\documentclass[%
reprint,
superscriptaddress,
amsmath,amssymb,
aps,
floatfix]{revtex4-2}

\usepackage{graphicx}
\usepackage{dcolumn}
\usepackage{bm}
\usepackage[hidelinks,colorlinks=true,linkcolor=blue,citecolor=blue,urlcolor=blue]{hyperref}
\usepackage[export]{adjustbox}

\begin{document}

\preprint{Universal fluctuation spectrum of Vlasov--Poisson turbulence}

\title{Universal fluctuation spectrum of Vlasov--Poisson turbulence}

\author{Michael L. Nastac}
\email{michael.nastac@physics.ox.ac.uk}
\affiliation{Rudolf Peierls Centre for Theoretical Physics, University of Oxford, Clarendon Laboratory, Parks Road, Oxford OX1 3PU, UK}
\affiliation{St. John’s College, Oxford OX1 3JP, UK}

\author{Robert J. Ewart}
\affiliation{Rudolf Peierls Centre for Theoretical Physics, University of Oxford, Clarendon Laboratory, Parks Road, Oxford OX1 3PU, UK}
\affiliation{Balliol College, Oxford, OX1 3BJ, UK}
\affiliation{Department of Astrophysical Sciences, Princeton University, Peyton Hall, Princeton, NJ 08544, United States}

\author{James Juno}
\affiliation{Princeton Plasma Physics Laboratory, Princeton, New Jersey 08540, USA}

\author{Michael Barnes}
\affiliation{Rudolf Peierls Centre for Theoretical Physics, University of Oxford, Clarendon Laboratory, Parks Road, Oxford OX1 3PU, UK}
\affiliation{University College, Oxford OX1 4BH, UK}

\author{Alexander~A.~Schekochihin}
\affiliation{Rudolf Peierls Centre for Theoretical Physics, University of Oxford, Clarendon Laboratory, Parks Road, Oxford OX1 3PU, UK}
\affiliation{Merton College, Oxford OX1 4JD, UK}

\date{\today}%

\begin{abstract}

The `thermal' fluctuation spectrum of the electric field arising due to particle noise in a quiescent Vlasov--Poisson plasma was derived in the 1960s. Here, we derive the universal fluctuation spectrum of the electric field, at Debye and sub-Debye scales, for a turbulent Vlasov--Poisson plasma. This spectrum arises from what is likely to be the `final cascade'—a universal regime to be encountered at the extreme small-scale end of any turbulent cascade in a nearly collisionless plasma. The cascaded invariant is the generalized (negative) entropy $C_2$, which is the quadratic Casimir invariant of the particle distribution function. $C_2$ cascades to small scales in both position and velocity space via both linear and nonlinear phase mixing, in such a way that the time scales associated with the two processes are `critically balanced' at every scale. We construct a phenomenological scaling theory of the fluctuation spectrum of $C_2$ in $(k, s)$ space—and hence of the electric field in $k$ space—in arbitrary dimensions, where $k$ and $s$ are the spatial and velocity wavenumbers, respectively. The electric-field spectrum is sufficiently steep for the nonlinear mixing to be controlled by the largest-scale electric fluctuations, and so the $C_2$ cascade resembles the Batchelor cascade of a passive scalar, albeit in the kinetic phase space. The conclusions of our theory are corroborated by direct numerical simulations of a forced 1D-1V plasma. We predict that the cascade is ultimately terminated at the (spatial) wavenumber where the turbulent electric-field spectrum gives way to the fluctuation spectrum associated with discrete particle noise; this transition scale is set by the amplitude of the turbulent electric fields and the plasma parameter. The characteristic time scale for this small-scale cutoff to be reached is the dynamical time of phase-space mixing times a logarithmic factor in the plasma parameter—this is the first concrete demonstration of this property of Vlasov--Poisson turbulence, analogous to how fluid turbulence dissipates energy at a rate independent (or nearly independent) of molecular diffusion. A key corollary is that the phase-space cascade enables the irreversibility of particle heating to be achieved on collisionless time scales. The mean distribution function arising from this process (in the case of forced 1D-1V turbulence) is shown to be derivable from quasilinear considerations and is not a Maxwellian, but rather resembles a flat top. Furthermore, in the presence of the phase-space cascade at Debye and sub-Debye scales—a scenario that may be ubiquitous—standard collisional plasma theory ceases to be valid. This calls for the development of new collision operators suited to such turbulent environments.

\end{abstract}

\maketitle

\section{Introduction} \label{intro}

Plasmas, even in equilibrium, have fluctuations. The fluctuation spectrum of the electric field in a quiescent Vlasov--Poisson plasma (in a Maxwellian equilibrium, or in an equilibrium that is stable and spatially homogeneous) was derived in the 1960s \cite{rostoker1961fluctuations,hubbard1961friction,klimontovich1962theory} and quickly became a textbook result \cite{klimontovich2013statistical,krall1973principles, pitaevskii2012physical}. Physically, this is the fluctuation spectrum of discrete particle noise, arising from the steady-state balance of the emission and absorption (Landau damping \cite{landau1946vibrations}) of Langmuir waves. In the years following this result, there were several attempts to compute the fluctuation spectrum in a turbulent Vlasov--Poisson plasma and, more generally, to construct a theory of Vlasov--Poisson turbulence \cite{Kadomtsev1965plasmaturbulence,dupree1966perturbation, orszag1967model, ichimaru1968theory, weinstock1969formulation,ichimaru1970theory, dupree1972theory, cook1973electric, misguich1975re, pelletier1975fluctuation,misguich1978kinetic, dubois1978direct}, but no universal theory emerged.

In this work, we revisit Vlasov--Poisson turbulence, working in the high-frequency, small-scale limit where the electrons are dynamical and the ions provide a stationary, neutralizing background. Within this approximation, we derive a universal turbulent fluctuation spectrum at Debye and sub-Debye scales. To find the fluctuation spectrum of the electric field, it turns out to be necessary to derive the fluctuation spectrum of the particle distribution function in both position and velocity space. Physically, this fluctuation spectrum represents a turbulent state in phase space, characterized by a turbulent cascade of fluctuations from injection to dissipation scales in phase space \cite{schekochihin2008gyrokinetic,schekochihin2009astrophysical,eyink2018cascades, nastac2023phase}. The main focus of this paper is to develop a theory of this cascade and hence derive the fluctuation spectrum of the turbulent electric field.

Perhaps the most influential early work on phase-space turbulence was by Dupree \cite{dupree1972theory} (see also \cite{dupree1970theory,kadomtsev1971theory}), who proposed that Vlasov--Poisson turbulence was composed of small-scale phase-space structures that he called `clumps,' or `granulations.' This idea inspired many follow-ups, in the context of unmagnetized \cite{misguich1978clumps, boutros1981theory, misguich1982relative, boutros1982theory, tetreault1983growth} and magnetized \cite{dupree1978role, biglari1988theory, kosuga2011relaxation, kosuga2014ion} plasma turbulence, as well as of energetic-particle dynamics in fusion devices \cite{berk1997spontaneous, berk1999spontaneous, lesur2013nonlinear, wang2013hole,  lilley2014formation, bierwage2021effect}. The clump theory also sparked criticism \cite{krommes1986comments,zhang1988comments, krommes1997clump}, primarily questioning the validity of the approximations that went into Dupree's calculations. While phase-space clumps continue to be a topic of interest \cite{diamond2010modern, kosuga2017role, lesur2020nonlinear}, their theoretical foundation is still obscure, and a definitive demonstration of their existence remains elusive, despite early attempts \cite{berman1983observation,berman1985simulation} to observe them in simulations.

In a completely separate strand of research, Schekochihin et al. \cite{schekochihin2008gyrokinetic,schekochihin2009astrophysical} proposed that magnetized plasma turbulence, within the gyrokinetic approximation, is a cascade of free energy \cite{krommes1994role, schekochihin2008gyrokinetic, schekochihin2009astrophysical, abel2013multiscale} through phase space. At sub-Larmor scales, the cascade occurs via so-called nonlinear perpendicular phase mixing, which was first identified as a mechanism of generating fine-scale phase-space structure in magnetized plasmas by Dorland and Hammett \cite{dorland1993gyrofluid}. The notion of phase-space cascades has continued to underlie theories of magnetized plasma turbulence \cite{plunk2010two, barnes2011critically, kunz2015inertial, schekochihin2016phase, kunz2018astrophysical, schekochihin2019constraints}, and these cascades have now been measured not only in numerical simulations \cite{tatsuno2009nonlinear, parker2016suppression,pezzi2018velocity,kawazura2019thermal,meyrand2019fluidization,cerri2018dual,kawazura2020ion,zhou2023electron} but also in nature using spacecraft data \cite{servidio2017magnetospheric,wu2023ion} and in laboratory plasmas \cite{kawamori2013verification,kawamori2022evidence}. The physics of phase-space cascades in plasmas outside the gyrokinetic regime is a topic of ongoing research \cite{servidio2017magnetospheric,cerri2018dual,pezzi2018velocity,eyink2018cascades,adkins2018solvable,nastac2023phase,celebre2023}.

In this paper, we propose a phase-space cascade theory of Vlasov--Poisson turbulence at Debye and sub-Debye scales. Namely, we envision a turbulent state in which the particle distribution function consists of an ensemble of phase-space `eddies,' or `blobs' (cf. clumps and granulations \cite{dupree1972theory, diamond2010modern}). As particles stream ballistically and get accelerated by turbulent electric fields, phase-space eddies stretch and fold in both position and velocity space, and the invariant \cite{knorr1977time, servidio2017magnetospheric,zhdankin2022generalized,nastac2023phase,lesur2013nonlinear,diamond2010modern}
\begin{equation}
    C_2 = \frac{1}{V} \iint \mathrm{d} \mathbf{x} \mathrm{d} \mathbf{v} \, \frac{f^2}{2},
\end{equation}
where $f$ is the particle distribution function and $V$ is the system volume, cascades from large to small phase-space scales. The cascaded invariant $C_2$ is conserved by collisionless Vlasov dynamics and dissipated by collisional velocity-space diffusion (or phase-space coarse-graining), so it is a generalized (negative) entropy of the distribution function. Phase-space turbulence is thus a mechanism by which the system becomes time irreversible even though collisions are rare, due to rapid dissipation of fluctuations at the small-scale end of the cascade.

Such a phase-space cascade of $C_2$ was recently studied for a passive plasma species in one spatial and one velocity dimension (1D-1V), accelerated by an externally imposed Gaussian random electric field that was white in time and had a specified spatial correlation function \cite{nastac2023phase}. This model was the plasma-kinetic analogue of the Kraichnan model of passive advection \cite{kraichnan1968small,falkovich2001particles}. In the Kraichnan model, the exponent of the electric-field spectrum is a free parameter. When this spectrum is shallow, the electric field is multiscale, and the mixing of phase-space eddies is local in scale, similar to the mixing of fluid scalars by Kolmogorov inertial-range turbulence \cite{kolmogorov1941b,obukhov1949structure,corrsin1951spectrum,frisch1995turbulence}. When the electric-field spectrum is steep, the electric field is effectively single-scale, and the phase-space mixing of the distribution function is dominated by the largest-scale electric fluctuations. This regime, also studied in \cite{adkins2018solvable}, is analogous to Batchelor turbulence in fluids \cite{batchelor1959small}, where, in the so-called viscous-convective range, a passive scalar is advected by an effectively single-scale flow (i.e., by the fluid motions at the viscous scale). In \cite{nastac2023phase}, we solved for the wavenumber fluctuation spectrum of $C_2$ (in phase space) in the Kraichnan model exactly, and also constructed a phenomenological theory that reproduced the asymptotic scalings of the exact spectrum. The chief assumption of that theory was that $C_2$ underwent a constant-flux cascade in both position and velocity space and that the phase-space eddy turnover time was determined by the ‘critical balance’ between the linear and nonlinear time scales, analogously to the critical-balance conjecture in magnetohydrodynamic turbulence \cite{goldreich1995toward,schekochihin2022mhd}.

In this paper, we extend these ideas to the Vlasov--Poisson turbulence in arbitrary dimensions, where the electric field is self-consistently determined by Poisson's equation, which in general contains both an external forcing and the effect of the distribution function on the field. This gives us a phenomenological scaling theory for the fluctuation spectrum of $C_2$ in phase space and of the electric field, which we test via direct numerical simulations of a forced 1D-1V plasma. The self-consistent Poisson coupling of the electric-field fluctuations to the electron-density fluctuations turns out to put the turbulence into the Batchelor regime. In the same way that Batchelor turbulence describes turbulent mixing at the very smallest possible scales of fluid scalars \cite{batchelor1959small, sreenivasan2019turbulent} in the viscous-convective range, the turbulent regime that we find here describes kinetic turbulence at Debye and sub-Debye scales, which are the very smallest possible scales of a kinetic plasma. Such small-scale kinetic Batchelor turbulence may be the universal `final cascade' in kinetic-plasma-turbulence systems, such as the solar wind and Earth's magnetosheath, and indeed everywhere else.

The rest of this paper is organized as follows. In Section \ref{C2_sect}, we motivate the existence of phase-space turbulence in the Vlasov--Poisson system and explain why $C_2$ is the invariant around which we construct a cascade theory. In Section \ref{phenom_theory}, we derive the fluctuation spectra of the electric field and of $C_2$ in arbitrary dimensions.  In Section \ref{forced_turbulence}, we corroborate our theory using direct numerical simulations of a 1D-1V plasma stirred by an external electric field. In Section \ref{thermo_turb}, we analyze theoretically and numerically the thermodynamics of the forced 1D-1V plasma. We find that the mean particle distribution function develops a broadening flat-top shape as particles are energized by the external field. In Section \ref{noise_turbulence_interplay}, we discuss how the turbulent cascade is ultimately terminated at small scales, whereby the fluctuation spectrum of turbulence transitions to that of discrete particle noise. Finally, in Section \ref{discussion}, we conclude by summarizing our results and discussing their implications. 

The main text is supplemented by several appendices. In Appendix \ref{PT_hd}, we supplement the turbulence theory of Section \ref{phenom_theory} by showing that scaling regimes of Vlasov--Poisson turbulence outside the Batchelor regime are inconsistent with a critically-balanced phase-space cascade of $C_2$. In Appendix \ref{PU_app}, we discuss the phase-space asymmetry of the distribution function due to the presence of phase-unmixing modes, which are a generalization of the plasma-echo effect \cite{gould1967plasma,malmberg1968plasma}. In Appendix \ref{LC_PSE_section}, we discuss the dynamical origin of kinetic Batchelor turbulence by considering the Lagrangian chaos of particle trajectories in phase space. In Appendix \ref{QL_app}, we derive and solve a quasilinear diffusion equation for the mean particle distribution function, whose solution we compare, successfully, to the mean particle distribution function obtained from our direct numerical simulations in Section \ref{thermo_turb}. Finally, in Appendix \ref{BD_app}, we discuss an early theory of Vlasov--Poisson turbulence above the Debye scale proposed by Biglari and Diamond \cite{biglari1988cascade,biglari1989clouds} (we rule out the validity of their approach below the Debye scale, where our theory applies). As a corollary of this discussion, we propose a new scaling regime for low-frequency electrostatic turbulence in which it is the ion distribution function that undergoes a phase-space cascade above the Debye scale, leading to scaling laws different from either those associated with our sub-Debye-scale cascade or those proposed by Biglari and Diamond for the electron-only turbulence.

\section{Vlasov--Poisson turbulence} \label{C2_sect}

We consider the Vlasov equation for an electrostatic plasma in $d$ dimensions (where $d = 1, 2$, or $3$):
\begin{equation} \label{vlasov}
    \frac{\partial f}{\partial t} + \textbf{v} \cdot \nabla f - \frac{e}{m}  \textbf{E}_{\mathrm{tot}} \cdot \nabla_{\mathbf{v}} f = C[f],
\end{equation}
where $f(\mathbf{x},\mathbf{v},t)$ is the electron distribution function, $C[f]$ is the collision operator, $e$ is the elementary charge, $m$ is the electron mass, and
\begin{equation} \label{E_tot}
    \textbf{E}_{\mathrm{tot}} = \textbf{E} + \textbf{E}_{\mathrm{ext}}
\end{equation}
is the total electric field, which comprises of a self-consistent field $\textbf{E}$ and an externally imposed electric field $\textbf{E}_{\mathrm{ext}}$. The self-consistent electric field is determined by Gauss's law,
\begin{equation} \label{gauss}
    \nabla \cdot \textbf{E} = 4 \pi e \left(\overline{n} - \int \mathrm{d} \mathbf{v} f \right),
\end{equation}
or equivalently, Poisson's equation for the electrostatic potential $\varphi$, where $\textbf{E} = - \nabla \varphi$. We assume the ions to be a stationary, neutralizing background with a constant number density $\overline{n}$.

The characteristic frequency and length scale of the Vlasov--Poisson system are the Langmuir frequency $ \omega_{\mathrm{pe}} = \sqrt{4 \pi e^2 \overline{n} /m}$, and the electron Debye length $ \lambda_{\mathrm{De}} = v_{\mathrm{the}}/\omega_{\mathrm{pe}} = \sqrt{T_{\mathrm{e}}/4 \pi e^2  \overline{n}}$, where $v_{\mathrm{the}} = \sqrt{T_{\mathrm{e}}/m}$ is the electron thermal velocity and $T_{\mathrm{e}}$ is the (mean) electron temperature, defined as
\begin{equation}
    T_{\mathrm{e}} = \frac{1}{d} \frac{1}{\overline{n} V} \iint \mathrm{d} \mathbf{x} \mathrm{d} \mathbf{v} \, m |\mathbf{v}-\mathbf{u}|^2 f, 
\end{equation}
where $V$ is the system's volume and
\begin{equation} \label{u}
    \mathbf{u}(\mathbf{x}, t) = \frac{1}{n(\mathbf{x}, t)} \int \mathrm{d} \mathbf{v} \, \mathbf{v} f(\mathbf{x},\mathbf{v},t)
\end{equation}
is the bulk flow velocity.

We are interested in Vlasov--Poisson turbulence on spatial scales comparable to and smaller than the Debye length, as this is where the Langmuir waves begin to be strongly damped, losing their fluid-like behavior as they interact with particles in the bulk of the electron distribution function. 

To probe a universal turbulent state, as one does in fluid-dynamical systems, we imagine the plasma to be stirred by an external random electric field $\textbf{E}_{\mathrm{ext}} = - \nabla \varphi_{\mathrm{ext}}$. Spatially, we assume that this external field is statistically homogeneous, isotropic, and localized in wavenumber space at the Debye scale, viz., $k_{\mathrm{f}} \lambda_{\mathrm{De}} \sim 1$, where $k_{\mathrm{f}}$ is the wavenumber shell where the forcing is concentrated. Temporally, we assume $\textbf{E}_{\mathrm{ext}}$ to have a finite correlation time $\tau_{\mathrm{corr}} \sim \omega_{\mathrm{pe}}^{-1}$. This forcing can be be conceptualized as a source of turbulent electric fluctuations from large scales, which are then processed at the Debye and sub-Debye scales considered here. Indeed, while we are explicitly considering turbulence driven by external forcing in this paper, ultimately, we predict that the fluctuation spectra that we derive below are universal and should also describe turbulence driven in other ways, e.g., by an instability. 

Throughout this paper, we shall consider a periodic system with length $L$ and volume $V = L^{d}$. For any dimensionality $d$, we interpret $f(\mathbf{x},\mathbf{v},t)$ as a 3D-3V distribution function with the non-participating dimensions (in both position and velocity space) integrated out; $f$~therefore has dimensions of $(\text{length}^{-d}) \times (\text{velocity}^{-d})$ and is assumed to integrate to the total number of particles~$N$. Notationally, this leaves \eqref{vlasov} unchanged, but  turns \eqref{gauss}~into
\begin{equation} \label{gauss2}
   \frac{e}{m} \nabla \cdot \textbf{E} = -\omega_{\mathrm{pe}}^2 \frac{\delta n}{n_0},
\end{equation}
where $n_0 \equiv \overline{n} L^{3-d}$ is the mean $d$-dimensional number density, and $\delta n = n - n_0$, where $n$ is the total ($d$-dimensional) number density of electrons. We have introduced the plasma frequency in \eqref{gauss2} for convenience of tracking dimensions of physical quantities.

Before continuing, let us discuss the nature of the collision operator $C[f]$. For all sections of this paper, except for Section \ref{noise_turbulence_interplay}, we assume the collision operator to be of the Fokker--Planck type, e.g., the Landau collision operator \cite{landau1936transport}. Strictly speaking, in the presence of high-frequency (Langmuir) fluctuations, modeling finite-$N$ effects in this way is not rigorous. To derive the Landau collision operator \cite{landau1936transport} (or, e.g., the Lenard-Balescu collision operator \cite{lenard1960bogoliubov,balescu1960irreversible}), according to Bogoliubov's hypothesis \cite{Bogolyubov1946}, requires a separation of time scales between the slow evolution of the single-particle distribution function $f$ and the fast evolution of pair correlations \cite{krall1973principles, klimontovich2013statistical, pitaevskii2012physical}. When $f$ contains high-frequency (Langmuir) fluctuations, no such scale separation exists, and one must in general simultaneously solve for the coupled single-particle and two-particle distribution functions (see, e.g., \cite{oberman1962high} for such a calculation).

Besides temporal scale separation, the validity of collision operators also hinges upon spatial scale separation. The distinction between collisionless and collisional physics lies in a coarse-graining of fluctuations at a length scale intermediate between the Debye length $\lambda_{\mathrm{D}}$ and the interparticle spacing $\Delta r \sim n_0^{-1/d}$, which are guaranteed to be asymptotically separated by the largeness of the plasma parameter $N_{\mathrm{D}} = n_0 \lambda^d_{\mathrm{D}}$ (the number of particle of particles in a Debye sphere), which is the fundamental large parameter in a weakly coupled plasma \cite{klimontovich2013statistical}. Sub-Debye turbulence blurs this distinction.

While keeping these caveats in mind, we expect that, like in fluid turbulence \cite{frisch1995turbulence}, the characteristics of the turbulence above the dissipation scale will be largely insensitive to the precise form of dissipation, and only the structure of the `dissipation range' will depend on the nature of collisions (finite-$N$ effects). To the end of understanding the `inertial range' of sub-Debye Vlasov--Poisson turbulence, we will treat the Vlasov equation with a Fokker--Planck collision operator as an adequate model in Sections \ref{C2_sect}-\ref{thermo_turb}. Then, in Section \ref{noise_turbulence_interplay}, we will theorize about how dissipation is ultimately achieved in real plasmas. 

\subsection{Electric fluctuations} \label{elc_fluc}

We are interested in electric fluctuations in the plasma, which can be characterized by the wavenumber spectrum
\begin{equation} \label{P_E}
  P_{\mathbf{E}}(\mathbf{k}) = \frac{e^2}{m^2} \frac{1}{V} \int \mathrm{d} \mathbf{r} \, e^{-i \mathbf{k} \cdot \mathbf{r}} \, \overline{\mathbf{E}(\mathbf{x}_1) \cdot \mathbf{E}(\mathbf{x}_2)},
\end{equation}
where $\mathbf{k}$ is the spatial wavevector, $\mathbf{x}_1$ and $\mathbf{x}_2$ are two points in position space, and $\overline{(...)}$ denotes ensemble averaging, e.g., averaging with respect to realizations of the stochastic forcing, or the initial conditions of the system. We assume that the electric field is statistically homogeneous, so the correlation function \eqref{P_E} depends only on the increment $\mathbf{r} = \mathbf{x_1}-\mathbf{x_2}$.

For a plasma in a Maxwellian equilibrium, the fluctuation spectrum of the electric field is \cite{rostoker1961fluctuations, klimontovich1962theory, hubbard1961friction}
\begin{equation} \label{TS}
    P_{\mathbf{E}}(\mathbf{k}) = \frac{8 \pi e^2}{V m^2} \frac{T_{\mathrm{e}}}{2} \frac{1}{1+(k \lambda_{\mathrm{De}})^2}.
\end{equation}
The shell-averaged version of this spectrum, in $d$ dimensions, is
\begin{equation} \label{TS_SA}
    \langle P_{\mathbf{E}} \rangle_{\mathrm{SA}}(k)  = \frac{S_d}{(2 \pi)^d} \frac{8 \pi e^2}{ m^2} \frac{T_{\mathrm{e}}}{2} \frac{k^{d-1}}{1+(k \lambda_{\mathrm{De}})^2},
\end{equation}
where $S_d$ is the area of a unit sphere in $d$ dimensions ($2$ in 1D, $2 \pi$ in 2D, and $4 \pi$ in 3D). Physically, this thermal spectrum is due to discrete particle noise. The motion of particles excites Langmuir waves, which then get Landau damped, returning their energy to particles; \eqref{TS} is the result of this continuous emission and absorption of these Langmuir waves. Above the Debye scale, each mode has an equal amount of energy, $T_{\mathrm{e}}/2$ (statistical-mechanical equipartition), whereas below the Debye scale, the spectrum is steepened. The sub-Debye scaling arises from (undressed) discrete particle noise and therefore also applies to `thermal' fluctuations in quiescent non-Maxwellian plasmas \cite{rostoker1961fluctuations}; we will explicitly show this in Section \ref{noise_turbulence_interplay}.

Of course, virtually all (nearly) collisionless plasmas of interest are turbulent and far from thermal equilibrium. This then begs the question: what is the fluctuation spectrum in a turbulent Vlasov--Poisson plasma? The essential difference between a turbulent system and an equilibrium one is that the former constantly receives energy and processes it into heat whereas the latter merely distributes its energy as fairly as possible between its degrees of freedom. Intuition from fluid dynamics suggests that progress can be made by formulating a Kolmogorov-like theory of plasma turbulence as a constant-flux energy cascade \cite{kolmogorov1941b,frisch1995turbulence}. The energy of the electric field, however, is not an invariant of the Vlasov--Poisson system. The total energy of (self-consistent) electric fields and particles is, but, while it can be injected into the plasma by the external forcing, there is no energy sink in this system (collisions conserve energy).

It is also not obvious how to determine the characteristic time scale of the turbulence. In fluid-dynamical systems, it is often given by the dynamical time associated with nonlinear advection by the turbulent flow. In the Vlasov--Poisson system, according to \eqref{gauss2}, any small-scale structure of the electric field comes from that of the electron density perturbations $\delta n$. This can only be found by solving the Vlasov equation for the distribution function $f$ and then taking its zeroth velocity moment. In \eqref{vlasov}, we can extract the dynamical times associated with the linear phase mixing due to particle streaming and the nonlinear mixing due to particle interaction with the electric field. Crucially, both of these times \textit{a priori} involve both position- and velocity-space scales. Consequently, no cascade phenomenology that coarse-grains over the velocity degrees of freedom in the distribution function can account for dynamical means by which inherently kinetic, small-scale structure is generated in the Vlasov--Poisson system.

Thus, Vlasov--Poisson turbulence is not simply a cascade of electric energy to small spatial scales. Nevertheless, we will find in the following sections that there is an invariant of the Vlasov equation around which we can construct a turbulence theory that simultaneously takes into account the velocity-space dynamics of the distribution function and predicts the electric-field fluctuation spectrum.

\subsection{Casimir invariants}

The nearly collisionless Vlasov equation \eqref{vlasov}, with or without the collision operator, conserves the particle number and momentum, as well as the total energy of particles and fields (in the absence of forcing). In addition to these conservation laws, in the absence of collisions, \eqref{vlasov} also conserves an infinite family of so-called Casimir invariants \cite{ye1992action}. Namely, for any smooth function $g(f)$, it is straightforward to show that the collisionless Vlasov equation, assuming a spatial domain without boundaries (e.g., a periodic box), conserves the quantity
\begin{equation} \label{casimir_invariants}
    G[f] = \iint \mathrm{d} \mathbf{x} \mathrm{d} \mathbf{v} \, g(f).          
\end{equation}
The conservation of these Casimir invariants is a consequence of the collisionless dynamics being the advection of the distribution function by the 2$d$-dimensional phase-space flow $(\mathbf{v}, (-e/m)\mathbf{E}_{\mathrm{tot}})$, which is incompressible and, therefore, phase-space-volume preserving.

As we discussed in the previous section, to formulate the theory of Vlasov--Poisson turbulence as a constant-flux cascade, we need a cascaded invariant that contains information both about the electric field and the velocity-space structure of the distribution function. Among the infinite family of invariants \eqref{casimir_invariants}, the quadratic invariant \cite{knorr1977time, servidio2017magnetospheric,zhdankin2022generalized,nastac2023phase,lesur2013nonlinear,diamond2010modern}
\begin{equation} \label{C2}
    C_2 = \frac{1}{V} \iint \mathrm{d} \mathbf{x} \mathrm{d} \mathbf{v} \, \frac{f^2}{2}
\end{equation}
will turn out to be a good vehicle for such a theory. It has, up to constant prefactors, also been called `enstrophy'  \cite{knorr1977time,servidio2017magnetospheric}, `phasestrophy' \cite{diamond2010modern,lesur2013nonlinear}, and `$f$-strophy' \cite{schekochihin2022lectures}.

\subsection{Mean fields plus fluctuations} \label{C2_subsection}

We define the mean distribution as $f_0 \equiv \overline{f}$. We assume that the distribution function is statistically homogeneous in position space, and hence $f_0$ does not depend on $\mathbf{x}$. We further adopt the ergodic hypothesis, which makes ensemble and spatial averages equivalent. To justify our focus on $C_2$, we note that if we decompose the distribution function into its mean and fluctuating parts,~viz., 
\begin{equation}
    f(\mathbf{x},\mathbf{v},t) = f_0(\mathbf{v},t) + \delta \! f(\mathbf{x},\mathbf{v},t),
\end{equation}
then $C_2$ splits into
\begin{equation}
    C_2 = C_{2,0} + \delta C_2,
\end{equation}
where
\begin{equation}
    C_{2,0} = \int \mathrm{d} \mathbf{v} \, \frac{f_0^2}{2}, \qquad \delta C_2 = \frac{1}{V} \iint \mathrm{d} \mathbf{x} d \mathbf{v} \, \frac{\delta \! f^2}{2}. 
\end{equation}
This property also holds for any phase-space coarse-graining (low-pass filtering) of the distribution function (cf. spatial low-pass filtering in hydrodynamics \cite{frisch1995turbulence}). Up to weight factors in the integrand of \eqref{C2}, $C_2$ is the only Casimir invariant that has this property (aside, trivially, from $C_1 = \iint \mathrm{d} \mathbf{x} \mathrm{d} \mathbf{v} \, f = N$, the number of particles). The additivity of $C_2$ is useful, as it will allow us to quantify, in a physically suggestive way, the exchange of $C_2$ between the mean distribution and fluctuations.

The Fourier spectrum of $\delta \! f$ is
\begin{equation} \label{spec_fourier}
    P_{\delta \! f}(\mathbf{k}, \mathbf{s}) = \frac{1}{(2 \pi)^{d} V} \iint \mathrm{d} \mathbf{r} \mathrm{d} \mathbf{u} \, e^{-i(\mathbf{k} \cdot \mathbf{r}-\mathbf{s} \cdot \mathbf{u})} C_{\delta \! f}(\mathbf{r}, \mathbf{u}),
\end{equation}
where $\mathbf{s}$ is the velocity wavevector, and the two-point correlation function of $\delta \! f$ is
\begin{equation} \label{corr_func}
    C_{\delta \! f}(\mathbf{r}, \mathbf{u}) = \int \mathrm{d} \mathbf{U} \, \frac{1}{2} \overline{\delta \! f(\mathbf{x}_1,\mathbf{v}_1) \, \delta \! f(\mathbf{x}_2,\mathbf{v}_2)},
\end{equation}
where $(\mathbf{x}_1, \mathbf{v}_1)$ and $(\mathbf{x}_2, \mathbf{v}_2)$ are two points in phase space. By statistical homogeneity, the correlation function \eqref{corr_func} depends only on the increment $\mathbf{r} = \mathbf{x_1}-\mathbf{x_2}$ and not on the mean position $\mathbf{R} = (\mathbf{x}_1 + \mathbf{x}_2)/2$. Because $f_0$ has an energy scale, the velocity space cannot be homogeneous; however, we can still integrate the full correlation function over the average velocity $\mathbf{U} = (\mathbf{v_1} + \mathbf{v_2})/2$ to obtain a homogeneous integrated correlation function that depends on velocity only via increment $\mathbf{u} = \mathbf{v_1} - \mathbf{v_2}$. The spectrum \eqref{spec_fourier} is related to $\delta C_2$ via Parseval's theorem:
\begin{equation} \label{parseval}
 \frac{V}{(2 \pi)^d} \iint \mathrm{d} \mathbf{k}  \mathrm{d} \mathbf{s} \, P_{\delta \! f}(\mathbf{k}, \mathbf{s}) = \delta C_2.
\end{equation}

We can relate the spectrum of $\delta C_2$ to the electric-field spectrum \eqref{P_E}. When $\mathbf{s} = \mathbf{0}$, the velocity integrals in \eqref{spec_fourier} and \eqref{corr_func} reduce \eqref{spec_fourier} to the Fourier spectrum of the density fluctuations $\delta n$. Using this result together with~\eqref{gauss2}, we have that
\begin{equation} \label{s=0_C2}
   2 (2 \pi)^{d} P_{\delta 
 \! f}(\mathbf{k}, \mathbf{s} = \mathbf{0}) = P_{\delta n}(\mathbf{k}) = \frac{n_0^2}{\omega_{\mathrm{pe}}^4} k^2 P_{\mathbf{E}}(\mathbf{k}), 
\end{equation}
where $P_{\mathbf{E}}(\mathbf{k})$ is defined by \eqref{P_E} and
\begin{equation} \label{density_spec}
    P_{\delta n}(\mathbf{k}) = \frac{1}{V} \int \mathrm{d} \mathbf{r} \, e^{-i \mathbf{k} \cdot \mathbf{r}} \, \overline{\delta n(\mathbf{x}_1) \delta n(\mathbf{x}_2)}.
\end{equation}
As we show in the following sections, computation of $P_{\delta \! f}(\mathbf{k}, \mathbf{s} = \mathbf{0})$ requires knowledge of $P_{\delta \! f}(\mathbf{k}, \mathbf{s})$ over the whole $(\mathbf{k}, \mathbf{s})$ space. This can be anticipated in light of the discussion in Section \ref{elc_fluc}, where we noted that the small-scale structure in \eqref{vlasov} in both position and velocity space is generated by intertwined processes.

\subsection{\texorpdfstring{$C_2$}{C2} budget: dissipation and injection} \label{C2_budget}

We find in the following sections that the distribution function undergoes mixing in phase space and therefore develops sharp phase-space gradients. The velocity-space gradients of the distribution function are ultimately limited by collisions. This is because the collision operator is a Fokker--Planck operator, whose dominant part when velocity-space gradients are large is velocity-space diffusion. Importantly, velocity-space diffusion negative-definitely dissipates $C_2$ (and, generally, any convex functional of the distribution function). To capture this property in the simplest possible way, we neglect electron-ion collisions and model electron-electron collisions as
\begin{equation} \label{coll_op_diffusion}
C[f] = \nu \, v_{\mathrm{the}}^2 \nabla^2_{\mathbf{v}} \delta 
 \! f, 
\end{equation}
where $\nu$ is the (velocity-independent) collision frequency, and we have anticipated that only $\delta \! f$ will develop fine velocity-space scales on dynamical time scales (in our simulations, we nevertheless utilize a Fokker--Planck operator that contains both drag and diffusion and acts on the full distribution function).

Using the decomposition $C_2 = C_{2,0} + \delta C_2$, we can therefore write a budget equation for $\delta C_2$:
\begin{equation} \label{budget_C2}
    \frac{d \delta C_2}{d t} = -\frac{d C_{2,0}}{dt} - \frac{\nu \, v_{\mathrm{the}}^2}{V} \iint \mathrm{d} \mathbf{x} \mathrm{d} \mathbf{v} \, \left | \nabla_{\mathbf{v}} \delta 
 \! f \right   |^2. 
\end{equation}
To compute $d C_{2,0}/dt$, we need the evolution equation for~$f_0$, which can be found by ensemble averaging the Vlasov equation \eqref{vlasov}:
\begin{equation} \label{df0dt}
    \frac{\partial f_0}{\partial t} = \frac{e}{m} \nabla_{\mathbf{v}} \cdot \left[ \overline{\mathbf{E}_{\mathrm{tot}} \delta 
 \! f} \right],
\end{equation}
whence, using \eqref{C2},
\begin{equation} \label{dC20dt}
    \frac{d C_{2,0}}{dt} = -\frac{e}{m} \int \mathrm{d} \mathbf{v} \,  \left( \nabla_{\mathbf{v}} f_0 \right) \, \cdot \left[ \overline{ \left(\mathbf{E}_{\mathrm{tot}} \right) \delta 
 \! f} \right].
\end{equation}
\textit{A priori}, \eqref{dC20dt} is not sign-definite. However, we find in Section \ref{thermo_turb} that the external electric field energizes particles and heats $f_0$ in such a way that $C_{2,0}$ decreases in time. Therefore, the first term on the right-hand side of~\eqref{budget_C2} is positive and so $\delta C_2$ is injected by the decrease of~$C_{2,0}$ and dissipated by collisions; indeed this is the only way to have a (quasi)steady state for $\delta C_2$. We show in the following sections that the injection and dissipation are at disparate scales; hence, we have a turbulence where a flux of $\delta C_2$ is transferred across scales.

\section{Phenomenological theory of phase-space cascade} \label{phenom_theory}

In this section, we construct a phenomenological theory of Vlasov--Poisson turbulence in phase space, reminiscent of cascade theories of hydrodynamic turbulence \cite{kolmogorov1941b,frisch1995turbulence}.

Let us suppose that the system reaches a quasi-steady-state balance between injection and dissipation, viz., $d \delta C_2/dt \simeq 0$. Then the budget \eqref{budget_C2} becomes
\begin{equation} \label{flux_def}
    \varepsilon \equiv -\frac{d C_{2,0}}{dt} = \frac{\nu \, v_{\mathrm{the}}^2}{V} \iint \mathrm{d} \mathbf{x} \mathrm{d} \mathbf{v} \, \left | \nabla_{\mathbf{v}} \delta 
 \! f \right   |^2.
\end{equation}
The assumption of a quasi-steady state means that the $\delta \! f$ fluctuations are to evolve faster than $f_0$, so $\varepsilon \simeq \mathrm{const}$. Physically, the injection of $\delta C_2$ is due to turbulent electric fields distorting $f_0$. To see this, consider the evolution equation for $\delta \!f$:
\begin{align} \label{vlasov_delta_f}
        &\frac{\partial \delta \! f}{\partial t} + \textbf{v} \cdot \nabla \delta \! f - \frac{e}{m}\mathbf{E}_{\mathrm{tot}} \cdot \nabla_{\mathbf{v}} \delta \! f \nonumber \\  &=  \frac{e}{m} \mathbf{E}_{\mathrm{tot}} \cdot \nabla_{\mathbf{v}} f_0 + \nu \, v_{\mathrm{the}}^2 \nabla^2_{\mathbf{v}} \delta 
 \! f,
\end{align}
where we have modeled collisions using \eqref{coll_op_diffusion} and dropped the term $\partial f_0 / \partial t$ on the basis that $f_0$ evolves slowly compared to $\delta \! f$. Manifestly, fluctuations of $\delta \! f$ are sourced by velocity-space gradients of $f_0$ that are aligned with the electric field, in the same way that, e.g., in fluids, temperature fluctuations are sourced by the position-space gradients of the mean temperature via its advection by turbulent flows. In velocity space, the injection of $\delta C_2$ is at the scales of $f_0$, which we assume are $u \sim v_{\mathrm{the}}$. On the other hand, collisional dissipation must necessarily be at much smaller scales when $\nu$ is small enough. This is because the velocity-space gradients of $\delta \! f$ must be large to compensate for the smallness of $\nu$. Thus, the injection and dissipation are concentrated at large and small spatial scales, respectively (we shall confirm this \textit{a posteriori}).

Like in any prototypical turbulent system, to bridge these two disparate scales, there will be an `inertial range' of scales $(r,u)$ in phase space through which the flux $\varepsilon$ of $\delta C_2$ passes, unaffected directly by forcing or collisional dissipation. Below, we determine the structure of $\delta \! f$ and the electric field in this inertial range.

\subsection{Symmetries} \label{inhomog_aniso}

First, we discuss the symmetries (or lack thereof) of the phase-space turbulence. We assume $\delta \! f$ to be statistically homogeneous in space; consequently, $\mathbf{E}$ and $\delta n$ are also statistically homogeneous in space. Because of the underlying mean distribution $f_0$, we cannot, however, assume $\delta \! f$ to be homogeneous in velocity space. Nevertheless, locally in velocity space, we expect $\delta \! f$ to be approximately homogeneous at sufficiently small velocity-space scales, in the sense that the full two-point correlation function of $\delta \! f$ will depend more sensitively on the velocity increment $\mathbf{u}$ than on the average velocity $\mathbf{U}$. By analyzing the spectrum \eqref{spec_fourier} of the correlation function \eqref{corr_func} integrated over $\mathbf{U}$, we are probing only the small-scale, homogeneous structure of the distribution function.

We also assume that $\delta \! f$ is statistically isotropic in both position and velocity space. The fields $\mathbf{E}$ and $\delta n$ are, therefore, also statistically isotropic in space. In general, statistical isotropy of $\delta \! f$ in phase space implies that the spectrum \eqref{spec_fourier} depends on the wavenumber magnitudes $k = |\mathbf{k}|$, $s = |\mathbf{s}|$, and on the angle between the two wavenumbers, viz., $\theta = \cos^{-1} (\mathbf{k} \cdot \mathbf{s}/k s)$. It would be simplifying if the spectrum \eqref{spec_fourier} did not depend on $\theta$; physically, this would mean that the distribution function were symmetric with respect to reflections in $\mathbf{x}$ and $\mathbf{v}$ separately (note that, by parity symmetry, the distribution function is necessarily symmetric with respect to simultaneous reflections in $\mathbf{x}$ and $\mathbf{v}$). However, in fact, if the spectrum \eqref{spec_fourier} were independent of $\theta$, then there would be no cascade. To understand this requires a slight detour. In Appendix \ref{PU_app}, we demonstrate this result by explicitly computing the flux of $\delta C_2$ in $s$ space. Here, we provide an intuitive explanation.

Physically, the reason for this asymmetry is that linear phase mixing, which causes the distribution function to develop fine-scale structure in velocity space, is a reversible process. The reversed process, phase unmixing \cite{schekochihin2016phase,adkins2018solvable,nastac2023phase}, was made famous by the `plasma-echo' effect \cite{gould1967plasma,malmberg1968plasma}. Phase unmixing can be understood in spectral space:
the Fourier transform of the free-streaming operator $\mathbf{v} \cdot \nabla$ in \eqref{vlasov} is $\mathbf{k} \cdot \partial_{\mathbf{s}}$, so phase mixing is an advection in $\mathbf{s}$ space with `velocity' $\mathbf{k}$. Explicitly, the spectrum \eqref{spec_fourier} satisfies the following equation, derived in Appendix \ref{PU_app}: 
\begin{equation} \label{E_delta_f_eq}
    \frac{\partial P_{\delta \! f} }{\partial t} + \mathbf{k} \cdot \frac{\partial P_{\delta \! f}}{\partial \mathbf{s}} + \overline{\mathcal{N}}_{\mathbf{k},\mathbf{s}} = \overline{\mathcal{S}}_{\mathbf{k},\mathbf{s}} - 2 \nu v_{\mathrm{the}}^2 s^2 P_{\delta \! f},
\end{equation}
where $\overline{\mathcal{N}}_{\mathbf{k},\mathbf{s}}$ and $\overline{\mathcal{S}}_{\mathbf{k},\mathbf{s}}$ are the nonlinear and source terms given by \eqref{NL_term_ks} and \eqref{S_term_ks}, respectively. Spectral modes with $\mathbf{k} \cdot \mathbf{s} > 0$ are advected from low to high $s$ and are, therefore, phase-mixing modes; modes with $\mathbf{k} \cdot \mathbf{s} < 0$ are advected from high to low $s$ and are, therefore, phase-unmixing modes.

If the spectrum $P_{\delta \! f}$ were a function only of $k$ and $s$, there would be no distinction between phase-mixing and phase-unmixing modes. The resulting velocity-space flux of $\delta C_2$ to small scales would then be zero, since phase mixing and phase unmixing would exactly cancel; this is explicitly shown in Appendix \ref{PU_app}. Therefore, phase-space asymmetry is necessary for there to be a phase-space cascade. In particular, the contributions to the velocity-space flux from the phase-mixing modes must outweigh those from the phase-unmixing modes; i.e., the system must have a preference for structures that have an alignment between $\mathbf{k}$ and $\mathbf{s}$. This is an assumption that we must put into our theory.

Despite $\delta \! f$ lacking symmetry in this way, we posit that the spectrum will have the same asymptotic scalings at large $k$ and $s$, in both the phase-mixing and phase-unmixing regions of phase space. More specifically, we assume that the angular structure of the spectrum, i.e., the difference between the spectrum in the two regions, is a higher-order correction (in $\mathbf{k}$ and $\mathbf{s}$) beyond the scalings that we compute in our phenomenological theory. This hypothesis is motivated by the Vlasov--Kraichnan model, which indeed has this property \cite{nastac2023phase}. In Section \ref{forced_turbulence}, we confirm that this holds also in our self-consistent Vlasov--Poisson system.

Because we do not attempt to predict the precise angular structure of the fluctuation spectrum, our analysis simplifies to predicting the $(k, s)$ dependence of the fluctuation spectrum shell-averaged in both $\mathbf{k}$ and $\mathbf{s}$, which we denote by $\langle P_{\delta \! f}(k,s) \rangle_{\mathrm{SA}}$. We include the `density-of-states' factors~$\sim V k^{d-1} s^{d-1}$ in the definition of $\langle P_{\delta \! f}  \rangle_{\mathrm{SA}} (k, s)$, so that Parseval's theorem \eqref{parseval} becomes
\begin{equation} \label{parseval_SA}
    \int^{\infty}_{0} \mathrm{d} k \, \int^{\infty}_{0} \mathrm{d} s \langle P_{\delta \! f}  \rangle_{\mathrm{SA}} (k, s) = \delta C_2.
\end{equation}

\subsection{Linear phase mixing and nonlinear mixing} \label{lpm_nm}

The $\delta \! f$ fluctuations are `eddies', or blobs, in phase space (cf. clumps and granulations \cite{dupree1972theory,diamond2010modern}); $\delta C_2$ is transferred to smaller scales as $\delta \! f$ is stretched and folded in phase space. This is accomplished by two mechanisms, as can be seen in \eqref{vlasov}: linear phase mixing due to ballistic particle streaming, which develops fine velocity-space structure in the distribution function on a characteristic time scale
\begin{equation} \label{tau_linear}
    \tau_{\mathrm{l}} \sim \frac{r}{u},
\end{equation}
and nonlinear mixing due to particle acceleration by electric fields, which develops fine position-space structure in the distribution function on a characteristic time scale
\begin{equation} \label{NL_time}
    \tau_{\mathrm{nl}} \sim \frac{u}{(e/m) \delta E_r} \sim \frac{u}{\gamma^2 r^{\beta}},
\end{equation}
where $\delta E_r$ is the electric field increment across scale $r$. Here $\gamma$, which has dimensions $(\text{length}^{(1-\beta)/2}) \times (\text{time}^{-1})$,  sets the effective `shearing rate' of the electric field, and $\beta$ controls the spatial roughness of the electric field, which we will determine self-consistently. Also note that the linear and nonlinear time scales in principle have directional dependence. For example, $\delta E_r$ should be interpreted as an increment of the electric field projected in the direction of the velocity gradient of $\delta \! f$. However, since the electric field is assumed statistically isotropic and we are not diagnosing the angular structure of $\delta \! f$, this distinction does not matter at the level of our scaling theory (and likewise for the angular structure imprinted by linear phase mixing).

If $\beta < 1$, the electric field is multiscale (i.e., rougher than a simple Taylor expansion in $r$), and the shearing of phase-space eddies at scale $r$ is dominated by the electric field at that scale, akin to Kolmogorov fluid turbulence \cite{kolmogorov1941b}. If $\beta = 1$, however, the electric field is smooth in space, and the shearing of phase-space eddies is instead dominated by the largest scales of the field, similar to Batchelor turbulence of fluid passive scalar fields \cite{batchelor1959small}. In this case, $\gamma$ in \eqref{NL_time} must be redefined to include contributions from both the external field (which, we assume is concentrated at large scales) and the self-consistent field, viz.,
\begin{equation} \label{gamma_def_Batchelor}
    \frac{e}{m} \delta E_{\mathrm{tot}, r} \sim \gamma^2 r.
\end{equation}

In what follows, we will assume $\beta = 1$ and \textit{a posteriori} show that this assumption is consistent. In Appendix \ref{PT_hd}, we show that trying to construct a scaling theory with $\beta < 1$ leads to a contradiction. Therefore, any scaling theory of self-consistent Vlasov--Poisson turbulence must have $\beta = 1$.

The remainder of this section will use the above intuition of linear phase mixing and nonlinear advection to construct a phenomenological theory of the $\delta C_2$ cascade. Before continuing, we advise a reader seeking a more thorough motivation for the dynamical origin of this cascade to look for it in Appendix \ref{LC_PSE_section}. There, we motivate the existence of phase-space eddies by showing that the interplay of free streaming and acceleration by spatially smooth electric fields causes Lagrangian particle trajectories to become chaotic in phase space, viz., that nearby particles separate exponentially. This Lagrangian chaos causes blobs (eddies) of $\delta \! f$ to be stretched in phase space; this process generates sharp phase-space gradients in $\delta \! f$, and is the mechanism by which $\delta C_2$ cascades to small scales. The quasi-steady turbulent state that our theory describes is reached over many correlation times of the electric field and many local stretching events, after which the distribution function is thoroughly mixed in phase space.

\subsection{Critical balance} \label{CB_section}

To make progress, we must ascertain a relationship between scales $r$ and $u$. We conjecture that the turbulence satisfies a critical balance in phase space between the linear and nonlinear time scales \cite{nastac2023phase}, viz., the phase-space eddy-turnover time $\tau_{\mathrm{c}}$, which is the typical time it takes for $\delta C_2$ to be transferred across phase-space scales $(r,u)$, is given by
\begin{align} \label{CB}
& \tau_{\mathrm{c}} \sim \tau_{\mathrm{l}} \sim \tau_{\mathrm{nl}} \implies \tau_{\mathrm{c}} \sim \gamma^{-1}.
\end{align}
This relation is analogous to the conjecture of critical balance in magnetohydrodynamic (MHD) turbulence \cite{schekochihin2022mhd,goldreich1995toward} and amounts to stating that phase-space eddies are most efficiently mixed in both position and velocity space simultaneously. In other words, critical balance \eqref{CB} implies that phase-space eddies have the aspect ratio
\begin{equation} \label{CB_aspect_ratio}
    r \sim \gamma^{-1} u.
\end{equation}
In Appendix \ref{LC_PSE_section}, we show that critical balance is a natural feature of the Lagrangian stretching of phase-space eddies. In brief, any initial structure for which $\tau_{\mathrm{l}} < \tau_{\mathrm{nl}}$ will be linearly phase-mixed to smaller $u$ and, therefore, smaller~$\tau_{\mathrm{nl}}$, and any one for which $\tau_{\mathrm{nl}} < \tau_{\mathrm{l}}$ will be nonlinearly phase-mixed to smaller $r$ and, therefore, smaller $\tau_{\mathrm{l}}$---phase-space dynamics push $\delta \! f$ toward critical balance at every scale.

We seek the shell-averaged fluctuation spectrum of $\delta C_2$ in $(k, s)$ space as a product of power laws, with the exponents changing across the critical-balance line $k \sim \gamma s$~\eqref{CB_aspect_ratio}:
\begin{equation} \label{F_ks}
\langle P_{\delta \! f}  \rangle_{\mathrm{SA}} (k, s) \propto
\begin{cases} 
      k^{d-1} s^{-a} , \qquad \quad & k \ll \gamma s, \\
      s^{d-1} k^{-b}, \qquad \quad & k \gg \gamma s.
   \end{cases}
\end{equation}
The goal is now to find the exponents $a$ and $b$, whereas the factors of $k^{d-1}$ and $s^{d-1}$ in \eqref{F_ks} arise simply from the wavenumber Jacobian in $d$ dimensions. To argue that these are the proper large-scale power laws, we also suppose that the non-shell-averaged spectrum is a product of power laws, based on the assumption that the angular structure of the spectrum is just an order-unity effect that ensures that the system accommodates a constant-flux cascade, as discussed in Section \ref{inhomog_aniso}. Then, the non-shell-averaged spectrum should have the scalings of \eqref{F_ks} divided by the wavenumber Jacobian $\propto k^{d-1} s^{d-1}$, i.e.,
\begin{equation} \label{F_ks_not_SA}
 P_{\delta \! f} (\mathbf{k}, \mathbf{s}) \propto
\begin{cases} 
       s^{-(d - 1 + a)} , \qquad \quad & k \ll \gamma s, \\
       k^{-(d - 1 + b)}, \qquad \quad & k \gg \gamma s.
   \end{cases}
\end{equation}

To argue that the $k \ll \gamma s$ asymptotic of \eqref{F_ks_not_SA} is independent of $k$, note that, under the critical-balance conjecture \eqref{CB}, the position-space correlation scale of the turbulence at a given velocity-space scale $u$ is simply~\eqref{CB_aspect_ratio}, $r \sim \gamma^{-1} u$. At long position-space scales, $k \ll \gamma s$, \eqref{F_ks_not_SA} (for fixed $s$) measures correlations between points in position space that are separated by distances greater than the correlation scale. Such points are effectively uncorrelated, so the spectrum at these wavenumbers is that of white noise in position space and is, therefore, independent of~$k$ (cf. thermal-equipartition spectra in hydrodynamics \cite{kraichnan1973helical,alexakis2019thermal,hosking2023emergence}).

To justify that the $k \gg \gamma s$ asymptotic of \eqref{F_ks_not_SA} is independent of $s$, a similar argument to the one above can be made, with $k$ and $\gamma s$ interchanged. The same result can also be deduced by noting that $P_{\delta \! f} (\mathbf{k}, \mathbf{s} = \mathbf{0})$, up to a constant factor, is the density-fluctuation spectrum, as per \eqref{s=0_C2}. In order for this to be finite, the $s \rightarrow 0$ limit of~\eqref{F_ks_not_SA} must depend only on $k$ and not on~$s$.

\subsection{Constant flux in velocity space}
 \label{const_flux_vspace}

To find the exponent $a$ in \eqref{F_ks}, we assume that $\delta C_2$ undergoes a constant-flux cascade in velocity space. Using $\tau_{\mathrm{c}} \sim \gamma^{-1}$, we have 
\begin{align} \label{spec_s_const_flux}
    & \frac{v_{\mathrm{the}}^d \delta \! f_{u}^2}{\tau_{\mathrm{c}}} \sim \varepsilon \implies \nonumber \\  & v_{\mathrm{the}}^d \delta \! f_{u}^2 \sim \varepsilon \gamma^{-1} = \textrm{const} \iff \int \mathrm{d} k \, \langle P_{\delta \! f}  \rangle_{\mathrm{SA}} (k, s) \propto s^{-1},
\end{align}
where $\delta \! f_{u}$ is the typical increment of $\delta \! f$ across scale $u$ and the flux $\varepsilon$ was defined in \eqref{flux_def}. We can also calculate the 1D $s$ spectrum by integrating~\eqref{F_ks} over $k$; if we assume $b > 1$ (which we will check \textit{a posteriori}), then the dominant contribution comes from integrating up to the critical-balance line:

\begin{align} \label{spec_s_CB_line}
\int \mathrm{d} k \, \langle P_{\delta \! f}  \rangle_{\mathrm{SA}} (k, s) & \sim      \int_{0}^{\gamma s} \mathrm{d} k \, \langle P_{\delta \! f}  \rangle_{\mathrm{SA}} (k, s) \nonumber \\  & \propto \int_{0}^{\gamma s} \mathrm{d} k \, k^{d-1} s^{-a} \propto s^{d - a}.
\end{align} 
Equating \eqref{spec_s_const_flux} and \eqref{spec_s_CB_line} yields $a = d + 1$.

\subsection{Constant flux in position space} \label{const_flux_ps_1D_1V}

To find the exponent $b$ in \eqref{F_ks}, we assume $\delta C_2$ undergoes a constant-flux cascade in position space. Using~$\tau_{\mathrm{c}} \sim \gamma^{-1}$, we have 
\begin{align} \label{spec_k_const_flux}
    & \frac{v_{\mathrm{the}}^d \delta \! f_{r}^2}{\tau_{\mathrm{c}}} \sim \varepsilon \implies \nonumber \\  & v_{\mathrm{the}}^d \delta \! f_{r}^2 \sim \varepsilon \gamma^{-1} = \textrm{const} \iff \int \mathrm{d} s \, \langle P_{\delta \! f}  \rangle_{\mathrm{SA}} (k, s) \propto k^{-1},
\end{align}
where $\delta \! f_{r}$ is the typical increment of $\delta \! f$ across scale $r$. To compute the 1D $k$ spectrum from the 2D spectrum, we can assume that the dominant contribution comes from integrating \eqref{F_ks} over $s$ up to the critical-balance line:
\begin{align} \label{spec_k_CB_line}
    \int \mathrm{d} s \, \langle P_{\delta \! f}  \rangle_{\mathrm{SA}} (k, s) & \sim \int_{0}^{\gamma^{-1} k} \mathrm{d} s \,  \langle P_{\delta \! f}  \rangle_{\mathrm{SA}} (k, s) \nonumber \\  & \propto \int_{0}^{\gamma^{-1} k} \mathrm{d} s \, s^{d-1} k^{-b} \propto k^{d-b}.
\end{align}
Equating \eqref{spec_k_const_flux} and \eqref{spec_k_CB_line} yields $b = d + 1$. The same result can, in fact, be inferred by requiring that the two asymptotics in  \eqref{F_ks} and \eqref{F_ks_not_SA} match along the critical-balance line $k \sim \gamma s$.

\subsection{Universal fluctuation spectrum} \label{UFS}

The full 2D fluctuation spectrum of $\delta C_2$ is, therefore,
\begin{equation} \label{F_ks_solved}
\langle P_{\delta \! f} \rangle_{\mathrm{SA}}(k,s) \sim \varepsilon 
\begin{cases} 
      k^{d-1} \left(\gamma s\right)^{-d-1} , & k \ll \gamma s, \\
      \left(\gamma s \right)^{d-1} k^{-d-1} , & k \gg \gamma s.
   \end{cases}
\end{equation}
The coefficients in front of \eqref{F_ks_solved} can be inferred by using the relationship \eqref{parseval_SA} between the spectrum and $\delta C_2$ and enforcing that 1D integrals over $k$ or $s$ of the 2D spectrum satisfy the constant-flux relations in position and velocity space, viz., \eqref{spec_s_const_flux} and \eqref{spec_k_const_flux}.

We can now use \eqref{F_ks_solved} to compute the fluctuation spectrum of the electric field. Combining \eqref{s=0_C2} with the $\mathbf{s} = \mathbf{0}$ asymptotic of $\delta C_2$ fluctuation spectrum shell-averaged only in $\mathbf{k}$ [so removing the $s^{d-1}$ factor in \eqref{F_ks_solved}] yields
\begin{equation} \label{Ek}
    \langle P_{\mathbf{E}}\rangle_{\mathrm{SA}} (k)  \sim \frac{\varepsilon \, \omega_{\mathrm{pe}}^4 \gamma^{d-1}}{ n_0^2} \, k^{-d-3}.
\end{equation}
Together, \eqref{F_ks_solved} and \eqref{Ek} are the universal fluctuation spectra of Vlasov--Poisson plasma turbulence.

In all dimensions, the fluctuation spectrum of the electric field is safely steeper than $k^{-3}$, which is the threshold beyond which any field with a steeper spectrum has the linear increments \eqref{gamma_def_Batchelor} \cite{frisch1995turbulence}. This validates our assumption that the turbulence is in the Batchelor regime. Indeed, the 1D $k$ spectrum of $\delta C_2$ integrated over $s$ is $\propto k^{-1}$, the same as the Batchelor spectrum \cite{batchelor1959small} for a passive scalar advected by a large-scale flow. Because this cascade occurs in the kinetic phase space and because, by critical balance \eqref{CB}, phase-space eddies have aspect ratio \eqref{CB_aspect_ratio}, $r \sim \gamma^{-1} u$, the 1D $s$ spectrum of $\delta C_2$ integrated over $k$ \eqref{spec_s_const_flux} also has the Batchelor exponent,~$s^{-1}$.

We observe that the scalings of the $\delta C_2$ spectrum \eqref{F_ks_solved} are exactly the same as in the Batchelor regime \cite{batchelor1959small} of the plasma-kinetic Vlasov--Kraichnan model \cite{adkins2018solvable,nastac2023phase} (in 1D-1V), in which the electric field is externally imposed to be a Gaussian random field that is smooth in space (single-scale in~$k$) and white in time. However, as discussed in Section~\ref{intro}, the external field is arbitrary in the Kraichnan model, and choosing a spatially rough external field (multiscale in $k$) leads to different scalings \cite{nastac2023phase}. In the self-consistent case considered here, the Batchelor regime, consistent with the electric-field spectrum \eqref{Ek}, is the only possible scaling regime for a critically-balanced constant-flux phase-space cascade. Indeed, in Appendix~\ref{PT_hd}, we show that a Vlasov--Poisson plasma cannot self-consistently support a spatially rough electric field.

\subsection{Outer scale} \label{outer_scale}

In Section \ref{C2_sect}, we stated that the phase-space cascade begins at the Debye scale, but so far in our scaling theory, we have made no mention of the outer scale of the cascade. Using our knowledge of the structure of turbulence in the inertial range, we can now specify what that scale is.

Because $\delta \! f$ is sourced via velocity-space gradients of $f_0$, as can be seen in \eqref{vlasov_delta_f}, the outer scale in velocity space, $u_0 \sim 1/s_0$, is just the velocity-space scale of $f_0$ \footnote{Note that by assuming \eqref{outer_scale_u0}, we are ignoring the possibility that the cascade is driven by $\delta \! f$ at scales larger than the outer scale $r_0$. If these fluctuations were associated with phase-space turbulence at larger scales (e.g., the gyrokinetic entropy cascade \cite{schekochihin2008gyrokinetic, schekochihin2009astrophysical, tatsuno2009nonlinear, plunk2010two}), they would likely have $u_0 < v_{\mathrm{the}}$. How the phase-space cascade is modified in the presence of an injection with fine-scale velocity-space structure is an interesting question for future work, which, along with the other considerations discussed in Section \ref{implications_dissipation}, is important for determining how the phase-space cascade described in this paper connects to turbulence at scales above $r_0$}, viz.,
\begin{equation} \label{outer_scale_u0}
    u_0 \sim v_{\mathrm{the}}.
\end{equation}

We can estimate the outer scale of the cascade in position space as the infrared cutoff of the Batchelor approximation, viz., that the outer-scale electric fields dominate the phase-space mixing of the distribution function. Using \eqref{Ek}, we conclude that the mean-squared fluctuation amplitude of the self-consistent field at the outer scale $r_0 \sim 1/k_0$ is
\begin{align} \label{Efield_os}
      \frac{e^2}{m^2} \overline{E_{r_0}^2} \sim \int^{\infty}_{k_0} \mathrm{d} k  \langle P_{\mathbf{E}} \rangle_{\mathrm{SA}} (k)  & \sim \frac{ \varepsilon \, \omega^{4}_{pe} \gamma^{d-1} r^{d+2}_{0}}{ n^2_0 }.
\end{align}
Meanwhile, by the definition of $\gamma$ in \eqref{gamma_def_Batchelor}, the mean-squared fluctuation amplitude of the total (external plus self-consistent) electric field  at the outer scale is
\begin{equation} \label{fluct_amp_tot}
    \frac{e^2}{m^2} \overline{E_{\mathrm{tot}, r_0}^2} \sim \gamma^4 r_{0}^{2}.
\end{equation}
Clearly, we must have $ \overline{E_{\mathrm{tot}, r_0}^2} \gtrsim \overline{E_{r_0}^2}$. To produce a useful constraint on $r_0$, we must relate $\varepsilon$ and $\gamma$.

Since $C_{2,0}$ sources $\delta C_2$ at the cascade rate $\tau_{\mathrm{c}} \sim \gamma^{-1}$,
\begin{equation} \label{epsilon_twiddle_general}
    \varepsilon = - \frac{d C_{2,0}}{d t} \sim \frac{C_{2,0}}{\tau_{\mathrm{c}}} \sim \frac{n^2_0 \gamma}{v_{\mathrm{the}}^d}.
\end{equation}
Using \eqref{epsilon_twiddle_general} in \eqref{Efield_os} and comparing the latter to \eqref{fluct_amp_tot}, we get
\begin{equation} \label{amp_ineq}
    \frac{\overline{E_{r_0}^2}}{\overline{E_{\mathrm{tot}, r_0}^2}} \sim \omega^4_{pe} \gamma^{d-4} v_{\mathrm{the}}^{-d} r^{d}_0 \lesssim 1,
\end{equation}
which, using $v_{\mathrm{the}} \sim \omega_{\mathrm{pe}} \lambda_{\mathrm{De}}$, implies
\begin{equation} \label{os_estimate}
    \frac{r_0}{\lambda_{\mathrm{De}}}  \lesssim \left( \frac{\gamma}{\omega_{\mathrm{pe}}} \right)^{(4-d)/d}.
\end{equation}
Another constraint on the outer scale is that we expect the injection to satisfy critical balance. Indeed, our outer-scale injection is a model for $C_2$ arriving from larger scales, and if it arrives by dynamical, organic means, it should naturally be critically balanced. Then, $r_0 \sim \gamma^{-1} u_0$, as per \eqref{CB_aspect_ratio}. Inserting this relation into \eqref{os_estimate}, with $u_0 \sim v_{\mathrm{the}}$, yields
\begin{equation} \label{CB_outer_scale_constraint}
\gamma \gtrsim \omega_{\mathrm{pe}}.
\end{equation}
The constraints \eqref{os_estimate} and \eqref{CB_outer_scale_constraint} are naturally satisfied if $\gamma \sim \omega_{\mathrm{pe}}$ and $r_0 \sim \lambda_{\mathrm{De}}$. The phase-space cascade, starting at the Debye scale, will then extend into the sub-Debye range $r \ll \lambda_{\mathrm{De}}$.

\subsection{Dissipation scale} \label{dissipation_cutoffs}

Fluctuations, after being injected at the outer scale, will cascade through the inertial range, to the dissipation range at large wavenumbers, where the spectra \eqref{F_ks_solved} and \eqref{Ek} are cut off by collisions. To find the velocity-space cutoff, we can balance the cascade time \eqref{CB}, $\tau_{\mathrm{c}} \sim \gamma^{-1}$, with the time scale of collisional velocity-space diffusion, which is, from \eqref{coll_op_diffusion},
\begin{equation}
   \tau_{\nu} \sim \nu^{-1} \, \frac{u^2}{v_{\mathrm{the}}^2}. 
\end{equation}
This gives us the collisional velocity-space scale:
\begin{equation} \label{u_nu}
    \frac{u_{\nu}}{v_{\mathrm{the}}} \sim \mathrm{Do}^{-1/2},
\end{equation}
where the Dorland number \cite{schekochihin2009astrophysical,schekochihin2008gyrokinetic, plunk2010two, tatsuno2009nonlinear}
\begin{equation} \label{dorland}
    \mathrm{Do} \equiv (\nu \tau_{\mathrm{c}})^{-1} = \frac{\gamma}{\nu}
\end{equation}
is the kinetic-plasma-turbulence analogue of the Reynolds number. When $\mathrm{Do} \gg 1$, there exists an inertial range of velocity-space scales between which $\delta C_2$ is cascaded from the outer scale $u_0 \sim v_{\mathrm{the}}$ to the dissipation scale $u_{\nu}$ \eqref{u_nu}.

To estimate the position-space cutoff, we can combine the estimate \eqref{u_nu} with the critical-balance relation \eqref{CB_aspect_ratio} between position-space and velocity-space scales. This implies that the smallest position-space scale satisfies 
\begin{equation} \label{r_nu}
    \frac{r_{\nu}}{\lambda_{\mathrm{De}}} \sim \frac{\omega_{\mathrm{pe}}}{\gamma} \, \mathrm{Do}^{-1/2}.
\end{equation}
If the outer scale satisfies critical balance \eqref{CB_aspect_ratio}, $r_0 \sim \gamma^{-1} v_{\mathrm{the}}$, \eqref{r_nu} becomes
\begin{equation} \label{r_nu2}
    \frac{r_{\nu}}{r_0} \sim \mathrm{Do}^{-1/2}.
\end{equation}
Thus, if $\mathrm{Do} \gg 1$, there also exists an inertial range of position-space scales between which $\delta C_2$ is cascaded from the outer scale $r_0$ to the dissipation scale $r_{\nu}$. This dissipation cutoff can be thought of as the Kolmogorov scale \cite{kolmogorov1941b} of phase-space turbulence. Since collisions are diffusive only in velocity space, it is a nontrivial result due to the collisionless dynamics, which intertwine position and velocity space, that such a cutoff exists in position space.

\subsection{Efficiency of phase-space mixing} \label{ph_mixing_eff}

Let us compute the characteristic time scale $\tau_{\mathrm{d}}$ on which phase-space eddies are processed by the turbulent cascade from the injection scale to the collisional dissipation scale. We can estimate $\tau_{\mathrm{d}}$ by dividing the total $\delta C_2$ stored in phase space in steady state by its injection rate $\varepsilon$. Integrating the 1D $k$ spectrum \eqref{spec_k_const_flux}, $\propto k^{-1}$, from $k_0 \sim r^{-1}_0 \sim \gamma /v_{\mathrm{the}}$ up to $k_{\nu} \sim r^{-1}_{\nu} \sim \gamma^{3/2} / (v_{\mathrm{the}} \nu^{1/2})$ [see \eqref{r_nu2}] yields
\begin{equation} \label{delta_C2_amp}
    \delta C_2 \sim \varepsilon \gamma^{-1} \log \frac{k_{\nu}}{k_{0}} \sim \varepsilon \gamma^{-1} \log \frac{\gamma}{\nu} = \varepsilon \gamma^{-1} \log \mathrm{Do}.
\end{equation}
Therefore,
\begin{equation} \label{tau_dissp}
    \tau_{\mathrm{d}} \sim \frac{\delta C_2}{\varepsilon} \sim \gamma^{-1} \log \mathrm{Do},
\end{equation}
which implies efficient phase-space mixing. Physically, the logarithmic dependence of \eqref{tau_dissp} on the Dorland number \eqref{dorland} arises from the fact that phase-space eddies are stretched exponentially in time at the rate $\gamma$, until they have developed phase-space gradients on the collisional scales $(r_{\nu}, u_{\nu})$ (this argument is given in greater detail in Appendix \ref{LC_PSE_section}). Such mixing is analogous to fluid-dynamical passive scalar mixing in the Batchelor regime, where $\tau_{\mathrm{d}}$ has the same form as \eqref{tau_dissp} but with~$\gamma$ replaced by the mixing rate associated with the advecting flow and~$\nu$ replaced by the rate of molecular diffusivity~\cite{batchelor1959small}.

The time scale \eqref{tau_dissp} should be contrasted to a linear system with only phase mixing, where \cite{su1968collisional,zocco2011reduced,kanekar2015fluctuation,banik2024relaxation}
\begin{equation} \label{tau_dissp_linear}
    \tau_{\mathrm{d}} \sim \frac{1}{\nu^{1/3}(k v_{\mathrm{the}})^{2/3}}
\end{equation}
for a linear perturbation with wavenumber $k$. The linear time scale \eqref{tau_dissp_linear} is asymptotically much longer than the nonlinear time scale \eqref{tau_dissp} in the small-$\nu$ limit, although, notably, both are much shorter than the collisional time~$\nu^{-1}$.

Note that because the cascade time $\tau_{\mathrm{c}} \sim \gamma^{-1}$ is independent of scale, $\tau_{\mathrm{d}} \rightarrow \infty$ as $\nu \rightarrow 0$, but only weakly (logarithmically with $\nu$). In other words, formally, in the limit $\nu \rightarrow 0$, there is no `dissipative anomaly' \cite{eyink2006onsager} for any finite time, but there is a dissipative anomaly if one instead takes $t \rightarrow \infty$ first and then $\nu \rightarrow 0$. In this way, Batchelor-like turbulence \cite{batchelor1959small} is distinct from Kolmogorov-like turbulence \cite{kolmogorov1941b}, where there is a dissipative anomaly in finite time, viz., $\tau_{\mathrm{c}} \rightarrow 0$ as $r \rightarrow 0$ and $\tau_{\mathrm{d}} \rightarrow \textrm{const}$, in the limit of small viscosity (or collision frequency) \cite{frisch1995turbulence,schekochihin2008gyrokinetic, tatsuno2009nonlinear, eyink2018cascades, zhdankin2022generalized, nastac2023phase}. However, in practice, for any system with a finite value of $\nu$, the phase-space cascade is still very efficient at dissipating $\delta C_2$ in a way nearly independent of $\nu$.

Finally, we ask the reader doubting the validity of a Fokker--Planck collision operator as the dissipation mechanism of the sub-Debye cascade to stick with us till Section \ref{noise_turbulence_interplay}. There, we will argue that in real plasmas, the turbulent cascade is not terminated by (collisional) velocity-space diffusion but rather by the takeover of turbulent fluctuations by discrete particle noise. As a consequence, the estimates \eqref{delta_C2_amp} and \eqref{tau_dissp} will be replaced by \eqref{delta_C2_amp_noise} and \eqref{tau_dissp_noise}, respectively.

\section{Numerical experiments: forced turbulence} \label{forced_turbulence}

In this section, we test the turbulence theory developed in Section \ref{phenom_theory} via direct numerical simulations of a forced 1D-1V system.

\subsection{Numerical setup} \label{numerical_setup}

We solve the Vlasov equation \eqref{vlasov} in 1D-1V ($d = 1$) using the continuum, discontinuous Galerkin Vlasov-Maxwell solver in the Gkeyll simulation framework \cite{juno2018discontinuous,9355299}. We solve for the electron distribution function with the initial condition
\begin{equation} \label{initial_condition}
    f(x,v,t = 0) = \frac{n_0}{\sqrt{2 \pi}v_{\mathrm{th},0}} \, e^{-v^2/2v_{\mathrm{th},0}^2},
\end{equation}
where $v_{\mathrm{th},0} = \sqrt{T_0/m}$, and $T_0$ is the initial temperature. Ions are treated as a stationary, neutralizing background. In position space, the domain is periodic, with box size $L = 2 \pi \lambda_{D, 0}$, where $\lambda_{D, 0} = v_{\mathrm{th},0}/\omega_{\mathrm{pe}}$ is the electron Debye length at $t = 0$. In velocity space, the distribution function has the extent $\pm 10 v_{\mathrm{th},0}$ with zero-flux boundary conditions. We neglect collisions between ions and electrons, and we model electron-electron collisions with a nonlinear Dougherty-Fokker--Planck collision operator \cite{dougherty1964model, hakim2020conservative},
\begin{equation} \label{col_op_FP}
C[f] = \nu \frac{\partial}{\partial v} \left[ \left(v-u \right)f + v_{\mathrm{th}}^2 \frac{\partial f}{\partial v} \right].    
\end{equation}
where $\nu$ is the (velocity-independent) collision frequency, $u(x,t)$ is the bulk flow velocity \eqref{u}, and $v_{\mathrm{th}}(x,t) = \sqrt{T(x,t)/m}$ is the local-in-space thermal velocity \footnote{Note that $v_{\mathrm{th}}$ in \eqref{col_op_FP} is the thermal velocity of the full distribution function $f$ and, therefore, depends not just on time but also on space. Throughout the paper, we use $v_{\mathrm{the}}$ to denote the thermal velocity of $f_0$, which is generally time-dependent but independent of space. However, for spatially homogeneous turbulence, we expect $v_{\mathrm{the}} \sim v_{\mathrm{th}}$, so this distinction, while important for~\eqref{col_op_FP} to satisfy exactly its necessary conservation laws, is unimportant, e.g., for dimensional estimates.}, with the local temperature determined from the energy density and particle number density as
\begin{equation}
    T(x, t) = \frac{1}{n} \int \mathrm{d} v \, m \left(v-u \right)^2 f.
\end{equation}
This collision operator contains both velocity-space drag and diffusion, and it can be shown to conserve the particle number density, momentum, and energy. The collision operator can also be shown to satisfy the H-theorem, viz., it drives the distribution function to a local-in-space Maxwellian. Further assuming that the local-in-space thermal velocity is comparable to the thermal velocity of $f_0$, viz., $v_{\mathrm{th}} \sim v_{\mathrm{the}}$, and that the variation of $\delta \! f$ with~$v$ is much greater than that of $f_0$, reduces \eqref{col_op_FP} to the diffusion operator \eqref{coll_op_diffusion}. For a detailed description of the numerical implementation of \eqref{col_op_FP} in Gkeyll, including its conservative properties, we refer the reader to \cite{hakim2020conservative}.

Electrons are accelerated by a self-consistent electric field $E$, that is obtained from Ampère's law using the electron current density computed from the particle distribution function \footnote{We solve Ampère's law for the electric field rather than Poisson's equation purely out of convenience of the simulation code that we are using. We ensure that the total volume-averaged current is zero at all time steps, which enforces that the volume-averaged electric field is zero at all time steps. Then, in a periodic box, the 1D-1V Vlasov-Ampère system is equivalent to the 1D-1V Vlasov--Poisson system.},
\begin{equation}
    \frac{\partial E}{\partial t} = -4 \pi e \int \mathrm{d} v v f. 
\end{equation}
There is also an external forcing field $E_{\mathrm{ext}}$, which drives turbulence in the system. Spatially, this external field is homogeneous and has a single wavenumber at the box scale, $k_{\mathrm{f}} = 2 \pi /L = \lambda_{D, 0}^{-1}$, which, in this case, is also the outer scale of the turbulence, viz., $k_{\mathrm{f}} = k_0$. Temporally, we model $E_{\mathrm{ext}}$ as an Ornstein–Uhlenbeck process \cite{kloeden2013stochastic}, i.e., a Gaussian random field with zero mean and the correlation function
\begin{equation} \label{ext_E_corr_tau}
    \frac{e^2}{m^2} \overline{E_{\mathrm{ext},k_{\mathrm{f}}}(t)E^*_{\mathrm{ext},k_{\mathrm{f}}}(t')} = \frac{D}{2 \tau_{\mathrm{corr}}} e^{-|t-t'|/\tau_{\mathrm{corr}}},
\end{equation}
where $D$ is a constant diffusion coefficient and $\tau_{\mathrm{corr}}$ is the correlation time of the field.

As discussed in Section \ref{C2_sect}, this forcing can be conceptualized as a source of turbulent electric fluctuations originating from turbulence at scales larger than our system size. We are interested here in how they are processed through the Debye scale and into phase space. Therefore, in our simulations, we set $\tau_{\mathrm{corr}} = \omega_{\mathrm{pe}}^{-1}$ and pick~$D$ so that the energy density of the external electric field (ensemble-averaged over realizations of the stochastic forcing and volume-averaged),
\begin{equation} \label{energy_density_ext_field}
\overline{W}_{\mathrm{ext}} \equiv \frac{\langle \overline{E_{\mathrm{ext}}^2} \rangle_{V}}{8 \pi}, 
\end{equation}
satisfies $\overline{W}_{\mathrm{ext}} = (0.1) K_0$, where $K_0 = n_0 T_0 /2$ is the initial kinetic-energy density of the electrons and $\langle ... \rangle_{V} \equiv V^{-1} \int \mathrm{d} \mathbf{x} (...)$. This amplitude is chosen so that the ratio of the external field's correlation time to its associated phase-space eddy-turnover time is of order unity---a property expected of self-consistently generated turbulent fields. Using \eqref{ext_E_corr_tau}, the root-mean-squared spatial gradient of external field (multiplied by the charge-to-mass ratio) is $\gamma^2 = (e/m)\sqrt{\overline{ |\partial E_{\mathrm{ext}}/ \partial x|^2}} = k_{\mathrm{f}} (D/ \tau_{\mathrm{corr}})^{1/2}.$ For the above forcing parameters, $\gamma^2 \simeq 1.12 \omega_{\mathrm{pe}}^2$. Therefore, $\tau_{\mathrm{c}} \sim \gamma^{-1} \simeq 0.94 \omega_{\mathrm{pe}}^{-1}$ and~$\tau_{\mathrm{corr}}/\tau_{\mathrm{c}} \simeq 1.06$, as desired.

We perform simulations across a range of collision frequencies, varying from $\nu = 10^{-4} \omega_{\mathrm{pe}}$ down to $\nu = 10^{-6} \omega_{\mathrm{pe}}$. The simulation domains are discretized into phase-space grid cells, in which the distribution function and the electric field are represented by piecewise quadratic polynomial elements. To determine the grid resolution, note that the smallest phase-space scales are $\Delta x = L/N_x$ and $\Delta v = L_v/N_v$, where $N_x$ is the number of position-space cells, $N_v$ is the number of velocity-space cells, and $L_v$ is the `box size' in velocity space. To resolve a critically-balanced cascade properly, \eqref{CB} suggests that an optimal grid spacing should satisfy $\Delta x \sim \gamma^{-1} \Delta v$, which implies $N_x/N_v \sim \gamma^{1/2} (L/L_v) $. For our simulations, where $L_v = 20 v_{\mathrm{th},0}$, $N_x/N_v \simeq 1/3$.  Of course, this estimate involves an order-unity constant that we do not know \textit{a priori}. Empirically, we found that $N_x \times N_v = 1024 \times 2048$ ($N_x/N_v = 1/2$) yielded resolved results. In terms of grid spacing, this resolution corresponds to $\Delta x \simeq 6 \times 10^{-3} \lambda_{D, 0}$ and $\Delta v \simeq 10^{-2} v_{\mathrm{th},0}$.

Before continuing, we remark on a subtlety regarding the injection and dissipation scales in our numerical simulations. In position space, the injection scale is the box scale $r_0 = 2 \pi \lambda_{\mathrm{D},0}$. As the external field heats the plasma, the Debye length $\lambda_{\mathrm{De}} = v_{\mathrm{the}}/\omega_{\mathrm{pe}}$ grows, yet, since the box size is fixed, the outer scale remains equal to the initial Debye length. Therefore, the inertial range in position space shrinks with time, since replacing $\lambda_{\mathrm{De}}$ with $\lambda_{D, 0}$ in the estimate \eqref{r_nu} gives $r_{\nu}/r_{0} \sim (\lambda_{\mathrm{De}}/\lambda_{D, 0}) \, \omega_{\mathrm{pe}} \gamma^{-1} \, \mathrm{Do}^{-1/2}$, which grows as $\lambda_{\mathrm{De}}/\lambda_{D, 0}$ does. Furthermore, while the system is initially critically balanced at the outer scale, it is not at later times. Note that the forcing satisfies $\gamma \sim \omega_{\mathrm{pe}}$, so, initially, $r_0 \sim \gamma^{-1} u_0 $ because $\gamma^{-1} v_{\mathrm{th},0} \sim \lambda_{\mathrm{D},0} \sim r_0$. As the plasma is heated, the velocity-space outer scale $v_{\mathrm{the}}$ grows, even though $\gamma$ and $r_0$ stay fixed; the injection is therefore out of critical balance ($\tau_{\rm{l}} <  \tau_{\rm{nl}})$. However, as can be seen in Fig.~\ref{fig:f0_heating}, the kinetic energy of particles will grow only by roughly a factor of ten by the end of our simulations, so the ratios of the final thermal velocity and Debye length to the initial ones are only $v_{\mathrm{the}}/v_{\mathrm{th}, 0} \sim \lambda_{\mathrm{De}}/\lambda_{D, 0} \sim 3$. The required modifications to the dissipation estimates derived in Section \ref{dissipation_cutoffs} are, therefore, order-unity effects and can be ignored.

\subsection{Phase-space eddies} \label{numeric_PS_eddies}

\begin{figure*}
	\centering
	\includegraphics[width=\textwidth]{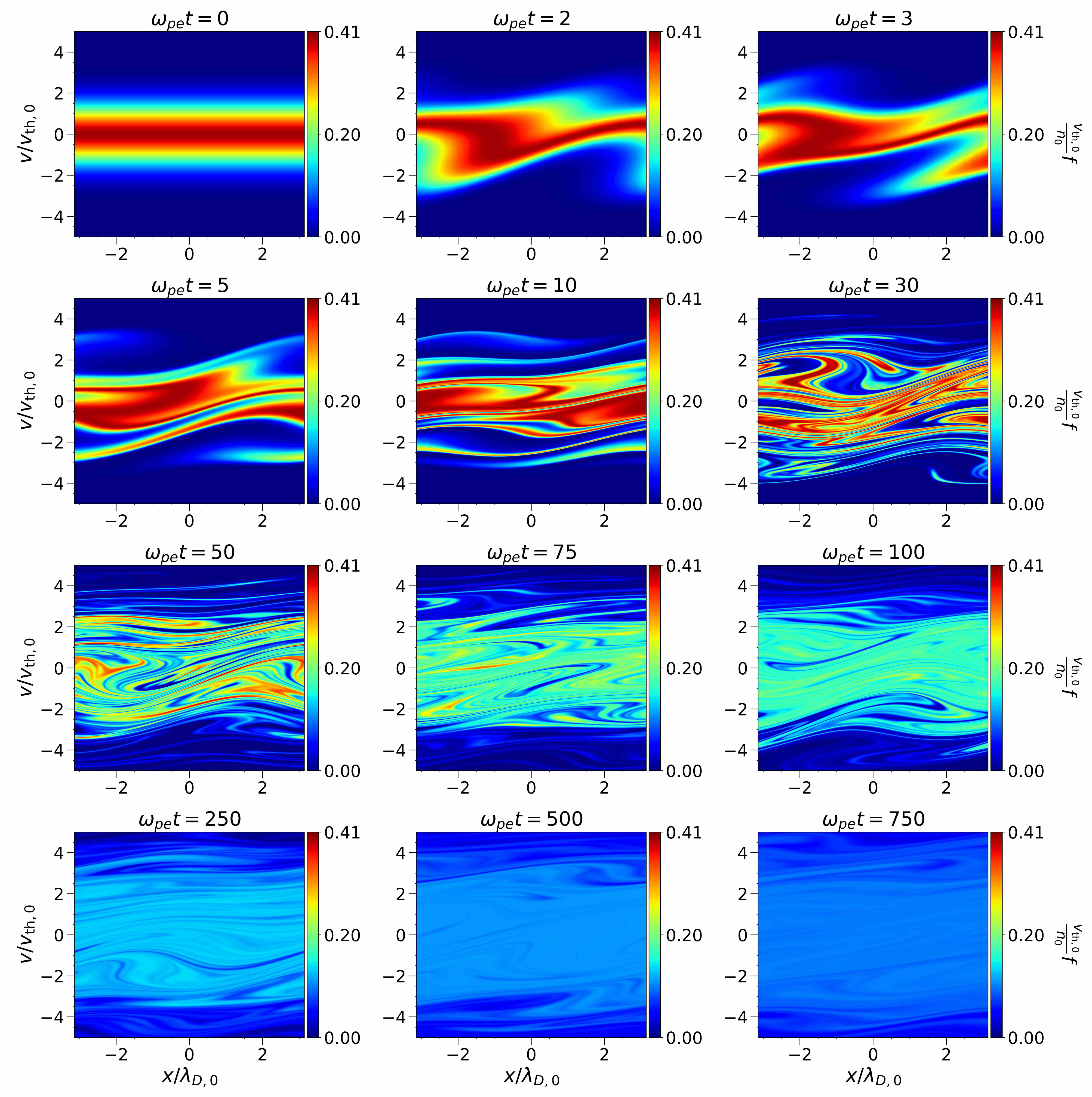}
	\caption{\label{fig:ps_eddy_numerical} Contour plots of the distribution function at a sequence of consecutive times, as indicated, from a numerical simulation with collision frequency $\nu = 10^{-5} \omega_{\mathrm{pe}}$.}
\end{figure*}

\begin{figure*}
	\centering
	\includegraphics[width=\textwidth]{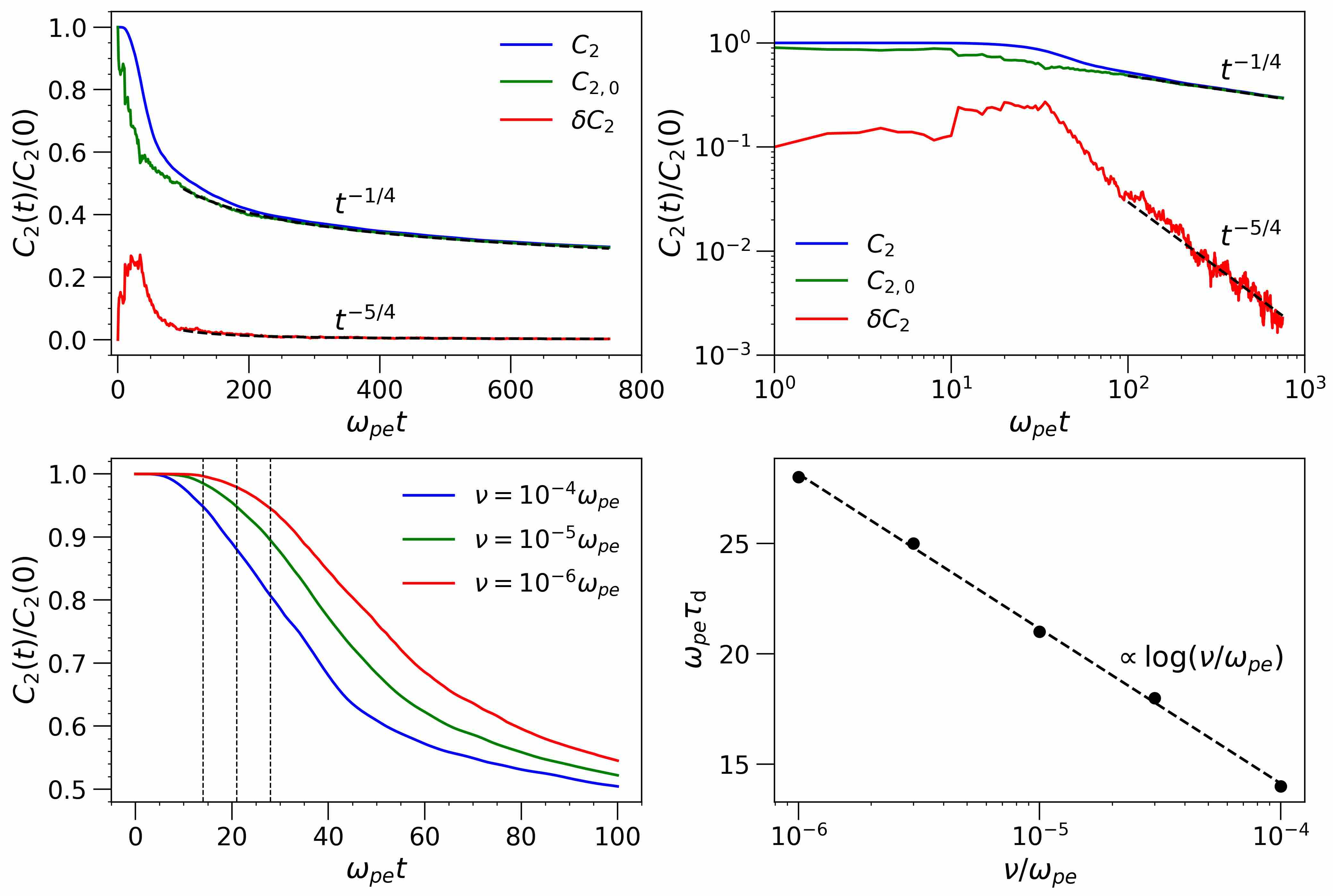}
 	\caption{\label{fig:C2_plots} Top left panel: $C_2$, $C_{2,0}$, and $\delta C_2$ versus time, shown alongside the theoretical predictions $C_{2,0} \propto t^{-1/4}$ [see \eqref{C20_time}] and $\delta C_2 \propto t^{-5/4}$ [see \eqref{decay_laws_deltaC2}]. This data is from a numerical simulation with collision frequency $\nu = 10^{-5} \omega_{\mathrm{pe}}$. The prefactor of the dashed curve $\propto t^{-1/4}$ is set by the analytical prediction \eqref{C20_time} (but with $\alpha \rightarrow \tilde{\alpha}$, where $\tilde{\alpha} = 0.84 \alpha$, with $\alpha$ computed from the forcing parameters of the simulation; see discussion in Section \ref{Stochastic_heating_numerics}). The prefactor of the dashed curve $\propto t^{-5/4}$ is obtained from a fit. Top right panel: same as top left panel, but with logarithmic spacing on the axes. Bottom left panel: $C_2$ versus time for simulations with three different collision frequencies. The vertical dashed lines indicate the time where $C_2$ is $95\%$ of its original value, which is a proxy for the time $\tau_{\mathrm{d}}$ that it takes collisions to be activated, predicted to be \eqref{tau_dissp}. From left to right, the vertical lines correspond to $ \omega_{pe }\tau_{\mathrm{d}}$ for the simulations with $\nu = 10^{-4} \omega_{\mathrm{pe}}$, $\nu = 10^{-5} \omega_{\mathrm{pe}}$, and $\nu = 10^{-6} \omega_{\mathrm{pe}}$, respectively. Bottom right panel: $ \omega_{pe }\tau_{\mathrm{d}}$ [using the proxy from the bottom left panel] versus $\nu /\omega_{\mathrm{pe}}$. The dashed line is $\propto \log (\nu/\omega_{\mathrm{pe}})$. In addition to cases plotted in the bottom left panel, we have also calculated $\tau_{\mathrm{d}}$ for simulations with $\nu = 3 \times 10^{-5} \omega_{\mathrm{pe}}$ and $\nu = 3 \times 10^{-6} \omega_{\mathrm{pe}}$. This confirms the logarithmic dependence \eqref{tau_dissp}. }
\end{figure*}

To illustrate the evolution of our phase-space turbulence, we show in Fig.~\ref{fig:ps_eddy_numerical} contour plots of the distribution function at a sequence of times. The distribution function starts as a Maxwellian with no spatial structure, i.e., $\delta \! f = 0$. As particles are moved around by the fluctuating electric field, $\delta \! f$ fluctuations are sourced by the velocity-space gradients of $f_0$, via \eqref{dC20dt} and \eqref{vlasov_delta_f}.  At early times, total $C_2$ is conserved, so the growth of $\delta C_2$ is at the expense of the depletion of $C_{2,0}$. This can be seen in Fig.~\ref{fig:C2_plots}, which show $C_2$, decomposed into $C_{2,0}$ and $\delta C_2$, versus time.

The $\delta \! f$ fluctuations, which are phase-space blobs, or eddies, undergo stretching and folding in phase space,  which generates small-scale structure in the distribution function. To understand stretching, we refer the reader to Appendix \ref{LC_PSE_section}, where we analyze the relation between Lagrangian chaos and phase-space turbulence. Here we give an abridged summary. At times short compared to the autocorrelation time of the electric field (which is the prescribed $\tau_{\mathrm{corr}}$ for the external field and the cascade time $\tau_{\mathrm{c}}$ for the self-consistent field; $\tau_{\mathrm{c}} \sim \tau_{\mathrm{corr}}$ in our case), two nearby Lagrangian particles with phase-space coordinates ($x_1, v_1)$ and $(x_2, v_2)$ see a difference in the total electric field $(e/m) \delta E_{\mathrm{tot},r} \sim \gamma^2 r$, where $\gamma$ is a local-in-space, constant-in-time shearing rate associated with the combined external and self-consistent fields (initially, there is only an external field, but later there is also a self-consistent field). The trajectory of the particle separation $(r,u) = (x_1-x_2, v_1-v_2)$ therefore satisfies the equations $(\dot{r}, \dot{u}) = (u, \gamma^2 r)$, which has the solution $(r, u) \propto (e^{\gamma t}, \gamma e^{\gamma t})$. There will be regions in space where $\gamma^2 < 0$, in which particles are trapped ($\gamma$ is imaginary). Trapped particles' trajectories cycle in phase space, causing phase-space eddies to rotate (but not stretch). In contrast, when $\gamma^2 > 0$, particle trajectories will separate exponentially, causing eddies to be stretched into thin filaments. This stretching is in both position and velocity space, which is clear from the fact that $(u, \gamma^2 r)$ is a superposition of two linear shear flows in phase space, viz., $(u, 0)$ and $(0, \gamma^2 r)$. The latter flow tilts eddies in the $r$ direction, while the former tilts eddies in the $u$ direction. The combined effect of the two flows orients eddies along the critical-balance line $u \sim \gamma r$.

Once the length of a filamented eddy reaches the scale of variation of the electric field (so that the expansion $\delta E_{\mathrm{tot},r} \sim \gamma^2 r$ is no longer valid), the eddy begins to rotate and fold, because two ends of it are now sheared by different values of $\gamma$. After one autocorrelation time of the field, phase-space eddies are effectively sheared by a new realization of the field, and the above process repeats. After many autocorrelation times, the distribution function becomes thoroughly mixed in phase space. This is manifest in Fig.~\ref{fig:ps_eddy_numerical}: the distribution function becomes more homogenized as time progresses.

The phase-space mixing process described above is collisionless, but according to \eqref{tau_dissp}, collisions are only negligible for a time~$ \tau_{\mathrm{d}} \sim \gamma^{-1} \log (\gamma/\nu)$, after which they start smoothing (diffusively) the fine-scale structure of the distribution function. In Fig.~\ref{fig:C2_plots} (bottom left panel), we show $C_2$ versus time for three cases of different collision frequencies, which shows that the time when collisions are first activated increases as $\nu$ decreases. In Fig.~\ref{fig:C2_plots} (bottom right panel), we show that the time when $C_2$ conservation is broken is consistent with a linear scaling with respect to $\log \nu$, in agreement with our prediction \eqref{tau_dissp}.

Once collisions are activated, $C_2$ decays \textit{ad infinitum}. In the left and right panels of Fig.~\ref{fig:C2_plots}, we show that, at long times, the decay laws
\begin{equation} \label{decay_laws}
    C_{2,0} \propto t^{-1/4}, \quad \delta C_2 \propto t^{-5/4}
\end{equation}
agree well with the simulation data. We will derive the power law for $C_{2,0}$ in Section \ref{Stochastic_heating} [see \eqref{C20_time}], but for the time being, we note that the decay law for $\delta C_2$ follows from that of $C_{2,0}$. In a saturated turbulent state, from \eqref{delta_C2_amp} and \eqref{flux_def}, we get
\begin{equation} \label{decay_laws_deltaC2}
    \delta C_2 \propto \varepsilon = -\frac{d C_{2,0}}{d t} \propto t^{-5/4},
\end{equation}
in agreement with our simulations.

For finite $\nu$, the time $\tau_{\mathrm{d}} \sim \gamma^{-1} \log (\gamma/\nu)$ is short compared to the long time over which $\varepsilon$ decays. The system is therefore in a quasi-steady turbulent state in which fast $\delta \! f$ fluctuations are sourced by a slowly evolving $f_0$, at a rate that is approximately balanced by the rate of collisional dissipation. This approximation becomes better as time progresses, because the decay of $\varepsilon$ becomes ever slower at long times. Conceptually, this turbulent state is analogous to fluid passive-scalar turbulence forced by a mean scalar gradient. In the latter case, a fixed mean profile allows for a truly steady-state turbulence \cite{yeung2002schmidt,donzis2010batchelor} (here, this would correspond to a frozen-in-time $f_0$), but a slowly evolving mean profile can still yield a quasi-steady-state turbulence \cite{schekochihin2004diffusion}, similar to our case.

\subsection{Spectra} \label{spectra_test}

\begin{figure}
	\centering
	\includegraphics[width=\linewidth]{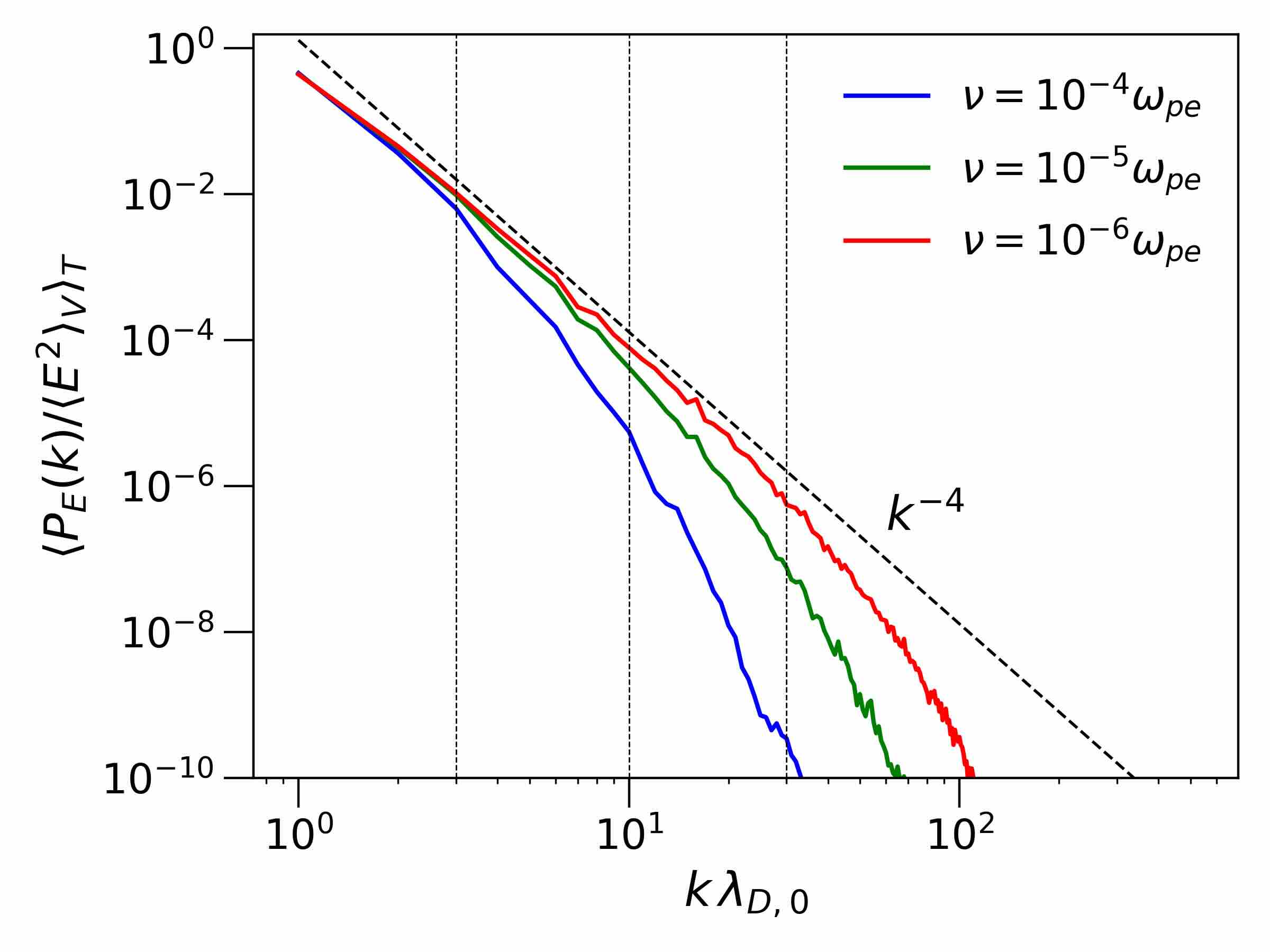}
	\caption{\label{fig:Ek_spec} Electric field spectrum versus time for three simulations with different collision frequencies. The spectra are time averaged over the interval $(30 \omega_{\mathrm{pe}}^{-1}, 250 \omega_{\mathrm{pe}}^{-1})$, where, at each time in the average, the spectra are normalized by the total electric energy density, $\langle E^2 \rangle_{V}$. The dashed vertical lines correspond to the dissipation cutoffs predicted in Section \ref{dissipation_cutoffs}; from left to right, these correspond to the $\nu = 10^{-4} \omega_{\mathrm{pe}}$, $\nu = 10^{-5} \omega_{\mathrm{pe}}$, and $\nu = 10^{-6} \omega_{\mathrm{pe}}$ cases, respectively. The cutoff for the simulation with $\nu = 10^{-4} \omega_{\mathrm{pe}}$ was determined by eye; the cutoffs for the other two cases are scaled from this reference value according to the scaling $k_{\nu} \propto \nu^{-1/2}$ \eqref{r_nu}. }
\end{figure}

\begin{figure*}
	\centering
	\includegraphics[width=\textwidth]{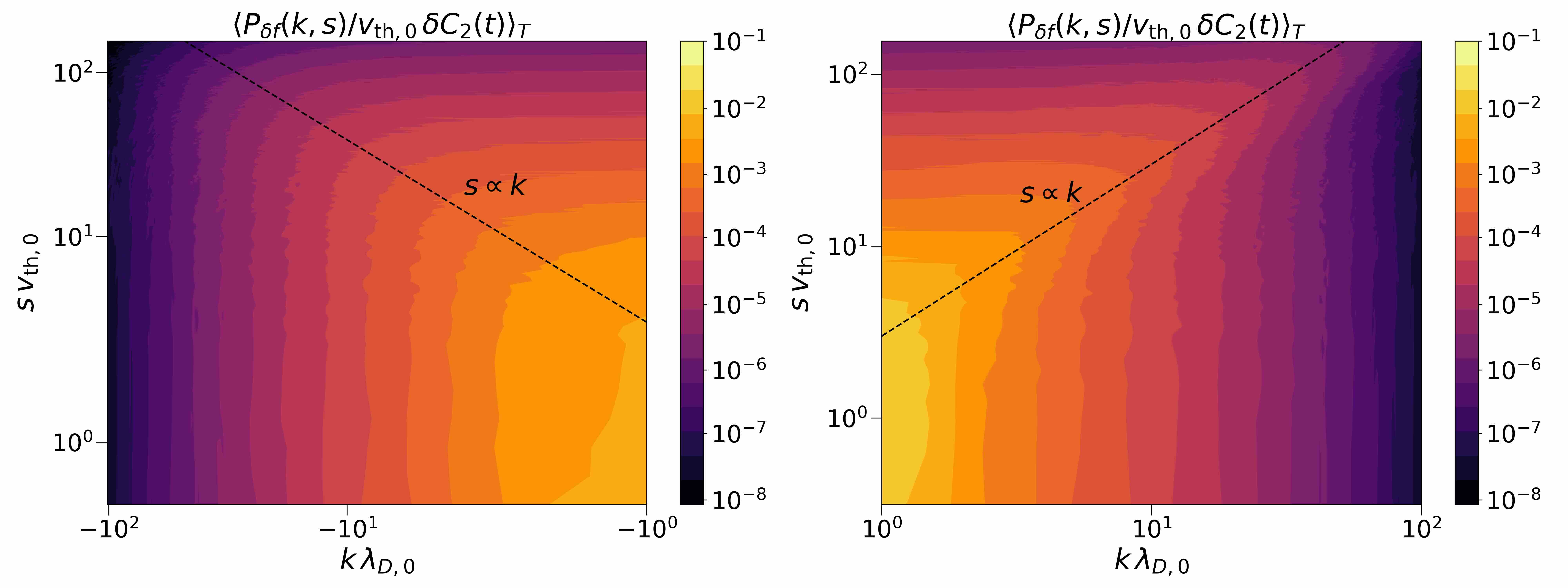}
	\caption{\label{fig:C2_2D_spec} 2D $\delta C_2$ spectrum from the simulation with $\nu = 10^{-6} \omega_{\mathrm{pe}}$, for $s > 0$ and (left panel) $k < 0$, (right panel) $k > 0$, averaged over the time interval $(30 \omega_{\mathrm{pe}}^{-1}, 250 \omega_{\mathrm{pe}}^{-1})$, where, at each time, the spectrum is normalized by the total~$\delta C_2$. The black dashed line is the critical-balance line, $s \sim \gamma^{-1} k$, separating the regions where the nonlinear mixing dominates ($s \gg \gamma^{-1} k$) and those where the linear phase mixing dominates ($s \ll \gamma^{-1} k$).}
\end{figure*}

\begin{figure*}
	\centering
	\includegraphics[width=\textwidth]{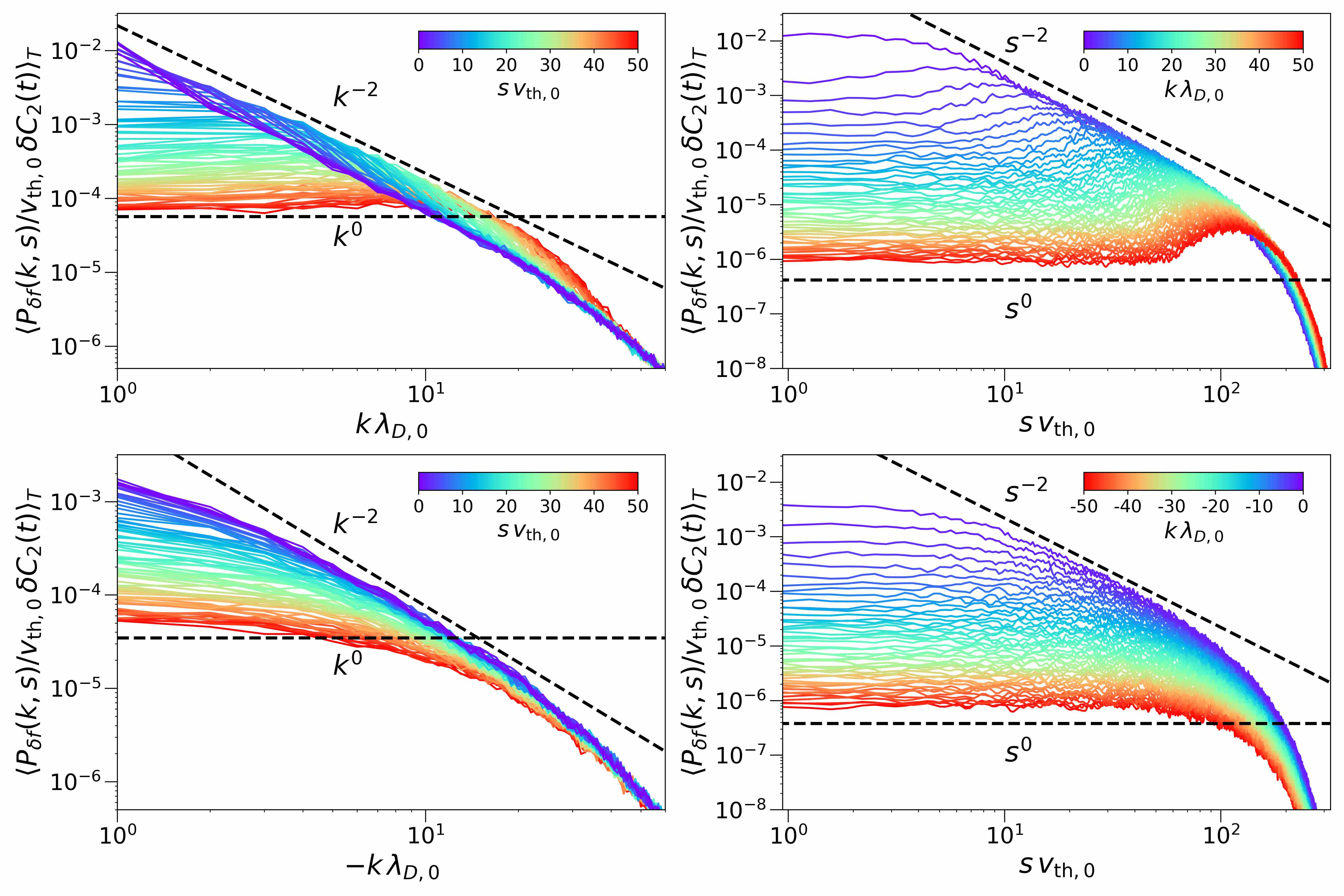}
	\caption{\label{fig:C2_1D_cuts} Top left panel: 1D cuts of the 2D $\delta C_2$ spectrum shown in Fig.~\ref{fig:C2_2D_spec}, as a function of $k > 0$, for a sequence of values of $s$ (color bar). Bottom left panel: the same as the top left panel but for~$k < 0$. Top right panel: 1D cuts again, but this time as a function of $s$, for a sequence of values of $k > 0$. Bottom right panel: the same as the top right panel but for $k < 0$. }
\end{figure*}

To confirm that this is kinetic Batchelor turbulence, we check the scalings \eqref{F_ks_solved} and \eqref{Ek}. To produce spectra from our numerical simulations, rather than compute ensemble averages over many simulations with different realizations of the forcing, we time average spectra over a window of time when the plasma is in the quasi-steady turbulent state, after the peak of $\delta C_2$ evident in Fig.~\ref{fig:C2_plots} (top left and right panels). Within our time averages, we normalize the $\delta C_2$ spectra and electric-field spectra by the total $\delta C_2$ and total electric energy at each instance in time to account for the fact that the turbulence is decaying, as discussed in the previous section. Notationally, quantities that were defined with ensemble averages in Sections \ref{C2_sect} and \ref{phenom_theory}, viz., the spectra $P_{\delta \! f}(k,s)$ and $P_{E}(k)$, will correspond to the same quantities but not ensemble averaged. Time averaging will be denoted by $\langle ... \rangle_{T}$.

In Fig.~\ref{fig:Ek_spec}, we plot the electric-field spectrum. The spectrum matches well the theoretical prediction \eqref{Ek} that $P_{E}(k) \propto k^{-4}$. As the collision frequency decreases, the inertial range extends to smaller scales in a way consistent with the estimate \eqref{r_nu}, viz., the cutoff wavenumber scales as $k_{\nu} \propto \nu^{-1/2}$. This is a demonstration of the remarkable property of phase-space turbulence that, in addition to $\delta C_2$, even `fluid-dynamical' quantities like the electric field have inertial ranges whose extent is controlled by collisions \cite{schekochihin2008gyrokinetic,schekochihin2009astrophysical,tatsuno2009nonlinear}, in the same way that, e.g., viscosity determines the dissipation scale of hydrodynamic turbulence \cite{frisch1995turbulence}. 

In Fig.~\ref{fig:C2_2D_spec}, we plot the time-averaged 2D $\delta C_2$ spectrum. The structure of the spectrum is clearly demarcated by the critical-balance line, $s \sim \gamma^{-1} k$. In Fig. \ref{fig:C2_1D_cuts}, we show 1D cuts of the 2D spectrum. The top left and bottom left panels show the spectrum for $k > 0$ and $k < 0$, respectively, at a sequence of values of $s$. These agree with the predicted scalings of~$k^{-2}$ at high $k$ and $k^{0}$ at low $k$, as given in \eqref{F_ks_solved}. Likewise, the top right and bottom right panels show the spectrum as a function of $s$ at a sequence of values of $k > 0$ and $k < 0$, respectively, following the predicted scalings of $s^{-2}$ at high $s$ and $s^{0}$ at low $s$.

\begin{figure*}
	\centering
	\includegraphics[width=\textwidth]{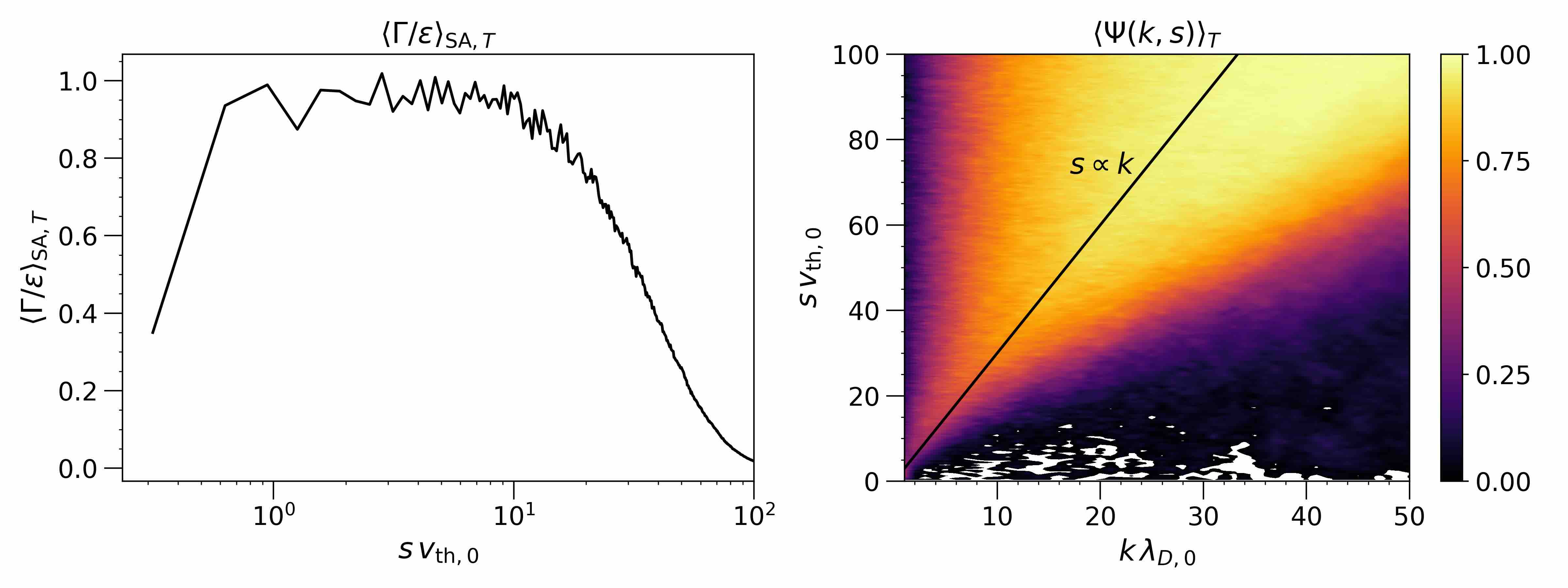}
	\caption{\label{fig:PU_flux} Left panel: shell-averaged (in $k$ and $s$) $s$ flux, as defined in \eqref{gamma_s_1D-1V_main}, for the simulation with $\nu = 10^{-5} \omega_{\mathrm{pe}}$. The flux is averaged over the time interval $(250 \omega_{\mathrm{pe}}^{-1}, 500 \omega_{\mathrm{pe}}^{-1})$, where at each time, we divide the flux by the $\delta C_2$ injection rate $\varepsilon = - dC_{2,0}/dt$, computed using the analytical expression \eqref{C20_time} (but with $\alpha \rightarrow \tilde{\alpha}$, where $\tilde{\alpha} = 0.84 \alpha$, with $\alpha$ computed from the forcing parameters of the simulation; see discussion in Section \ref{Stochastic_heating_numerics}). Right panel: contour plot of $\langle\Psi(k,s)\rangle_{T}$ [see \eqref{Lambda}] in $(k,s)$ space, for the same simulation.  The average again is done over the time interval $(250 \omega_{\mathrm{pe}}^{-1}, 500 \omega_{\mathrm{pe}}^{-1})$. The black line is the critical-balance line,~$s \sim \gamma^{-1} k$. }
\end{figure*}

The existence of a populated spectrum at both $k > 0$ and $k < 0$ is a demonstration that the turbulent state involves the coexistence of phase-mixing and phase-unmixing modes, respectively. Of course, the injected flux at large scales cannot be processed to dissipation scales in the phase-unmixing region. As we discuss in Section \ref{inhomog_aniso} and show explicitly in Appendix \ref{PU_app}, in order to have a constant-flux cascade of $\delta C_2$, the contributions to the $s$ flux integrated over $k$ from the phase-mixing modes must outweigh the corresponding contributions from the phase-unmixing modes. More precisely, the quantity
\begin{equation} \label{gamma_s_1D-1V_main}
	\langle \Gamma \rangle_{\mathrm{SA}} = 2  \sum_k \, \left[ P_{\delta \! f}(k, s) - P_{\delta \! f}(-k, s) \right]
\end{equation}
must be positive [note $s > 0$ in \eqref{gamma_s_1D-1V_main}; the $s < 0$ contribution is accounted for by the factor of two in front of the sum]. We show in Fig.~\ref{fig:PU_flux} (left panel) that the flux~\eqref{gamma_s_1D-1V_main} (time-averaged and normalized) is indeed positive. Furthermore, in an inertial range of velocity-space scales, this flux is approximately constant and equal to the injected flux $\varepsilon = -d C_{2,0}/dt$, consistent with a net constant-flux velocity-space cascade.

\begin{figure*}
	\centering
	\includegraphics[width=\textwidth]{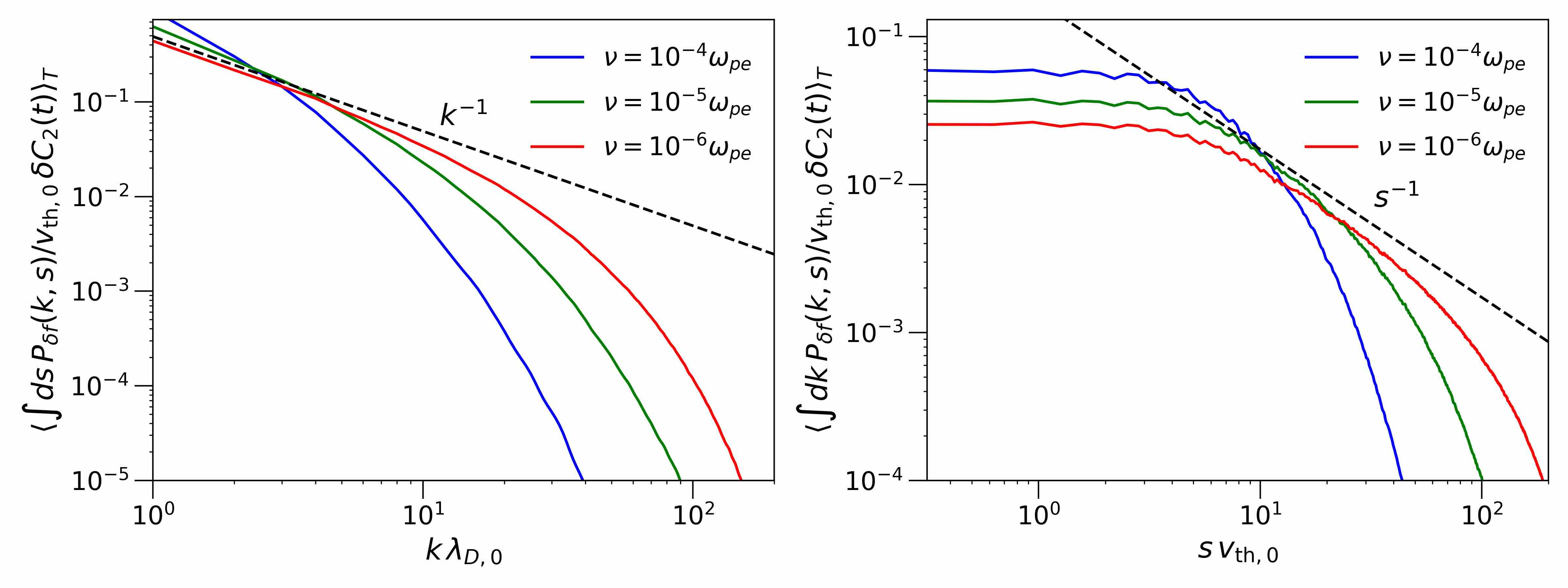}
	\caption{\label{fig:C2_1D_spec} Left panel: 1D $\delta C_2$ spectrum in $k$ obtained by integrating over all $s$ the 2D spectrum from Fig.~\ref{fig:C2_2D_spec}. Here it is shown from simulations with $\nu = 10^{-4} \omega_{\mathrm{pe}}$, $10^{-5} \omega_{\mathrm{pe}}$, and $ 10^{-6} \omega_{\mathrm{pe}}$. Right panel: the same as the left panel, but for the 1D $\delta C_2$ spectrum in $s$ integrated over all $k$. }
\end{figure*}

It is instructive to compare the phase-mixing and phase-unmixing fluxes locally in $(k,s)$ space. The competition between the two processes is encoded in the sign of the integrand of \eqref{gamma_s_1D-1V_main}, which is positive where phase mixing dominates and negative where phase-unmixing dominates. A useful quantity to measure this competition~is
\begin{equation} \label{Lambda}
 \Psi(k,s) = \frac{ P_{\delta \! f}(k,s) -  P_{\delta \! f}(-k,s)}{ P_{\delta \! f}(k,s) +  P_{\delta \! f}(-k,s)},
\end{equation}
where $k > 0$ and $s \geq 0$. A state with only phase mixing modes has $\Psi = 1$, while a state with only phase-unmixing modes has $\Psi = -1$. For reference, in the Vlasov--Kraichnan model, $\eqref{Lambda}$ is everywhere positive in the inertial range \cite{adkins2018solvable,nastac2023phase}, except along the line $s = 0$, where the $s$ flux vanishes (since, in this part of phase space, fluctuations are cascaded in position space to high~$k$).

In Fig.~\ref{fig:PU_flux} (right panel), we plot $\langle \Psi(k,s)\rangle_{T}$, which is mostly positive and is $\mathcal{O}(1)$ in the critical-balance region, from which the dominant contributions to the integrated flux \eqref{gamma_s_1D-1V_main} comes. This is in agreement with the Kraichnan model and is consistent with the positivity of the flux in Fig.~\ref{fig:PU_flux} (left panel). Note, however, that there are some patches in the high-$k$ and low-$s$ part of phase space where~$\langle \Psi(k,s)\rangle_{T}$ is negative. In these regions (shown in white),~$\langle\Psi(k,s) \rangle_{T} > -0.1$. We do not have any physical reason to expect such patches, and we believe that the presence of these regions is a numerical artifact.

Finally, in Fig.~\ref{fig:C2_1D_spec}, we plot the 1D $\delta C_2$ spectra in $k$ and~$s$, integrated over $s$ and $k$, respectively. The numerically obtained spectra are consistent with the theoretical predictions, $\int \mathrm{d} s P_{\delta \! f}(k,s) \propto k^{-1}$ and $\int \mathrm{d} k P_{\delta \! f}(k,s) \propto s^{-1}$.

\section{Thermodynamics of turbulent plasma} \label{thermo_turb}

Now that we have diagnosed the small-scale structure of the phase-space turbulence, we analyze the evolution and structure of the mean distribution $f_0$---the slowly evolving `equilibrium' of our turbulent plasma. This is the velocity-space equivalent of turbulent-transport theory: evolution of the mean macroscopic distribution subject to microscopic fluctuations.

The question of whether collisionless equilibria of turbulent plasmas have a degree of universality was first posed decades ago \cite{kadomtsev1970collisionless,dupree1972theory} and remains a topic of ongoing interest (see, e.g., \cite{diamond2010modern,ewart2022,ewart2023nonthermal,ewart2024relaxation,banik2024relaxation} for recent advances). Rather than attempting to construct a universal theory, our goal in this section is more modest: to explain the evolution and structure of $f_0$ in the situation of forced 1D-1V turbulence, as considered in our numerical simulations in Section \ref{forced_turbulence}. Specifically, in Section \ref{Stochastic_heating}, we develop an analytical theory of stochastic heating, which we then validate against our numerical results in Section \ref{Stochastic_heating_numerics}. In Section \ref{decay_irreversible}, we apply this theory to determine how the $\delta C_2$ flux \eqref{flux_def} and the energy of electric fluctuations decay in time. The implications of these results for the broader question of universal collisionless equilibria in turbulent plasmas are discussed in Section \ref{turb_coll_op}.

\subsection{Stochastic heating: theory} \label{Stochastic_heating}

Particles accelerated by stochastic fields undergo non-resonant energization; this phenomenon is known as stochastic heating \cite{sturrock1966stochastic,chandran2010perpendicular,verscharen2019multi,cerri2021stochastic}. Analytically, stochastic heating is normally modeled either phenomenologically, by estimating diffusion coefficients using scalings of the turbulent electric fields, or using quasilinear theory. Here, we take the quasilinear approach.

In Appendix \ref{QL_app}, we compute the correlator between the electric field and $\delta \! f$ in \eqref{df0dt} by assuming that it is dominated by the forcing scale $k_{\mathrm{f}}$ and that the $\delta \! f$ evolution is linear. This leads to the following evolution equation for $f_0$:
\begin{equation} \label{QL_evol}
    \frac{\partial f_0}{\partial t} =  \frac{\partial}{\partial v} D_{\mathrm{eff}}(v) \frac{\partial f_0}{\partial v},
\end{equation}
with the effective diffusion coefficient
\begin{equation} \label{D_eff}
    D_{\mathrm{eff}}(v) = \frac{D}{\left[(k_{\mathrm{f}} v \, \tau_{\mathrm{corr}} )^2 + 1 \right] |\epsilon(k_{\mathrm{f}} v, k_{\mathrm{f}})|^2},
\end{equation}
where $\epsilon(k_{\mathrm{f}} v, k_{\mathrm{f}})$ is the dielectric function, defined in \eqref{dielectric_omega}, and $D$ is the constant diffusion coefficient introduced in \eqref{ext_E_corr_tau}.

To understand \eqref{QL_evol}, first consider the case of passive particles accelerated by a white-noise field, viz., $\epsilon = 1$ and $\tau_{\mathrm{corr}} = 0$. Then, $D_{\mathrm{eff}}(v) = D$, and the solution of \eqref{QL_evol} for a Maxwellian initial condition is simply a spreading Maxwellian:
\begin{equation} \label{f0_diffusion}
    f_0 = \frac{n_0}{\sqrt{2 \pi} v_{\mathrm{the}}}e^{-v^2/2 v_{\mathrm{the}}^2}, \quad v_{\mathrm{the}} = \sqrt{v_{\mathrm{th},0}^2 + 2 D t},
\end{equation}
with a kinetic-energy density that grows linearly in time:
\begin{equation} \label{K_diffusion}
     K \equiv \int \mathrm{d} v \frac{m v^2}{2} f_0 = \frac{n_0 T_0}{2} + n_0 m D t.
\end{equation}
Particles random-walk in velocity space, so $f_0$ is heated diffusively at a constant rate $\propto D$. As the distribution function broadens, its $C_{2,0}$ decreases: using \eqref{f0_diffusion}, we find
\begin{equation} \label{C20_diffusion}
    C_{2,0} = \int \mathrm{d} v \frac{f_0^2}{2} = \frac{n_0^2}{4 \sqrt{\pi} v_{\mathrm{the}}},
\end{equation}
with $v_{\mathrm{the}}$ given in \eqref{f0_diffusion}. The long-time decay law of $C_{2,0}$ is thus
\begin{equation}
    C_{2,0} \propto t^{-1/2}.
\end{equation}

The presence of self-consistent fields `dresses' the diffusion coefficient \eqref{D_eff} with a factor of $|\epsilon|^{-2}$, reflecting the fact that above the Debye length, the plasma strives to maintain quasineutrality and shield electic fields. In doing so, Langmuir waves are generated, with phase speeds satisfying $v_p/v_{\mathrm{the}} \sim 1/(k_{\mathrm{f}} \lambda_{\mathrm{De}})$. Initially, $k_{\mathrm{f}} \lambda_{\mathrm{De}} = 1$, so the Langmuir waves interact with the bulk of the distribution, and $ \epsilon$ has some order-unity variation in velocity space. As the distribution heats, the Debye length grows, as $\lambda_{\mathrm{De}} = v_{\mathrm{the}}/\omega_{\mathrm{pe}}$, and at long times, the system is pushed into the regime where $k_{\mathrm{f}} \lambda_{\mathrm{De}} \gg 1$, so $|\epsilon| \simeq 1$ (see Appendix \ref{QL_app}). Physically, below the Debye length, the plasma cannot respond to shield the external electric field, since the self-consistent electric fields are strongly Landau-damped. Since we focus on the regime where $k_{\mathrm{f}} \lambda_{\mathrm{De}}$ is never small, the dressing is never qualitatively important---but interesting and distinct behavior can be found from quasilinear theory in the presence of large-scale forcing, $k_{\mathrm{f}} \lambda_{\mathrm{De}} \ll 1$ \cite{banik2024universal}. 

Let us now examine the effect of a finite correlation time. From \eqref{D_eff}, it is clear that the presence of a finite $\tau_{\mathrm{corr}}$ affects the diffusion coefficient depending on the size of $k_{\mathrm{f}} v_{\mathrm{the}} \tau_{\mathrm{corr}}$, where we have estimated the typical velocity by $v_{\mathrm{the}}$.  If $k_{\mathrm{f}} v_{\mathrm{the}} \tau_{\mathrm{corr}} \ll 1$, assuming $k_{\mathrm{f}} \lambda_{\mathrm{De}} \gtrsim 1$ so that $\epsilon \approx 1$, then $D_{\mathrm{eff}}(v) \approx D$, and the solution (\ref{f0_diffusion}-\ref{K_diffusion}) approximately holds. However, $v_{\mathrm{the}}$ grows as the plasma is heated, and for any finite $\tau_{\mathrm{corr}}$, inevitably the system is pushed into the  $k_{\mathrm{f}} v_{\mathrm{the}} \tau_{\mathrm{corr}} \gg 1$ regime, in which $D_{\mathrm{eff}}(v) \simeq D/(k_{\mathrm{f}} v \tau_{\mathrm{corr}})^{2}$. In this regime, assuming also $|\epsilon| \approx 1$, \eqref{df0dt} has a similarity solution
\begin{equation} \label{flat-top}
    f_0 = \frac{n_0}{\Gamma(1/4) (\alpha t)^{1/4}} e^{-v^4/ 16 \alpha t}, \quad \alpha = \frac{D}{(k_{\mathrm{f}} \tau_{\mathrm{corr}})^2},
\end{equation}
where $\Gamma$ is the Gamma function. We derive \eqref{flat-top} in Appendix \ref{QL_app}. The solution \eqref{flat-top} is a `flat-top' distribution, with a relatively flatter core than a Maxwellian. The kinetic-energy density associated with this distribution grows as
\begin{equation} \label{K_subdiffusion}
     K = \frac{2 \, \Gamma(3/4)}{\Gamma(1/4)} m n_0 (\alpha t)^{1/2} \simeq 0.68 \, m n_0 (\alpha t)^{1/2}.
\end{equation}
The heating is subdiffusive, $K \propto t^{1/2}$, slower than in the diffusive regime \eqref{K_diffusion}. Physically, since the ratio of the correlation time $\tau_{\mathrm{corr}}$ to the typical streaming time $(k_{\mathrm{f}} v_{\mathrm{the}})^{-1}$ is large, viz., $k_{\mathrm{f}} v_{\mathrm{the}} \tau_{\mathrm{corr}} \gg 1$, the particles are now fast enough to be able to sample the spatial structure of the electric field before the field decorrelates, resulting in partial averaging-out of the stochastic acceleration. The particles therefore gain relatively less energy than in the $k_{\mathrm{f}} v_{\mathrm{the}} \tau_{\mathrm{corr}} \ll 1$ regime, which leads to slower heating.

This slower heating also manifests in a slower decrease in $C_{2,0}$. Using the expression \eqref{flat-top}, we have
\begin{equation} \label{C20_time}
    C_{2,0} = \int \mathrm{d} v \frac{f_0^2}{2} = \frac{n_0^2}{2^{5/4} \Gamma(1/4) (\alpha t)^{1/4}} \propto t^{-1/4},
\end{equation}
whereas $C_{2,0} \propto t^{-1/2}$ is the diffusive regime, as per \eqref{C20_diffusion}.

\begin{figure*}
	\centering
	\includegraphics[width=\textwidth]{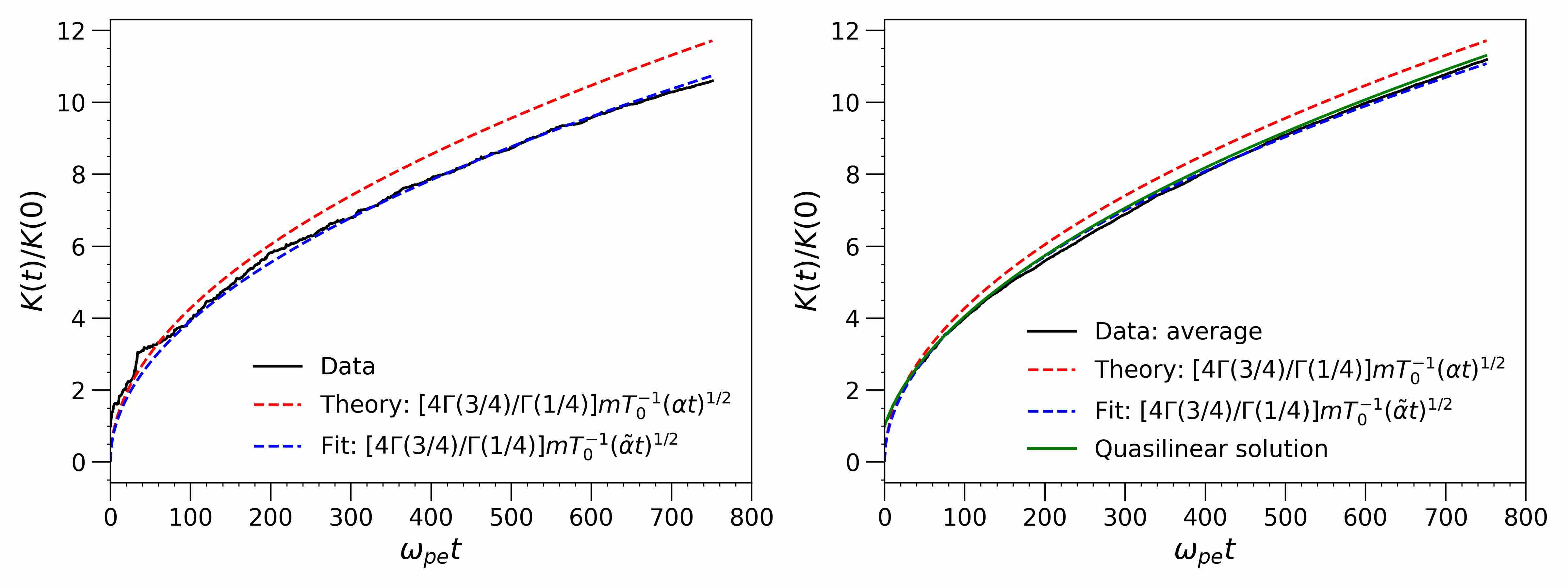}
	\caption{\label{fig:f0_heating} Left panel: kinetic-energy density of the particle distribution function versus time from the simulation with $\nu = 10^{-5} \omega_{\mathrm{pe}}$. The red dashed curve is the theoretical asymptotic \eqref{K_subdiffusion} with $\alpha = D/(k_{\mathrm{f}} \tau_{\mathrm{corr}})^2$ [as defined in \eqref{flat-top}] with $D$, $k_{\mathrm{f}}$, and $\tau_{\mathrm{corr}}$ taken as the forcing parameters used in the simulation. The blue dashed curve is \eqref{K_subdiffusion} but with $\alpha \rightarrow \tilde{\alpha}$, where $\tilde{\alpha}$ is statistically fit to the simulation data. We fit $\log  K(t)/K(0) = (1/2) \log t + \log A$, where $A = (4 \Gamma(3/4)/\Gamma(1/4)) m T_0^{-1} \tilde{\alpha}^{1/2}$, to the simulation data in the range $t \in (250 \omega_{\mathrm{pe}}^{-1}, 750 \omega_{\mathrm{pe}}^{-1})$. The fit parameter $\tilde{\alpha}$ satisfies $\tilde{\alpha} \simeq 0.84 D/(k_{\mathrm{f}} \tau_{\mathrm{corr}})^2$. Right panel: the black solid curve is the average kinetic energy versus time taken from an ensemble average of ten simulations with different realizations of the stochastic forcing (different seed of the random number generator). Each simulation was run with the same forcing parameters and initial condition as described in Section \ref{numerical_setup}, the collision frequency $\nu = 10^{-5} \omega_{\mathrm{pe}}$, and the grid resolution $N_x \times N_v = 512 \times 1024$. The dashed red curve is the theoretical asymptotic~\eqref{K_subdiffusion}. The green solid curve is the kinetic energy versus time obtained by directly solving the quasilinear equation \eqref{QL_evol} with $|\epsilon| = 1$. To produce this solution, velocity space was discretized on a grid of extent $\pm 10 v_{\mathrm{th}, 0}$, with zero-flux boundary conditions. Time stepping was done with Euler's method, and the diffusion operator was discretized using a second-order central-difference scheme. The initial condition was taken to be the Maxwellian \eqref{initial_condition}, and the forcing parameters were set to the values used in our Vlasov simulations. The blue dashed curve is a fit to the average heating curve, done in the same way as in the right panel. The fitting parameter here is $\tilde{\alpha} \simeq 0.89 D/(k_{\mathrm{f}} \tau_{\mathrm{corr}})^2$.}
\end{figure*}

\subsection{Stochastic heating: numerical simulations} \label{Stochastic_heating_numerics}

We now test this theory against our numerical simulations. In what follows, $f_0$ as computed from simulation data is taken to be the volume-averaged distribution function (over the entire box), i.e., $f_0 \equiv \langle f \rangle_{V}$.

\begin{figure*}
	\centering
	\includegraphics[width=\textwidth]{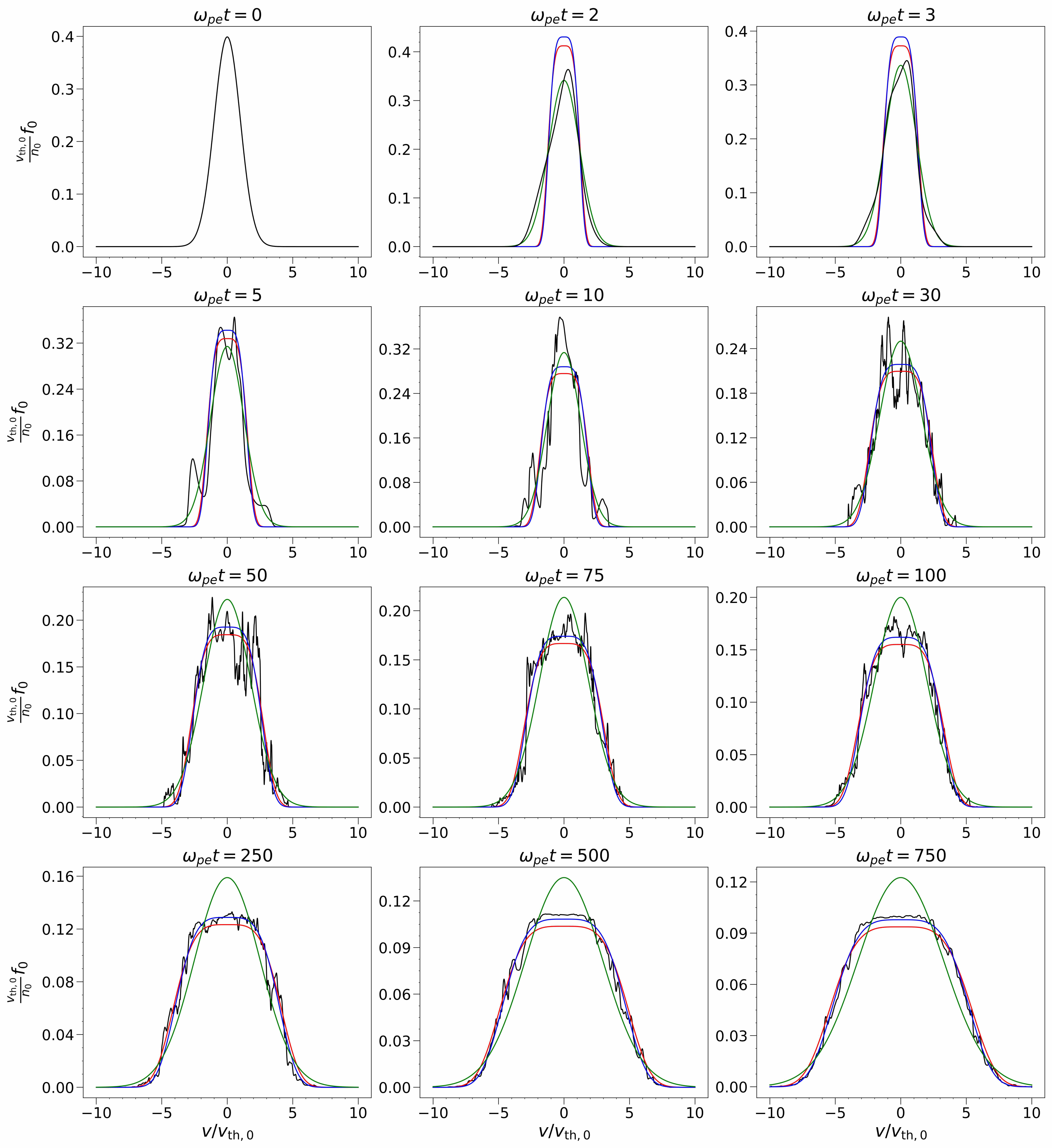}
	\caption{\label{fig:f0_plots} Evolution of $f_0$ through a set of consecutive times in the numerical simulation with collision frequency $\nu = 10^{-5} \omega_{\mathrm{pe}}$. The black curves are the numerically obtained volume-averaged distribution function at each simulation time. The red curves are the profile \eqref{flat-top} with $\alpha = D/(k_{\mathrm{f}} \tau_{\mathrm{corr}})^2$, where $D$, $k_{\mathrm{f}}$, and $\tau_{\mathrm{corr}}$ are taken as the forcing parameters used in the simulation. The blue curves are the same profile but with $\alpha \rightarrow \tilde{\alpha} \simeq 0.84 D/(k_{\mathrm{f}} \tau_{\mathrm{corr}})^2$. The parameter $\tilde{\alpha}$ corresponds to the statistical fit of the evolution of the kinetic-energy density in the left panel of Fig.~\ref{fig:f0_heating}. The green curves are Maxwellian fits, with thermal velocity at each time chosen so that the kinetic energy of the Maxwellian is equal to that of the true distribution function. }
\end{figure*}

In the left panel of Fig.~\ref{fig:f0_heating}, we plot the kinetic-energy density versus time for the simulation with $\nu = 10^{-5} \omega_{\mathrm{pe}}$. Initially, $k_{\mathrm{f}} v_{\mathrm{the}} \tau_{\mathrm{corr}} = 1$, so there is no asymptotic regime in time where we would predict the heating to be purely diffusive according to \eqref{K_diffusion}. At long times, the kinetic-energy density grows in a way that is consistent with the subdiffusive scaling $\propto t^{1/2}$ predicted in \eqref{K_subdiffusion}. However, the heating rate is not consistent with the precise quasilinear prediction. Namely, the red curve, which corresponds to~\eqref{K_subdiffusion} with $\alpha = D/(k_{\mathrm{f}} \tau_{\mathrm{corr}})^2$ [as defined in \eqref{flat-top}] with $D$, $k_{\mathrm{f}}$, and $\tau_{\mathrm{corr}}$ taken as the forcing parameters used in the simulation, overestimates the heating. Instead, if we take $\alpha \rightarrow \tilde{\alpha}$ in \eqref{K_subdiffusion} and treat $\tilde{\alpha}$ as a fitting parameter, we find that $\tilde{\alpha} \simeq 0.84 D/(k_{\mathrm{f}} \tau_{\mathrm{corr}})^2$ matches the data much better and indeed remarkably well (the blue curve). This agreement is corroborated by the evolution of $f_0$, which we show in Fig.~\ref{fig:f0_plots}. Initially, $f_0$ is a Maxwellian, but at long times, $f_0$ tends to a distribution that is certainly non-Maxwellian and that agrees well with the flat-top distribution \eqref{flat-top} with $\alpha$ replaced by the fitting parameter $\tilde{\alpha}$.

The discrepancy between the expected $\alpha$ and the fit $\tilde{\alpha}$ that matches the data much better is not a failure of quasilinear theory in describing the evolution of $f_0$ but is rather due to the approximate solution \eqref{flat-top} to~\eqref{QL_evol} being subject to quantitatively non-negligible corrections. To obtain \eqref{flat-top}, we approximated the diffusion coefficient \eqref{D_eff} as $D_{\mathrm{eff}}(v) \simeq D/(k_{\mathrm{f}} v \tau_{\mathrm{corr}})^{2}$. If we restore `$+1$' in the denominator of \eqref{D_eff}, the heating rate is reduced. We solved the quasilinear equation \eqref{QL_evol} numerically with $D_{\mathrm{eff}}(v)$ given by \eqref{D_eff}, taking $|\epsilon| = 1$ and adopting the forcing parameters used in our Vlasov simulations. The kinetic-energy density of the resulting solution is plotted in the right panel of Fig.~\ref{fig:f0_heating} (green solid line). We also ran an ensemble of ten simulations with different realizations of the stochastic forcing; the average heating curve from this ensemble is also plotted in the right panel of Fig.~\ref{fig:f0_heating} (black solid line). The two curves match extremely well, and are manifestly below the theoretical prediction \eqref{K_subdiffusion}. Fitting $\tilde{\alpha}$ to the average heating curve now yields $\tilde{\alpha} \simeq 0.89 D/(k_{\mathrm{f}} \tau_{\mathrm{corr}})^2$. To see why there is a roughly $10\%$ discrepancy, note that at the final time plotted in Fig.~\ref{fig:f0_heating}, the kinetic-energy has grown from its initial amount by a factor of approximately $10$, so $(k_{\mathrm{f}} v_{\mathrm{the}} \tau_{\mathrm{corr}})^2 \approx 10$. The $+1$ in the denominator of~\eqref{D_eff} thus amounts to an approximately $ 10\%$ correction.

That there is agreement between our simulations and quasilinear theory should perhaps be puzzling, since the system is clearly nonlinear, given the evidence of phase-space turbulence (and, therefore, nonlinear mode coupling) gathered in Sections \ref{numeric_PS_eddies} and \ref{spectra_test}. An argument as to why it is not contradictory that the system exhibits both nonlinear and quasilinear behavior is as follows. The quasilinear approximation amounts to neglecting the nonlinear acceleration term in the $\delta \! f$ equation. A dimensional estimate of the ratio of this term to the linear acceleration term, which is retained, at some phase-space scale $(r,u)$~is
\begin{equation} \label{ratio_NL_L}
    R_{r,u} = \frac{\partial \delta \! f_{r, u}/ \partial v}{\partial f_0/ \partial v} \sim \frac{\delta \! f_{r,u}}{f_0} \frac{v_{\mathrm{the}}}{u} ,
\end{equation}
where $\delta \! f_{r,u}$ is the characteristic amplitude of $\delta \! f$ at scale $(r,u)$. The nonlinearity is important if $\delta \! f$ has a large amplitude and/or sharp velocity gradients. Since the energy injection is done by an external field at the outer scale $k_0 = k_{\mathrm{f}} = \lambda_{\mathrm{D},0}^{-1}$ and since this is Batchelor turbulence, viz., it is the largest-scale mode of the electric field that predominantly stirs $\delta \! f$, it is reasonable to approximate the right-hand side of \eqref{df0dt} with the term that comes from the correlation between the $k_{\mathrm{f}}$ components of the electric field and $\delta \! f$ [explicitly, the right-hand-side of \eqref{df0dt} decomposes into a wavenumber sum, as can be seen in \eqref{QL_corr}]. Then, we only have to estimate \eqref{ratio_NL_L} for $r \sim \lambda_{\mathrm{D},0}$. Because $\delta \! f$ fluctuations are injected by the velocity-space gradients of $f_0$, a natural estimate for the dominant velocity-space scale of $\delta \! f$ is $u \sim v_{\mathrm{the}}$. This implies that the validity of quasilinear theory is determined solely by the size of $\delta \! f_{r,u}/f_0$ at the outer scale. A reasonable proxy for that is \footnote{Implicit in using $\delta C_2$ as a proxy for the amplitude of $\delta \! f$ is that the logarithmic dependence of \eqref{delta_C2_amp} on $\nu$ is order-unity. For finite $\nu$, this approximation is reasonable and essentially captures how the amplitude of $\delta \! f$ depends on the flux $\varepsilon$.}
\begin{equation}
    \frac{\delta \! f_{r,u}}{f_0} \sim \left ( \frac{\delta C_2}{C_{2,0}} \right)^{1/2} \propto t^{-1/2},
\end{equation}
where we have combined \eqref{C20_time}, \eqref{delta_C2_amp}, and \eqref{flux_def} to get the $t^{-1/2}$ scaling. Hence,
\begin{equation}
    R_{r,u} \propto t^{-1/2}
\end{equation}
for $(r,u) \sim (\lambda_{\mathrm{D},0}, v_{\mathrm{the}}) $. This estimate also applies to the regime where $f_0$ is the diffusing Maxwellian \eqref{f0_diffusion}, or indeed to any situation where $C_{2,0}$ decays as a power law in time and $\delta C_2$ is in a forced quasi-steady state. This analysis suggests that, if the plasma enters a regime where the time evolution of $f_0$ is quasilinear, it will stay in that regime, with the quasilinear approximation becoming better with time. Therefore, a quasilinear outer scale can be compatible with a nonlinear cascade at small scales. 

\subsection{Decay of electric fluctuations} \label{decay_irreversible}

As we discussed at the end of Section \ref{numeric_PS_eddies}, even though the turbulent state that we have been considering is forced, i.e., energy is injected, the turbulence is decaying with respect to $C_2$, as can be seen in Fig.~\ref{fig:C2_plots}. Physically, the external field can inject energy into the plasma, but it cannot extract entropy from it.

\begin{figure}
	\centering
	\includegraphics[width=\linewidth]{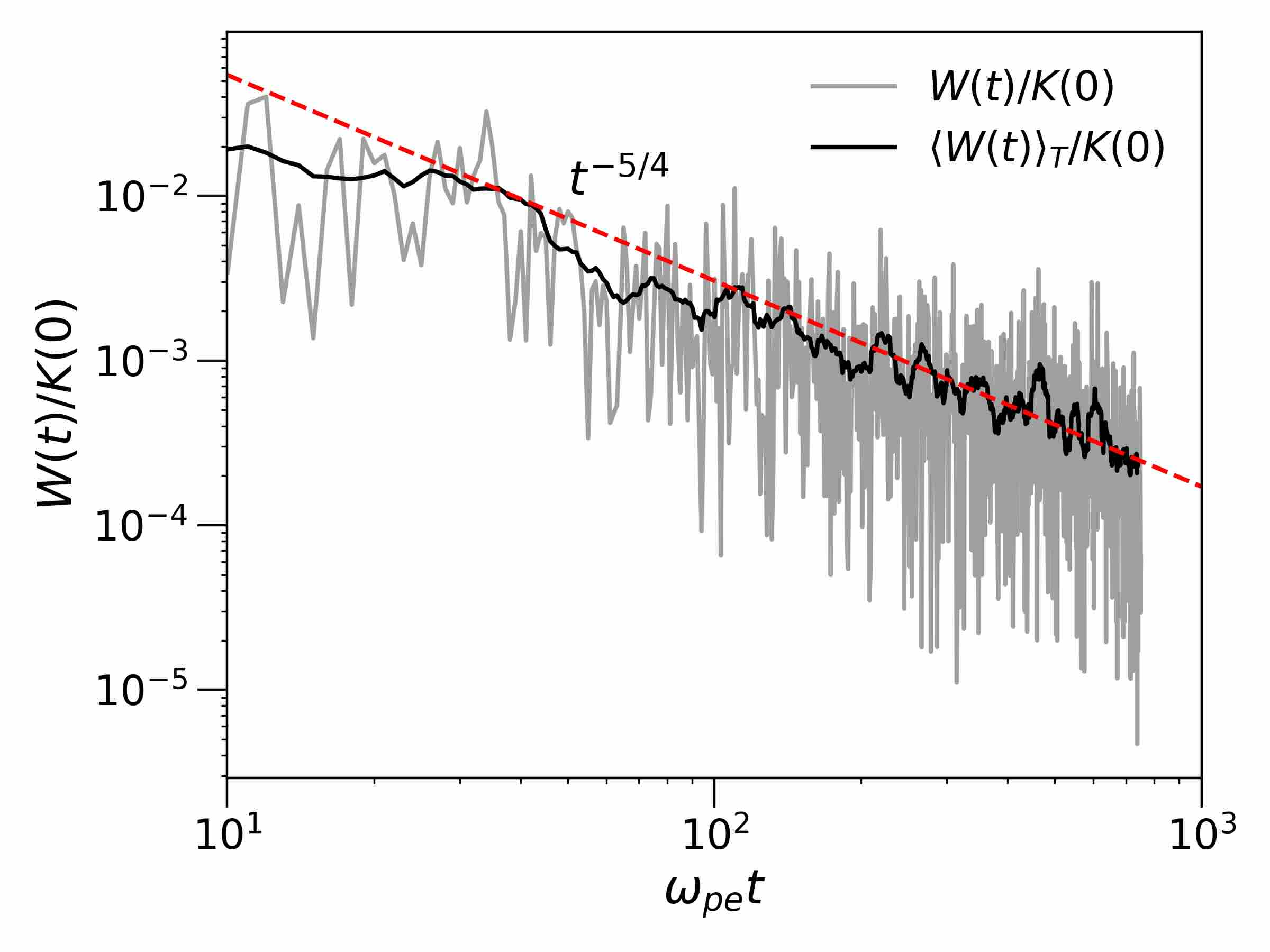}
	\caption{\label{fig:W_decaying} The gray curve is the energy density of the electric field, $W = \langle E^2/8 \pi \rangle_{V}$, versus time. The black curve is a time average of the gray curve: at each time $t$, the energy density is averaged over the time interval $(t-T/2, t+T/2)$, where $T = 20 \omega_{\mathrm{pe}}^{-1}$. The red dashed line is $\propto t^{-5/4}$. }
\end{figure}

The decay of $\delta C_2$ is set by the decay of the flux $\varepsilon = - d C_{2,0}/dt \propto t^{-5/4}$, since $\delta C_2 \propto \varepsilon$, according to \eqref{delta_C2_amp}. The fact that $\varepsilon$ decays also immediately implies that the electric-energy density decays in time. Using \eqref{Ek} and Parseval's theorem, we have
\begin{align} \label{electric_energy_decay}
    W &\equiv \frac{\langle E^2 \rangle_{V}}{8 \pi}  = \frac{1}{8 \pi} \int \mathrm{d} k  \, \frac{m^2}{e^2}  \langle P_{\mathbf{E}} \rangle_{\mathrm{SA}} (k) \nonumber \\ & \sim \frac{ m^2 \varepsilon \, \omega^{4}_{pe} \gamma^{d-1}}{e^2 n^2_0 k^{d+2}_{0} } \propto t^{-5/4},
\end{align}
where we have approximated the integral over $k$ by the contribution from the (assumed constant) outer-scale wavenumber $k_0$ and ignored factors of order unity. In Fig.~\ref{fig:W_decaying}, we plot $W$ versus time: it indeed decays in a way that is consistent with the power law \eqref{electric_energy_decay}.

Energy is continuously injected into the plasma, yet the electric energy decays in time due to the decaying nature of the phase-space turbulence. In the long-time limit, all of the injected energy is necessarily deposited into the kinetic energy of the particles. Phase-space turbulence is therefore the mechanism by which the particle energization becomes irreversible in time, i.e., it becomes `heating' in the thermodynamic sense. This heating occurs on collisionless time scales,~$\nu t \ll 1$.

In \cite{nastac2023phase}, we argued that phase-space turbulence enables the irreversibility of stochastic heating of a passive species accelerated by a random, external electric field. This is because stochastic heating transfers $C_{2,0}$ into $\delta C_2$, which then cascades to small phase-space scales and rapidly gets dissipated by collisions (or, rather, finite-$N$ effects; see Section \ref{noise_turbulence_interplay}). This breaking of $C_2$ conservation renders the heating irreversible, analogously to how, in a fluid passive scalar advected by a turbulent flow, the irreversibility of the turbulent diffusion of the mean scalar profile is enabled by the cascade of scalar fluctuations to small scales and their subsequent dissipation by molecular diffusivity \cite{schekochihin2004diffusion, falkovich2001particles, sreenivasan2019turbulent}. This argument also applies to the present case with external and self-consistent fields. However, in the latter case, phase-space turbulence also explains why the self-consistent electric fields decay in time, which is not \textit{a priori} obvious. On the other hand, in the passive case, since there are no self-consistent fields, the turbulent cascade is unnecessary to work out the system's energy budget: all of the injected energy is necessarily deposited into particles' kinetic~energy.

\section{Discrete particle noise} \label{noise_turbulence_interplay}

In our turbulence calculations in Sections \ref{C2_sect}-\ref{forced_turbulence}, we ignored electric fluctuations arising from discrete particle noise, as encapsulated by the thermal spectrum \eqref{TS}. In this section, we examine the interplay between turbulence and discrete particle noise, and, as a result, infer how the phase-space cascade is ultimately terminated at small scales.

\subsection{Fluctuation spectra of discrete particle noise}

The Vlasov--Poisson system \eqref{vlasov}-\eqref{gauss} captures the effect of thermal (interparticle) electric fluctuations on the single-particle distribution function via the collision operator, but does not contain the thermal fluctuations themselves, which have been coarse-grained away. These fluctuations are, however, encoded in the Klimontovich distribution function \cite{klimontovich2013statistical}:
\begin{equation} \label{Klim}
    F(\mathbf{x}, \mathbf{v}, t) = \sum_{i = 1}^{N} \delta^{2d}\left(\mathbf{Q} -\mathbf{Q}_i(t) \right),
\end{equation}
where $\mathbf{Q} = (\mathbf{x}, \mathbf{v})$, and $\mathbf{Q}_i(t) = (\mathbf{x}_i(t), \mathbf{v}_i(t))$ is the phase-space coordinate of the~$i^{\textrm{th}}$ particle at time $t$. The single-particle distribution function $f(\mathbf{x}, \mathbf{v}, t)$ is obtained by averaging the Klimontovich distribution function, viz.,
\begin{equation} \label{full_F}
    F(\mathbf{x}, \mathbf{v}, t) = f(\mathbf{x}, \mathbf{v}, t) + \delta \! F(\mathbf{x}, \mathbf{v}, t), \quad f = \left \langle F \right \rangle,
\end{equation}
where $\langle ... \rangle$ denotes ensemble averaging over realizations of different initial conditions for particle motion, or coarse-graining over a length scale $\ell$ such that~$\Delta r \sim n_0^{-1/d} \ll \ell \ll \lambda_{\mathrm{De}}$. For the purposes of our discussion, we assume that these two averaging procedures produce the same outcome.

We are interested in sub-Debye-scale electric fluctuations. The $k \lambda_{\mathrm{De}} \gg 1$ limit of \eqref{TS}, in fact, does not require the plasma to be in Maxwellian equilibrium and can be shown to hold for a plasma in any stable, spatially homogeneous equilibrium $f_0(\mathbf{v})$ \cite{rostoker1961fluctuations}. This is because the electric-field spectrum at sub-Debye scales is dominated by ‘bare’ thermal fluctuations arising from uncorrelated particles. Indeed, near equilibrium, particle correlations are significant only at the Debye scale and above, where thermal fluctuations are `dressed' by Debye shielding, and the exact form of the spectrum depends on $f$ \cite{rostoker1961fluctuations,morrison2008fluctuation}.

The sub-Debye electric-field spectrum can be calculated directly from the $\delta \! F$ fluctuations. The two-point correlation function of  $\delta \! F$ is
\begin{align} \label{Klim_corr_step1}
    & \left \langle \delta \! F(\mathbf{x}_1, \mathbf{v}_1, t) \delta \! F(\mathbf{x}_2, \mathbf{v}_2, t) \right \rangle \nonumber \\ & = \left \langle  F(\mathbf{x}_1, \mathbf{v}_1, t) F(\mathbf{x}_2, \mathbf{v}_2, t) \right \rangle -  f(\mathbf{x}_1, \mathbf{v}_1, t) f(\mathbf{x}_2, \mathbf{v}_2, t).
\end{align}
We can compute $\langle  F(\mathbf{x}_1, \mathbf{v}_1, t) F(\mathbf{x}_2, \mathbf{v}_2, t) \rangle $ from \eqref{Klim}:
\begin{align} \label{Klim_corr_step2}
& \langle  F(\mathbf{x}_1, \mathbf{v}_1, t) F(\mathbf{x}_2, \mathbf{v}_2, t) \rangle  \nonumber \\& = \left \langle \sum_{i = 1}^{N} \delta^{2d}\left(\mathbf{Q}_1 -\mathbf{Q}_i(t) \right) \delta^{2d}\left(\mathbf{Q}_1 -\mathbf{Q}_2\right) \right \rangle  \nonumber \\ & + \left \langle  \sum_{i = 1}^{N} \sum_{j \neq i}^{N} \delta^{2d}\left(\mathbf{Q}_1 -\mathbf{Q}_i(t) \right) \delta^{2d}\left(\mathbf{Q}_2 -\mathbf{Q}_j(t) \right) \right \rangle \nonumber \\ & = f \left(\mathbf{Q}_1, t \right) \delta^{2d}\left(\mathbf{Q}_1 -\mathbf{Q}_2\right) + \frac{N-1}{N}  f_2 \left(\mathbf{Q}_1, \mathbf{Q}_2, t \right),
\end{align}
where we have isolated the $i = j$ component of the double sum in the second line, and $f_2$ is the two-particle distribution function \cite{klimontovich2013statistical}. For uncorrelated particles,~$f_2 \left(\mathbf{Q}_1, \mathbf{Q}_2 \right) = f(\mathbf{Q}_1) f(\mathbf{Q}_2)$. Inserting \eqref{Klim_corr_step2} into \eqref{Klim_corr_step1}, we obtain, to lowest order in~$1/N$ \cite{klimontovich2013statistical},
\begin{align} \label{Klim_corr}
	  \langle \delta \! F(\mathbf{x}_1, & \mathbf{v}_1, t) \delta \! F(\mathbf{x}_2, \mathbf{v}_2, t) \rangle \nonumber \\ & \simeq f (\mathbf{x}_1, \mathbf{v}_1, t) \delta^{d}\left(\mathbf{x}_1 -\mathbf{x}_2 \right) \delta^{d}\left(\mathbf{v}_1 -\mathbf{v}_2\right).
\end{align}
The phase-space spectrum of these fluctuations, defined similarly to \eqref{spec_fourier}, is 
\begin{equation} \label{spec_ps_noise}
    P_{\delta \! F}(\mathbf{k}, \mathbf{s}) = \frac{n_0}{2 (2 \pi)^d V}.
\end{equation}
The electric-field spectrum corresponding to \eqref{spec_ps_noise} is then [cf. \eqref{s=0_C2}]
\begin{equation} \label{UFS_noise}
    P_{\mathbf{E}}(\mathbf{k}) =  \frac{\omega_{\mathrm{pe}}^4}{N} k^{-2},
\end{equation}
which agrees with the spectrum \eqref{TS} corresponding to a Maxwellian equilibrium, in the limit $k \lambda_{\mathrm{De}} \gg 1$. The shell-averaged versions of \eqref{spec_ps_noise} and \eqref{UFS_noise} are
\begin{gather}
    \langle  P_{\delta \! F} \rangle_{\mathrm{SA}}(k,s) = \frac{S^2_d \, n_0}{2 (2 \pi)^{2d}} k^{d-1} s^{d-1}, \label{spec_ps_noise_SA} \\
    \langle P_{\mathbf{E}} \rangle_{\mathrm{SA}}(k) = \frac{S_d}{(2 \pi)^d}  \frac{\omega_{\mathrm{pe}}^4}{n_0} k^{d-3}. \label{UFS_noise_SA}
\end{gather}
These are the universal sub-Debye fluctuation spectra of discrete particle noise.

\subsection{Transition from turbulence to noise}

The noise spectrum \eqref{UFS_noise_SA} is much shallower than the turbulence spectrum \eqref{Ek}. The wavenumber $k_{\mathrm{the}}$ at which they meet is given by balancing the two expressions:
\begin{equation} \label{kth_def}
    k_{\mathrm{the}} \sim \left( \frac{\varepsilon \gamma^{d-1}}{n_0} \right)^{1/2d} \sim \frac{n_0^{1/2d} \gamma^{1/2}}{v^{1/2}_{\mathrm{the}}},
\end{equation}
where we have used \eqref{epsilon_twiddle_general} to write $\varepsilon$ in terms of $n_0$, $v_{\mathrm{the}}$, and $\gamma$. With respect to the Debye length, this transition happens when
\begin{equation} \label{kth}
    k_{\mathrm{the}} \lambda_{\mathrm{De}} \sim \left(n_0 \lambda_{\mathrm{De}}^d \right)^{1/2d} \left( \frac{\gamma}{\omega_{\mathrm{pe}}} \right)^{1/2} = N_{\mathrm{D}}^{1/2d} \left( \frac{\gamma}{\omega_{\mathrm{pe}}} \right)^{1/2},
\end{equation}
where $N_{\mathrm{D}} = n_0 \lambda_{\mathrm{De}}^d$ is the plasma parameter. 

In order for the turbulence spectrum \eqref{Ek} to show itself above the noise floor \eqref{UFS_noise_SA}, the transition wavenumber~\eqref{kth} must be much larger than the outer-scale one $k_0$. If fluctuations at the outer scale are critically balanced (see Section \ref{outer_scale}), then $k_0 \lambda_{\mathrm{De}} \sim \gamma / \omega_{\mathrm{pe}}$. This implies
\begin{equation} \label{kth_k0}
    \frac{k_{\mathrm{the}}}{k_0} \sim N_{\mathrm{D}}^{1/2d} \left( \frac{\omega_{\mathrm{pe}}}{\gamma} \right)^{1/2},
\end{equation}
which safely satisfies $k_{\mathrm{the}}/k_0 \gg 1$ if $\gamma \sim \omega_{\mathrm{pe}}$ and $N_{\mathrm{D}} \gg 1$.

Note that the length scale of the transition \eqref{kth_def} is guaranteed to be larger than the interparticle spacing $\Delta r \sim n_0^{-1/d}$:
\begin{equation} \label{kth_delta_r}
    k_{\mathrm{the}} \Delta r \sim N_{\mathrm{D}}^{-1/2d} \left( \frac{\gamma}{\omega_{\mathrm{pe}}} \right)^{1/2} \ll 1,
\end{equation}
again provided that $\gamma \sim \omega_{\mathrm{pe}}$ and $N_{\mathrm{D}} \gg 1$. This condition is equivalent to $n_0 r_{\mathrm{the}}^{d} \gg 1$, where $r_{\mathrm{the}} \sim k_{\mathrm{the}}^{-1}$, viz., the number of particles contained inside `cells' of length $r_{\mathrm{the}}$ is large. This implies that a kinetic-theoretic description of the plasma is still valid at least down to the thermal cutoff scale $r_{\mathrm{the}}$.

\subsection{Invalidity of the collision operator} \label{invalid_col_op}

In Sections \ref{C2_sect}-\ref{forced_turbulence}, we assumed that fluctuations at the small-scale end of the cascade were dissipated by collisional (velocity-space) diffusion, described by a Fokker--Planck operator (e.g., the Landau operator \cite{landau1936transport}). As discussed in Section \ref{C2_sect}, the validity of modelling finite-$N$ effects on $f$ by such a collision operator hinges on there being a scale separation between collisionless (turbulence) and collisional (noise) dynamics, both temporally and spatially.

Temporally, collective turbulent fluctuations have frequencies of order $\omega_{\mathrm{pe}}$ or larger (see Section \ref{outer_scale}). This is the same time scale over which Debye-shielding clouds form (whose interactions make up the `kicks' that lead to collisional velocity-space diffusion), so there is no scale separation in time between the turbulence and the noise. Thus, formally, the encapsulation of the effect of fluctuations on the single-particle distribution function via the Landau (or Lenard-Balescu) collision operator is~invalid~\cite{landau1936transport,Bogolyubov1946, lenard1960bogoliubov, balescu1960irreversible}.

Even if we ignore the lack of time-scale separation and assume that there is such a thing as a Coulomb-collision operator in a turbulent plasma, clean spatial scale separation between the turbulence and noise would require that the collisional cutoff \eqref{r_nu} is reached before the transition scale \eqref{kth}. The ratio of the two wavenumbers is
\begin{equation}
    \frac{k_{\mathrm{the}}}{k_{\nu}} \sim N_{\mathrm{D}}^{1/2d} \left( \frac{\nu}{\omega_{\mathrm{pe}}} \right)^{1/2} \frac{\omega_{\mathrm{pe}}}{\gamma}.
\end{equation}
For $d = 2$ and $d = 3$, $\nu/\omega_{\mathrm{pe}} \sim N^{-1}_{D}$ \cite{klimontovich2013statistical}. Then, for~$\gamma \sim \omega_{\mathrm{pe}}$,
\begin{equation} \label{kth_knu_3D}
     \frac{k_{\mathrm{the}}}{k_{\nu}} \sim N_{\mathrm{D}}^{(1-d)/2d} \ll 1.
\end{equation}
For $d = 1$, one must go to higher order to derive the collision operator, and $\nu/\omega_{\mathrm{pe}} \sim N^{-2}_{D}$ (see, e.g., \cite{fouvry2022kinetic}). This yields
\begin{equation} \label{kth_knu_1D}
     \frac{k_{\mathrm{the}}}{k_{\nu}} \sim N_{\mathrm{D}}^{-3/2} \ll 1.
\end{equation}

Note that, by critical balance \eqref{CB}, the thermal cutoff in velocity space is
\begin{equation} \label{sth}
    s_{\mathrm{the}} \sim \frac{k_{\mathrm{the}}}{\gamma} \sim \left( \frac{\varepsilon }{n_0 \gamma^{d+1}} \right)^{1/2d},
\end{equation}
which satisfies [cf. \eqref{kth}]
\begin{equation}
    s_{\mathrm{the}} v_{\mathrm{the}} \sim N_{\mathrm{D}}^{1/2d} \left( \frac{\gamma}{\omega_{\mathrm{pe}}} \right)^{3/2} \gg 1,
\end{equation}
provided $\gamma \sim \omega_{\mathrm{pe}}$ and $N_{\mathrm{D}} \gg 1$. Compared to the collisional cutoff $s_{\nu} \sim k_{\nu}/\gamma$ [see \eqref{u_nu} and \eqref{r_nu2}], $s_{\mathrm{the}}/s_{\nu}$ satisfies the same scalings as \eqref{kth_knu_3D} and~\eqref{kth_knu_1D} in $d = 2, 3$ and $d = 1$, respectively.

Thus, the turbulence is dissipated not necessarily by a collision operator in Landau's form \cite{landau1936transport}, but rather by being swamped by the particle noise. The phase-space spectrum of the full $F$ \eqref{full_F} is, therefore,
\begin{align} \label{F_ks_turb_noise}
&\langle P_{F} \rangle_{\mathrm{SA}}(k, s) \sim  \nonumber \\ &
\begin{cases} 
     \varepsilon k^{d-1} \left(\gamma s \right)^{-d-1},  \, & k \ll \gamma s, \, s \ll s_{\mathrm{the}},  \\
      \varepsilon \left(\gamma s \right)^{d-1} k^{-d-1}, \, & k \gg \gamma s, \, k \ll k_{\mathrm{the}}, \\
      n_0 k^{d-1} s^{d-1}, \,  &k \gg k_{\mathrm{the}}, \, \, \text{and/or} \, \, s \gg s_{\mathrm{the}}. 
   \end{cases}
\end{align}
Likewise, the full fluctuation spectrum of the total (turbulent and noise) electric field is
\begin{equation} \label{Ek_turb_noise}
\langle P_{\mathrm{\mathbf{E}}} \rangle_{\mathrm{SA}}(k) \sim \frac{\omega_{\mathrm{pe}}^4}{n_0} 
\begin{cases} 
      \displaystyle \frac{\varepsilon \, \gamma^{d-1}}{ n_0} \, k^{-d-3} , & k \ll k_{\mathrm{the}}, \\ \\
       k^{d-3} , & k \gg k_{\mathrm{the}}.
   \end{cases}
\end{equation}
In our continuum simulations presented in Section \ref{forced_turbulence}, there are no thermal fluctuations, so the transition from the turbulence to noise spectra in \eqref{F_ks_turb_noise} and \eqref{Ek_turb_noise} is not observed. Rather, the tubulent fluctuations are dissipated by a Fokker--Planck collision operator. However, the turbulence-to-noise transition has recently been observed in particle-in-cell simulations of the 1D-1V (electron-only) two-stream instability~\cite{ewart2024relaxation}.

\subsection{Renormalized collision rate} \label{dissp_cg}

It is an open question what is the `collision operator' that replaces Landau's \cite{landau1936transport} in the presence of the sub-Debye cascade, i.e., what is the effective field theory of a turbulent plasma. We discuss on this further in Section \ref{turb_coll_op}, but here we estimate how the efficiency of phase-space mixing (cf. Section \ref{ph_mixing_eff}) is modified by the presence of noise and offer a simple prescription to capture the dissipation scales \eqref{kth_def} and \eqref{sth} with a velocity-space diffusion operator.

In the presence of dissipation via the turbulence-to-noise transition, the total $\delta C_2$ stored in phase space (of fluctuations above the noise floor) is found by replacing $k_{\nu}$ \eqref{r_nu} by $k_{\mathrm{the}}$ \eqref{kth_def} in \eqref{delta_C2_amp}. Using \eqref{kth_k0}, this gives
\begin{equation} \label{delta_C2_amp_noise}
    \delta C_2 \sim \varepsilon \gamma^{-1} \log \frac{k_{\mathrm{the}}}{k_{0}} \sim \varepsilon \gamma^{-1} \log \left[ N_{\mathrm{D}}^{1/2d} \left( \frac{\omega_{\mathrm{pe}}}{\gamma} \right)^{1/2} \right].
\end{equation}
Therefore, the characteristic time over which turbulent fluctuations are processed by the cascade and dissipated~is
\begin{equation} \label{tau_dissp_noise}
     \tau_{\mathrm{d}} \sim \frac{\delta C_2}{\varepsilon} \sim \gamma^{-1} \log \mathrm{Do}_{\mathrm{the}},
\end{equation}
where we have defined the `thermal Dorland number'
\begin{equation} \label{nu_the}
    \mathrm{Do}_{\mathrm{the}} \equiv \frac{\gamma}{\nu_{\mathrm{the}}}, \quad \nu_{\mathrm{the}} \equiv \frac{\gamma^{2}}{\omega_{\mathrm{pe}} N_{\mathrm{D}}^{1/d}}.
\end{equation}
Roughly speaking, $\nu_{\mathrm{the}}$ is the `renormalized' collision rate that, if used as the diffusion coefficient in a velocity-space diffusion operator, would provide a cutoff of the turbulence spectrum at $k_{\mathrm{the}}$ \eqref{kth_def} and $s_{\mathrm{the}}$ \eqref{sth} rather than at $k_{\nu}$ \eqref{r_nu} and $s_{\nu}$ \eqref{u_nu}, respectively. Because $k_{\mathrm{the}} \ll k_{\nu}$ and $s_{\mathrm{the}} \ll s_{\nu}$, this rate is greater than that of Coulomb collisions. Indeed, taking $\gamma \sim \omega_{\mathrm{pe}}$, we have that
\begin{equation} \label{nu_the_comp}
\nu_{\mathrm{the}} \sim \frac{\omega_{\mathrm{pe}}}{N_{\mathrm{D}}^{1/d}} \gg \nu \sim  
\begin{cases} 
     \omega_{\mathrm{pe}}/N_{\mathrm{D}}, \qquad  d = 2, 3,  \\
      \omega_{\mathrm{pe}}/N_{\mathrm{D}}^2, \qquad  d = 1.
   \end{cases}
\end{equation}
Although continuum simulations without thermal fluctuations cannot accurately capture the turbulence-to-noise transitions in \eqref{F_ks_turb_noise} and \eqref{Ek_turb_noise}, the turbulent fluctuation spectra \eqref{F_ks_solved} and \eqref{Ek} can be correctly cut off at \eqref{kth_def} and~\eqref{sth} by adopting $\nu_{\mathrm{the}}$ as the collision rate.

For both \eqref{tau_dissp} and \eqref{tau_dissp_noise}, assuming $\gamma \sim \omega_{\mathrm{pe}}$, the characteristic time scale of turbulent dissipation satisfies~$\tau_{\mathrm{d}} \sim \gamma^{-1} \log N_{\mathrm{D}}$, up to order-unity factors. While the nature of the dissipation range predicted here is quite different from what it would be in the presence of a Fokker--Planck collision operator and in the absence of thermal fluctuations (as calculated in Sections \ref{dissipation_cutoffs} and \ref{ph_mixing_eff}), the conclusion that the distribution function will be rapidly mixed to small scales by the inertial-range dynamics of the phase-space cascade remains unchanged.

\section{Discussion} \label{discussion}

\subsection{Summary}

The key result of this work is a theory of the universal fluctuation spectrum of the electric field \eqref{Ek} at Debye and sub-Debye scales in Vlasov--Poisson turbulence. The spectrum is $\propto k^{-d-3}$, where $d$ is the dimensionality. This fluctuation spectrum arises from a state of strong phase-space turbulence, in which $C_2$, the quadratic Casimir invariant of the (electron) distribution function, is the cascaded invariant. We have also derived scalings for the fluctuation spectrum of $\delta C_2$ in phase space using the critical-balance conjecture \eqref{CB}, which gives the relationship between position- and velocity-space scales. Because the electric-field spectrum is so steep, the phase-space mixing of the distribution function at small scales is controlled by the electric fields at the outer (Debye) scale, similarly to Batchelor turbulence \cite{batchelor1959small} of a passive scalar in fluid dynamics.

We have confirmed our turbulence theory numerically via continuum simulations of forced 1D-1V turbulence. We have also found that at long times, the mean distribution function evolves a broadening flat-top shape, which agrees well with our quasilinear prediction \eqref{flat-top}. We have worked out theoretically, and confirmed numerically, the decay law of the electric-energy density. As time goes on, the energy injected by the external forcing is converted into the (internal) kinetic energy of particles, with less and less energy retained in the fluctuating fields. Phase-space turbulence is the mechanism by which this particle heating becomes irreversible. We have shown that this irreversibility is achieved on collisionless time scales, $\nu t \ll 1$ [see \eqref{tau_dissp}].

In our simulations, the turbulence is dissipated by Fokker--Planck collisions. However, in Section \ref{noise_turbulence_interplay}, we argue that standard collisional theory \cite{landau1936transport,lenard1960bogoliubov,balescu1960irreversible,Bogolyubov1946} is, in fact, not valid for turbulent, nearly collisionless plasmas exhibiting the sub-Debye cascade. We theorize that turbulent fluctuations in real plasmas are `dissipated' by the takeover of thermal fluctuations at small scales, viz., the transition from the turbulence spectra to the spectra of discrete particle noise, as given in \eqref{F_ks_turb_noise} and \eqref{Ek_turb_noise}. The spatial wavenumber $k_{\mathrm{the}}$ \eqref{kth_def} where this transition occurs is controlled by the plasma parameter and the ratio of the turbulent mixing rate $\gamma$ to the plasma frequency $\omega_{\mathrm{pe}}$. When $\gamma \sim \omega_{\mathrm{pe}}$, the transition scale is asymptotically below the Debye length and asymptotically above the scale of interparticle spacing. The effect of the turbulence is to renormalize the effective collision rate upward, to \eqref{nu_the_comp}.

The characteristic time scale for turbulent fluctuations to be dissipated is the time scale of turbulent mixing,~$\gamma^{-1}$, multiplied by the logarithm of the Dorland number [see \eqref{tau_dissp} and \eqref{tau_dissp_noise}], which is the plasma-kinetic analogue of the Reynolds number. When $\gamma \sim \omega_{\mathrm{pe}}$, this time scale, up to order-unity factors, reduces to the turbulent mixing time multiplied by the logarithm of the plasma parameter. Phase-space mixing is, therefore, fast, and the dynamics of what are nominally considered `collisionless' plasmas are in fact collisionless for only a short time before dissipation via finite-particle-number effects is activated.

\subsection{The final dissipation range of kinetic plasma turbulence} \label{implications_dissipation}

We predict that kinetic Batchelor turbulence is the `final cascade'—a universal regime to be encountered at the extreme small-scale end of any turbulent cascade in a nearly collisionless plasma. This is because any turbulent fluctuations that reach the Debye scale can be efficiently mixed in phase space via the phase-space entropy cascade [assuming $\mathrm{Do}_{\mathrm{the}} \gg 1$; see \eqref{nu_the}], leading to rapid activation of `collisional' dissipation ($\tau_{\mathrm{d}} \sim \gamma^{-1} \log N_{\mathrm{D}}$) and irreversible conversion of field energy into the internal energy of particles.

Testing the universality of our theory requires validating it in circumstances beyond just the 1D-1V forced case that we simulated. Indeed, recent particle-in-cell simulations of the nonlinear saturation of the two-stream instability \cite{ewart2024relaxation} confirmed the scalings of the phase-space spectrum \eqref{F_ks_solved} and the electric-field spectrum \eqref{Ek} at Debye and sub-Debye scales. These simulations contained thermal noise produced by the finite number of simulated macroparticles \cite{birdsall2018plasma}, and the turbulence-to-noise transition was also observed, verifying the theory developed in Section \ref{noise_turbulence_interplay}. Beyond demonstrating the turbulence-to-noise transition, this work shows that the sub-Debye phase-space cascade can be self-consistently excited by an instability and does not require external forcing, supporting the expectation that the turbulent regime proposed in this paper is a universal feature of kinetic plasma~turbulence.

An important direction for future work is to test our theory in 2D-2V and 3D-3V and to understand how the sub-Debye cascade connects to turbulence above the Debye length. In electrostatic plasmas (or plasmas in which magnetic fields are dynamically subdominant to electric fields), the sub-Debye cascade may be the ultimate `sink' of small-scale fluctuations, e.g., in Langmuir turbulence \cite{zakharov1972collapse,goldman1984strong,robinson1997nonlinear}, i.e., turbulence of coupled Langmuir and ion-acoustic waves. This type of turbulence has been studied extensively in beam-plasma and laser-plasma laboratory experiments, as well as in space and astrophysical plasmas (see, e.g., \cite{goldman1984strong,robinson1997nonlinear} and references therein); notably, it is thought to play a significant role in explaining the coherent emission from solar type III radio bursts \cite{che2017electron,reid2021fine}. In 2D and 3D, strong Langmuir turbulence at $k \lambda_{\mathrm{D}} \ll 1$ is thought to lead to Langmuir collapse \cite{zakharov1972collapse}, in which localized caverns filled with Langmuir fluctuations explosively collapse to scales as small as the Debye length, whereupon the energy of the Langmuir fluctuations is thought to be deposited into electron heat. One might expect that this happens via the cascade studied~here.

We expect the sub-Debye cascade to exist also in magnetized plasmas (assuming the Debye length is much smaller than the electron gyroradius, which is true in many plasmas of interest \cite{thorne2017modern}), e.g., in the solar wind and the Earth's magnetosheath. It is yet to be understood generically how high-frequency (Langmuir) fluctuations are generated at the Debye scale to initiate the phase-space cascade. As discussed above, solar type III radio bursts are known to produce Langmuir fluctuations \cite{che2017electron,reid2021fine}, but the question remains how high-frequency fluctuations may be excited outside of these special circumstances. The prevailing paradigm of magnetized astrophysical plasma turbulence is that the turbulence is dominated by low-frequency (Alfv{\'e}nic and compressive) fluctuations \cite{quataert1999turbulence,de2007chromospheric,chen2016recent,kawazura2020ion, schekochihin2022mhd, schekochihin2009astrophysical}, which, within the gyrokinetic approximation \cite{schekochihin2009astrophysical,kunz2015inertial,kunz2018astrophysical}, stay asymptotically below the cyclotron and plasma frequencies even at length scales below the ion and electron gyroscales. However, in practice, e.g., in the solar wind, the gyrokinetic approximation is thought to break down somewhere in between the ion and electron gyroscales \cite{howes2008model}, at which point high-frequency (ion-cyclotron) fluctuations can be excited. How these turbulent fluctuations may reach the Langmuir frequency at the Debye scale and initiate the phase-space cascade predicted by our theory is not~known.

Another feature to reconcile is the fact that magnetized plasma turbulence is anisotropic \cite{schekochihin2009astrophysical, schekochihin2022mhd}, while, at sub-Debye scales, the turbulence should be isotropic (particle trajectories are not significantly deflected by magnetic fields at these scales). Understanding the transition from a low-frequency, anisotropic turbulent state at super-Debye scales to a high-frequency, isotropic turbulent state at sub-Debye scales is crucial to developing a complete theory of kinetic plasma turbulence in phase~space.

The sub-Debye cascade ultimately terminates at small scales when turbulent fluctuations are overtaken by thermal noise. This interplay between turbulence and noise is reminiscent of that in molecular fluids, where recent results \cite{bandak2022dissipation,mcmullen2022navier,bell2022thermal} suggest that thermal noise can alter the dissipation range of Navier-Stokes turbulence at length scales much larger than the molecular mean free path. In typical incompressible turbulent flows, such as the atmospheric boundary layer, the transition to the thermal noise spectrum occurs at a length scale comparable to the Kolmogorov scale. However, in kinetic plasmas, the transition scale between the turbulence and noise spectra is asymptotically larger than the ‘Kolmogorov scale’ associated with dissipation via Coulomb collisions [see \eqref{kth_knu_3D} and \eqref{kth_knu_1D}]. Moreover, standard collisional theory breaks down in the presence of the sub-Debye cascade (see Section \ref{turb_coll_op}). Kinetic plasma turbulence differs fundamentally from fluid turbulence in this regard.

\subsection{Comparison to the clump theory} \label{clump_comp}

At the time when Dupree proposed his clump theory \cite{dupree1972theory}, it was not known that kinetic plasmas undergo phase-space cascades. It is, therefore, not surprising that no mention of cascades was made in the original clump theory. Yet, arguably, Dupree was on the right track, leading to Vlasov--Poisson Batchelor turbulence. On a qualitative level, his clumps are similar to the phase-space eddies in our theory. Moreover, a key aspect of the clump theory is that clumps have a `lifetime' given by the time that it takes two nearby particles to separate exponentially in phase space, which is a logarithmic function of their initial separation. Dupree posited the existence of a quasi-steady turbulent state in which the amplitude of the two-point correlation function of the perturbed distribution function $\delta \! f$ is set by the clump lifetime. Indeed, as discussed in Appendix \ref{LC_PSE_section}, exponential separation of Lagrangian particle trajectories in phase space is precisely the dynamical mechanism of kinetic Batchelor turbulence, just like in fluid Batchelor turbulence \cite{batchelor1959small,falkovich2001particles}. Furthermore, the amplitude of $ \delta C_2$ in steady state is set by the logarithm of the scale separation between the outer and dissipation scales of the cascade [see \eqref{delta_C2_amp} and~\eqref{delta_C2_amp_noise}]. Roughly speaking, this aligns with Dupree's prediction, if one interprets the smallest separation between two particles as the phase-space dissipation scale, \eqref{kth_def} and \eqref{sth}. We further discuss the relationship between our theory and Dupree's clump theory \cite{dupree1972theory} in Appendix~\ref{LC_PSE_section}.

\subsection{Relation to phase-space holes} \label{generalizations}

Beyond clumps, Dupree and others \cite{dupree1982theory,dupree1983growth, berman1983observation, berman1985simulation,biglari1988cascade,biglari1989clouds} also proposed a somewhat alternative picture of Vlasov turbulence, in which it consisted of an ensemble of Bernstein-Greene-Kruskal (BGK) modes \cite{bernstein1957exact}, or `phase-space (electron) holes' \cite{roberts1967nonlinear, morse1969one, berk1970phase, schamel1986electron,ghizzo1988stability, hutchinson2017electron, hutchinson2024kinetic}, which could interact, e.g., via hole collisions and mergers. In our terms, phase-space holes are long-lived, rotating phase-space eddies. Our simulations reported in Section \ref{forced_turbulence} did not have phase-space holes. Because they were in a regime with $\tau_{\mathrm{corr}} \sim \tau_{\mathrm{nl}}$, particles were trapped as often as they were detrapped, so phase-space eddies had little opportunity to stay coherent (become a phase-space hole) before being sheared apart. A turbulent state with many phase-space holes did, however, emerge in the nonlinear saturation of the 1D-1V two-stream instability \cite{roberts1967nonlinear,ewart2024relaxation}. Nevertheless, Ewart et al.~\cite{ewart2024relaxation} found that the small-scale structure at sub-Debye scales of such a turbulence was well described by our theory of the phase-space cascade. We believe the reason for this is as follows. First of all, the long-lived existence of phase-space holes themselves is not a manifestation of a phase-space cascade. We argued in Section \ref{numeric_PS_eddies} and Appendix \ref{LC_PSE_section} that the physical mechanism of the cascade is the Lagrangian stretching of phase-space eddies caused by nearby particles exponentially separating. Trapped particles do not separate, so, while they cause eddies to rotate, they do not refine the small-scale structure of the eddies. Instead, having formed, the phase-space holes can stir the ambient plasma as they move around, effectively acting as a (Debye-scale) forcing that can excite the sub-Debye cascade. Furthermore, when phase-space holes form, as well as when they collide and merge, it has been observed in numerical simulations that the distribution function undergoes filamentation and subsequent collisional dissipation \cite{galeotti2005asymptotic,califano2006vlasov,carril2023formation}. Filamentation is essentially a phase-space stretching process, so we expect that such filamentation at sub-Debye scales corresponds to a kinetic cascade of $C_2$ as described in our paper.

What our theory does not describe, however, is the phase-space structure above the Debye scale in a turbulent system with many phase-space holes. How to quantify the structure of this type of phase-space turbulence in a meaningful way is an open problem. An attempt to solve this problem was a relatively unnoticed (including by us, until well after our theory was completed) work by Biglari and Diamond \cite{biglari1988cascade,biglari1989clouds}. They proposed that holes serve as the fundamental turbulent structures governing the fluctuation spectra, with their dynamics at each scale dictated by a localized form of the `hole instability' \cite{dupree1982theory}. Based on this principle, they developed a cascade theory to describe the formation of a hierarchy of holes at different scales. In their theory, the phase-space cascade was a process transferring $C_2$ from larger-scale holes to smaller-scale ones, down to a collisional cutoff. This approach led to a set of scalings that are reproduced, in our language, in Appendix \ref{BD_app}, where we also show that the same approach would be inconsistent if applied below the Debye scale. Above the Debye scale, it is not \textit{a priori} ruled out, but, as far as we are aware, the resulting scalings have not been observed in a numerical simulation. This may be because their validity depends on a number of physical assumptions, which, while not obviously invalid, are at least questionable—we discuss them in Appendix \ref{BD_app}. We also present there our own take on turbulence above the Debye scale, which is a hybrid of the Biglari and Diamond's scheme for taking account of Debye shielding and of our own prescription for a phase-space cascade. An intriguing case of somewhat similar scalings (even if somewhat different physics) turns out to be a new scaling regime of ion phase-space turbulence, also above the Debye scale—this is worked out in Appendix \ref{IS_sect}. Regardless of whether the Biglari--Diamond cascade is realizable in a real Vlasov--Poisson system, their work must be acknowledged as a conceptual advance beyond Dupree's original clump theory and was prescient, long predating the later studies of phase-space cascades \cite{schekochihin2008gyrokinetic,schekochihin2009astrophysical} in the context of gyrokinetic turbulence. 

\subsection{Collision operators and mean distributions in the presence of turbulence} \label{turb_coll_op}

A significant implication of this work is that in plasmas that are turbulent at Debye and sub-Debye scales, standard collisional theory \cite{landau1936transport,balescu1960irreversible,lenard1960bogoliubov,Bogolyubov1946} is not valid. Fundamentally, this is because the encapsulation of finite-$N$ effects into the Landau \cite{landau1936transport} (or Lenard-Balescu \cite{lenard1960bogoliubov,balescu1960irreversible}) collision operator is based on the assumption that the fluctuation spectrum of the electric field at Debye and sub-Debye scales is solely that of discrete particle noise \eqref{UFS_noise_SA} (this is the plasma analogue of Boltzmann's Stosszahlansatz~\cite{Boltzmann}). When there is a phase-space cascade, the turbulent fluctuation spectrum \eqref{Ek_turb_noise} supersedes the noise spectrum at wavenumbers below the transition wavenumber~\eqref{kth_def}. We argued in Section \ref{dissp_cg} that as a result, the effective collision rate in such a plasma would be renormalized upwards, to \eqref{nu_the_comp}. The natural question is then, what is the collision operator in the presence of these turbulent fluctuations? This question is not new, and indeed was a key preoccupation of many early theories of Vlasov--Poisson turbulence (see, e.g., \cite{dupree1972theory,misguich1978kinetic,misguich1978clumps, dubois1978direct}). Developing such a theory requires understanding the structure of turbulence, and, as discussed in Section \ref{intro}, no definitive theory of fluctuations emerged from the early work on this subject, so progress stalled. In light of the fluctuation spectra \eqref{F_ks_turb_noise} and \eqref{Ek_turb_noise}, we believe that this topic can be fruitfully revisited. We leave the task of developing such a theory to future work, but here, we offer some speculations on the problem.

The derivation of collision operators is a coarse-graining procedure \cite{klimontovich2013statistical}, and, analogously to the theory of turbulent viscosity in hydrodynamics \cite{davidson2015turbulence}, it is an arbitrary choice over which scales to coarse-grain, with eliminated modes at any cutoff scale renormalizing the `effective collisionality' felt by fluctuations above that scale. That being said, we think that there are naturally two special scales over which coarse-graining should provide good effective descriptions of the plasma, viz., the thermal scale \eqref{kth} and the Debye scale.  

To coarse-grain over thermal fluctuations at $k > k_{\mathrm{the}}$ (and $s > s_{\mathrm{the}}$) would yield a kinetic description of the plasma down to essentially the smallest possible scales where such a description can still be valid [cf. \eqref{kth_delta_r}]. One would have to apply this coarse-graining procedure systematically to the Klimontovich equation \eqref{Klim} and hence derive an effective equation for the single-particle distribution function.

To coarse-grain at the Debye scale is to derive a `turbulent collision operator,' which, in effect, would replace the collision operators of Landau \cite{landau1936transport} or Lenard and Balescu \cite{lenard1960bogoliubov,balescu1960irreversible}. This collision operator would describe how turbulent fluctuations at Debye and sub-Debye scales provide an effective collisionality of the plasma at super-Debye scales. Intuitively, the `turbulent collision rate,' $\nu_{\mathrm{turb}}$, of such an effective collision operator should essentially be equal to the rate of turbulent mixing, $\gamma$. This can be explicitly estimated as follows. Using \eqref{gamma_def_Batchelor}, \eqref{CB}, and assuming that the turbulent collisionality is dominated by the outer-scale fields and that these fields satisfy the critical-balance relation \eqref{CB_aspect_ratio}, viz., $r_0 \sim \gamma^{-1} v_{\mathrm{the}}$, we have
\begin{equation}
    \nu_{\mathrm{turb}} \sim v_{\mathrm{the}}^{-2} \frac{e^2}{m^2} \delta E^2_{r_0} \tau_{\mathrm{c}} \sim v_{\mathrm{the}}^{-2} \gamma^4 r_0^2 \gamma^{-1} \sim \gamma.
\end{equation}
We argued in Section \ref{outer_scale} that $\gamma \sim \omega_{\mathrm{pe}}$, so the relaxation of the coarse-grained distribution function due to turbulence will be extremely fast compared to (Coulomb) collisional relaxation ($\nu_{\mathrm{turb}} \gg \nu$), similarly to the way in which turbulence enhances mixing rates in fluids \cite{davidson2015turbulence,sreenivasan2019turbulent}.

Unlike Coulomb collisions, turbulence need not drive the distribution function toward a Maxwellian. In the example of forced 1D-1V turbulence considered in Section \ref{forced_turbulence}, we found that the time evolution of the mean distribution function was well described by the quasilinear equation \eqref{QL_evol}, and, at long times, had the form of the broadening flat-top \eqref{flat-top}. In this case, the operator in~\eqref{QL_evol} is the collision operator. Note that because the self-consistent fields decay in time [see Section \eqref{decay_irreversible}], while the fluctuation amplitude of the external field is fixed, the long-time evolution of $f_0$ is dictated by the external field rather than the self-consistent fields. 

In contrast, the simulations of the two-steam instability by \cite{ewart2024relaxation} involved only self-consistent fields, and the mean (volume-averaged) distribution function in their simulations evolved toward a non-thermal distribution with a power-law tail in energy with exponent $-2$. This distribution function is a maximum-entropy state of Lynden-Bell statistical mechanics \cite{lynden1967statistical} derived by \cite{ewart2023nonthermal}, who showed that a broad class of Lynden-Bell equilibria would have this form. While the Lynden-Bell theory as originally devised \cite{lynden1967statistical, ewart2023nonthermal} assumed perfect conservation of phase-space volume [conservation of all Casimir invariants \eqref{casimir_invariants}], \cite{ewart2024relaxation} proposed that universal Lynden-Bell equilibria could, in fact, be reached not despite the presence of turbulence, which breaks phase-space volume conservation, but rather thanks to it. In essence, they showed that, as fluctuations were continuously processed and dissipated by the turbulent cascade, the Casimirs of the distribution function \eqref{casimir_invariants} continuously changed, and, consequently, so did the Lynden-Bell equilibrium to which the system strove to relax, in such a way that this equilibrium fell into the universality class with the aforementioned power-law tail. Although quasilinear collision operators that have Lynden-Bell equilibria as their fixed points have been constructed \cite{kadomtsev1970collisionless,chavanis2021kinetic,ewart2022}, a theory of turbulent collisions that would capture the evolution of the Casimirs observed in \cite{ewart2024relaxation} remains to be developed.

While both the forced and the two-stream systems exhibit sub-Debye cascades as predicted by our theory in Section \ref{phenom_theory}, the structure of the mean distribution function is vastly different between the two cases. We think that this is because the distortions of $f_0$ by phase-space turbulence is dominated by fluctuations at the outer scale of the turbulence (roughly the Debye scale or above). In our simulations, these are phase-space eddies rapidly mixed by the external electric field, while, in the simulations of \cite{ewart2024relaxation}, the dynamics at the outer scale is controlled by the interactions and merging of many phase-space holes \footnote{An alternative explanation for the power-law tail in energy with exponent $-2$ could be that the phase-space holes effectively act as a large-scale forcing at $k_{\mathrm{f}} \lambda_{\mathrm{De}} \gtrsim 1$. According to the quasilinear theory of \cite{banik2024universal}, who considered the $k_{\mathrm{f}} \lambda_{\mathrm{De}} \ll 1$ limit of the dressed diffusion equation \eqref{QL_evol_general}, such forcing dynamically generates the power-law tail, akin to how the flat top \eqref{flat-top} in Section~\ref{thermo_turb} emerges in the $k_{\mathrm{f}} \lambda_{\mathrm{De}} \gg 1$ regime. If the mean distribution function in the simulations of \cite{ewart2024relaxation} is indeed governed by quasilinear theory, it must also be reconciled with how this quasilinear behavior is consistent with the presence of phase-space turbulence (cf. the discussion at the end of Section \ref{Stochastic_heating_numerics}).}. It seems that the presence of these long-lived, coherent structures may support the system's ability to relax toward the Lynden-Bell equilibria with power-law tails \cite{ewart2023nonthermal}. This dichotomy suggests that a more detailed understanding of phase-space turbulence at the outer scale of the cascade is needed to make progress toward constructing a theory of turbulent collisions.

\subsection{Implications for general plasma physics and beyond}

Beyond the Vlasov--Poisson system, our results suggest that a complete understanding of the thermodynamics (what is $f_0$ and how are particles irreversibly energized?) and turbulence (what is the small-scale structure of $\delta \! f$ and the electromagnetic fields?) of nearly collisionless plasmas requires understanding not just energy but also entropy (in our case, quadratic Casimir) budgets. This is known to be true in the gyrokinetic approximation, where the energy of the fields is intertwined with the entropy of $\delta \! f$ to form the free energy invariant, which is the cascaded invariant of gyrokinetic turbulence \cite{schekochihin2008gyrokinetic,schekochihin2009astrophysical,abel2013multiscale,kunz2015inertial,kunz2018astrophysical}.

Outside the gyrokinetic regime, typical paradigms of nearly collisionless plasma turbulence remain largely fluid-dynamical \cite{marino2023scaling}. Yet, in our system, the scaling of the electric-field spectrum came not from an energy cascade but instead was a byproduct of the $C_2$ cascade. Thus, $C_2$ appears to be the relevant cascaded invariant around which one should construct new theories of kinetic (Vlasov) turbulence.

Beyond plasmas, the Vlasov--Poisson system also describes self-gravitating systems, where the constituent particles are, e.g., stars or dark matter particles, and the field is gravitational rather than electric \cite{binney2011galactic,rampf2021cosmological, hamilton2024kinetic}. It is then natural to expect that phase-space turbulence also exists in gravitational-kinetic systems. Indeed, a theory of gravitational turbulence in cold dark matter was recently developed by \cite{ginat2024cosmological}, highlighting both the intriguing similarities and important differences with our theory of phase-space turbulence in plasmas.

\section*{Acknowledgements}

This paper is dedicated to the memory of~Bill Dorland (1965-2024).

We are thankful to U. Banik, F. Califano, G. Celebre, B. Chandran, W. Clarke, P. Dellar, B. Dorland, H. Fetsch, T. Fülöp, Y. B. Ginat, C. Hamilton, G. Hammett, D. Hosking, M. Kunz, T. Passot, L. Richard, W. Sengupta, J. Squire, D. Uzdensky, D. Verscharen, and L. Wilson for helpful discussions related to this work. M.L.N was supported by a Clarendon Scholarship. R.J.E was supported by a UK EPSRC studentship. J. Juno was supported by the U.S. Department of Energy under Contract No. DE-AC02-09CH1146 via an LDRD grant. The work of A.A.S was supported in part by grants from STFC (ST/W000903/1) and EPSRC (EP/R034737/1), as well as by the Simons Foundations via a Simons Investigator award. This research was also supported in part by the NSF grant PHY-2309135 to the Kavli Institute for Theoretical Physics (KITP) and benefited from interactions with the participants of the program “Interconnections between the Physics of Plasmas and Self-gravitating Systems.” The simulations performed for this paper were done on the Stellar cluster at Princeton University and the Frontera cluster at the Texas Advanced Computing Center at the University of Texas at Austin. Frontera access was made possible by the NSF award OAC-1818253.

\appendix

\section{Incompatibility of spatially rough electric fields with a critically-balanced phase-space cascade} \label{PT_hd}

In this appendix, we show that a critically balanced constant-flux phase-space cascade in the Vlasov--Poisson system is incompatible with self-consistent electric fields that are not spatially smooth but rather only Hölder continuous, i.e., have increments that scale as
\begin{equation} \label{E_r_increment}
    \frac{e}{m}\delta E_r \sim \gamma^2 r^{\beta},
\end{equation}
where $0 < \beta < 1$ is the Hölder exponent. The increments and shell-averaged fluctuation spectrum of such an electric field obey the following relation \cite{frisch1995turbulence}:
\begin{equation} \label{relation_inc_spec}
    \delta E_r \propto r^{\beta} \iff  \langle P_{\mathbf{E}} \rangle_{\mathrm{SA}} (k) \propto k^{-(2\beta+1)}.
\end{equation}
Note that our analysis in this appendix does not \textit{a priori} exclude the possibility of $\beta \leq 0$, for which the relations \eqref{E_r_increment} and \eqref{relation_inc_spec} can still hold provided that $^{``}\delta E_r^"$ is carefully defined (e.g., using low-pass filtering \cite{eyink1995besov}). As we show below, negative values of $\beta$ are not relevant in the Vlasov--Poisson system \eqref{vlasov}-\eqref{gauss}. However, they do appear in the scaling regime of ion turbulence above the Debye scale that is derived in Appendix \ref{IS_sect}.

\subsection{Critical balance}

For an electric field satisfying \eqref{E_r_increment}, the critical-balance condition~\eqref{CB} yields a scale-dependent phase-space eddy-turnover time:
\begin{align} \label{CB_line_rough}
    & \tau_{\mathrm{c}} \sim \tau_{\mathrm{l}} \sim \tau_{\mathrm{nl}} \implies u \sim \gamma \, r^{(1+\beta)/2},  \nonumber \\  & \tau_{\mathrm{c}} \sim  \gamma^{-1}r^{(1-\beta)/2} \sim \gamma^{-2/(1+\beta)} u^{(1-\beta)/(1+\beta)}.
\end{align}
Similarly to our theory in Section \ref{CB_section}, we expect the shell-averaged fluctuation spectrum of $\delta C_2$ to be a product of power laws, with exponents changing across the critical-balance line \eqref{CB_line_rough}:
\begin{equation} \label{F_ks_hd}
\langle P_{\delta \! f} \rangle_{\mathrm{SA}}(k, s) \propto
\begin{cases} 
      k^{d-1} s^{-a} , \qquad & k \ll \left(\gamma s \right)^{2/(1+\beta)}, \\
      s^{d-1} k^{-b}, \qquad & k \gg \left(\gamma s \right)^{2/(1+\beta)},
   \end{cases}
\end{equation}
where the goal is to find the exponents $a$ and $b$, whereas the factors of $k^{d-1}$ and $s^{d-1}$ arise from the wavenumber Jacobian in $d$ dimensions.
The arguments presented in Section \ref{CB_section} that established these as the appropriate large-scale power laws naturally extend to the rough-field case [with the critical-balance condition~\eqref{CB} replaced by~\eqref{CB_line_rough}].

\subsection{Self-consistency}

The exponent $b$ can be related to the parameter $\beta$ by enforcing self-consistency via Poisson's equation. Note that the fluctuation spectrum of $\delta C_2$ shell-averaged over just spatial wavenumbers evaluated at $\mathbf{s} = \mathbf{0}$ scales as $\propto k^{-b}$, as can be seen in \eqref{F_ks_hd}. Then, since the fluctuation spectrum of the electric field is proportional to the $\delta C_2$ spectrum evaluated at $\mathbf{s} = \mathbf{0}$ divided by $k^2$, as per \eqref{s=0_C2}, we have that the shell-averaged electric-field spectrum scales as
\begin{equation} \label{self_consistent_spec}
    \langle P_{\mathbf{E}} \rangle_{\mathrm{SA}} (k) \propto k^{-b-2}.
\end{equation}
Combining \eqref{self_consistent_spec} and \eqref{relation_inc_spec} implies that an electric field self-consistently generated by $\delta \! f$ satisfies
\begin{equation}
b = 2\beta-1.
\end{equation}

\subsection{Constant flux in position space} \label{contradiction}

Now we assume that there is a constant-flux cascade of $\delta C_2$ in position space:
\begin{align} \label{spec_k_const_flux_hd}
    & \frac{v_{\mathrm{the}}^d \delta \! f_{r}^2}{\tau_{\mathrm{c}}} \sim \varepsilon \implies \nonumber \\  & \delta \! f_{r} \propto r^{(1-\beta)/4} \iff \int \mathrm{d} s \, \langle P_{\delta \! f} \rangle_{\mathrm{SA}}(k,s) \propto k^{-(3-\beta)/2}.
\end{align}
To find $\beta$, we demand that integrating the 2D spectrum~\eqref{F_ks_hd} over $s$ yield the same scaling as in \eqref{spec_k_const_flux_hd}. Assuming that the dominant contribution to the integral comes from integrating \eqref{F_ks_hd} from $s = 0$ up to the critical-balance line $s \sim \gamma^{-1}k^{(1+\beta)/2}$, we get that
\begin{align} \label{spec_k_CB_line_hd}
    &\int_{0}^{\gamma^{-1}k^{(1+\beta)/2}} \mathrm{d} s \,  \langle P_{\delta \! f} \rangle_{\mathrm{SA}}(k,s) \nonumber \\  & \propto \int_{0}^{\gamma^{-1}k^{(1+\beta)/2}} \mathrm{d} s \, s^{d-1} k^{-(2\beta - 1)} \propto k^{[d+2-(4-d) \beta]/2}.
\end{align}
Equating \eqref{spec_k_const_flux_hd} and \eqref{spec_k_CB_line_hd} yields $\beta = (d+5)/(5-d)$. This evaluates to $\beta =  3/2, 7/3,$ and  $4$ for $d = 1, 2,$ and $3$, respectively. This is a contradiction, as we assumed $\beta < 1$ in order to have a scale-dependent cascade time \eqref{CB_line_rough}. Indeed, if $\beta > 1$, the cascade time \eqref{CB_line_rough} gets longer at smaller scales. This implies that the shearing of $\delta \! f$ at small scales is dominated not by local electric fields at those scales but rather by the largest-scale fields, i.e., $\beta = 1$ and the turbulence is in the Batchelor regime. Therefore, spatially rough electric fields (with Hölder exponent $\beta < 1$) are incompatible with a critically balanced constant-flux phase-space cascade of $\delta C_2$ in the Vlasov--Poisson system.

\subsection{An equivalent scaling argument} \label{argument_rw}

Here, we give an alternative calculation ruling out non-Batchelor scaling regimes, without the use of wavenumber spectra. Beyond providing conceptual clarity, recasting the argument in this different language will also prove useful in the comparison of our theory to the Biglari--Diamond theory \cite{biglari1988cascade,biglari1989clouds} in Appendix \ref{BD_app}.

From Gauss's law \eqref{gauss}, the electric-field increment across scale $r$ can be estimated as
\begin{equation} \label{E_r_Gauss}
    \frac{e}{m}\delta E_r \sim \omega_{\mathrm{pe}}^2 \frac{\delta n_r}{n_0} r,
\end{equation}
where $\delta n_r$ is the density increment across the same scale. To estimate $\delta n_r$ in terms of ~$\delta \! f_r$, we posit that the velocity integral in \eqref{gauss} accumulates as a product of $d$ (Brownian) random walks in velocity space, where the random walk in each velocity dimension has a step size $\sim u$ and number of steps $\sim v_{\mathrm{the}}/u$, and $u$ is the velocity scale corresponding, via critical balance \eqref{CB_line_rough}, to the spatial scale~$r$. This yields
\begin{equation} \label{delta_n_r}
    \delta n_r \sim v_{\mathrm{the}}^d \left( \frac{u}{v_{\mathrm{the}}} \right)^{d/2} \delta \! f_r.
\end{equation}
This is equivalent to the argument in Section~\ref{CB_section} that the $s$ dependence of the spectrum in the $k \gg \gamma s$ limit is that of a $d$-dimensional white noise in velocity space [see~\eqref{F_ks}].

Assuming that the electric field has the increments \eqref{E_r_increment} and invoking \eqref{CB_line_rough} gives $u \propto r^{(\beta+1)/2}$. Furthermore, the constant-flux relation \eqref{spec_k_const_flux_hd} implies that~$\delta \! f_r \propto r^{(1-\beta)/4}$. Combining these scalings with \eqref{delta_n_r} and inserting the result back into \eqref{E_r_Gauss} gives
\begin{equation} \label{E_r_rough_accumulation}
    \delta E_r \propto r^{1+[1+\beta)d + (1-\beta)]/4}.
\end{equation}
Demanding consistency, we equate the exponent in \eqref{E_r_rough_accumulation} to $\beta$, obtaining~$\beta = (d+5)/(5-d)$, which is precisely the contradictory scaling for $\beta$ that we found in Appendix~\ref{contradiction}.

Note that if we insert $\beta = 1$ into \eqref{delta_n_r} and \eqref{E_r_rough_accumulation}, we~get
\begin{equation} \label{increments_ho_Batchelor}
    \delta n_r \propto r^{d/2}, \quad \delta E_r \propto r^{(d+2)/2}.
\end{equation}
Because the electric-field spectrum \eqref{Ek} is steeper than~$k^{-3}$, two-point increments of the electric field satisfy $\delta E_r \propto r$ in any dimension. However, if one computes increments using more than two points so that the lowest-order terms in the Taylor expansion cancel, it is possible to relate the scalings of spectra steeper than $k^{-3}$ directly to the scalings of multi-point structure functions (see, e.g., \cite{cho2009simulations}). If the increments $\delta E_r$ in \eqref{increments_ho_Batchelor} are interpreted as three-point (or higher) increments in $d = 1$ or four-point (or higher) increments in $d = 2, 3$ \cite{cho2009simulations}, they directly translate into the spectrum \eqref{Ek} via the relationship $\delta E_r \propto r^{\mu} \iff \langle P_{\mathbf{E}}\rangle_{\mathrm{SA}}(k) \propto k^{-2 \mu +1}$.

The (multi-point) increments \eqref{increments_ho_Batchelor} have turned out to be consistent with our conclusion that the turbulence is in the Batchelor regime because the local-in-scale shearing of $\delta \! f$ by the electric fluctuations at small scales described by \eqref{increments_ho_Batchelor} is sub-dominant to the shearing of $\delta \! f$ by the largest-scale fields, captured by the two-point increment $\delta E_r \propto r$. This is clear from the fact that the scale-dependent cascade time \eqref{CB_line_rough} gets longer at smaller scales when $\beta  > 1$.

\section{Phase-space asymmetry} \label{PU_app}

In this appendix, we show that phase-space asymmetry in the distribution function is necessary for a plasma to exhibit a velocity-space cascade. In Fourier space, this implies that the fluctuation spectrum \eqref{spec_fourier} depends not just on the wavenumbers' magnitudes, $k$ and $s$, but also on the angle between them, $\theta = \cos^{-1} ( \mathbf{k} \cdot \mathbf{s}/k s)$. 

To demonstrate this, we need to derive the evolution equation for the spectrum \eqref{spec_fourier}. To do this, we derive the evolution equation for the correlation function \eqref{corr_func} and then Fourier transform it. Let us write an evolution equation for the product $\delta \! f(\mathbf{x_1},\mathbf{v_1}) \delta \! f(\mathbf{x_2},\mathbf{v_2})$, make the coordinate transformations $(\mathbf{x_1},\mathbf{x_2}) \rightarrow (\mathbf{r}, \mathbf{R}) = (\mathbf{x_1}-\mathbf{x_2}, (\mathbf{x_1} + \mathbf{x_2})/2) $ and $(\mathbf{v_1},\mathbf{v_2}) \rightarrow (\mathbf{u}, \mathbf{U}) = (\mathbf{v_1}-\mathbf{v_2}, (\mathbf{v_1} + \mathbf{v_2})/2) $, integrate over the average velocity $\mathbf{U}$, and then ensemble average. This yields
\begin{equation} \label{corr_eq}
    \frac{\partial C_{\delta \! f}}{\partial t} + \mathbf{u} \cdot \nabla_{\mathbf{r}} C_{\delta \! f} + \overline{\mathcal{N}}(\mathbf{r}, \mathbf{u}) = \overline{\mathcal{S}}(\mathbf{r}, \mathbf{u}) + 2 \nu v_{\mathrm{the}}^2 \nabla^2_{\mathbf{u}} C_{\delta \! f},
\end{equation}
where the nonlinear term is
\begin{align}
 \mathcal{N} &=  \frac{1}{2} \int \mathrm{d} \mathbf{U} \,  \left[\mathbf{E}_{\mathrm{tot}}(\mathbf{R} + \mathbf{r}/2) -\mathbf{E}_{\mathrm{tot}}(\mathbf{R} - \mathbf{r}/2) \right] \nonumber \\ & \cdot \frac{\partial}{\partial \mathbf{u}} \left[\delta \! f(\mathbf{R} + \mathbf{r}/2,\mathbf{U} + \mathbf{u}/2) \delta \! f(\mathbf{R} - \mathbf{r}/2,\mathbf{U} - \mathbf{u}/2) \right],
\end{align}
and the source term, which comes from the second term on the right-hand side of \eqref{vlasov_delta_f}, is
\begin{align}
    &\mathcal{S} = \frac{1}{2}\int \mathrm{d} \mathbf{U} \, \nabla_{\mathbf{u}} \cdot \big[f_0(\mathbf{U} + \mathbf{u}/2) \times \nonumber \\ &\mathbf{E}_{\mathrm{tot}}(\mathbf{R} + \mathbf{r}/2) \delta \! f(\mathbf{R} - \mathbf{r}/2,\mathbf{U} - \mathbf{u}/2) -f_0(\mathbf{U} - \mathbf{u}/2) \times \nonumber \\ &\mathbf{E}_{\mathrm{tot}}(\mathbf{R} - \mathbf{r}/2) \delta \! f(\mathbf{R} + \mathbf{r}/2,\mathbf{U} + \mathbf{u}/2) \big].
\end{align}
Note that the nonlinear and source terms are only independent of the average position $\mathbf{R}$ after ensemble averaging. Also, while we are assuming that finite-$N$ effects can be modelled by the collisional velocity-space diffusion operator \eqref{coll_op_diffusion}, everything derived below also holds if the dissipation scale is controlled by the transition to the noise spectrum, as derived in Section \ref{noise_turbulence_interplay}---we would just have to replace the collisional scale $u_{\nu}$ \eqref{u_nu} by the thermal cutoff scale $u_{\mathrm{the}} \sim s_{\mathrm{the}}^{-1}$ \eqref{sth}.

Fourier transforming \eqref{corr_eq} yields \eqref{E_delta_f_eq}, where
\begin{equation} \label{NL_term_ks}
    \overline{\mathcal{N}}_{\mathbf{k},\mathbf{s}} = \frac{1}{(2 \pi)^{d} V} \iint \mathrm{d} \mathbf{r} \mathrm{d} \mathbf{u} \, e^{-i(\mathbf{k} \cdot \mathbf{r}-\mathbf{s} \cdot \mathbf{u})} \overline{\mathcal{N}}(\mathbf{r}, \mathbf{u}),
\end{equation}
\begin{equation} \label{S_term_ks}
    \overline{\mathcal{S}}_{\mathbf{k},\mathbf{s}} = \frac{1}{(2 \pi)^{d} V} \iint \mathrm{d} \mathbf{r} \mathrm{d} \mathbf{u} \, e^{-i(\mathbf{k} \cdot \mathbf{r}-\mathbf{s} \cdot \mathbf{u})} \overline{\mathcal{S}}(\mathbf{r}, \mathbf{u}).
\end{equation}
To interpret \eqref{E_delta_f_eq}, first note that the spectrum is sourced by the term $\overline{\mathcal{S}}_{\mathbf{k},\mathbf{s}}$, which physically represents distortion of $f_0$ by electric fields. \textit{A priori}, this term could have a broad support in wavenumber space. However, if $f_0$ has small velocity-space gradients, e.g., if it is the flat-top distribution \eqref{flat-top}, as found in our simulations, we can reasonably assume the source predominantly injects fluctuations at low $s$. Likewise, since the external field is concentrated at the outer scale, $k_{\mathrm{f}} \lambda_{\mathrm{De}} \sim 1$, and the self-consistent field has a steep power law \eqref{Ek} in wavenumber space, we can also reasonably assume that the injection is predominantly at low $k$.

Fluctuations sourced at low wavenumbers are then coupled to other wavenumbers via linear phase mixing [the second term in \eqref{E_delta_f_eq}] and nonlinear mode coupling [the third term in \eqref{E_delta_f_eq}]. As discussed in Section \ref{inhomog_aniso}, the phase-mixing term can transfer $\delta C_2$ to both small and large velocity-space scales: phase-mixing perturbations with $\mathbf{k} \cdot \mathbf{s} > 0$ are advected from high to low $s$, while phase-unmixing perturbations with $\mathbf{k} \cdot \mathbf{s} < 0$ are advected from low to high $s$ \cite{nastac2023phase, adkins2018solvable, schekochihin2016phase}. Without nonlinearity, a system sourced at low wavenumbers only has phase-mixing modes, which propagate to high $\mathbf{s}$ until they are damped by collisions. However, nonlinearity, which couples Fourier modes with different $\mathbf{k}'s$, can couple phase-mixing modes to create phase-unmixing modes. Indeed, we saw in Section \ref{spectra_test} that the system in the fully developed turbulent state consists of a broadly populated spectrum of both phase-mixing and phase-unmixing modes.

Intuitively, a net velocity-space cascade from injection to dissipation scales requires the flux of $\delta C_2$ carried by phase-mixing modes (from low to high $s$) to outweigh the flux of $\delta C_2$ carried by phase-unmixing modes (from high to low $s$) in the inertial range. To demonstrate this explicitly, we first show how a constant-flux velocity-space cascade manifests in spectral space. Assume a steady state (on the fast time scales of  $\delta \! f$, where $f_0$ and all average quantities can be considered to evolve slowly), integrate \eqref{E_delta_f_eq} over $\mathbf{k}$, and consider wavenumbers $\mathbf{s}$ such that dissipation is negligible, viz., $s u_{\nu} \ll 1$, where $u_{\nu}$ \eqref{u_nu} is the velocity-space collisional cutoff derived in Section \ref{dissipation_cutoffs}. The nonlinear term vanishes when integrated over $\mathbf{k}$, and we get
\begin{equation} \label{eq_s_flux_int_k}
    \frac{\partial}{\partial \mathbf{s}} \cdot \mathbf{\Gamma} = S(\mathbf{s}),
\end{equation}
a balance between the $\mathbf{s}$-space flux
\begin{equation} \label{def_s_flux}
    \mathbf{\Gamma}(\mathbf{s}) = \frac{V}{(2 \pi)^d} \int \mathrm{d} \mathbf{k} \, \mathbf{k} \,  P_{\delta \! f}(\mathbf{k},\mathbf{s})
\end{equation}
and the source
\begin{equation}
    S(\mathbf{s}) = \frac{V}{(2 \pi)^d} \int \mathrm{d} \mathbf{k} \, \overline{\mathcal{S}}_{\mathbf{k},\mathbf{s}}.
\end{equation}
For $s v_{\mathrm{the}} \gg 1$, we can approximate $S(\mathbf{s}) \simeq \varepsilon \delta(\mathbf{s})$. Then, the flux shell-averaged over angles in $\mathbf{s}$ is radial and constant:
\begin{equation} \label{s_flux_SA}
    \langle \mathbf{\Gamma} \rangle_{\mathrm{SA}} = \varepsilon \, \hat{\mathbf{s}}.
\end{equation}
This solution holds in the inertial range $1/v_{\mathrm{the}} \ll s \ll 1/u_{\nu}$ and implies a constant-flux velocity-space cascade of $\delta C_2$. Note that this result was demonstrated in the 1D-1V Kraichan model in \cite{nastac2023phase}, but here we demonstrate that a steady-state $\delta C_2$ velocity-space cascade, under the above assumptions, is a generic feature of the Vlasov--Poisson system in an arbitrary number of dimensions. Also note that, formally, $\mathbf{\Gamma}$ could have angular structure. For example, for a delta-function source, the flux can contain arbitrary divergence-free circulations added to it in the inertial range and still satisfy \eqref{eq_s_flux_int_k}. However, we do not have to worry about this structure, as it would integrate to zero after shell averaging and, therefore, not contribute to the net flux of $\delta C_2$ to high $s$.

Now suppose that $P_{\delta \! f}(\mathbf{k},\mathbf{s})$ were symmetric in phase space. Then, since the integrand in \eqref{def_s_flux} is odd in $\mathbf{k}$, $\mathbf{\Gamma}(\mathbf{s})= 0$. Thus, $\delta \! f$ must have statistical asymmetry in phase space in order for the integral in \eqref{def_s_flux} to be non-zero and hence for there to be a velocity-space cascade.

This required asymmetry is explicitly related to the asymmetry in the amplitudes of the phase-mixing and phase-unmixing modes. Indeed, taking the dot product of \eqref{s_flux_SA} with $\hat{\mathbf{s}}$, we have
\begin{equation} \label{gamma_SA_dot}
    \frac{V}{(2 \pi)^d} \int \mathrm{d} \mathbf{k} \, \frac{\mathbf{k} \cdot \mathbf{s}}{s} \,  \langle P_{\delta \! f} \rangle_{\mathrm{SA}, s}(\mathbf{k},s) = \varepsilon,
\end{equation}
where the shell-averaging in \eqref{gamma_SA_dot} is only in $\mathbf{s}$. For \eqref{s_flux_SA} to be positive, the positive contributions to the integral in \eqref{gamma_SA_dot} from phase-mixing modes, with $\mathbf{k} \cdot \mathbf{s} > 0$, must be larger than negative contributions to the integral from phase-unmixing modes, with $\mathbf{k} \cdot \mathbf{s} < 0$.

These results demonstrate that the asymmetry required for a velocity-space cascade is a dependence of $P_{\delta \! f}$ on the angle $\theta = \cos^{-1} ( \mathbf{k} \cdot \mathbf{s}/k s)$. Barring any asymmetry set by the outer scale, since there are no special directions in position space or velocity space, we expect there to be no other angular structure (i.e., no anisotropy) in the inertial-range spectrum.

\section{Lagrangian chaos and phase-space eddies} \label{LC_PSE_section}

In this section, we argue that the dynamical mechanism of kinetic Batchelor turbulence is the stretching of phase-space blobs (eddies) due to the exponential separation of the phase-space trajectories of nearby Lagrangian particles, accelerated by spatially smooth electric fields. This mechanism is analogous to how, in the Batchelor regime of passive scalar turbulence, a fluid-dynamical passive scalar undergoes filamentation in position space due to the Lagrangian chaos of the trajectories of particles advected by spatially smooth flow velocities \cite{batchelor1959small, falkovich2001particles}.

The problem set-up is as follows. Consider test particles (electrons) accelerated solely by an external electric field, viz., $\mathbf{E}_{\mathrm{tot}} = \mathbf{E}_{\mathrm{ext}}$. Then, a particle labeled by index $n$ and having Lagrangian phase-space coordinates $(\mathbf{X}_n(t), \mathbf{V}_n(t))$ satisfies the equations of motion
\begin{equation} \label{Lagrangian_EOM}
    \frac{d \mathbf{X}_n}{d t} = \mathbf{V}_n(t), \quad \frac{d \mathbf{V}_n}{d t} = -\frac{e}{m} \mathbf{E}_{\mathrm{ext}}(\mathbf{X}_n(t),  t). 
\end{equation}
We wish to calculate the two-particle separation in phase space, viz.,
\begin{equation}
   \left(\mathbf{r}(t), \mathbf{u}(t) \right) = \left(\mathbf{X}_1(t)-\mathbf{X}_2(t), \mathbf{V}_1(t)-\mathbf{V}_2(t)\right), 
\end{equation}
which satisfies
\begin{equation}
    \frac{d \mathbf{r}}{d t} = \mathbf{u}, \quad \frac{d \mathbf{u}}{d t} = -\frac{e}{m} \left[\mathbf{E}_{\mathrm{ext}}(\mathbf{X}_1(t), t) - \mathbf{E}_{\mathrm{ext}}(\mathbf{X}_2(t), t) \right].
\end{equation}
For position-space separations smaller than the scale of variation of the electric field, which we assume to be smooth in space, we can Taylor expand the $i^{\mathrm{th}}$ component of the electric field as
\begin{equation} \label{elc_increment}
    -\frac{e}{m} \left [E_{\mathrm{ext}}^{i}(\mathbf{X_1}(t),t) - E_{\mathrm{ext}}^{i}(\mathbf{X_2}(t),t) \right] \simeq \sigma^{i}_{j} r^{j},
\end{equation}
where $\mathbf{r} = \mathbf{X}_1 - \mathbf{X}_2$, and
\begin{equation} \label{sigma}
   \sigma^{i}_{j} = -\frac{e}{m}\frac{\partial E^{i}_{\mathrm{ext}}}{\partial x_{j}}(\mathbf{X}_2(t), t)
\end{equation}
is the Lagrangian `rate of strain' tensor. Then the separations satisfy
\begin{equation} \label{drdt_dudt}
    \frac{d r^{i}}{d t} = u^{i}, \quad \frac{d u^{i}}{d t} = \sigma^{i}_{j} r^{j}.
\end{equation}

To make further progress, we must specify $\sigma$. Our aim is to be pedagogical rather than exhaustive, so, rather than attempt to prove general statements about particles accelerated by general fields (and hence general $\sigma$), we consider two instructive cases: passive particles accelerated by electric fields that are either static in time or white-noise in time. The static field is an approximation of a finite-time-correlated field on time scales $t \ll \tau_{\mathrm{corr}}$, while the white-noise field is an approximation of a finite-time-correlated field on time scales $t \gg \tau_{\mathrm{corr}}$. These two approaches are inspired by the classic works on fluid-dynamical passive-scalar mixing by Batchelor \cite{batchelor1959small} and Kraichnan \cite{kraichnan1968small}, respectively. After analyzing these two extreme cases, we will make a case for how to interpret the situation with finite-time-correlated electric fields (externally imposed and/or self-consistent), present, e.g., in our numerical simulations in Section~\ref{forced_turbulence}, and, expectedly, in most relevant situations in nature.

Before continuing, we remark that, while our main aim in this appendix is to understand collisionless dynamics, we will invoke collisional velocity-space diffusion as a mechanism of dissipation. However, as we argue in Section \ref{noise_turbulence_interplay}, in reality, we do not predict phase-space turbulence to be terminated at small scales by a Fokker--Planck collision operator, but rather by the takeover of discrete particle noise. The simplest (albeit crude) fix to account for this is to replace the diffusive cutoffs $r_{\nu}$ \eqref{r_nu} and $u_{\nu}$ \eqref{u_nu} with the thermal cutoffs $r_{\mathrm{the}} \sim k_{\mathrm{the}}^{-1}$ \eqref{kth_def} and $u_{\mathrm{the}} \sim s_{\mathrm{the}}^{-1}$ \eqref{sth}. Even so, insofar as the calculations performed in this appendix serve to provide intuition on the physics of phase mixing in the inertial range, the choice of dissipation mechanism does not affect any of the key results.

\subsection{Static electric field} \label{static_Batchelor}

First, we consider the case of an electric field that is static in time. While \eqref{drdt_dudt} is easily solvable for arbitrary time-independent $\sigma_{ij}$ in arbitrary dimensions, we consider the 1D-1V case for simplicity and set $\sigma = \mathrm{const}$, where $\sigma$ is proportional to the local gradient of the electric field: $\sigma = (e/m)\varphi_{\mathrm{ext}}''(x_2)$. There are three distinct regimes: $\sigma = 0$, $\sigma < 0$, and $\sigma > 0$. In what follows, we denote $\gamma = \sqrt{|\sigma|}$.

When $\sigma = 0$, the solution to \eqref{drdt_dudt} with initial separation $(r(0), u(0)) = (r_{0}, u_{0})$ is
\begin{align} \label{ru_fs}
    &r(t) = r_{0} + u_0 t, \nonumber \\ &u(t) = u_0.
\end{align}
Particles move ballistically along their initial trajectories, and any initial separation in their velocities leads to separation in position space that grows linearly in time.

When $\sigma < 0$, the particle trajectories are
\begin{align} \label{ru_osc}
    &r(t) = r_{0} \cos( \gamma t) + \gamma^{-1} u_{0} \sin( \gamma t), \nonumber \\ &u(t) = -\gamma r_{0} \sin( \gamma t) + u_{0} \cos( \gamma t).
\end{align}
The electrostatic potential at $x_2$ is concave, so nearby particles are trapped; their trajectories oscillate in phase space. 

When $\sigma > 0$, the particle trajectories are  
\begin{align} \label{ru_exp}
    &r(t) = \frac{r_0 + \gamma^{-1} u_0}{2} e^{\gamma t} +  \frac{r_0-\gamma^{-1} u_0}{2} e^{-\gamma t}, \nonumber \\ &u(t) = \frac{\gamma r_0 + u_0}{2} e^{\gamma t} -  \frac{\gamma r_0- u_0}{2} e^{-\gamma t}.
\end{align}
This is a toy model of Lagrangian chaos; any initial phase-space separation is exponentially amplified on a time scale $\sim \gamma^{-1}$, where $\gamma$ is the positive Lyapunov exponent of the phase-space flow $(u, \gamma^2 r)$.

In higher dimensions, one can substitute $\mathbf{r} = \hat{\mathbf{r}} e^{\lambda t}$ into~\eqref{drdt_dudt}, which gives
\begin{equation} \label{eigenvalue_problem}
    \lambda^2 \hat{\mathbf{r}} = \sigma \cdot \hat{\mathbf{r}},
\end{equation}
so the spectrum of Lyapunov exponents is set by the eigenvalues of the rate-of-strain matrix $\sigma$. Depending on the Lyapunov exponents and corresponding eigenvectors \eqref{eigenvalue_problem}, particle trajectories can either move ballistically or oscillate and/or exponentially separate. Different dimensions can be also be coupled. For example, two particles that initially coincide in one direction can still exponentially separate in that direction if they have initial separations in another direction. We do not explicitly consider the 2D-2V and 3D-3V cases here, but the calculations can be straightforwardly generalized to higher dimensions.

\subsubsection{Deformation of phase-space eddies} \label{static_Batchelor_eddies}

The solutions \eqref{ru_fs}, \eqref{ru_osc}, and \eqref{ru_exp} describe how individual particles separate. Using these solutions, we can compute how their separation causes a distribution of particles to deform in phase space. For simplicity, we restrict ourselves to 1D-1V. Combining \eqref{corr_eq} and \eqref{elc_increment}, in the absence of a source and collisions, the equation for the $\delta \! f$ correlator \eqref{corr_func} is
\begin{equation} \label{corr_eq_static}
    \frac{\partial C_{\delta \! f}}{\partial t} + u\frac{\partial C_{\delta \! f}}{\partial r} + \sigma r \frac{\partial C_{\delta \! f}}{\partial u} = 0.
\end{equation}
Note that $(r,u)$ here denotes the Eulerian phase-space separation, rather than the Lagrangian one discussed in the previous section. Also note that the Vlasov equation~\eqref{vlasov} for the distribution function $f(x,v,t)$ in $(x,v)$ space and for a linear electric field profile, $E = \sigma x$, is locally the same as \eqref{corr_eq_static}.

We can solve \eqref{corr_eq_static} via the method of characteristics. For the initial condition $C_{\delta \! f}(r, u, t = 0) = g(r, u)$, the solution at later time is $C_{\delta \! f}(r, u, t) = g(r_0, u_0)$, where $(r_0, u_0)$ is given by the inversion of the characteristics \eqref{ru_fs}, \eqref{ru_osc}, and \eqref{ru_exp}, viz., for $\sigma = 0$,
\begin{align} \label{ru_fs0}
    &r_0 = r - u t, \nonumber \\ &u_0 = u,
\end{align}
for $\sigma < 0$,
\begin{align} \label{ru_osc0}
    &r_0 = r \cos( \gamma t) - \gamma^{-1} u \sin( \gamma t), \nonumber \\ & u_0 = u \cos( \gamma t) + \gamma r \sin( \gamma t),
\end{align}
and for $\sigma > 0$,
\begin{align} \label{ru_exp0}
    &r_0 = \frac{r- \gamma^{-1}u}{2}e^{\gamma t} + \frac{r + \gamma^{-1}u}{2}e^{-\gamma t}, \nonumber \\ &u_0 = \frac{u- \gamma r}{2}e^{\gamma t} + \frac{u + \gamma r}{2}e^{-\gamma t}.
\end{align}

\begin{figure*}
	\centering
    	\includegraphics[width=\textwidth]{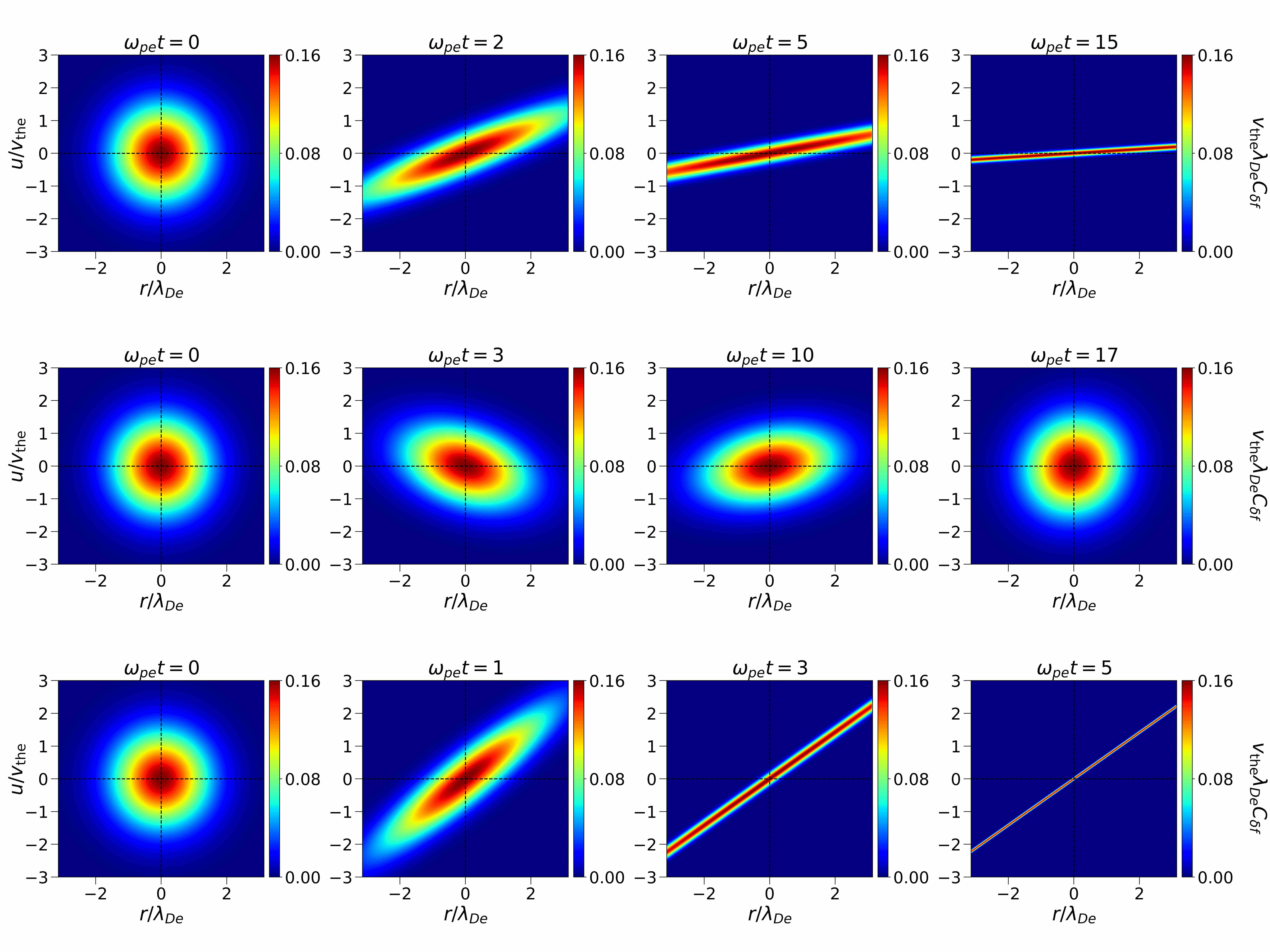}
	\caption{\label{fig:ps_eddy_shearing} Sequences in time of contour plots of $C_{\delta \! f}$, for (top row) $\sigma = 0$, (middle row) $\sigma = -(1/2) \omega_{\mathrm{pe}}^2$, (bottom row) $\sigma = (1/2) \omega_{\mathrm{pe}}^2$. The correlation function $C_{\delta \! f}$ is solved for by combining the initial condition \eqref{blob_IC} with the characteristics \eqref{ru_fs0}, \eqref{ru_osc0}, and \eqref{ru_exp0}, corresponding to the top, middle, and bottom rows, respectively.}
\end{figure*}

Consider the initial condition
\begin{equation} \label{blob_IC}
    g(r,u) = \frac{1}{2 \pi \, v_{\mathrm{the}} \lambda_{\mathrm{De}}} e^{-r^2/2 \lambda_{\mathrm{De}}^2 - u^2/2 v_{\mathrm{the}}^2},
\end{equation}
which represents a phase-space blob, or eddy, of size $r \sim \lambda_{\mathrm{De}}$ and $u \sim v_{\mathrm{the}}$ centered at $(r, u) = (0, 0)$. Note that, for simplicity, we consider the domain to be infinite in position space rather than periodic. We have also normalized $g(r,u)$ so that its integral over phase space is unity. In Fig.~\ref{fig:ps_eddy_shearing}, we plot the solution $C_{\delta \! f}(r, u, t) = g(r_0, u_0)$, with $(r_0, u_0)$ given by \eqref{ru_fs0}, \eqref{ru_osc0}, and \eqref{ru_exp0}, for the above three cases taken at a sequence of times.

In Fig.~\ref{fig:ps_eddy_shearing} (top row), we plot the case with $\sigma = 0$. The phase-space eddy is sheared in velocity space by the phase-space shear flow $(u, 0)$, which causes the eddy to deform into a thin, elongated filament that tends to align with the $r$ axis. The distribution function consequently develops fine-scale structure in velocity space. This process, linear phase mixing, underlies linear Landau damping \cite{landau1946vibrations}.

In Fig.~\ref{fig:ps_eddy_shearing} (middle row), we plot the case with $\sigma < 0$. The phase-space flow $(u, \sigma r)$ causes the eddy to rotate. Up to oscillations set by $\gamma$ in \eqref{ru_osc0}, trapped particles do not separate. Therefore, the phase-space eddy is not stretched, and the distribution function does not develop fine-scale structure.

In Fig.~\ref{fig:ps_eddy_shearing} (bottom row), we plot the case with $\sigma > 0$. The phase-space flow $(u, \sigma r)$ is the combination of the shear flows $(u, 0)$ and $(0, \sigma r)$. The flow $(0, \sigma r)$, corresponding to nonlinear phase mixing by the electric field, would alone deform the eddy into a thin, elongated filament that is aligned with the $u$ axis, causing the distribution function to develop sharp gradients in position space. Together with the flow $(u,0)$ corresponding to linear phase mixing, the eddy is filamented in both position and velocity space, and is stretched along the line $u = \gamma r$, where $g(r_0, u_0)$ is maximized at long times.

This interplay of linear and nonlinear phase mixing in the $\sigma > 0$ case underlies the physics of critical balance built into our turbulence theory. In Section \ref{CB_section}, we justified critical balance by arguing that phase-space mixing should occur simultaneously in both position and velocity space. Here, critical balance is a built-in feature of phase-space stretching. A phase-space eddy is naturally sheared in both position and velocity space, fundamentally because of the coupling between them, viz., any initial particle separation in phase space leads to exponential separation in both variables. Furthermore, the notion that phase-space eddies have aspect ratio $u \sim \gamma r$ manifests in the tilting of the eddy along the line $u = \gamma r$.

The shearing of the phase-space eddy is limited by collisions. To estimate heuristically the time scale on which collisions are activated, we can solve
\begin{equation} \label{coll_decay_C}
    \frac{\partial C_{\delta \! f}}{\partial t} \sim   \nu v_{\mathrm{the}}^2 \frac{\partial^2 C_{\delta \! f}}{\partial u^2},
\end{equation}
where, on the right-hand side of \eqref{coll_decay_C}, we insert the collisionless solution $C_{\delta \! f} = g(r_0, u_0)$. For the linear phase-mixing ($\sigma = 0$) case, we have
\begin{equation}
     \nu v_{\mathrm{the}}^2 \frac{\partial^2 C_{\delta \! f}}{\partial u^2}  \sim \nu v_{\mathrm{the}}^2 \frac{\partial^2 g}{\partial r_0^2} \left(\frac{\partial r_0}{\partial u} \right)^2 \sim -\nu \frac{v_{\mathrm{the}}^2}{\lambda_{\mathrm{De}}^2} t^2 C_{\delta \! f},
\end{equation}
where the $u_0$ derivatives are subdominant since $u_0$ does not depend on time [see \eqref{ru_fs0}]. Then, solving \eqref{coll_decay_C} yields
\begin{equation}
    C_{\delta \! f} \sim  \exp{\left(-\frac{\nu v_{\mathrm{the}}^2}{3 \lambda_{\mathrm{De}}^2} t^3\right)},
\end{equation}
which is dissipated away on a time scale $\tau_{\mathrm{d}} \propto \nu^{-1/3}$ \cite{su1968collisional,kanekar2015fluctuation,zocco2011reduced,banik2024relaxation}, in agreement with \eqref{tau_dissp_linear}.

For the $\sigma > 0$ case, note that after a Lyapunov time $\sim \gamma^{-1}$, the derivatives  $(\partial r_0/\partial u)$ and $ (\partial u_0/\partial u)$ are $ \propto e^{\gamma t}$.  Therefore, the term contributing to the collisional diffusion involving $(\partial r_0/\partial u)^2$ scales as
\begin{align} \label{du_exp}
     \nu v_{\mathrm{the}}^2 \frac{\partial^2}{\partial u^2} C_{\delta \! f} & \sim \nu v_{\mathrm{the}}^2  \frac{\partial^2 g}{\partial r_0^2} \left(\frac{\partial r_0}{\partial u} \right)^2  \sim -\nu \frac{v_{\mathrm{the}}^2}{\lambda_{\mathrm{De}}^2} \frac{e^{2 \gamma t}}{4 \gamma^2 } C_{\delta \! f}.
\end{align}
The other terms involving derivatives of $u_0$ have the same functional form. Then, inserting \eqref{du_exp} into \eqref{coll_decay_C} yields
\begin{equation}
	C_{\delta \! f} \sim  \exp{\left(-\frac{\nu v_{\mathrm{the}}^2}{8 \lambda_{\mathrm{De}}^2 \gamma^{3}} e^{2 \gamma t} \right)},
\end{equation}
which is dissipated away on a time scale $\tau_{\mathrm{d}} \propto |\log \nu|$ [cf.~\eqref{tau_dissp}], asymptotically faster than the linear-phase mixing time scale when $\nu$ is small. Physically, the time that it takes for two nearby particles to separate exponentially scales logarithmically with their initial separation, as can be calculated from \eqref{ru_exp}. The smallest possible phase-space separations that are not smeared by collisional diffusion are given by balancing the collision time $\tau_{\nu} \sim \nu^{-1} u^2/v_{\mathrm{the}}^2$ with the stretching time $\sim \gamma^{-1}$, alongside the critical-balance relation \eqref{CB_aspect_ratio}. This gives $r_{\nu}, u_{\nu} \propto \nu^{1/2}$ and hence $\tau_{\mathrm{d}} \propto |\log \nu|$, in agreement with the estimates in Sections \ref{dissipation_cutoffs} and \ref{ph_mixing_eff}.

\subsubsection{Phase-space spectrum} \label{static_Batchelor_spec}

The collisional estimates found at the end of the previous section for the $\sigma > 0$ regime agree with those computed for the phase-space cascade in Section \ref{dissipation_cutoffs}, where the cascade time is just the shearing time, viz., $\tau_{\mathrm{c}} \sim \gamma^{-1}$. In fact, this filamentation is just an `elementary step' of the phase-space cascade. Let us work out the $\delta C_2$ spectrum associated with it.

We Fourier transform \eqref{corr_eq_static} in position and velocity space, and bring back the source and collision terms, as in \eqref{E_delta_f_eq}, viz.,
\begin{equation} \label{spec_eq_static}
     \frac{\partial P_{\delta \! f} }{\partial t} + k \frac{\partial P_{\delta \! f}}{\partial s} + \gamma^2 s \frac{\partial P_{\delta \! f}}{\partial k}  = \mathcal{S}_{k,s} - 2 \nu v_{\mathrm{the}}^2 s^2 P_{\delta \! f}.
\end{equation}
We only consider the case $\gamma^2 = \sigma > 0$. A crude model for a source localized at low $k$ (due to the spatial smoothness of the electric field) and low $s$ (due to characteristic velocity-space scale of $f_0$ being $v_{\mathrm{the}}$) is
\begin{equation}
	\mathcal{S}_{k,s} = \frac{\sqrt{2 \pi} \varepsilon v_{\mathrm{the}}}{L} e^{-s^2 v_{\mathrm{the}}^2/2} \delta(k),
\end{equation}
which satisfies $(L/ 2 \pi) \iint \mathrm{d} k \mathrm{d} s \, \mathcal{S}_{k,s} = \varepsilon$.

We assume that \eqref{spec_eq_static} reaches a steady state, in which the $\delta C_2$ injection at low $(k,s)$ by the source is balanced by the dissipation at high $(k,s)$ by collisions, viz.,
\begin{equation}
    \varepsilon = \frac{ \nu \, v_{\mathrm{the}}^2 L}{\pi} \iint \mathrm{d} k \, \mathrm{d} s \, s^2 P_{\delta \! f}.
\end{equation}
We solve for the spectrum away from the collisional dissipation range, where it satisfies
\begin{equation} \label{spec_eq_IR_ss}
    k \frac{\partial P_{\delta \! f}}{\partial s} + \gamma^2 s \frac{\partial P_{\delta \! f}}{\partial k}  = \frac{\sqrt{2 \pi} \varepsilon v_{\mathrm{the}}}{L} e^{-s^2 v_{\mathrm{the}}^2/2} \delta(k).
\end{equation}
Because $\delta \! f$ is real, $P_{\delta \! f}(k, s) = P_{\delta \! f}(-k, -s)$. Therefore, we only need to solve \eqref{spec_eq_IR_ss} in the upper half-plane, $-\infty < k < +\infty$ and $s \geq 0$.

Away from $k = 0$, the solution of \eqref{spec_eq_IR_ss} is of the form
\begin{equation} \label{h_pm}
    P_{\delta \! f}(k, s) =  \begin{cases} 
     h_{+}(s^2 -\gamma^{-2} k^2) , & k > 0, \\
      h_{-}(s^2 -\gamma^{-2} k^2) , & k < 0,
   \end{cases}
\end{equation}
where $h_{\pm}$ are functions to be determined. Let us integrate \eqref{spec_eq_IR_ss} over a small region $(-\delta, +\delta)$ around $k = 0$ and then take $\delta \rightarrow 0^{+}$. In this limit, the free-streaming term vanishes but the acceleration and source terms are finite, which yields the following boundary condition:
\begin{equation} \label{static_BC}
    \gamma^2 s \left[ h_{+}(s^2) - h_{-}(s^2)  \right] = \frac{ \sqrt{2 \pi} \varepsilon v_{\mathrm{the}}}{L} e^{-s^2 v_{\mathrm{the}}^2/2}.
\end{equation}
Likewise, we can integrate \eqref{spec_eq_IR_ss} over a vanishingly small region around $s = 0$. In this limit, the acceleration and source terms vanish but the free-streaming term is finite.  Also applying the reality condition $P_{\delta \! f}(k, s) = P_{\delta \! f}(-k, -s)$ to express the spectrum in the lower half-plane in terms of the spectrum in the upper half-plane, we obtain
\begin{equation} \label{static_BC2}
    k \left[ h_{+}(-\gamma^{-2} k^2) - h_{-}(-\gamma^{-2} k^2)  \right] = 0,
\end{equation}
which implies $h_{+} = h_{-}$ whenever $s^2 -\gamma^{-2} k^2 < 0$.

To solve \eqref{spec_eq_IR_ss} with the boundary conditions \eqref{static_BC} and \eqref{static_BC2},  note that the characteristics of \eqref{spec_eq_IR_ss} satisfy $ds/dt = k$ and $dk/dt = \gamma^2 s$, which have solutions
\begin{align} \label{sk_exp}
    &s(t) = \frac{s_0 + \gamma^{-1} k_0}{2} e^{\gamma t} +  \frac{s_0-\gamma^{-1} k_0}{2} e^{-\gamma t}, \nonumber \\ &k(t) = \frac{\gamma s_0 + k_0}{2} e^{\gamma t} -  \frac{\gamma s_0- k_0}{2} e^{-\gamma t},
\end{align}
for initial conditions $s(0) = s_0$ and $k(0) = k_0$. `Particles' initially sourced at low $(k,s)$ will spread exponentially fast to high $(k,s)$, with trajectories at long times satisfying $s \sim \gamma^{-1} k$. This is just the spectral manifestation of critically balanced stretching of phase-space eddies, as calculated in Appendix \ref{static_Batchelor_eddies}. This calculation suggests that at long times, the inertial-range spectrum will only be populated in the region where $k s > 0$, viz., the phase-mixing region. Therefore, it is reasonable to impose $h_{-} = 0$. Then, \eqref{static_BC} yields
\begin{equation} \label{approx_sol}
    h_{+}(s^2) = \frac{ \sqrt{2 \pi} \varepsilon v_{\mathrm{the}}}{L} \frac{e^{-s^2 v_{\mathrm{the}}^2/2}}{\gamma^2 s},
\end{equation}
which, together with \eqref{h_pm} and \eqref{static_BC2}, gives the inertial-range spectrum
\begin{equation} \label{spec_max}
    P_{\delta \! f}(k, s) = \frac{ \sqrt{2 \pi} \varepsilon v_{\mathrm{the}}}{L} \frac{e^{-|s^2 -\gamma^{-2} k^2| v_{\mathrm{the}}^2/2}}{\gamma^2 |s^2 -\gamma^{-2} k^2|^{1/2}} H \left(s^2 -\gamma^{-2} k^2 \right)
\end{equation}
for $k > 0$ and $P_{\delta \! f}(k, s) = 0$ for $k < 0$. Here, $H \left(s^2 -\gamma^{-2} k^2 \right)$ is the Heaviside step function (defined with the convention that $H(0) = 1/2$), which ensures that the boundary condition \eqref{static_BC2} is satisfied. In the large $v_{\mathrm{the}}$ limit, the Gaussian in \eqref{spec_max} can be approximated as a delta function, so
\begin{align} \label{spec_static}
     P_{\delta \! f}(k, s) &\simeq \frac{ \pi \varepsilon }{L} \frac{\delta \left(|s^2 -\gamma^{-2} k^2|^{1/2} \right)}{\gamma^2 |s^2 -\gamma^{-2} k^2|^{1/2}} \nonumber \\ &= \frac{ \pi \varepsilon }{L} \frac{\delta(s- \gamma^{-1} k)}{\gamma |k|}.
\end{align}
Note that the delta function in the first line of \eqref{spec_static} is in fact equal to the term in the second line plus a term~$\propto \delta(s + \gamma^{-1} k)$. However, we have dropped this latter term, since it would only yield a non-zero contribution in the phase-unmixing region $k < 0$, where the spectrum vanishes ($h_{-} = 0$). Also note that \eqref{spec_static} satisfies the reality condition $P_{\delta \! f}(k, s) = P_{\delta \! f}(-k, -s)$, so that \eqref{spec_static} is the spectrum in the full $(k, s)$ plane.

The spectrum \eqref{spec_static} is sharply concentrated along the critical-balance line $s = \gamma^{-1} k$, and has the integrated 1D spectra,
\begin{equation} \label{static_1D_k}
    \int \mathrm{d} s \, P_{\delta \! f}(k, s) = \frac{ \pi \varepsilon}{L \gamma} \, |k|^{-1}
\end{equation}
and
\begin{equation} \label{static_1D_s}
       \int \mathrm{d} k \, P_{\delta \! f}(k, s) = \frac{ \pi \varepsilon}{L \gamma}|s|^{-1},
\end{equation}
in agreement with the Batchelor scalings \eqref{spec_k_const_flux} and \eqref{spec_s_const_flux}, respectively. These scalings correspond to a constant flux of $\delta C_2$ processed from injection to dissipation scales in both position and velocity space. Indeed, defining $\hat{\nabla} = \left(\partial/ \partial s, \partial/\partial k \right)$ and the flux
\begin{equation}
    \mathbf{\Xi} = \left(\Xi^{s}, \Xi^{k} \right) = \left( k P_{\delta \! f}, \gamma^2 s P_{\delta \! f}  \right),
\end{equation}
\eqref{spec_eq_static} can be written in the flux-gradient form
\begin{equation} \label{spec_eq_static_flux}
     \frac{\partial P_{\delta \! f} }{\partial t} + \hat{\nabla} \cdot \mathbf{\Xi}  = \mathcal{S}_{k,s} - 2 \nu v_{\mathrm{the}}^2 s^2 P_{\delta \! f}.
\end{equation}
Using the solution \eqref{spec_static}, we find that the steady-state integrated fluxes in the inertial range satisfy the constant-flux relations \cite{nastac2023phase}
\begin{gather}
    \frac{L}{2 \pi} \int \mathrm{d} k \, \Xi^{s} = \frac{\varepsilon}{2} \, \mathrm{sgn}(s), \\
    \frac{L}{2 \pi} \int \mathrm{d} s \, \Xi^{k} = \frac{\varepsilon}{2} \, \mathrm{sgn}(k).
\end{gather}

That such a simple model recovers the above power laws suggests that these scalings are simply a spectral manifestation of phase-space stretching.  Indeed, the original derivation of the Batchelor spectrum in fluid-dynamical passive scalars was done in an analogous way, viz., by considering position-space stretching of a passive scalar advected by a velocity field that is static in time and smooth in space \cite{batchelor1959small}.

Despite this agreement with the 1D spectra, aside from capturing the dominant structure along the critical-balance line $s = \gamma^{-1} k$, the solution \eqref{spec_static} does not capture the high-$k$ or high-$s$ asymptotics of the 2D spectrum~\eqref{F_ks_solved}. Furthermore, the spectrum \eqref{spec_static} is devoid of any structure in the phase-unmixing region $ks < 0$. Fundamentally, this is because a static field with a single stretching rate $\gamma$ can only stretch eddies in one direction in phase space. Roughly speaking, to fill the $(k,s)$ plane requires a distribution of stretching rates; indeed, in the turbulence theory in Section \ref{phenom_theory}, $\gamma$ is an effective average stretching rate. In the following section, we examine the opposite limit of the static-field case, considering particles that are accelerated by an electric field that is a white noise in time, and, therefore, inherently possesses a statistical distribution of stretching rates. The phase-space mixing of the distribution function induced by this field leads to a fluctuation spectrum with the same scalings as~\eqref{F_ks_solved} \cite{adkins2018solvable,nastac2023phase}.

\subsection{White-noise electric field: Kraichnan model} \label{Kraichnan}

In this section, we consider the phase-space separations of particles accelerated by a random electric field that is smooth in space. In particular, consider a potential external electric field, $\mathbf{E}_{\mathrm{ext}} = -\nabla \varphi_{\mathrm{ext}}$, which is a Gaussian random variable with zero mean. Spatially, we take the field to be smooth and statistically homogeneous and isotropic. Temporally, we take it to be a white noise. Then, its correlation tensor is
\begin{equation} \label{corr_func_elc}
   \frac{e^2}{m^2} \overline{E_{\mathrm{ext}}^{i}(\mathbf{x}, t) E_{\mathrm{ext}}^{j}(\mathbf{x'}, t')} = 2 \, \kappa^{ij}(\mathbf{r}) \delta(t-t').
\end{equation}
where $\delta$ is a Dirac delta function, $\mathbf{r} = \mathbf{x}-\mathbf{x'}$, and the correlation matrix $\kappa^{ij}$, at small separations, satisfies \cite{ schekochihin2002spectra}
\begin{equation} \label{kappa_tensor}
    \kappa^{ij}(\mathbf{r}) \simeq \kappa_0 \delta^{ij} - \frac{1}{2} \kappa_2 r^2 \left(\delta^{ij} + 2 \hat{r}^i \hat{r}^j \right),
\end{equation}
where $\delta^{ij}$ is the Kronecker delta, $\hat{\mathbf{r}} = \mathbf{r}/r$, and $\kappa_0$ and $\kappa_2$ are positive constants. Note that $\kappa_2$ controls the effective shearing rate of the electric field: indeed, the field's increments satisfy
\begin{align} \label{kraichnan_increment}
    & \frac{e^2}{m^2} \overline{\left[E_{\mathrm{ext}}^{i}(\mathbf{x}, t)-E_{\mathrm{ext}}^{i}(\mathbf{x'}, t)\right] \left[ E_{\mathrm{ext}}^{j}(\mathbf{x}, t')-E_{\mathrm{ext}}^{j}(\mathbf{x'}, t') \right]} \nonumber \\ &= 2 \left[\kappa^{ij}(\mathbf{0})-\kappa^{ij}(\mathbf{r}) \right] \delta(t-t') \nonumber \\ &\simeq \kappa_2 r^2 \left(\delta^{ij} + 2 \hat{r}^i \hat{r}^j \right) \delta(t-t').
\end{align}

Passive acceleration of particles by random electric fields with the above properties is a plasma-kinetic analogue of the Kraichnan model of passive advection in fluid dynamics \cite{kraichnan1968small,falkovich2001particles}. Phase-space turbulence in the plasma-kinetic Kraichnan model was studied in 1D-1V by \cite{adkins2018solvable,nastac2023phase}, where it was found that when the stochastic electric field is smooth in space, the $\delta C_2$ spectrum has the Batchelor scalings \eqref{F_ks_solved}. These works  were focused on the spectral structure of the distribution function, and the interpretation of the cascade was done in wavenumber space rather than in position and velocity space, in terms of the interplay between phase mixing, phase unmixing, and nonlinear mode coupling. We show below that in the Kraichnan model, nearby particles separate exponentially in phase space. This result clarifies that the physical mechanism of the $\delta C_2$ cascade in terms of phase-space-eddy stretching developed in Section \ref{numeric_PS_eddies} and Appendix \ref{static_Batchelor_eddies} also applies to the Kraichnan model.

\subsubsection{Derivation of separation transition probability distribution function}

Because the electric field is stochastic, rather than computing trajectories in single realizations of it, we deal with the probability distribution of particles averaged over these realizations. The general object of interest is the conditional transition probability distribution function to find $N$ particles at phase-space coordinates $(\underline{\mathbf{x}}, \underline{\mathbf{v}})$ at time $t >0$, given the particles had phase-space coordinates $(\underline{\mathbf{x}_0}, \underline{\mathbf{v}_0})$ at the initial time $t = 0$:
\begin{equation} \label{P2_avg}
    P_N( \underline{\mathbf{x}}, \underline{\mathbf{v}}, t; \underline{\mathbf{x}_0}, \underline{\mathbf{v}_0}, 0) = \overline{p_N},
\end{equation}
where the $N$-particle transition probability distribution before ensemble averaging is
\begin{align} \label{P2_unavg}
    p_N( \underline{\mathbf{x}}, \underline{\mathbf{v}}, t; \underline{\mathbf{x}_0}, \underline{\mathbf{v}_0}, 0)& \nonumber \\  = \prod_{n=1}^{N}  \delta^{d}(\mathbf{x}_{n}-&\mathbf{X}_{n}(t; 0, \mathbf{x}_{n, 0}, \mathbf{v}_{n, 0})) \nonumber \\ \times  \delta^{d}(\mathbf{v}_{n}-&\mathbf{V}_{n}(t; 0, \mathbf{x}_{n, 0}, \mathbf{v}_{n, 0})).
\end{align}
where we have denoted $\underline{\mathbf{x}} = (\mathbf{x}_1, ..., \mathbf{x}_N)$, $\underline{\mathbf{v}} = (\mathbf{v}_1, ..., \mathbf{v}_N) $, $\underline{\mathbf{x}_0} = (\mathbf{x}_{1,0}, ..., \mathbf{x}_{N,0})$, and $\underline{\mathbf{v}_0} = (\mathbf{v}_{1,0}, ..., \mathbf{v}_{N,0})$. In~\eqref{P2_unavg},~$\delta^{d}(\mathbf{...})$ is a $d$-dimensional Dirac delta function. We denote by $(\mathbf{X}_{n}(t; 0, \mathbf{x}_{n, 0}, \mathbf{v}_{n, 0} ), \mathbf{V}_{n}(t; 0, \mathbf{x}_{n, 0}, \mathbf{v}_{n, 0}))$ the phase-space coordinates of the Lagrangian particle's trajectories at time $t$ satisfying the equations of motion \eqref{Lagrangian_EOM} with initial conditions $(\mathbf{X}_{n}(0), \mathbf{X}_{n}(0)) = ( \mathbf{x}_{n, 0},  \mathbf{v}_{n, 0})$. 

Taking the time derivative of $p_N$ given by \eqref{P2_unavg} yields the Liouville equation
\begin{equation} \label{Liouville}
    \frac{\partial p_N}{\partial t} + \sum_{n=1}^{N} \mathbf{v}_n \cdot \nabla_{\mathbf{x_n}} p_N -\frac{e}{m} \sum_{n=1}^{N} \mathbf{E}_{\mathrm{ext}}(\mathbf{x}_n,  t) \cdot \nabla_{\mathbf{v_n}} p_N = 0.
\end{equation}
To derive \eqref{Liouville}, we have used the relation
\begin{align}
    &\frac{\partial }{\partial t} \delta^{d}(\mathbf{x}_{n}-\mathbf{X}_{n}(t; 0, \mathbf{x}_{n, 0}, \mathbf{v}_{n, 0})) \nonumber \\ &= -\frac{d \mathbf{X}_{n}}{dt} \cdot \nabla_{\mathbf{x_n}} \delta^{d}(\mathbf{x}_{n}-\mathbf{X}_{n}(t; 0, \mathbf{x}_{n, 0}, \mathbf{v}_{n, 0}))
\end{align}
and similarly for the time derivative of the velocity delta functions. We have also used the relation $a \, \delta (a-b) = b \, \delta (a-b)$ to replace the time derivatives of the Lagrangian phase-space coordinates with those of the Eulerian phase-space coordinates.

We now ensemble-average \eqref{Liouville} to obtain the evolution equation for $P_N$:
\begin{equation} \label{Liouville2}
    \frac{\partial P_N}{\partial t} + \sum_{n=1}^{N} \mathbf{v}_n \cdot \nabla_{\mathbf{x_n}} P_N -\frac{e}{m} \sum_{n=1}^{N} \overline{ \mathbf{E}_{\mathrm{ext}}(\mathbf{x}_n,  t) \cdot \nabla_{\mathbf{v_n}} p_N} = 0.
\end{equation}
To compute the ensemble average of the acceleration term in \eqref{Liouville2}, we employ the the Furutsu-Novikov theorem \cite{furutsu1964statistical,novikov1965functionals} for splitting correlators involving Gaussian random fields $\mathbf{E}_{\mathrm{ext}}$. Denoting all the variables that $\mathbf{E}_{\mathrm{ext}}$ depends on by $\mathbf{q}$ (space and time), this theorem states
\begin{align} \label {NF}
    &\overline{E^{i}_{\mathrm{ext}}(\mathbf{q}) F[\mathbf{E}_{\mathrm{ext}}]} \nonumber \\ &= \int \mathrm{d} \mathbf{q'} \overline{ E^{i}_{\mathrm{ext}}(\mathbf{q}) E^{j}_{\mathrm{ext}}(\mathbf{q}') } \, \overline{\frac{\delta \! F[\mathbf{E}_{\mathrm{ext}}]}{\delta E^{j}_{\mathrm{ext}}(\mathbf{q}')}},
\end{align}
where $F[\mathbf{E}_{\mathrm{ext}}]$ is a differentiable functional of $\mathbf{E}_{\mathrm{ext}}$, and the integration in \eqref{NF} is over all possible values of $\mathbf{q}'$. Combining \eqref{corr_func_elc} and \eqref{NF}, the desired correlator is
\begin{align} \label{correlator_Liouville}
   & \overline{E_{\mathrm{ext}}^{i}(\mathbf{x}_n, t) p_N (\underline{\mathbf{x}}, \underline{\mathbf{v}}, t)} \nonumber \\ &= \frac{m^2}{e^2} \int \mathrm{d} t' \int \mathrm{d} \mathbf{x}' \, 2 \kappa^{ij}(\mathbf{x}_n -\mathbf{x}') \delta(t-t') \overline{\frac{\delta p_N (\underline{\mathbf{x}}, \underline{\mathbf{v}}, t)}{\delta E^{j}_{\mathrm{ext}}(\mathbf{x}', t')}}.
\end{align}
Formally integrating \eqref{Liouville} in time yields
\begin{align} 
    p_N (\underline{\mathbf{x}}, \underline{\mathbf{v}}, t) = - & \int^{t}_{\textrm{the past}}  \mathrm{d} t'' \bigg[ (...) \nonumber \\ &-\frac{e}{m} \sum_{m=1}^{N}  E^{\ell}_{\mathrm{ext}}(\mathbf{x}_m, t'') \frac{\partial p_N}{\partial v_{m. \ell}}    \bigg], 
\end{align}

where $(...)$ represents terms that will vanish once we take the functional derivative in \eqref{correlator_Liouville}:
\begin{equation} \label{func_deriv}
    \overline{\frac{\delta p_N (\underline{\mathbf{x}}, \underline{\mathbf{v}}, t)}{\delta E^{j}_{\mathrm{ext}}(\mathbf{x}', t')}} = \frac{e}{m} \sum_{m=1}^{N} \frac{\partial P_N}{\partial v_{m, j}} \delta^{d}(\mathbf{x}_m - \mathbf{x}') H(t-t'),
\end{equation}
where the Heaviside step function $H(t-t')$ is defined with the convention that $H(0) = 1/2$. Inserting \eqref{func_deriv} into \eqref{correlator_Liouville} gives
\begin{equation}
    \overline{E_{\mathrm{ext}}^{i}(\mathbf{x}_n, t) p_N (\underline{\mathbf{x}}, \underline{\mathbf{v}}, t)} = \frac{m}{e} \sum_{m=1}^{N} \kappa^{ij}(\mathbf{x}_n-\mathbf{x}_m) \frac{\partial P_N}{\partial v_{m, j}}.
\end{equation}
Therefore, the final closed equation for $P_N$ is
\begin{align} \label{Liouville2_closed}
    \frac{\partial P_N}{\partial t} &+ \sum_{n=1}^{N} \mathbf{v}_n \cdot \nabla_{\mathbf{x_n}} P_N \nonumber \\  &- \sum_{n, m=1}^{N} \kappa^{ij}(\mathbf{x}_n-\mathbf{x}_m) \frac{\partial^2 P_N}{\partial v_{n, i} v_{m, j}}  = 0.
\end{align}

In this paper, we are only interested in the two-particle transition probability distribution, $P_2$. We leave analyzing $P_N$ to future work. Since we are interested in phase-space separations, we integrate $P_2$ over the average phase-space coordinates. Namely, we change variables $(\mathbf{x_1},\mathbf{x_2}) \rightarrow (\mathbf{r}, \mathbf{R}) = (\mathbf{x_1}-\mathbf{x_2}, (\mathbf{x_1} + \mathbf{x_2})/2) $ and $(\mathbf{v_1},\mathbf{v_2}) \rightarrow (\mathbf{u}, \mathbf{U}) = (\mathbf{v_1}-\mathbf{v_2}, (\mathbf{v_1} + \mathbf{v_2})/2) $ and define the separation transition probability as
\begin{equation}
    \mathcal{P}_2(\mathbf{r}, \mathbf{u}, t; \mathbf{r}_0, \mathbf{u}_0, t_0)  = \int \mathrm{d} \mathbf{R} \, \mathrm{d} \mathbf{U} \, P_2. 
\end{equation}
From \eqref{Liouville2_closed}, $\mathcal{P}_2$ satisfies
\begin{equation} \label{P2_sep_arbit}
    \frac{\partial \mathcal{P}_2}{\partial t} + \mathbf{u} \cdot \nabla_{\mathbf{r}} \mathcal{P}_2 -2 \left[\kappa^{ij}(\mathbf{0}) - \kappa^{ij}(\mathbf{r})  \right] \frac{\partial^2 \mathcal{P}_2}{\partial u_i u_j} = 0
\end{equation}
with initial condition $\mathcal{P}_2(\mathbf{r}, \mathbf{u}, t_0; \mathbf{r}_0, \mathbf{u}_0, 0) = \delta^{d}(\mathbf{r} -\mathbf{r}_0) \delta^{d}(\mathbf{u} -\mathbf{u}_0)$. For position-space separations $r$ smaller than the variation scale of the electric field, such that the Taylor expansion \eqref{kappa_tensor} is valid, \eqref{P2_sep_arbit} becomes
\begin{equation} \label{P2_sep_final}
    \frac{\partial \mathcal{P}_2}{\partial t} + \mathbf{u} \cdot \nabla_{\mathbf{r}} \mathcal{P}_2 - \kappa_2 r^2 \left( \delta^{ij} + 2 \hat{r}^i \hat{r}^j \right) \frac{\partial^2 \mathcal{P}_2}{\partial u_i u_j} = 0.
\end{equation}

\subsubsection{Analysis of separation transition probability distribution function} \label{particle_sep_Kraichnan}

Explicitly solving \eqref{P2_sep_final} for the full distribution $\mathcal{P}_2$ is difficult. However, we can easily solve for moments of~$\mathcal{P}_2$. We define averages of functions $h(r,u)$ with respect to $\mathcal{P}_2$~as
\begin{equation}
    \mathbb{E} \left[ h(\mathbf{r}, \mathbf{u}) \right] \equiv \int \mathrm{d} \mathbf{r} \, \mathrm{d} \mathbf{u} \, \mathcal{P}_2 \, h(\mathbf{r}, \mathbf{u}).
\end{equation}
Note that, by construction in \eqref{P2_unavg}, $\mathbb{E}[1] = 1$.

Let us examine the first and second moments. From~\eqref{P2_sep_final}, the first moments satisfy the closed equations
\begin{align} \label{avg_sep}
    &\frac{d}{d t} \mathbb{E}[\mathbf{r}] = \mathbb{E}[\mathbf{u}], \nonumber \\ & \frac{d}{d t} \mathbb{E}[\mathbf{u}] = 0.
\end{align}
These equations have the solution
\begin{equation}
    \mathbb{E}[\mathbf{r}] = \mathbf{r}_0 + \mathbf{u}_0 t, \quad \mathbb{E}[\mathbf{u}] = \mathbf{u}_0. 
\end{equation}
For the white-noise external field satisfying \eqref{corr_func_elc}, the rate-of-strain tensor $\sigma^{i}_{j}$ in the separation equations \eqref{drdt_dudt} is temporally a white noise with zero mean, so that the mean velocity separation $\mathbb{E}[\mathbf{u}]$ cannot change. However, the mean separation in position space grows linearly in time at the rate $\mathbf{u}_0$.

 The second moments satisfy the closed hierarchy of equations:
\begin{align} \label{Dupree_hierarchy}
    &\frac{d}{d t} \mathbb{E}[|\mathbf{r}|^2] = 2 \, \mathbb{E}[\mathbf{r} \cdot \mathbf{u}], \nonumber \\ & \frac{d}{d t} \mathbb{E}[\mathbf{r} \cdot \mathbf{u}] = \mathbb{E}[|\mathbf{u}|^2], \nonumber \\ & \frac{d}{d t} \mathbb{E}[|\mathbf{u}|^2] = 6 \kappa_2 \, \mathbb{E}[|\mathbf{r}|^2],
\end{align}
which reduces to
\begin{align} \label{Dupree_eq}
    &\frac{d^3}{d t^3} \mathbb{E}[|\mathbf{r}|^2] = 12 \kappa_2 \, \mathbb{E}[|\mathbf{r}|^2].
\end{align}
The solution of \eqref{Dupree_eq} is
\begin{align} \label{Dupree_sol}
    &\mathbb{E}[|\mathbf{r}|^2] = c_1 e^{\lambda t} \nonumber \\ &+ e^{-(\lambda/2) t} \left[c_2 \cos \left( \frac{\sqrt{3}}{2} \lambda t \right) + c_3 \sin \left( \frac{\sqrt{3}}{2} \lambda t \right) \right],
\end{align}
where the Lypaunov exponent $\lambda = (12 \kappa_2)^{1/3}$, and
\begin{align}
    &c_1 = \frac{\lambda^2 r_0^2 + 2 \lambda \, \mathbf{r}_0 \cdot \mathbf{u}_0 + 2 u_0^2}{3 \lambda^2}, \nonumber \\ &c_2 = \frac{2 \left(\lambda^2 r_0^2 - \lambda \, \mathbf{r}_0 \cdot \mathbf{u}_0 - u_0^2 \right) }{3 \lambda^2}, \nonumber \\ &c_3 =  \frac{2 \left(\lambda \, \mathbf{r}_0 \cdot \mathbf{u}_0 - u_0^2\right)}{\sqrt{3} \lambda^2}.
\end{align}
For any non-zero initial separation $(\mathbf{r_0}, \mathbf{u_0})$, $c_1 > 0$ (the only non-zero initial conditions that satisfy $c_1 \leq 0$ are complex and hence unphysical). Therefore, after a time $\lambda t \gtrsim 1$, the first term in \eqref{Dupree_sol} dominates, and the mean-squared particle positions separate exponentially, viz., $\mathbb{E}[|\mathbf{r}|^2] \propto e^{\lambda t}$. Via \eqref{Dupree_hierarchy}, $\mathbb{E}[\mathbf{r} \cdot \mathbf{u}]$ and $\mathbb{E}[|\mathbf{u}|^2]$ also grow exponentially. Therefore, like in the static field case in Appendix \ref{static_Batchelor}, the particle trajectories are chaotic in phase space.

Dupree \cite{dupree1972theory} in his clump theory also studied particle separations in phase space. In fact, he derived the 1D-1V versions of \eqref{P2_sep_final} and \eqref{Dupree_eq}. His derivation was not explicitly for the Kraichnan field \eqref{corr_func_elc}, but he used a series of approximations that in the end amounted to assuming the field to be Kraichnan, viz., an externally imposed field that is spatially smooth and white-noise in time. Dupree's equation [the 1D-1V version of \eqref{P2_sep_final}] was subsequently studied by other authors (see, e.g., \cite{misguich1982relative,suzuki1984scaling,pecseli1990phase}). Dupree's main conclusion from \eqref{Dupree_eq} was that particles formed so-called phase-space clumps only for as long as the particles were not exponentially separated. Namely, after $\lambda t \gtrsim 1$, the time it took for $\mathbb{E}[|\mathbf{r}|^2]$ to reach an arbitrary separation $r^2_{\mathrm{ref}}$ was the `clump lifetime'
\begin{equation} \label{clump_lifetime}
    \tau_{\mathrm{cl}} = \frac{1}{\lambda} \log \left( \frac{r^2_{\mathrm{ref}}}{c_1 }\right),
\end{equation}
the time scale on which a phase-space clump could stay coherent. In the following section, we will discuss previous results on phase-space turbulence in the Kraichnan model and how they relate to Dupree's arguments.

\subsubsection{Relationship between particle separation and the phase-space cascade}

The exponential separation of particle trajectories is precisely the physical mechanism of the phase-space cascade in the Kraichnan model. Making this connection concrete requires some additional formalism, viz., the fact that $P_N$ is the Green's function of the $N$-point correlation function of the particle distribution function. To show this, note that in a single realization of the electric field, given an initial distribution function $f_0(\mathbf{x}_0, \mathbf{v}_0)$ at time $t = 0$, the distribution function $f(\mathbf{x}, \mathbf{v}, t)$ at time $t > 0$ is given by \cite{cardy2008non}
\begin{align} \label{f_lagrang}
    f(\mathbf{x}, \mathbf{v}, t) = f_0(\mathbf{X}_{n,0}(0; t, \mathbf{x}, \mathbf{v}), &\mathbf{V}_{n,0}(0; t, \mathbf{x}, \mathbf{v})) \nonumber \\ = \int \mathrm{d} \mathbf{x}_0 \mathrm{d} \mathbf{v}_0 f_0(\mathbf{x}_0, \mathbf{v}_0) &\delta^{d}(\mathbf{x}_0 -\mathbf{X}_{n,0}(0; t, \mathbf{x}, \mathbf{v})) \nonumber \\ \times \, \delta^{d}(\mathbf{v}_0 &-\mathbf{V}_{n,0}(0; t, \mathbf{x}, \mathbf{v})),
\end{align}
where $(\mathbf{X}_{n,0}(0; t, \mathbf{x}, \mathbf{v}), \mathbf{V}_{n,0}(0; t, \mathbf{x}, \mathbf{v}))$ are the phase-space coordinates of the Lagrangian particle trajectories at the initial time, found by solving the equations of motion \eqref{Lagrangian_EOM} backward in time, starting from $(\mathbf{X}_{n,0}(t), \mathbf{V}_{n,0}(t) = (\mathbf{x}, \mathbf{v})$. Using \eqref{f_lagrang}, we have that the ensemble-averaged, $N$-point correlation function of the distribution function is given by
\begin{align}
     C_{f, N}(\underline{\mathbf{x}}, \underline{\mathbf{v}}, t) = &\overline{f(\mathbf{x}_1, \mathbf{v}_1, t) \cdot \cdot \cdot f(\mathbf{x}_N, \mathbf{v}_N, t)} \nonumber \\ = \int \mathrm{d} \underline{\mathbf{x}_0} \mathrm{d} \underline{\mathbf{v}_0} f_0(\mathbf{x}_{1, 0},& \mathbf{v}_{1, 0}, 0) \times \cdot \cdot f_0(\mathbf{x}_N, \mathbf{v}_N, 0) \nonumber \\ & \cdot P_N(\underline{\mathbf{x}_0}, \underline{\mathbf{v}_0}, 0; \underline{\mathbf{x}}, \underline{\mathbf{v}}, t).
\end{align}
Because the stochastic electric field is statistically invariant with respect to time inversion, the propagator $P_N$ has the symmetry $P_N(\underline{\mathbf{x}_0}, \underline{\mathbf{v}_0}, 0; \underline{\mathbf{x}}, \underline{\mathbf{v}}, t) = P_N(\underline{\mathbf{x}}, \underline{\mathbf{v}}, t; \underline{\mathbf{x}_0}, \underline{\mathbf{v}_0}, 0)$. Physically, this means that nearby Lagrangian particle trajectories exponentially separate backward and forward in time. Furthermore, this symmetry implies $C_{f,N}$ satisfies the same equation as $P_N$, viz., \eqref{Liouville2_closed}. In particular, the two-point correlation function integrated over the average phase-space coordinates $(\mathbf{R}, \mathbf{U})$ satisfies
\begin{equation} \label{Cf_closed}
    \frac{\partial \mathcal{C}_{f, 2}}{\partial t} + \mathbf{u} \cdot \nabla_{\mathbf{r}} \mathcal{C}_{f, 2} - \kappa_2 r^2 \left( \delta^{ij} + 2 \hat{r}^i \hat{r}^j \right) \frac{\partial^2 \mathcal{C}_{f, 2}}{\partial u_i u_j} = 0,
\end{equation}
same as \eqref{P2_sep_final}.

We do not formally solve \eqref{Cf_closed} in this paper; instead, we opt for `twiddle algebra.' Like in the time-evolution equation \eqref{corr_eq_static} for the correlation function in the static-field case, linear phase mixing shears phase-space eddies in velocity space, on a time scale $\tau_{\mathrm{l}} \sim r/u$. For the nonlinear term, however, unlike in the static-field case, where the electric field linearly shears eddies in position space, the Markovian character of the stochastic electric field turns the nonlinear term into a velocity-space diffusion operator, with a diffusion tensor $\propto r^2$. This causes position-space mixing of eddies on a time scale $\tau_{\mathrm{nl}} \sim u^2/\kappa_2 r^2$. If we assume that these two processes lead to a phase-space cascade of $C_2$ in which the phase-space mixing is critically balanced, viz., the cascade time satisfies $\tau_{\mathrm{c}} \sim \tau_{\mathrm{l}} \sim \tau_{\mathrm{nl}}$, then $\tau_{\mathrm{c}} \sim \kappa^{-1/3}$, independent of scale. Following the analysis of Section \ref{phenom_theory}, but without invoking Poisson's equation for self-consistency, yields the same scalings for the spectrum [which here is the Fourier transform of $\mathcal{C}_{f, 2}$, defined analogously as \eqref{spec_fourier}] that we found for the Vlasov--Poisson system, viz., \eqref{F_ks_solved}. Indeed, the steady-state inertial-range spectrum in the 1D-1V Kraichnan model was exactly solved for in \cite{adkins2018solvable,nastac2023phase}, and it satisfies the same asymptotic scalings as in \eqref{F_ks_solved}. Furthermore, \cite{nastac2023phase} showed that this spectrum satisfies the critical balance and constant-flux relations in phase space on which we have built our phenomenological turbulence theory in this paper.

The works \cite{adkins2018solvable,nastac2023phase} interpreted the phase-space cascade in wavenumber space. Fourier transforming \eqref{Cf_closed} yields an equation for the spectrum:
\begin{equation} \label{spec_eq_Kraichnan}
    \frac{\partial P_{f}}{\partial t} + \mathbf{k} \cdot \frac{\partial P_{f}}{\partial \mathbf{s}} - \kappa_2 s^2 \left( \delta^{ij} + 2 \hat{s}^i \hat{s}^j \right) \frac{\partial^2 P_{f}}{\partial k_i k_j} = 0.
\end{equation}
In \eqref{spec_eq_Kraichnan}, linear phase mixing manifests as advection of the spectrum in $\mathbf{s}$ at rate $\mathbf{k}$, and nonlinear phase mixing manifests as diffusion of the spectrum in $\mathbf{k}$ space, with a turbulent-diffusion coefficient $\sim \kappa_2 s^2$. The above analysis shows that the interplay of these two processes is simply the spectral manifestation of mixing of phase-space eddies. This mixing encodes the exponential separation of nearby particle trajectories in phase space, as calculated in Appendix \ref{particle_sep_Kraichnan}. Indeed, the cascade time $\tau_{\mathrm{c}} \sim \kappa_2^{-1/3}$, up to an order-unity constant, is simply the inverse of the Lypaunov exponent of the particle separation in \eqref{Dupree_sol}. Therefore, just as was the case in the static-field case considered in Appendix \ref{static_Batchelor}, the physical mechanism of the $C_2$ cascade in the Kraichnan model in the Batchelor regime is the Lagrangian chaos of particle orbits in phase space.

Before discussing the case with electric fields that are finite-time-correlated and self-consistent, we briefly return to the relevance of Dupree's notion of the clump lifetime \eqref{clump_lifetime}. In \eqref{clump_lifetime}, using $\lambda \sim \tau_{\mathrm{c}}^{-1}$ and equating $r_{\mathrm{ref}}$ with the outer scale of the turbulence and $(r_0, u_0)$ with the dissipation scale $(r_{\nu}, u_{\nu})$ [or $(r_{\mathrm{the}}, u_{\mathrm{the}})$], we find that the clump lifetime $\tau_{\mathrm{cl}}$ is just $\tau_{\mathrm{d}}$ [as given in \eqref{tau_dissp} or \eqref{tau_dissp_noise}], viz., the time that it takes for the fluctuations injected at the outer scale to cascade down to the dissipation scale. While Dupree did not argue in terms of a cascade of $C_2$, he had the right intuition that phase-space eddies (clumps in his language) can stay coherent for as long as particles do not exponentially separate.

\subsection{Finite-time-correlated and self-consistent electric fields} \label{FTC}

Given the behavior of particle trajectories and the phase-space structure of the distribution function for the static and white-noise electric field cases, we can now piece together what happens in a system with finite-time-correlated electric fields.

Starting from the static-field case, note that once an eddy is stretched in position space to a shape as long as the variation scale of the electric field, the Taylor expansion \eqref{elc_increment} is no longer valid, and the two ends of the eddy are sheared by local electric-field variations with different values of the strain $\sigma$ (strain tensor $\sigma^{i}_{j}$ in 2D-2V and 3D-3V). This leads to rotation and stretching of the ends of the eddy with different aspect ratios, depending on the local value of $\sigma$. Furthermore, after a correlation time of the electric field, $\tau_{\mathrm{corr}}$, the field everywhere in space is `renewed.’ The local stretching rates are now different, and the above process resets and repeats.

We hypothesize that after a time $t \gg \tau_{\mathrm{corr}}$ during which the distribution function has undergone many sequences of local stretching ($\sigma > 0$) and rotation ($\sigma < 0$) events, the distribution function becomes thoroughly mixed in phase space. Indeed, this is what we observe in our numerical simulations in Section \ref{numeric_PS_eddies}.

With regard to the phase-space spectrum, the Kraichnan model \cite{adkins2018solvable, nastac2023phase} correctly reproduces the scaling \eqref{F_ks_solved}, but the spectrum from the static-field model, while capturing the integrated 1D spectra \eqref{static_1D_k} and \eqref{static_1D_s}, lacks any structure in the $(k,s)$ plane away from the critical-balance line $s = \gamma^{-1} k$. This is intuitive, given that the solution \eqref{static_1D_s} is the spectrum associated with a single stretching event (a single `step' in the cascade), while the spectrum in the Kraichnan model arises from many stretching events and, therefore, fills the entire~$(k,s)$ plane.

We do not attempt any explicit calculations of phase-space stretching and turbulent statistics of particles accelerated by finite-time correlated fields (beyond our turbulence theory in Section \ref{phenom_theory} and Appendix \ref{PT_hd}). Quite general results on the statistics of stretching in the Batchelor regime of fluid passive-scalar advection can be obtained \cite{chertkov1995, balkovsky1999universal, falkovich2001particles} using random dynamical systems theory \cite{furstenberg1960products, oseledec1968multiplicative}. An interesting avenue for future work would be to pursue analogous calculations for the passive acceleration of particles in the plasma-kinetic Batchelor regime, which would help bridge the gap between the static-field and white-noise-field cases.

Note that the calculations presented in this appendix exclusively involved particles that are passively accelerated by external fields. Strictly speaking, these calculations do not apply directly to phase-space turbulence with self-consistent fields. However, the intuition gained from these calculations still sheds light on self-consistent turbulence. This is because, as demonstrated by the theory in Section \ref{phenom_theory} and Appendix~\ref{PT_hd}, phase-space turbulence in the Vlasov–Poisson system naturally falls into the Batchelor regime.

Finally, we remark that in self-consistent turbulence, whether phase-space eddies rotate or stretch at a given point in space is related to the electron density fluctuation at that point. Namely, the Lagrangian rate-of-strain tensor [cf. \eqref{sigma}]
\begin{equation}
    \sigma^{i}_{j} = \frac{e}{m}\frac{\partial^2 \varphi}{\partial x_i \partial x_j}
\end{equation}
satisfies, via Poisson's equation \eqref{gauss}, 
\begin{equation}
    \mathrm{Tr}(\sigma) = \frac{e}{m}\nabla^2 \varphi = \frac{4 \pi e^2}{m} \delta n.
\end{equation}
In 1D, phase-space eddies are stretched in regions of over-density and rotated in regions of under-density. This explains why, e.g., electron holes \cite{roberts1967nonlinear, morse1969one, berk1970phase, schamel1986electron,ghizzo1988stability, hutchinson2017electron, hutchinson2024kinetic}, which have $\delta n < 0$, are considered rotating phase-space eddies in our framework. In 2D and 3D, the sum of the eigenvalues of $\sigma$ is determined by~$\delta n$, allowing for both stretching and rotation to occur in different directions at the same spatial point. In particular, when $\delta n > 0$, there must be at least one positive eigenvalue of $\sigma$---and hence at least one direction with stretching.

\section{Quasilinear evolution in the presence of a random external field} \label{QL_app}

In this section, we derive and solve the quasilinear equation \eqref{QL_evol} for the mean distribution function $f_0$ in the presence of a random external field. The solution is compared to our numerical simulations in Section \ref{Stochastic_heating_numerics}. Since our simulations are in 1D-1V, we restrict ourselves to one spatial and one velocity dimension, but this calculation can be straightforwardly generalized to higher dimensions.

\subsection{Derivation of quasilinear diffusion equation}

As in the main text, let $E_{\mathrm{tot}} = -\nabla \varphi_{\mathrm{tot}}$, where $\varphi_{\mathrm{tot}} = \varphi + \varphi_{\mathrm{ext}}$, with $\varphi$ the self-consistent electrostatic potential and $\varphi_{\mathrm{ext}}$ the external one. We will consider a forcing with an arbitrary number of Fourier modes and later restrict ourselves to a single Fourier mode at the box scale. Following \eqref{energy_density_ext_field}, we consider the Fourier modes of $\varphi_{\mathrm{ext}}$ to be statistically independent, finite-time-correlated Gaussian random fields with zero mean and corrrelation function
\begin{equation} \label{ext_phi_corr_tau}
    \frac{e^2}{m^2} \overline{\varphi_{\mathrm{ext},k}(t)\varphi^*_{\mathrm{ext},k'}(t')} = \frac{D_k}{k^2 \tau_{\mathrm{corr},k}} \delta_{k,k'} \, e^{-|\tau|/\tau_{\mathrm{corr},k}},
\end{equation}
where $D_k$ is the diffusion coefficient and $\tau_{\mathrm{corr},k}$ is the correlation time of wavenumber $k$.

We now Fourier transform in space and linearize \eqref{vlasov}, writing $f = f_0(v,t) + \delta \! f(x,v,t)$, where $f_0$ is the mean (ensemble-averaged) distribution function and $\delta \! f \ll f_0$ is the fluctuation, assumed small. We assume that the ensemble average is equivalent to a spatial average; $f_0$, therefore, does not depend on spatial position.

Using \eqref{gauss2}, we find
\begin{align} \label{QL_corr}
    \frac{\partial f_0}{\partial t} & =  \frac{\partial }{\partial v} \frac{e}{m} \sum_k i k \overline{ \varphi^*_{\mathrm{tot},k} \delta \! f_{k} } \nonumber \\ & = -\frac{\partial }{\partial v} \frac{e}{m} \sum_k k \operatorname{Im} \overline{ \varphi^*_{\mathrm{tot},k} \delta \! f_{k} },
\end{align}
\begin{equation} \label{delta_f_QL}
    \frac{\partial \delta \! f_{k}}{\partial t} + i k v \delta \! f_{k} = -i \frac{e}{m} k \varphi_{\mathrm{tot},k} \frac{\partial f_0}{\partial v},
\end{equation}
\begin{equation}
    \varphi_{\mathrm{tot},k} = -\frac{4 \pi e}{k^2} \int \mathrm{d} v \, \delta \! f_{k} + \varphi_{\mathrm{ext},k}.
\end{equation}
    Note that, to get the expression in the second line of~\eqref{QL_corr}, we split the wavenumber sum into two equal parts, took $k \rightarrow -k$ in one of them, and applied the reality conditions $\varphi^*_{\mathrm{tot},k} = \varphi_{\mathrm{tot},-k}$ and $\delta \! f^{*}_{k} = \delta \! f_{-k}$. Also, in~\eqref{delta_f_QL}, we dropped the $\partial f_0/\partial t$ term on the assumption that $f_0$ evolves slowly compared to $\delta \! f$. The goal now is to compute the correlation function in \eqref{QL_corr} to get a closed equation for~$f_0$.

We assume that there is no initial $\delta \! f_{k}$ and $f_0$ is linearly stable both initially and at all times during the quasilinear evolution. Under these assumptions, we can solve~\eqref{delta_f_QL} for $\delta \! f_{k}$ via a Fourier transform in time:
\begin{equation} \label{delta_f_omega}
    \delta \! f_{k} = -\frac{e}{m} \int \mathrm{d} \omega \, e^{-i \omega t} \, \varphi_{\mathrm{tot}, k, \omega} \frac{\partial f_0 /\partial v}{v- \omega/k -i 0^{+}},
\end{equation}
where the time-Fourier-transformed component of the electrostatic potential is
\begin{equation} \label{phi_omega}
    \varphi_{\mathrm{tot}, k, \omega} = \frac{\varphi_{\mathrm{ext},k,\omega}}{\epsilon(\omega,k)},
\end{equation}
and the dielectric function is
\begin{equation} \label{dielectric_omega}
    \epsilon(\omega,k) \equiv 1 - \frac{\omega_{\mathrm{pe}}^2}{k^2 n_0} \int \mathrm{d} v' \, \frac{\partial f_0 /\partial v'}{v'- \omega/k -i0^{+}}.
\end{equation}
In \eqref{delta_f_omega} and \eqref{dielectric_omega}, the frequency and velocity integrals must be taken along the Landau contour \cite{landau1946vibrations}, circumnavigating the pole at $v' = \omega/k$ from below (assuming $k > 0$; otherwise, the pole is circumnavigated from above). This prescription is enforced by the symbol $-i0^{+}$ (replaced by $+i0^{+}$ when $k < 0$).

Combining \eqref{delta_f_omega}, \eqref{phi_omega}, and \eqref{dielectric_omega}, we have
\begin{align} \label{corr_phi_delta_f}
    \overline{ \varphi^*_{\mathrm{tot},k} \delta \! f_{k} } = -\frac{e}{m} \iint \mathrm{d} \omega' \, \mathrm{d} \omega \, & e^{-i(\omega-\omega')t}  \frac{\overline{ \varphi^*_{\mathrm{ext},k,\omega'} \varphi_{\mathrm{ext},k,\omega}  }}{\epsilon^*(\omega',k) \epsilon(\omega,k)} \nonumber \\ & \times \frac{\partial f_0 /\partial v}{v- \omega/k -i0^{+}}.
\end{align}
Using \eqref{ext_phi_corr_tau}, we can compute the correlation function of the external potential in frequency space:
\begin{align} \label{corr_func_phi_ext_freq}
  & \frac{e^2}{m^2} \overline{ \varphi^*_{\mathrm{ext},k,\omega'} \varphi_{\mathrm{ext},k,\omega}  } \nonumber   \\ &= \frac{1}{(2 \pi)^2} \iint \mathrm{d} t' \, \mathrm{d} t e^{-i \omega' t'} e^{i \omega t} \overline{ \varphi^*_{\mathrm{ext},k}(t') \varphi_{\mathrm{ext},k}(t) } \nonumber \\ & = \frac{D_k }{(2 \pi)^2 k^2 \tau_{\mathrm{corr},k}}  \int \mathrm{d} T e^{i(\omega-\omega')T} \nonumber \\ & \times \int \mathrm{d} \tau \, e^{i (\omega + \omega') \tau/2-|\tau|/\tau_{\mathrm{corr},k}} \nonumber \\ & = \frac{D_{k} }{2 \pi k^2 \tau_{\mathrm{corr},k}} \delta(\omega-\omega') \int \mathrm{d} \tau \, e^{i \omega \tau-|\tau|/\tau_{\mathrm{corr},k}} \nonumber \\ &=  \frac{D_k}{\pi k^2} \frac{\delta(\omega-\omega ')}{(\omega \tau_{\mathrm{corr},k})^2 + 1},
\end{align}
where in the second line we changed variables $(t,t') \rightarrow (T,\tau) = ((t+t')/2,t-t')$. Combining \eqref{corr_phi_delta_f} and \eqref{corr_func_phi_ext_freq} gives
\begin{align} \label{corr_phi_delta_f_simpf}
    \overline{ \varphi^*_{\mathrm{tot}, k} \delta \! f_{k}} = -\frac{m}{e}\frac{D_k}{\pi k^2} \int \mathrm{d} \omega & \frac{1}{\left[(\omega \tau_{\mathrm{corr},k})^{2} + 1\right] |\epsilon(\omega, k)|^2}  \nonumber \\ & \times \frac{\partial f_0 /\partial v}{v- \omega/k -i0^{+}}.
\end{align}  
We need only the imaginary part of \eqref{corr_phi_delta_f_simpf} in \eqref{QL_corr}. It is given by the residue contribution in \eqref{corr_phi_delta_f_simpf} from the pole at $\omega = k v$:
\begin{equation}
    \operatorname{Im} \overline{ \varphi^*_{\mathrm{tot}, k} \delta \! f_{k}} = -\frac{m}{e} \frac{D_k}{k} \frac{1}{\left[(k v \, \tau_{\mathrm{corr},k})^{2} + 1\right] |\epsilon(k v, k)|^2} \frac{\partial f_0}{\partial v}. 
\end{equation}
Inserting this term into \eqref{QL_corr} yields the desired quasilinear diffusion equation:
\begin{equation} \label{QL_evol_general}
    \frac{\partial f_0}{\partial t} = \frac{\partial }{\partial v} D_{\mathrm{eff}}(v) \frac{\partial f_0}{\partial v},
\end{equation}
where the diffusion coefficient is
\begin{equation} \label{D_eff_general}
    D_{\mathrm{eff}}(v) = \sum_k \frac{D_k}{\left[(k v \, \tau_{\mathrm{corr},k})^{2} + 1\right] |\epsilon(k v, k)|^2}.
\end{equation}
If $\varphi_{\mathrm{ext}}$ has a single mode with $k = k_{\mathrm{f}}$, then $D_{\pm k} = D/2$, $\tau_{\mathrm{corr}, \pm k} = \tau_{\mathrm{corr}}$, and \eqref{QL_evol_general} and \eqref{D_eff_general} reduce to \eqref{QL_evol} and \eqref{D_eff}, respectively. Since this is the situation in our simulations, for the rest of this appendix, we will restrict ourselves to this case. The conclusions below can easily be generalized to an external forcing that has a spectrum of wavenumbers.

Before continuing, we remark that  `dressed diffusion equations' of similar form to (\ref{QL_evol_general}-\ref{D_eff_general}), which take into account the self-consistent response to forcing by an external stochastic field, have been derived and analyzed before in the kinetic theory of self-gravitating systems (see, e.g., \cite{chavanis2023secular} and references therein). In the plasma-physics literature, the only instance in which we are aware of such an equation being analyzed is \cite{banik2024universal}, where they focus on length scales much larger than the Debye length, while here we focus on length scales below the Debye length.

\subsection{Stochastic heating and self-similar flat-top distributions}

In Section \ref{Stochastic_heating}, we discussed the physics associated with the dielectric function $\epsilon$. There are naturally two asymptotic regimes where analytical progress is possible: $k_{\mathrm{f}} \lambda_{\mathrm{De}} \gg 1$ and $k_{\mathrm{f}} \lambda_{\mathrm{De}} \ll 1$. Here, we study the former regime, which is the regime accessed in the long-time limit of our simulations.

In the limit $k_{\mathrm{f}} \lambda_{\mathrm{De}} \gg 1$, since the velocity variation of $f_0$ is $v_{\mathrm{the}}$, the second term in the dielectric function \eqref{dielectric_omega} scales as
\begin{equation} \label{dielectric_estimate}
    \frac{\omega_{\mathrm{pe}}^2}{k^2 n_0} \int \mathrm{d} v' \, \frac{\partial f_0 /\partial v'}{v'- \omega/k -i0^{+}} \sim \frac{\omega_{\mathrm{pe}}^2}{k_{\mathrm{f}}^2 v_{\mathrm{the}}^2} = \frac{1}{k_{\mathrm{f}}^2 \lambda_{\mathrm{De}}^2},
\end{equation}
so $\epsilon = 1 + \mathcal{O}((k_{\mathrm{f}} \lambda_{\mathrm{De}})^{-2}) \simeq 1$ in this limit.  Note that in this approximation, there are no resonances where $\epsilon \simeq 0$, because Langmuir waves are strongly Landau-damped, and the roots of \eqref{dielectric_omega} are not close to the real axis.

As we discussed in Section \ref{Stochastic_heating}, if additionally $k_{\mathrm{f}} v_{\mathrm{the}} \, \tau_{\mathrm{corr}} \ll 1$, then \eqref{QL_evol} reduces to a diffusion equation with an approximately constant diffusion coefficient $D_{\mathrm{eff}}(v) \simeq D$. Then, $f_0$ is a spreading Maxwellian~\eqref{f0_diffusion}. However, as $f_0$ heats up, $v_{\mathrm{the}}$ grows, so the system is naturally pushed into the regime where $k_{\mathrm{f}} v_{\mathrm{the}} \, \tau_{\mathrm{corr}} \gg 1$. In this limit, $D_{\mathrm{eff}}(v) \simeq D/(k v \,\tau_{\mathrm{corr}})^2$, and $f_0$ satisfies
\begin{equation} \label{df0dt_FT}
\frac{\partial f_0}{\partial t} = \frac{\partial }{\partial v} \left(\frac{\alpha}{v^2} \frac{\partial f_0}{\partial v}  \right), \quad \alpha = \frac{D}{(k \tau_{\mathrm{corr}})^2}.
\end{equation}
This equation admits a similarity solution of the form
\begin{equation} \label{f0_ss_initial}
    f_0 = t^{-\zeta} \, g(\eta), \quad \eta = \frac{v^4}{\alpha t},
\end{equation}
where $\zeta$ can be inferred from the constraint that \eqref{f0_ss_initial} must conserve the particles' number density. Integrating~\eqref{f0_ss_initial} over velocity space, noting that \eqref{f0_ss_initial} is even in $v$, and changing variables $v \rightarrow \eta$ in the integral, we have
\begin{align} \label{ss_constraint}
    & \int^{+\infty}_{-\infty}  \mathrm{d} v \, f_0 = 2 \int^{\infty}_{0} \mathrm{d} v \, t^{-\zeta} \, g(\eta) \nonumber \\ &  = \frac{\alpha^{1/4}}{2} t^{1/4-\zeta} \int^{\infty}_{0} d \eta \, \eta^{-3/4} g(\eta) = n_0 = \mathrm{const}, 
\end{align}
which implies $\zeta = 1/4$.

Substituting \eqref{f0_ss_initial} into \eqref{df0dt_FT} gives us an ordinary differential equation for $g$:
\begin{equation} \label{g_eq}
    16 \eta \frac{d^2 g}{d \eta^2} + (\eta + 4)\frac{d g}{d \eta} + \frac{1}{4}g = 0.
\end{equation}
To solve this equation, let
\begin{equation} \label{h_def}
    h = \frac{d g}{d \eta} +  \frac{1}{16}g, 
\end{equation}
which, using \eqref{g_eq}, satisfies
\begin{equation} \label{h_eq}
    \frac{d h}{d \eta} \eta + \frac{1}{4} h = 0.
\end{equation}
Integrating this equation yields $h = c_1/\eta^{1/4}$, where $c_1$ is a constant. Inserting this solution into \eqref{h_def} and solving for $g$ yields
\begin{equation} \label{g_sol}
    g(\eta) = c_1 \int_{0}^{\eta} \mathrm{d} \eta' \, (\eta ')^{-1/4} e^{-(\eta-\eta ')/16} + g(0) e^{-\eta/16}.
\end{equation}

We must set $c_1 = 0$, because the term multiplying $c_1$ in \eqref{g_sol} would lead to a divergent density in \eqref{ss_constraint}. To see this, we compute its asymptotic at large $\eta$:
\begin{align} \label{g_two_int}
   & \int_{0}^{\eta} \mathrm{d} \eta' \, (\eta ')^{-1/4} e^{-(\eta-\eta ')/16} = \frac{1}{\eta^{1/4}} \int^{\eta}_{0} \mathrm{d} y \frac{e^{-y/16}}{(1-y/\eta)^{1/4}} \nonumber \\ & \approx \frac{1}{\eta^{1/4}} \int^{\infty}_{0} \mathrm{d} y e^{-y/16} = \frac{16}{\eta^{1/4}},
\end{align}
If we insert this asymptotic into \eqref{ss_constraint}, we get a logarithmic divergence. Therefore, we must have $c_1 = 0$.

Now, to find $g(0)$, we insert \eqref{g_sol} into \eqref{ss_constraint}, which yields $g(0) = n_0/\Gamma(1/4) \, \alpha^{1/4}$ and, therefore,
\begin{equation}
    g(\eta) = \frac{n_0}{\Gamma(1/4) \, \alpha^{1/4} } e^{-\eta/16},
\end{equation}
which, together with \eqref{f0_ss_initial}, yields \eqref{flat-top}.

We can also derive the heating profile \eqref{K_subdiffusion}. To do so, we combine the definition of kinetic-energy density in \eqref{K_diffusion} with \eqref{flat-top}, convert the integral over velocity space into one over only positive velocities, and make the variable change $v \rightarrow \sigma = v^4/16 \alpha t$:
\begin{align}
      K &= \int \mathrm{d} v \frac{m v^2}{2} f_0 = \frac{m n_0}{\Gamma(1/4) (\alpha t)^{1/4}} \int^{\infty}_{0} d \mathrm{d} \, v^2 e^{-\frac{v^4}{16 \alpha t}} \nonumber \\ &= \frac{2 m n_0 (\alpha t)^{1/2}}{\Gamma(1/4)} \int^{\infty}_{0} \mathrm{d} \sigma \, \sigma^{-1/4} e^{-\sigma} \nonumber \\ &= \frac{2 \, \Gamma(3/4)}{\Gamma(1/4)} m n_0 (\alpha t)^{1/2}.
\end{align}

Finally, we remark that it is well known in the theory of stochastic acceleration that the functional form of \eqref{ext_phi_corr_tau} controls the functional form of the diffusion coefficient~\eqref{D_eff_general} (in the case where collective effects are ignored, viz., $\epsilon = 1$), so changing the correlator \eqref{ext_phi_corr_tau} can lead to a myriad of different particle distribution functions \cite{sturrock1966stochastic,vanden1997some}. However, as far as we are aware, the flat-top distribution \eqref{flat-top} has never been published before.

\section{Phase-space turbulence above the Debye scale} \label{BD_app}

\subsection{Biglari--Diamond theory} \label{BD_above}

It has been known since the early simulations \cite{roberts1967nonlinear} of the two-stream instability in a 1D-1V Vlasov--Poisson plasma with dynamic electrons and stationary ions, and continues to be the case in the most recent ones \cite{ewart2024relaxation}, that strong phase-space turbulence above the Debye scale contains a dynamic population of phase-space holes. Biglari and Diamond \cite{biglari1988cascade,biglari1989clouds} proposed that such holes were the basic elementary turbulent structures whose dynamics determined the fluctuation spectra, and that this dynamics would be controlled, at each scale, by a local version of the hole instability \cite{dupree1982theory}.

In a simple model of this instability, a hole of width $u$ in velocity space and depth $\delta \! f_u$ added on top of a stable distribution goes unstable with the growth rate \cite{dupree1982theory}
\begin{equation} \label{dispersion_relation_hole}
  \gamma_{\mathrm{hole}}^2 = \frac{\omega_{\mathrm{pe}}^2 \delta \! f_u \, u}{n_0\epsilon_0}- \frac{\left(k u \right)^2}{4},
\end{equation}
where
\begin{equation} \label{epsilon_0}
    \epsilon_0 \equiv \epsilon(0, k) = 1 + \frac{1}{(k \lambda_{\mathrm{De}})^{2}}
\end{equation}
is the dielectric function \eqref{dielectric_omega} evaluated at $\omega = 0$. Biglari and Diamond assumed that a theory of turbulence could be obtained by  treating $\delta \! f_u$ as a typical perturbation at scale $u$ and marginalizing \eqref{dispersion_relation_hole}--they called this constraint `quasi-virialization,' a nomenclature that stemmed from the hole instability \eqref{dispersion_relation_hole} being an electrostatic plasma analogue of the gravitational Jeans instability \cite{jeans1902stability}. Using also $\epsilon_0 \approx (k \lambda_{\mathrm{De}})^{-2}$ at $k \lambda_{\mathrm{De}} \ll 1$, one gets then
\begin{align} \label{BD_s_spec}
     (k u)^2  &\sim \frac{\omega_{\mathrm{pe}}^2 \delta \! f_u \, u}{n_0 \epsilon_0} \sim k^2 \frac{v_{\mathrm{the}}^2}{n_0} \delta \! f_u \, u \nonumber \\  \implies \delta \! f_u &\sim \frac{n_0}{v_{\mathrm{the}}^2} u \iff \int \mathrm{d} k \, \langle P_{\delta \! f}  \rangle_{\mathrm{SA}} (k, s) \propto s^{-3}
\end{align}
(this scaling is not explicitly given in their paper but the logic of their argument demands it).

Finally, the velocity-space increments can be converted into position-space ones by assuming a constant-flux cascade of $C_2$ with the cascade time set by either of the two terms in \eqref{dispersion_relation_hole}  [which are, by \eqref{BD_s_spec}, comparable]: letting $\tau_{\mathrm{c}}^{-1} \sim k u \sim u/r$, one gets
\begin{equation} \label{CB_BD}
	\varepsilon  \sim \frac{v_{\mathrm{the}} \delta \! f^2_u}{\tau_{\mathrm{c}}} \sim \frac{n_0^2}{v_{\mathrm{the}}} \frac{u^3}{r}  \implies u \propto r^{1/3}, \, \tau_{\mathrm{c}} \propto r^{2/3}. 
\end{equation} 
Finally, for $r$ and $u$ related in this way,
\begin{equation} \label{BD_k_spec}
	\delta \! f_r \sim \delta \! f_u \propto r^{1/3} \iff \int \mathrm{d} s \, \langle P_{\delta \! f}  \rangle_{\mathrm{SA}} (k, s) \propto k^{-5/3},
\end{equation}
which is the headline result of \cite{biglari1988cascade,biglari1989clouds}.

The assumption of marginalization of \eqref{dispersion_relation_hole} (`quasi-virialization') is, of course, equivalent to the critical-balance conjecture. Indeed, physically, \eqref{dispersion_relation_hole} represents the competition between particle trapping, which is destabilizing, and particle streaming, which is stabilizing. In terms of the linear and nonlinear times, \eqref{dispersion_relation_hole} can be rewritten as
\begin{equation} \label{dispersion_relation_hole_time_scales}
   \gamma^2_{\mathrm{hole}} \sim \frac{1}{\tau_{\mathrm{nl, \mathrm{BD}}} \tau_{\mathrm{l}}}- \frac{1}{\tau_{\mathrm{l}}^2},
\end{equation}
so marginalizing it implies the critical balance
\begin{equation} \label{QV_CB}
    \tau_{\mathrm{l}} \sim \tau_{\mathrm{nl, \mathrm{BD}}},
\end{equation}
where $\tau_{\mathrm{l}}$ is again given by \eqref{tau_linear}, but the nonlinear time is calculated differently from \eqref{NL_time}, in two ways. First, the electric-field increment in the nonlinear time is `dressed' by a factor of $1/\epsilon_0$ to account for Debye shielding:  
\begin{equation} \label{NL_time_dressed}
    \tau_{\mathrm{nl}, \mathrm{BD}} \sim \frac{u}{(e/m) \delta E_r/\epsilon_0}.
\end{equation}
Secondly, $\delta E_r$ is related to the density increment $\delta n_r$ via Poisson's law \eqref{E_r_Gauss}, where $\delta n_r$ is now computed as
\begin{equation} \label{BD_integral_velocity_rule}
    \delta n_r \sim \int  \mathrm{d} v \, \delta \! f_r \sim u \, \delta \! f_r
\end{equation}
[note that $\delta \!f _r \sim \delta \!f_u$ as long as $u$ and $r$ are related via critical balance and the constant-flux cascade]. The prescription \eqref{BD_integral_velocity_rule} is, effectively, an assumption that $\delta \!f$ at each scale contains a single hole (or a finite number of isolated ones) of width $u$. This is, in fact, the only part of Biglari and Diamond's theory where the idea that the phase-space cascade is a cascade of holes is needed. The prescription \eqref{BD_integral_velocity_rule} is to be contrasted to our theory's \eqref{delta_n_r}, where the velocity-space integral of $\delta \! f$ accumulates as a random walk instead.

For the electric field, combining \eqref{BD_integral_velocity_rule} with \eqref{E_r_Gauss}, \eqref{CB_BD}, and \eqref{BD_k_spec}, one gets
\begin{equation} \label{BD_Ek}
   \delta E_r \propto r^{5/3}, \quad \langle P_{E}\rangle_{\mathrm{SA}} (k) \propto k^{-13/3}.
\end{equation}

Biglari and Diamond's scalings, \eqref{BD_s_spec}, \eqref{CB_BD}, \eqref{BD_k_spec}, and~\eqref{BD_Ek}, are formally consistent, in that a local-in-scale cascade from large to small $r$ is assumed and indeed it turns out that the cascade time gets shorter with decreasing scale [see \eqref{CB_BD}]. However, a number of physically significant questions hang over the validity of this whole approach. In a strongly nonlinear, turbulent system, is Debye shielding at $k \lambda_{\mathrm{De}} \ll 1$ correctly captured, even qualitatively, by dressing the electric field in \eqref{NL_time_dressed} with the dielectric function \eqref{epsilon_0} computed from linear theory? Are the density fluctuations in a hole-dominated turbulence correctly estimated by the prescription \eqref{BD_integral_velocity_rule}, i.e., by assuming just a few holes per scale? Perhaps most importantly, is the formation of holes of different sizes a large-to-small-scale cascade process at all? Indeed, in, e.g., numerical simulations of the nonlinear evolution of the two-stream instability \cite{roberts1967nonlinear,ewart2024relaxation}, it does not appear that smaller holes form as a result of breaking up of larger holes (see, however, \cite{carril2023formation})—rather, in the long-time limit of such simulations, holes merge until a single box-scale hole remains (whether this merging process can be understood in terms of standard turbulence-theory paradigms, e.g., an inverse cascade of some invariant, is an interesting question for future work). Biglari and Diamond conjectured that perhaps their theory did not hold in this long-time limit, but it was left unclear how one might work out some intermediate time scale on which it would apply. All these doubts are given some additional weight by the fact that, as far as we are aware, there is no numerical evidence so far to support their theory.

Two further  caveats concern the applicability of this approach to the `real world.' First, it remains to be seen whether a turbulence containing many phase-space holes can exist in more than one dimension, given that BGK modes require additional symmetries to exist in two and three dimensions \cite{ng2005bernstein,ng2006weakly} and are thus less likely to be stable. 
Secondly, if the frequencies of fluctuations drop below the plasma frequency above the Debye scale, it is no longer a good approximation to ignore ion dynamics. We shall offer some speculations on alternative approaches that address both of these concerns in Appendices \ref{rw_above} and \ref{IS_sect}.

\subsection{A cascade of holes below the Debye scale?} \label{BD_below}

While Biglari and Diamond's theory \cite{biglari1988cascade,biglari1989clouds} was intended to describe turbulence above the Debye scale, it is useful to consider what their approach would imply below the Debye scale and whether it is a viable alternative  to our theory. The difference in \eqref{NL_time_dressed} is that there is no Debye shielding at $k \lambda_{\mathrm{De}} \gg 1$, so $\epsilon_0 \sim 1$. Combining this with \eqref{E_r_Gauss} and \eqref{BD_integral_velocity_rule} gives us the nonlinear time, and hence cascade time:
\begin{equation}
    \tau_{\mathrm{c}} \sim \tau_{\mathrm{nl, \mathrm{BD}}} \sim \frac{n_0}{\omega_{\mathrm{pe}}^2 \delta \! f_r r}.
\end{equation}
Assuming, as always, a constant-flux cascade of $C_2$ in position space yields
\begin{align} \label{tau_spec_BD_below_Debye}
    & \frac{v_{\mathrm{the}} \delta \! f_{r}^2}{\tau_{\mathrm{c}}} \sim \varepsilon \implies \nonumber \\ & \delta \! f_{r} \propto r^{-1/3} \iff \int \mathrm{d} s \, \langle P_{\delta \! f} \rangle_{\mathrm{SA}}(k,s) \propto k^{-1/3}.
\end{align}
This is, indeed, the scaling that was obtained under the same assumptions by Diamond et al. \cite{diamond2010modern}. It implies, however, that $\tau_{\mathrm{c}} \propto r^{-2/3}$ gets longer at smaller scales, so, in fact, a local-in-scale cascade cannot occur—it will be overwhelmed by the mixing of $\delta \! f$ by outer-scale electric fields, i.e., the system will be in the Batchelor regime. This is the same argument by which, in Appendix \ref{PT_hd}, we ruled out the possibility of a local-in-scale cascade with non-smooth electric fields within our own theory.

\subsection{Phase-space cascade with Debye shielding} \label{rw_above}

Given the multiple caveats casting doubt on the notion of a cascade of holes, encoded in the prescription \eqref{BD_integral_velocity_rule} for the relationship between $\delta n_r$ and $\delta \! f_r$, we would like to examine what happens if we restore our own, random-walk prescription \eqref{delta_n_r} but keep Biglari and Diamond's assumption that Debye shielding at scales above $\lambda_{\mathrm{De}}$ can be accounted for by the factor of $\epsilon_0^{-1}$ in the nonlinear time~\eqref{NL_time_dressed}.

As in Appendix \ref{PT_hd}, let $(e/m) \delta E_r \sim \gamma^2 r^{\beta}$, where $\beta$ must be determined self-consistently. Critical balance~\eqref{QV_CB} above the Debye scale, where $\epsilon_0 \sim (r/\lambda_{\mathrm{De}})^2$,  implies that
\begin{align} \label{CB_appE_3}
    & u \sim \gamma \lambda_{\mathrm{De}} r^{(\beta-1)/2} \implies \nonumber \\ & \tau_{\mathrm{c}} \sim (\gamma \lambda_{\mathrm{De}})^{-1} r^{(3-\beta)/2} \sim (\gamma \lambda_{\mathrm{De}})^{-2/(\beta-1)} u^{(3-\beta)/(\beta-1)}.
\end{align}
Assuming a constant-flux cascade of $\delta C_2$ in position space gives
\begin{align} \label{spec_k_const_flux_above_debye_shielding_rw}
    & \frac{v_{\mathrm{the}}^d \delta \! f_{r}^2}{\tau_{\mathrm{c}}} \sim \varepsilon \implies \nonumber \\  &  \delta \! f_{r} \propto r^{(3-\beta)/4} \iff \int \mathrm{d} s \, \langle P_{\delta \! f} \rangle_{\mathrm{SA}}(k,s) \propto k^{-(5-\beta)/2}.
\end{align}
Combining \eqref{E_r_Gauss}, \eqref{delta_n_r}, \eqref{CB_appE_3}, and \eqref{spec_k_const_flux_above_debye_shielding_rw} implies that electric field increments scale as
\begin{equation} \label{E_inc_rw_above}
    \delta E_r \propto r^{[7-d + (d-1)\beta] /4},
\end{equation}
which, upon equating to $r^{\beta}$, implies
\begin{equation} \label{beta_appE}
    \beta = \frac{7-d}{5-d}.
\end{equation}
This evaluates to $\beta = 3/2, 5/3$, and $2$ for $d = 1, 2,$ and $3$, respectively. The electric-field spectrum is, therefore,
\begin{equation} \label{spec_E_super_Debye}
    \langle P_{\mathbf{E}}\rangle_{\mathrm{SA}} (k) \sim \frac{\varepsilon \omega_{\mathrm{pe}}^4 \left( \gamma \lambda_{\mathrm{De}} \right)^{d-1}}{n_0^2} k^{-(19-3d)/(5-d)},
\end{equation}
which is $\propto k^{-4}$, $k^{-13/3}$, and $k^{-5}$ for $d = 1, 2,$ and $3$, respectively. Interestingly, \eqref{spec_E_super_Debye} is the same as the sub-Debye electric-field spectrum \eqref{Ek} for $d = 1$, but not for $d = 2$ or $3$.

For $\beta$ given by \eqref{beta_appE}, \eqref{CB_appE_3} becomes
\begin{align} \label{CB_appE_beta}
	& u \sim \gamma \lambda_{\mathrm{De}} r^{1/(5-d)} \implies \nonumber \\ & \tau_{\mathrm{c}} \sim (\gamma \lambda_{\mathrm{De}})^{-1} r^{(4-d)/(5-d)} \sim (\gamma \lambda_{\mathrm{De}})^{d-5} u^{4-d}.
\end{align}
Thus, the cascade time gets shorter at smaller scales, so a local-in-scale cascade is consistent. Analogously to the calculations in Section \ref{phenom_theory}, it is now straightforward to show that the 2D phase-space spectrum is
\begin{align} \label{F_ks_shielding}
&\langle P_{\delta \! f} \rangle_{\mathrm{SA}}(k,s) \sim \nonumber \\ & \varepsilon
\begin{cases} 
     k^{d-1} \left(\gamma \lambda_{\mathrm{De}} s\right)^{-(1+d)(5-d)} , & k \ll  \left(\gamma \lambda_{\mathrm{De}} s \right)^{5-d}, \\
      \left(\gamma \lambda_{\mathrm{De}} s \right)^{d-1} k^{-(9-d)/(5-d)} , &  k \gg \left( \gamma \lambda_{\mathrm{De}} s \right)^{5-d}.
   \end{cases}
\end{align}
The corresponding 1D spectra
\begin{align} \label{shielding_1D_spectra}
	& \int \mathrm{d} s \, \langle P_{\delta \! f} \rangle_{\mathrm{SA}}(k,s) \sim \varepsilon \left( \gamma \lambda_{\mathrm{De}} \right)^{-1} \, k^{-(9-2d)/(5-d)}, \nonumber \\ & \int \mathrm{d} k \, \langle P_{\delta \! f} \rangle_{\mathrm{SA}}(k,s) \sim \varepsilon \left( \gamma \lambda_{\mathrm{De}} \right)^{d-5} \, s^{-(5-d)},
\end{align}
are consistent with \eqref{spec_k_const_flux_above_debye_shielding_rw} and a constant-flux cascade of~$\delta C_2$ in velocity space, respectively.

Like in the case of the Biglari--Diamond theory, there is currently no conclusive numerical evidence to support these scalings, but it is intriguing that the electric-field spectra reported in simulations of the 1D-1V two-stream instability \cite{ewart2024relaxation} seem to extend some distance into the $k \lambda_{\mathrm{De}} < 1$ range without interruption of the $k^{-4}$ scaling, as they would do if \eqref{spec_E_super_Debye} were correct.

\subsection{Ion turbulence above the Debye scale} \label{IS_sect}

\subsubsection{Ion kinetics and Boltzmann electrons}

Throughout this paper, we have limited ourselves to the high-frequency limit of the Vlasov--Poisson system, in which electrons are dynamic but ions are assumed to be stationary. Outside this limit, one must generally consider ion and electron dynamics at the same time. However, it is possible to consider the opposite limit of low-frequency fluctuations, in which only ion kinetics need to be solved for. Namely, consider fluctuations with frequencies $\omega \sim k \, v_{\mathrm{thi}}$, where $v_{\mathrm{thi}} = \sqrt{T_{\mathrm{i}}/m_{\mathrm{i}}}$ is ion thermal velocity, $T_{\mathrm{i}}$ is the ion temperature, and $m_{\mathrm{i}}$ is the ion mass. On these time scales, assuming $T_{\mathrm{i}} \sim T_{\mathrm{e}}$, since $v_{\mathrm{the}} \gg v_{\mathrm{thi}}$, the electron distribution function approximately satisfies
\begin{equation} \label{fast_fe}
	\mathbf{v} \cdot \nabla f_{\mathrm{e}} + \frac{e}{m} \nabla \varphi \cdot \nabla_{\mathbf{v}} f_{\mathrm{e}} = 0,
\end{equation}
ignoring collisions and external forcing. The solution to \eqref{fast_fe} is $f_{\mathrm{e}} = f_{\mathrm{e}}(\mathcal{E})$, any function of the particle energy	$\mathcal{E} = m_{\mathrm{e}} |\mathbf{v}|^2/2-e\varphi$. For simplicity, let
\begin{equation} \label{Maxwellian_electrons}
    f_{\mathrm{e}} = \frac{n_0}{(2 \pi v_{\mathrm{the}})^{3/2}}\exp{\left[-\frac{1}{T_{\mathrm{e}}}\left(\frac{m_{\mathrm{e}} |\mathbf{v}|^2}{2}-e\varphi\right)\right]}.
\end{equation}
The density of the electrons is then given by the `Boltzmann response':
\begin{equation} \label{Boltzmann_response}
    n_e = n_0 e^{e\varphi/T_{\mathrm{e}}} \approx n_0\left(1 + \frac{e \varphi}{T_{\mathrm{e}}}\right),
\end{equation}
assuming $e \varphi/T_{\mathrm{e}} \ll 1$ (the same follows for a general $f_{\mathrm{e}}(\mathcal{E})$ for a suitably defined $T_{\mathrm{e}}$). Above the Debye scale, $k \lambda_{\mathrm{De}, \mathrm{i}} \ll 1$, the right-hand side of Poisson's equation (assuming ion charge $e$)
\begin{equation}
	-\lambda_{\mathrm{De}}^2 \nabla^2 \frac{e \varphi}{T_{\mathrm{e}}} = \frac{n_{\mathrm{i}}-n_{\mathrm{e}}}{n_0}
\end{equation}
dominates over the left-hand side, and so the electrostatic potential is determined by quasineutrality and \eqref{Boltzmann_response}:
\begin{equation} \label{quasi-neut}
	\frac{\delta n_i}{n_0} = \frac{\delta n_e}{n_0} = \frac{e \varphi}{T_{\mathrm{e}}}.
\end{equation}
This closure can now be coupled to the Vlasov equation for the ion distribution function:
\begin{equation} \label{vlasov_i}
    \frac{\partial f_{\mathrm{i}}}{\partial t} + \textbf{v} \cdot \nabla f_{\mathrm{i}} + \frac{e}{m_{\mathrm{i}}}  \mathbf{E} \cdot \nabla_{\mathbf{v}} f_{\mathrm{i}} = 0,
\end{equation}
where $\mathbf{E} = -\nabla \varphi$.

We imagine turbulence in \eqref{vlasov_i}-\eqref{quasi-neut} to be excited by external forcing or an instability, driven at length scales much greater than the Debye scale and at frequencies much lower than the plasma frequency. In the former case, we would append an external potential to \eqref{vlasov_i}-\eqref{quasi-neut}, taking $\varphi \rightarrow \varphi + \varphi_{\mathrm{ext}}$. Alternatively, if one allows for electron dynamics beyond \eqref{Maxwellian_electrons}, one can derive a kinetic version of Zakharov’s equations \cite{zakharov1972collapse} (see \cite{schekochihin2022lectures} for a tutorial exposition). In this model, on ion time scales, the electron-density perturbation \eqref{quasi-neut} contains both the Boltzmann response \eqref{quasi-neut} and the ponderomotive potential, which is essentially an effective pressure due to Langmuir oscillations. This ponderomotive potential, which couples ion and electron dynamics, can also drive ion-scale turbulence.

\subsubsection{Ion phase-space cascade}

Let us seek scaling laws for a critically balanced, constant-flux phase-space cascade in the system \eqref{vlasov_i}-\eqref{quasi-neut}. As in Appendices \ref{PT_hd} and \ref{rw_above}, let $(e/m_{\mathrm{i}}) \delta E_r \sim \gamma^2 r^{\beta}$, where $\beta$ must be determined self-consistently. For such an electric field, the critical-balance condition \eqref{CB} yields the scale-dependent cascade time \eqref{CB_line_rough}. Similar to our theory for the electron turbulence, we seek a shell-averaged fluctuation spectrum of $\delta C_2$ in the form
\begin{equation} \label{F_ks_ions_rough}
	\langle P_{\delta \! f_{\mathrm{i}}} \rangle_{\mathrm{SA}}(k, s) \propto
	\begin{cases} 
		k^{d-1} s^{-a} , \qquad & k \ll \left(\gamma s \right)^{2/(1+\beta)}, \\
		s^{d-1} k^{-b}, \qquad & k \gg \left(\gamma s \right)^{2/(1+\beta)}.
	\end{cases}
\end{equation}

The key difference between \eqref{vlasov}-\eqref{gauss} and \eqref{vlasov_i}-\eqref{quasi-neut} is in the difference between Poisson's equation \eqref{gauss} and quasineutrality \eqref{quasi-neut}: the former implies $P_{\mathbf{E}}(\mathbf{k}) \propto k^{-2} P_{\delta n_{\mathrm{e}}}(\mathbf{k})$, while the latter implies $P_{\mathbf{E}}(\mathbf{k}) \propto k^{2} P_{\delta n_{\mathrm{i}}}(\mathbf{k})$. The electric field spectrum consistent with \eqref{F_ks_ions_rough} and \eqref{quasi-neut}, is, therefore,
 \begin{equation} \label{self_consistent_spec_ions}
 	\langle P_{\mathbf{E}} \rangle_{\mathrm{SA}} (k) \propto k^{2-b}.
 \end{equation}
 Combining \eqref{self_consistent_spec_ions} and \eqref{relation_inc_spec} implies that, if the electric field is self-consistently generated by $\delta \! f_{\mathrm{i}}$,
 \begin{equation}
 b = 2\beta + 3.
 \end{equation}
 
 Assuming a constant-flux cascade of $\delta C_2$ in position space and using \eqref{CB_line_rough} for the cascade time gives
 \begin{align} \label{spec_k_const_flux_ions}
 	& \frac{v_{\mathrm{thi}}^d \delta \! f_{\mathrm{i}, r}^2}{\tau_{\mathrm{c}}} \sim \varepsilon \implies \nonumber \\  &  \delta \! f_{\mathrm{i}, r} \propto r^{(1-\beta)/4} \iff \int \mathrm{d} s \, \langle P_{\delta \! f_{\mathrm{i}}} \rangle_{\mathrm{SA}}(k,s) \propto k^{-(3-\beta)/2}.
 \end{align}
 To find $\beta$, we demand that integrating the 2D spectrum~\eqref{F_ks_ions_rough} over $s$ yield the same scaling as in \eqref{spec_k_const_flux_ions}. Assuming that the dominant contribution to the integral comes from integrating \eqref{F_ks_ions_rough} from $s = 0$ up to the critical-balance line $s \sim \gamma^{-1}k^{(1+\beta)/2}$, we get that
 \begin{align} \label{spec_k_CB_line_ions}
 	&\int_{0}^{\gamma^{-1}k^{(1+\beta)/2}} \mathrm{d} s \,  \langle P_{\delta \! f_{\mathrm{i}}} \rangle_{\mathrm{SA}}(k,s) \propto k^{[d-6-(4-d)\beta]/2}.
 \end{align}
Equating \eqref{spec_k_const_flux_ions} and \eqref{spec_k_CB_line_ions} gives us
\begin{equation} \label{beta_ions}
	\beta = \frac{3-d}{5-d} \implies b = \frac{9-d}{5-d}.
\end{equation}

Thus, the electric fields are very rough:~$\delta E_r \propto r^{(3-d)/(5-d)}$, which is~$\propto r^{-1/2}$, $r^{-1/3}$, and $r^{0}$ for $d = 1, 2,$ and $3$, respectively. Their spectra are
\begin{equation} \label{spec_E_i}
	\frac{e^2}{T_{\mathrm{e}}^2} \langle P_{\mathbf{E}}\rangle_{\mathrm{SA}} (k) \sim \frac{\varepsilon \,  \gamma^{d-1}}{n_0^2} k^{-(d-1)/(5-d)},
\end{equation}
which is $\propto k^{0}$, $k^{-1/3}$, and $k^{-1}$ for $d = 1, 2,$ and $3$, respectively [here $P_{\mathbf{E}}$ is defined as in \eqref{P_E} but without the factor of $e^2/m_{\mathrm{e}}^2$].

Inserting \eqref{beta_ions} into \eqref{CB_line_rough} gives the relationship between position- and velocity-space scales and the cascade time:
\begin{align} \label{CB_line_rough_ions}
	u \sim \gamma \, r^{1/(5-d)} & \implies  \nonumber \\  & \tau_{\mathrm{c}} \sim  \gamma^{-1}r^{(4-d)/(5-d)} \sim \gamma^{d-5} u^{4-d}.
\end{align}
The cascade time gets shorter at smaller scales, so the ion cascade is local in scale.

To find the remaining exponent $a$ in \eqref{F_ks_ions_rough}, we can demand that the two asymptotics in \eqref{F_ks_ions} match along the critical balance line $k \sim  \left(\gamma s \right)^{5-d}$ \eqref{CB_line_rough_ions}. This yields $a = (1+d)(5-d)$, and, therefore,
\begin{equation} \label{F_ks_ions}
	\langle P_{\delta \! f_{\mathrm{i}}} \rangle_{\mathrm{SA}}(k,s) \sim\varepsilon
	\begin{cases} 
		 k^{d-1} \left( \gamma s \right)^{-(1+d)(5-d)} , & k \ll  \left(\gamma s \right)^{5-d}, \\
	 \left(\gamma s \right)^{d-1} k^{-(9-d)/(5-d)} , &  k \gg \left( \gamma s \right)^{5-d}.
	\end{cases}
\end{equation}
The high-$s$ asymptotic of \eqref{F_ks_ions} is easily shown to be consistent with a constant-flux cascade of $\delta C_2$ in velocity space.

The 2D phase-space spectrum \eqref{F_ks_ions} has the same scaling laws as \eqref{F_ks_shielding}. To see why, consider the electric-field increments in the ion system: using \eqref{quasi-neut}, they are
\begin{equation}
    \frac{e}{m_{\mathrm{i}}}\delta E_r \sim \frac{e}{m_{\mathrm{i}}} \frac{\delta \varphi_r}{r} \sim \frac{T_{\mathrm{e}} }{m_{\mathrm{i}}} \frac{\delta n_{i, r}}{n_0} r^{-1}.
\end{equation}
Therefore, the nonlinear time is
\begin{equation} \label{tau_nl_ions}
    \tau_{\mathrm{nl}} \sim \frac{u}{(e/m_{\mathrm{i}}) \delta E_r} \sim \frac{T_{\mathrm{i}}}{T_{\mathrm{e}}}\frac{n_0}{v_{\mathrm{thi}}^2} \frac{u \, r}{\delta n_{\mathrm{i}, r}}.
\end{equation}
This reduces to the cascade time in \eqref{CB_line_rough_ions} if one combines the ion analogue of the random-walk prescription \eqref{delta_n_r} and critical balance \eqref{CB_line_rough_ions}. With respect to the density fluctuations, up to order-unity factors, \eqref{tau_nl_ions} scales in exactly the same way as the dressed nonlinear time \eqref{NL_time_dressed} above the Debye scale: via \eqref{E_r_Gauss},
\begin{equation} \label{tau_NL_BD_density}
  \tau_{\mathrm{nl},\mathrm{BD}} \sim \frac{n_0}{v_{\mathrm{the}}^2} \frac{u \, r}{\delta n_{\mathrm{e}, r}}.
\end{equation}
The cascade theories in this section and in Appendix \ref{rw_above} thus give the same scalings for the $\delta C_2$ spectra \eqref{F_ks_shielding} and \eqref{F_ks_ions}, but the electric-field spectra \eqref{spec_E_i} and \eqref{spec_E_super_Debye} are different (by four powers of $k$) because of the difference between \eqref{quasi-neut} and Poisson's equation \eqref{gauss}.

We leave further investigation of the predictions \eqref{spec_E_i} and \eqref{F_ks_ions}, as well as of ion and multispecies phase-space turbulence more generally, to future work.

\bibliography{paper}

\providecommand{\noopsort}[1]{}\providecommand{\singleletter}[1]{#1}%
\begin{thebibliography}{179}%
\makeatletter
\providecommand \@ifxundefined [1]{%
 \@ifx{#1\undefined}
}%
\providecommand \@ifnum [1]{%
 \ifnum #1\expandafter \@firstoftwo
 \else \expandafter \@secondoftwo
 \fi
}%
\providecommand \@ifx [1]{%
 \ifx #1\expandafter \@firstoftwo
 \else \expandafter \@secondoftwo
 \fi
}%
\providecommand \natexlab [1]{#1}%
\providecommand \enquote  [1]{``#1''}%
\providecommand \bibnamefont  [1]{#1}%
\providecommand \bibfnamefont [1]{#1}%
\providecommand \citenamefont [1]{#1}%
\providecommand \href@noop [0]{\@secondoftwo}%
\providecommand \href [0]{\begingroup \@sanitize@url \@href}%
\providecommand \@href[1]{\@@startlink{#1}\@@href}%
\providecommand \@@href[1]{\endgroup#1\@@endlink}%
\providecommand \@sanitize@url [0]{\catcode `\\12\catcode `\$12\catcode
  `\&12\catcode `\#12\catcode `\^12\catcode `\_12\catcode `\%12\relax}%
\providecommand \@@startlink[1]{}%
\providecommand \@@endlink[0]{}%
\providecommand \url  [0]{\begingroup\@sanitize@url \@url }%
\providecommand \@url [1]{\endgroup\@href {#1}{\urlprefix }}%
\providecommand \urlprefix  [0]{URL }%
\providecommand \Eprint [0]{\href }%
\providecommand \doibase [0]{https://doi.org/}%
\providecommand \selectlanguage [0]{\@gobble}%
\providecommand \bibinfo  [0]{\@secondoftwo}%
\providecommand \bibfield  [0]{\@secondoftwo}%
\providecommand \translation [1]{[#1]}%
\providecommand \BibitemOpen [0]{}%
\providecommand \bibitemStop [0]{}%
\providecommand \bibitemNoStop [0]{.\EOS\space}%
\providecommand \EOS [0]{\spacefactor3000\relax}%
\providecommand \BibitemShut  [1]{\csname bibitem#1\endcsname}%
\let\auto@bib@innerbib\@empty
\bibitem [{\citenamefont {Rostoker}(1961)}]{rostoker1961fluctuations}%
  \BibitemOpen
  \bibfield  {author} {\bibinfo {author} {\bibfnamefont {N.}~\bibnamefont
  {Rostoker}},\ }\bibfield  {title} {\bibinfo {title} {Fluctuations of a plasma
  ({I})},\ }\href
  {https://iopscience.iop.org/article/10.1088/0029-5515/1/2/004} {\bibfield
  {journal} {\bibinfo  {journal}
  {\href{https://iopscience.iop.org/article/10.1088/0029-5515/1/2/004}{Nuclear
  Fusion}}\ }\textbf {\bibinfo {volume} {1}},\ \bibinfo {pages} {101} (\bibinfo
  {year} {1961})}\BibitemShut {NoStop}%
\bibitem [{\citenamefont {Hubbard}(1961)}]{hubbard1961friction}%
  \BibitemOpen
  \bibfield  {author} {\bibinfo {author} {\bibfnamefont {J.}~\bibnamefont
  {Hubbard}},\ }\bibfield  {title} {\bibinfo {title} {The friction and
  diffusion coefficients of the {F}okker-{P}lanck equation in a plasma},\
  }\href {https://doi.org/10.1098/rspa.1961.0017} {\bibfield  {journal}
  {\bibinfo  {journal}
  {\href{https://doi.org/10.1098/rspa.1961.0017}{Proceedings of the Royal
  Society A: Mathematical and Physical Sciences}}\ }\textbf {\bibinfo {volume}
  {260}},\ \bibinfo {pages} {114} (\bibinfo {year} {1961})}\BibitemShut
  {NoStop}%
\bibitem [{\citenamefont {Klimontovich}\ and\ \citenamefont
  {Silin}(1962)}]{klimontovich1962theory}%
  \BibitemOpen
  \bibfield  {author} {\bibinfo {author} {\bibfnamefont {Y.~L.}\ \bibnamefont
  {Klimontovich}}\ and\ \bibinfo {author} {\bibfnamefont {V.~P.}\ \bibnamefont
  {Silin}},\ }\bibfield  {title} {\bibinfo {title} {Theory of fluctuations of
  the particle distributions in a plasma},\ }\href@noop {} {\bibfield
  {journal} {\bibinfo  {journal} {Soviet Physics Journal of Experimental and
  Theoretical Physics}\ }\textbf {\bibinfo {volume} {15}},\ \bibinfo {pages}
  {199} (\bibinfo {year} {1962})}\BibitemShut {NoStop}%
\bibitem [{\citenamefont {Klimontovich}(1967)}]{klimontovich2013statistical}%
  \BibitemOpen
  \bibfield  {author} {\bibinfo {author} {\bibfnamefont {Y.~L.}\ \bibnamefont
  {Klimontovich}},\ }\href@noop {} {\emph {\bibinfo {title} {The Statistical
  Theory of Non-Equilibrium Processes in a Plasma}}}\ (\bibinfo  {publisher}
  {Pergamon Press},\ \bibinfo {address} {Oxford},\ \bibinfo {year}
  {1967})\BibitemShut {NoStop}%
\bibitem [{\citenamefont {Krall}\ and\ \citenamefont
  {Trivelpiece}(1973)}]{krall1973principles}%
  \BibitemOpen
  \bibfield  {author} {\bibinfo {author} {\bibfnamefont {N.~A.}\ \bibnamefont
  {Krall}}\ and\ \bibinfo {author} {\bibfnamefont {A.~W.}\ \bibnamefont
  {Trivelpiece}},\ }\href@noop {} {\emph {\bibinfo {title} {Principles of
  {P}lasma {P}hysics}}}\ (\bibinfo  {publisher} {McGraw-Hill},\ \bibinfo
  {address} {New York},\ \bibinfo {year} {1973})\BibitemShut {NoStop}%
\bibitem [{\citenamefont {Pitaevskii}\ and\ \citenamefont
  {Lifshitz}(1981)}]{pitaevskii2012physical}%
  \BibitemOpen
  \bibfield  {author} {\bibinfo {author} {\bibfnamefont {L.~P.}\ \bibnamefont
  {Pitaevskii}}\ and\ \bibinfo {author} {\bibfnamefont {E.}~\bibnamefont
  {Lifshitz}},\ }\href@noop {} {\emph {\bibinfo {title} {Physical Kinetics
  (Landau and Lifshitz’s Course of Theoretical Physics)}}},\ Vol.~\bibinfo
  {volume} {10}\ (\bibinfo  {publisher} {Pergamon Press},\ \bibinfo {address}
  {Oxford},\ \bibinfo {year} {1981})\BibitemShut {NoStop}%
\bibitem [{\citenamefont {Landau}(1965{\natexlab{a}})}]{landau1946vibrations}%
  \BibitemOpen
  \bibfield  {author} {\bibinfo {author} {\bibfnamefont {L.~D.}\ \bibnamefont
  {Landau}},\ }\bibfield  {title} {\bibinfo {title} {On the vibrations of the
  electronic plasma},\ }in\ \href@noop {} {\emph {\bibinfo {booktitle}
  {Collected Papers of L.D. Landau}}},\ \bibinfo {editor} {edited by\ \bibinfo
  {editor} {\bibfnamefont {D.}~\bibnamefont {ter Haar}}}\ (\bibinfo
  {publisher} {Pergamon Press},\ \bibinfo {address} {Oxford},\ \bibinfo {year}
  {1965})\ pp.\ \bibinfo {pages} {445--460}\BibitemShut {NoStop}%
\bibitem [{\citenamefont {Kadomtsev}(1965)}]{Kadomtsev1965plasmaturbulence}%
  \BibitemOpen
  \bibfield  {author} {\bibinfo {author} {\bibfnamefont {B.~B.}\ \bibnamefont
  {Kadomtsev}},\ }\href@noop {} {\emph {\bibinfo {title} {Plasma Turbulence}}}\
  (\bibinfo  {publisher} {Academic Press},\ \bibinfo {address} {New York},\
  \bibinfo {year} {1965})\BibitemShut {NoStop}%
\bibitem [{\citenamefont {Dupree}(1966)}]{dupree1966perturbation}%
  \BibitemOpen
  \bibfield  {author} {\bibinfo {author} {\bibfnamefont {T.~H.}\ \bibnamefont
  {Dupree}},\ }\bibfield  {title} {\bibinfo {title} {A perturbation theory for
  strong plasma turbulence},\ }\href {https://doi.org/10.1063/1.1761932}
  {\bibfield  {journal} {\bibinfo  {journal}
  {\href{https://doi.org/10.1063/1.1761932}{Physics of Fluids}}\ }\textbf
  {\bibinfo {volume} {9}},\ \bibinfo {pages} {1773} (\bibinfo {year}
  {1966})}\BibitemShut {NoStop}%
\bibitem [{\citenamefont {Orszag}\ and\ \citenamefont
  {Kraichnan}(1967)}]{orszag1967model}%
  \BibitemOpen
  \bibfield  {author} {\bibinfo {author} {\bibfnamefont {S.~A.}\ \bibnamefont
  {Orszag}}\ and\ \bibinfo {author} {\bibfnamefont {R.~H.}\ \bibnamefont
  {Kraichnan}},\ }\bibfield  {title} {\bibinfo {title} {Model equations for
  strong turbulence in a {V}lasov plasma},\ }\href
  {https://doi.org/10.1063/1.1762351} {\bibfield  {journal} {\bibinfo
  {journal} {\href{https://doi.org/10.1063/1.1762351}{Physics of Fluids}}\
  }\textbf {\bibinfo {volume} {10}},\ \bibinfo {pages} {1720} (\bibinfo {year}
  {1967})}\BibitemShut {NoStop}%
\bibitem [{\citenamefont {Ichimaru}\ and\ \citenamefont
  {Nakano}(1968)}]{ichimaru1968theory}%
  \BibitemOpen
  \bibfield  {author} {\bibinfo {author} {\bibfnamefont {S.}~\bibnamefont
  {Ichimaru}}\ and\ \bibinfo {author} {\bibfnamefont {T.}~\bibnamefont
  {Nakano}},\ }\bibfield  {title} {\bibinfo {title} {Theory of a turbulent
  stationary state of a plasma},\ }\href
  {https://doi.org/10.1103/PhysRev.165.231} {\bibfield  {journal} {\bibinfo
  {journal} {\href{https://doi.org/10.1103/PhysRev.165.231}{Physical Review}}\
  }\textbf {\bibinfo {volume} {165}},\ \bibinfo {pages} {231} (\bibinfo {year}
  {1968})}\BibitemShut {NoStop}%
\bibitem [{\citenamefont {Weinstock}(1969)}]{weinstock1969formulation}%
  \BibitemOpen
  \bibfield  {author} {\bibinfo {author} {\bibfnamefont {J.}~\bibnamefont
  {Weinstock}},\ }\bibfield  {title} {\bibinfo {title} {Formulation of a
  statistical theory of strong plasma turbulence},\ }\href
  {https://doi.org/10.1063/1.2163666} {\bibfield  {journal} {\bibinfo
  {journal} {\href{https://doi.org/10.1063/1.2163666}{Physics of Fluids}}\
  }\textbf {\bibinfo {volume} {12}},\ \bibinfo {pages} {1045} (\bibinfo {year}
  {1969})}\BibitemShut {NoStop}%
\bibitem [{\citenamefont {Ichimaru}(1970)}]{ichimaru1970theory}%
  \BibitemOpen
  \bibfield  {author} {\bibinfo {author} {\bibfnamefont {S.}~\bibnamefont
  {Ichimaru}},\ }\bibfield  {title} {\bibinfo {title} {Theory of strong
  turbulence in plasmas},\ }\href {https://doi.org/10.1063/1.1693117}
  {\bibfield  {journal} {\bibinfo  {journal}
  {\href{https://doi.org/10.1063/1.1693117}{Physics of Fluids}}\ }\textbf
  {\bibinfo {volume} {13}},\ \bibinfo {pages} {1560} (\bibinfo {year}
  {1970})}\BibitemShut {NoStop}%
\bibitem [{\citenamefont {Dupree}(1972)}]{dupree1972theory}%
  \BibitemOpen
  \bibfield  {author} {\bibinfo {author} {\bibfnamefont {T.~H.}\ \bibnamefont
  {Dupree}},\ }\bibfield  {title} {\bibinfo {title} {Theory of phase space
  density granulation in plasma},\ }\href {https://doi.org/10.1063/1.1693911}
  {\bibfield  {journal} {\bibinfo  {journal}
  {\href{https://doi.org/10.1063/1.1693911}{Physics of Fluids}}\ }\textbf
  {\bibinfo {volume} {15}},\ \bibinfo {pages} {334} (\bibinfo {year}
  {1972})}\BibitemShut {NoStop}%
\bibitem [{\citenamefont {Cook}\ and\ \citenamefont
  {Taylor}(1973)}]{cook1973electric}%
  \BibitemOpen
  \bibfield  {author} {\bibinfo {author} {\bibfnamefont {I.}~\bibnamefont
  {Cook}}\ and\ \bibinfo {author} {\bibfnamefont {J.~B.}\ \bibnamefont
  {Taylor}},\ }\bibfield  {title} {\bibinfo {title} {Electric field
  fluctuations in turbulent plasmas},\ }\href
  {https://doi.org/10.1017/S0022377800007388} {\bibfield  {journal} {\bibinfo
  {journal} {\href{https://doi.org/10.1017/S0022377800007388}{Journal of Plasma
  Physics}}\ }\textbf {\bibinfo {volume} {9}},\ \bibinfo {pages} {131}
  (\bibinfo {year} {1973})}\BibitemShut {NoStop}%
\bibitem [{\citenamefont {Misguich}\ and\ \citenamefont
  {Balescu}(1975)}]{misguich1975re}%
  \BibitemOpen
  \bibfield  {author} {\bibinfo {author} {\bibfnamefont {J.~H.}\ \bibnamefont
  {Misguich}}\ and\ \bibinfo {author} {\bibfnamefont {R.}~\bibnamefont
  {Balescu}},\ }\bibfield  {title} {\bibinfo {title} {Re-normalized
  quasi-linear approximation of plasma turbulence: {P}art 1. {M}odification of
  the {W}einstock weak-coupling limit},\ }\href
  {https://doi.org/10.1017/S0022377800025174} {\bibfield  {journal} {\bibinfo
  {journal} {\href{https://doi.org/10.1017/S0022377800025174}{Journal of Plasma
  Physics}}\ }\textbf {\bibinfo {volume} {13}},\ \bibinfo {pages} {385}
  (\bibinfo {year} {1975})}\BibitemShut {NoStop}%
\bibitem [{\citenamefont {Pelletier}\ and\ \citenamefont
  {Pomot}(1975)}]{pelletier1975fluctuation}%
  \BibitemOpen
  \bibfield  {author} {\bibinfo {author} {\bibfnamefont {G.}~\bibnamefont
  {Pelletier}}\ and\ \bibinfo {author} {\bibfnamefont {C.}~\bibnamefont
  {Pomot}},\ }\bibfield  {title} {\bibinfo {title} {Fluctuation spectrum in
  turbulent plasma},\ }\href {https://doi.org/10.1017/S0022377800009788}
  {\bibfield  {journal} {\bibinfo  {journal}
  {\href{https://doi.org/10.1017/S0022377800009788}{Journal of Plasma
  Physics}}\ }\textbf {\bibinfo {volume} {14}},\ \bibinfo {pages} {491}
  (\bibinfo {year} {1975})}\BibitemShut {NoStop}%
\bibitem [{\citenamefont {Misguich}\ and\ \citenamefont
  {Balescu}(1978{\natexlab{a}})}]{misguich1978kinetic}%
  \BibitemOpen
  \bibfield  {author} {\bibinfo {author} {\bibfnamefont {J.~H.}\ \bibnamefont
  {Misguich}}\ and\ \bibinfo {author} {\bibfnamefont {R.}~\bibnamefont
  {Balescu}},\ }\bibfield  {title} {\bibinfo {title} {Kinetic theory of binary
  correlations in turbulent plasmas},\ }\href
  {https://doi.org/10.1017/S0022377800023710} {\bibfield  {journal} {\bibinfo
  {journal} {\href{https://doi.org/10.1017/S0022377800023710}{Journal of Plasma
  Physics}}\ }\textbf {\bibinfo {volume} {19}},\ \bibinfo {pages} {147}
  (\bibinfo {year} {1978}{\natexlab{a}})}\BibitemShut {NoStop}%
\bibitem [{\citenamefont {DuBois}\ and\ \citenamefont
  {Espedal}(1978)}]{dubois1978direct}%
  \BibitemOpen
  \bibfield  {author} {\bibinfo {author} {\bibfnamefont {D.}~\bibnamefont
  {DuBois}}\ and\ \bibinfo {author} {\bibfnamefont {M.}~\bibnamefont
  {Espedal}},\ }\bibfield  {title} {\bibinfo {title} {Direct interaction
  approximation and plasma turbulence theory},\ }\href
  {https://iopscience.iop.org/article/10.1088/0032-1028/20/12/001} {\bibfield
  {journal} {\bibinfo  {journal}
  {\href{https://iopscience.iop.org/article/10.1088/0032-1028/20/12/001}{Plasma
  Physics}}\ }\textbf {\bibinfo {volume} {20}},\ \bibinfo {pages} {1209}
  (\bibinfo {year} {1978})}\BibitemShut {NoStop}%
\bibitem [{\citenamefont {Schekochihin}\ \emph {et~al.}(2008)\citenamefont
  {Schekochihin}, \citenamefont {Cowley}, \citenamefont {Dorland},
  \citenamefont {Hammett}, \citenamefont {Howes}, \citenamefont {Plunk},
  \citenamefont {Quataert},\ and\ \citenamefont
  {Tatsuno}}]{schekochihin2008gyrokinetic}%
  \BibitemOpen
  \bibfield  {author} {\bibinfo {author} {\bibfnamefont {A.~A.}\ \bibnamefont
  {Schekochihin}}, \bibinfo {author} {\bibfnamefont {S.~C.}\ \bibnamefont
  {Cowley}}, \bibinfo {author} {\bibfnamefont {W.}~\bibnamefont {Dorland}},
  \bibinfo {author} {\bibfnamefont {G.~W.}\ \bibnamefont {Hammett}}, \bibinfo
  {author} {\bibfnamefont {G.~G.}\ \bibnamefont {Howes}}, \bibinfo {author}
  {\bibfnamefont {G.~G.}\ \bibnamefont {Plunk}}, \bibinfo {author}
  {\bibfnamefont {E.}~\bibnamefont {Quataert}},\ and\ \bibinfo {author}
  {\bibfnamefont {T.}~\bibnamefont {Tatsuno}},\ }\bibfield  {title} {\bibinfo
  {title} {Gyrokinetic turbulence: a nonlinear route to dissipation through
  phase space},\ }\href {https://doi.org/10.1088/0741-3335/50/12/124024}
  {\bibfield  {journal} {\bibinfo  {journal}
  {\href{https://doi.org/10.1088/0741-3335/50/12/124024}{Plasma Physics and
  Controlled Fusion}}\ }\textbf {\bibinfo {volume} {50}},\ \bibinfo {pages}
  {124024} (\bibinfo {year} {2008})}\BibitemShut {NoStop}%
\bibitem [{\citenamefont {Schekochihin}\ \emph {et~al.}(2009)\citenamefont
  {Schekochihin}, \citenamefont {Cowley}, \citenamefont {Dorland},
  \citenamefont {Hammett}, \citenamefont {Howes}, \citenamefont {Quataert},\
  and\ \citenamefont {Tatsuno}}]{schekochihin2009astrophysical}%
  \BibitemOpen
  \bibfield  {author} {\bibinfo {author} {\bibfnamefont {A.~A.}\ \bibnamefont
  {Schekochihin}}, \bibinfo {author} {\bibfnamefont {S.~C.}\ \bibnamefont
  {Cowley}}, \bibinfo {author} {\bibfnamefont {W.}~\bibnamefont {Dorland}},
  \bibinfo {author} {\bibfnamefont {G.~W.}\ \bibnamefont {Hammett}}, \bibinfo
  {author} {\bibfnamefont {G.~G.}\ \bibnamefont {Howes}}, \bibinfo {author}
  {\bibfnamefont {E.}~\bibnamefont {Quataert}},\ and\ \bibinfo {author}
  {\bibfnamefont {T.}~\bibnamefont {Tatsuno}},\ }\bibfield  {title} {\bibinfo
  {title} {Astrophysical gyrokinetics: kinetic and fluid turbulent cascades in
  magnetized weakly collisional plasmas},\ }\href
  {https://doi.org/10.1088/0067-0049/182/1/310} {\bibfield  {journal} {\bibinfo
   {journal} {\href{https://doi.org/10.1088/0067-0049/182/1/310}{Astrophysical
  Journal Supplement Series}}\ }\textbf {\bibinfo {volume} {182}},\ \bibinfo
  {pages} {310} (\bibinfo {year} {2009})}\BibitemShut {NoStop}%
\bibitem [{\citenamefont {Eyink}(2018)}]{eyink2018cascades}%
  \BibitemOpen
  \bibfield  {author} {\bibinfo {author} {\bibfnamefont {G.~L.}\ \bibnamefont
  {Eyink}},\ }\bibfield  {title} {\bibinfo {title} {Cascades and dissipative
  anomalies in nearly collisionless plasma turbulence},\ }\href
  {https://doi.org/10.1103/PhysRevX.8.041020} {\bibfield  {journal} {\bibinfo
  {journal} {\href{https://doi.org/10.1103/PhysRevX.8.041020}{Physical Review
  X}}\ }\textbf {\bibinfo {volume} {8}},\ \bibinfo {pages} {041020} (\bibinfo
  {year} {2018})}\BibitemShut {NoStop}%
\bibitem [{\citenamefont {Nastac}\ \emph {et~al.}(2024)\citenamefont {Nastac},
  \citenamefont {Ewart}, \citenamefont {Sengupta}, \citenamefont
  {Schekochihin}, \citenamefont {Barnes},\ and\ \citenamefont
  {Dorland}}]{nastac2023phase}%
  \BibitemOpen
  \bibfield  {author} {\bibinfo {author} {\bibfnamefont {M.~L.}\ \bibnamefont
  {Nastac}}, \bibinfo {author} {\bibfnamefont {R.~J.}\ \bibnamefont {Ewart}},
  \bibinfo {author} {\bibfnamefont {W.}~\bibnamefont {Sengupta}}, \bibinfo
  {author} {\bibfnamefont {A.~A.}\ \bibnamefont {Schekochihin}}, \bibinfo
  {author} {\bibfnamefont {M.}~\bibnamefont {Barnes}},\ and\ \bibinfo {author}
  {\bibfnamefont {W.~D.}\ \bibnamefont {Dorland}},\ }\bibfield  {title}
  {\bibinfo {title} {Phase-space entropy cascade and irreversibility of
  stochastic heating in nearly collisionless plasma turbulence},\ }\href
  {https://doi.org/10.1103/PhysRevE.109.065210} {\bibfield  {journal} {\bibinfo
   {journal} {\href{https://doi.org/10.1103/PhysRevE.109.065210}{Physical
  Review E}}\ }\textbf {\bibinfo {volume} {109}},\ \bibinfo {pages} {065210}
  (\bibinfo {year} {2024})}\BibitemShut {NoStop}%
\bibitem [{\citenamefont {Dupree}(1970)}]{dupree1970theory}%
  \BibitemOpen
  \bibfield  {author} {\bibinfo {author} {\bibfnamefont {T.~H.}\ \bibnamefont
  {Dupree}},\ }\bibfield  {title} {\bibinfo {title} {Theory of resistivity in
  collisionless plasma},\ }\href {https://doi.org/10.1103/PhysRevLett.25.789}
  {\bibfield  {journal} {\bibinfo  {journal}
  {\href{https://doi.org/10.1103/PhysRevLett.25.789}{Physical Review Letters}}\
  }\textbf {\bibinfo {volume} {25}},\ \bibinfo {pages} {789} (\bibinfo {year}
  {1970})}\BibitemShut {NoStop}%
\bibitem [{\citenamefont {Kadomtsev}\ and\ \citenamefont
  {Pogutse}(1971)}]{kadomtsev1971theory}%
  \BibitemOpen
  \bibfield  {author} {\bibinfo {author} {\bibfnamefont {B.}~\bibnamefont
  {Kadomtsev}}\ and\ \bibinfo {author} {\bibfnamefont {O.}~\bibnamefont
  {Pogutse}},\ }\bibfield  {title} {\bibinfo {title} {Theory of beam-plasma
  interaction},\ }\href {https://doi.org/10.1063/1.1693356} {\bibfield
  {journal} {\bibinfo  {journal}
  {\href{https://doi.org/10.1063/1.1693356}{Physics of Fluids}}\ }\textbf
  {\bibinfo {volume} {14}},\ \bibinfo {pages} {2470} (\bibinfo {year}
  {1971})}\BibitemShut {NoStop}%
\bibitem [{\citenamefont {Misguich}\ and\ \citenamefont
  {Balescu}(1978{\natexlab{b}})}]{misguich1978clumps}%
  \BibitemOpen
  \bibfield  {author} {\bibinfo {author} {\bibfnamefont {J.~H.}\ \bibnamefont
  {Misguich}}\ and\ \bibinfo {author} {\bibfnamefont {R.}~\bibnamefont
  {Balescu}},\ }\bibfield  {title} {\bibinfo {title} {`{C}lumps' as enhanced
  correlations and turbulent {D}ebye screening},\ }\href
  {https://iopscience.iop.org/article/10.1088/0032-1028/20/8/006/} {\bibfield
  {journal} {\bibinfo  {journal}
  {\href{https://iopscience.iop.org/article/10.1088/0032-1028/20/8/006/}{Plasma
  Physics}}\ }\textbf {\bibinfo {volume} {20}},\ \bibinfo {pages} {781}
  (\bibinfo {year} {1978}{\natexlab{b}})}\BibitemShut {NoStop}%
\bibitem [{\citenamefont {Boutros-Ghali}\ and\ \citenamefont
  {Dupree}(1981)}]{boutros1981theory}%
  \BibitemOpen
  \bibfield  {author} {\bibinfo {author} {\bibfnamefont {T.}~\bibnamefont
  {Boutros-Ghali}}\ and\ \bibinfo {author} {\bibfnamefont {T.~H.}\ \bibnamefont
  {Dupree}},\ }\bibfield  {title} {\bibinfo {title} {Theory of two-point
  correlation function in a {V}lasov plasma},\ }\href
  {https://doi.org/10.1063/1.863265} {\bibfield  {journal} {\bibinfo  {journal}
  {\href{https://doi.org/10.1063/1.863265}{Physics of Fluids}}\ }\textbf
  {\bibinfo {volume} {24}},\ \bibinfo {pages} {1839} (\bibinfo {year}
  {1981})}\BibitemShut {NoStop}%
\bibitem [{\citenamefont {Misguich}\ and\ \citenamefont
  {Balescu}(1982)}]{misguich1982relative}%
  \BibitemOpen
  \bibfield  {author} {\bibinfo {author} {\bibfnamefont {J.~H.}\ \bibnamefont
  {Misguich}}\ and\ \bibinfo {author} {\bibfnamefont {R.}~\bibnamefont
  {Balescu}},\ }\bibfield  {title} {\bibinfo {title} {On relative spatial
  diffusion in plasma and fluid turbulences: clumps, {R}ichardson's law and
  intrinsic stochasticity},\ }\href
  {https://iopscience.iop.org/article/10.1088/0032-1028/24/3/007/} {\bibfield
  {journal} {\bibinfo  {journal}
  {\href{https://iopscience.iop.org/article/10.1088/0032-1028/24/3/007}{Plasma
  Physics}}\ }\textbf {\bibinfo {volume} {24}},\ \bibinfo {pages} {289}
  (\bibinfo {year} {1982})}\BibitemShut {NoStop}%
\bibitem [{\citenamefont {Boutros-Ghali}\ and\ \citenamefont
  {Dupree}(1982)}]{boutros1982theory}%
  \BibitemOpen
  \bibfield  {author} {\bibinfo {author} {\bibfnamefont {T.}~\bibnamefont
  {Boutros-Ghali}}\ and\ \bibinfo {author} {\bibfnamefont {T.~H.}\ \bibnamefont
  {Dupree}},\ }\bibfield  {title} {\bibinfo {title} {Theory of nonlinear
  ion-electron instability},\ }\href {https://doi.org/10.1063/1.863817}
  {\bibfield  {journal} {\bibinfo  {journal}
  {\href{https://doi.org/10.1063/1.863817}{Physics of Fluids}}\ }\textbf
  {\bibinfo {volume} {25}},\ \bibinfo {pages} {874} (\bibinfo {year}
  {1982})}\BibitemShut {NoStop}%
\bibitem [{\citenamefont {Tetreault}(1983)}]{tetreault1983growth}%
  \BibitemOpen
  \bibfield  {author} {\bibinfo {author} {\bibfnamefont {D.~J.}\ \bibnamefont
  {Tetreault}},\ }\bibfield  {title} {\bibinfo {title} {Growth rate of the
  clump instability},\ }\href {https://doi.org/10.1063/1.864100} {\bibfield
  {journal} {\bibinfo  {journal}
  {\href{https://doi.org/10.1063/1.864100}{Physics of Fluids}}\ }\textbf
  {\bibinfo {volume} {26}},\ \bibinfo {pages} {3247} (\bibinfo {year}
  {1983})}\BibitemShut {NoStop}%
\bibitem [{\citenamefont {Dupree}(1978)}]{dupree1978role}%
  \BibitemOpen
  \bibfield  {author} {\bibinfo {author} {\bibfnamefont {T.~H.}\ \bibnamefont
  {Dupree}},\ }\bibfield  {title} {\bibinfo {title} {Role of clumps in
  drift-wave turbulence},\ }\href {https://doi.org/10.1063/1.862286} {\bibfield
   {journal} {\bibinfo  {journal}
  {\href{https://doi.org/10.1063/1.862286}{Physics of Fluids}}\ }\textbf
  {\bibinfo {volume} {21}},\ \bibinfo {pages} {783} (\bibinfo {year}
  {1978})}\BibitemShut {NoStop}%
\bibitem [{\citenamefont {Biglari}\ \emph {et~al.}(1988)\citenamefont
  {Biglari}, \citenamefont {Diamond},\ and\ \citenamefont
  {Terry}}]{biglari1988theory}%
  \BibitemOpen
  \bibfield  {author} {\bibinfo {author} {\bibfnamefont {H.}~\bibnamefont
  {Biglari}}, \bibinfo {author} {\bibfnamefont {P.~H.}\ \bibnamefont
  {Diamond}},\ and\ \bibinfo {author} {\bibfnamefont {P.~W.}\ \bibnamefont
  {Terry}},\ }\bibfield  {title} {\bibinfo {title} {Theory of
  trapped-ion-temperature-gradient-driven turbulence and transport in
  low-collisionality plasmas},\ }\href {https://doi.org/10.1063/1.866542}
  {\bibfield  {journal} {\bibinfo  {journal}
  {\href{https://doi.org/10.1063/1.866542}{Physics of Fluids}}\ }\textbf
  {\bibinfo {volume} {31}},\ \bibinfo {pages} {2644} (\bibinfo {year}
  {1988})}\BibitemShut {NoStop}%
\bibitem [{\citenamefont {Kosuga}\ and\ \citenamefont
  {Diamond}(2011)}]{kosuga2011relaxation}%
  \BibitemOpen
  \bibfield  {author} {\bibinfo {author} {\bibfnamefont {Y.}~\bibnamefont
  {Kosuga}}\ and\ \bibinfo {author} {\bibfnamefont {P.~H.}\ \bibnamefont
  {Diamond}},\ }\bibfield  {title} {\bibinfo {title} {On relaxation and
  transport in gyrokinetic drift wave turbulence with zonal flow},\ }\href
  {https://doi.org/10.1063/1.3662428} {\bibfield  {journal} {\bibinfo
  {journal} {\href{https://doi.org/10.1063/1.3662428}{Physics of Plasmas}}\
  }\textbf {\bibinfo {volume} {18}},\ \bibinfo {pages} {122305} (\bibinfo
  {year} {2011})}\BibitemShut {NoStop}%
\bibitem [{\citenamefont {Kosuga}\ \emph {et~al.}(2014)\citenamefont {Kosuga},
  \citenamefont {Itoh}, \citenamefont {Diamond}, \citenamefont {Itoh},\ and\
  \citenamefont {Lesur}}]{kosuga2014ion}%
  \BibitemOpen
  \bibfield  {author} {\bibinfo {author} {\bibfnamefont {Y.}~\bibnamefont
  {Kosuga}}, \bibinfo {author} {\bibfnamefont {S.-I.}\ \bibnamefont {Itoh}},
  \bibinfo {author} {\bibfnamefont {P.~H.}\ \bibnamefont {Diamond}}, \bibinfo
  {author} {\bibfnamefont {K.}~\bibnamefont {Itoh}},\ and\ \bibinfo {author}
  {\bibfnamefont {M.}~\bibnamefont {Lesur}},\ }\bibfield  {title} {\bibinfo
  {title} {Ion temperature gradient driven turbulence with strong trapped ion
  resonance},\ }\href {https://doi.org/10.1063/1.4897179} {\bibfield  {journal}
  {\bibinfo  {journal} {\href{https://doi.org/10.1063/1.4897179}{Physics of
  Plasmas}}\ }\textbf {\bibinfo {volume} {21}},\ \bibinfo {pages} {102303}
  (\bibinfo {year} {2014})}\BibitemShut {NoStop}%
\bibitem [{\citenamefont {Berk}\ \emph {et~al.}(1997)\citenamefont {Berk},
  \citenamefont {Breizman},\ and\ \citenamefont
  {Petviashvili}}]{berk1997spontaneous}%
  \BibitemOpen
  \bibfield  {author} {\bibinfo {author} {\bibfnamefont {H.~L.}\ \bibnamefont
  {Berk}}, \bibinfo {author} {\bibfnamefont {B.~N.}\ \bibnamefont {Breizman}},\
  and\ \bibinfo {author} {\bibfnamefont {N.~V.}\ \bibnamefont {Petviashvili}},\
  }\bibfield  {title} {\bibinfo {title} {Spontaneous hole-clump pair creation
  in weakly unstable plasmas},\ }\href
  {https://doi.org/10.1016/S0375-9601(97)00523-9} {\bibfield  {journal}
  {\bibinfo  {journal}
  {\href{https://doi.org/10.1016/S0375-9601(97)00523-9}{Physics Letters A}}\
  }\textbf {\bibinfo {volume} {234}},\ \bibinfo {pages} {213} (\bibinfo {year}
  {1997})}\BibitemShut {NoStop}%
\bibitem [{\citenamefont {Berk}\ \emph {et~al.}(1999)\citenamefont {Berk},
  \citenamefont {Breizman}, \citenamefont {Candy}, \citenamefont {Pekker},\
  and\ \citenamefont {Petviashvili}}]{berk1999spontaneous}%
  \BibitemOpen
  \bibfield  {author} {\bibinfo {author} {\bibfnamefont {H.~L.}\ \bibnamefont
  {Berk}}, \bibinfo {author} {\bibfnamefont {B.~N.}\ \bibnamefont {Breizman}},
  \bibinfo {author} {\bibfnamefont {J.}~\bibnamefont {Candy}}, \bibinfo
  {author} {\bibfnamefont {M.}~\bibnamefont {Pekker}},\ and\ \bibinfo {author}
  {\bibfnamefont {N.~V.}\ \bibnamefont {Petviashvili}},\ }\bibfield  {title}
  {\bibinfo {title} {Spontaneous hole--clump pair creation},\ }\href
  {https://doi.org/10.1063/1.873550} {\bibfield  {journal} {\bibinfo  {journal}
  {\href{https://doi.org/10.1063/1.873550}{Physics of Plasmas}}\ }\textbf
  {\bibinfo {volume} {6}},\ \bibinfo {pages} {3102} (\bibinfo {year}
  {1999})}\BibitemShut {NoStop}%
\bibitem [{\citenamefont {Lesur}\ and\ \citenamefont
  {Diamond}(2013)}]{lesur2013nonlinear}%
  \BibitemOpen
  \bibfield  {author} {\bibinfo {author} {\bibfnamefont {M.}~\bibnamefont
  {Lesur}}\ and\ \bibinfo {author} {\bibfnamefont {P.~H.}\ \bibnamefont
  {Diamond}},\ }\bibfield  {title} {\bibinfo {title} {Nonlinear instabilities
  driven by coherent phase-space structures},\ }\href
  {https://doi.org/10.1103/PhysRevE.87.031101} {\bibfield  {journal} {\bibinfo
  {journal} {\href{https://doi.org/10.1103/PhysRevE.87.031101}{Physical Review
  E}}\ }\textbf {\bibinfo {volume} {87}},\ \bibinfo {pages} {031101} (\bibinfo
  {year} {2013})}\BibitemShut {NoStop}%
\bibitem [{\citenamefont {Wang}\ \emph {et~al.}(2013)\citenamefont {Wang},
  \citenamefont {Todo},\ and\ \citenamefont {Kim}}]{wang2013hole}%
  \BibitemOpen
  \bibfield  {author} {\bibinfo {author} {\bibfnamefont {H.}~\bibnamefont
  {Wang}}, \bibinfo {author} {\bibfnamefont {Y.}~\bibnamefont {Todo}},\ and\
  \bibinfo {author} {\bibfnamefont {C.~C.}\ \bibnamefont {Kim}},\ }\bibfield
  {title} {\bibinfo {title} {Hole-clump pair creation in the evolution of
  energetic-particle-driven geodesic acoustic modes},\ }\href
  {https://doi.org/10.1103/PhysRevLett.110.155006} {\bibfield  {journal}
  {\bibinfo  {journal}
  {\href{https://doi.org/10.1103/PhysRevLett.110.155006}{Physical Review
  Letters}}\ }\textbf {\bibinfo {volume} {110}},\ \bibinfo {pages} {155006}
  (\bibinfo {year} {2013})}\BibitemShut {NoStop}%
\bibitem [{\citenamefont {Lilley}\ and\ \citenamefont
  {Nyqvist}(2014)}]{lilley2014formation}%
  \BibitemOpen
  \bibfield  {author} {\bibinfo {author} {\bibfnamefont {M.~K.}\ \bibnamefont
  {Lilley}}\ and\ \bibinfo {author} {\bibfnamefont {R.~M.}\ \bibnamefont
  {Nyqvist}},\ }\bibfield  {title} {\bibinfo {title} {Formation of phase space
  holes and clumps},\ }\href {https://doi.org/10.1103/PhysRevLett.112.155002}
  {\bibfield  {journal} {\bibinfo  {journal}
  {\href{https://doi.org/10.1103/PhysRevLett.112.155002}{Physical Review
  Letters}}\ }\textbf {\bibinfo {volume} {112}},\ \bibinfo {pages} {155002}
  (\bibinfo {year} {2014})}\BibitemShut {NoStop}%
\bibitem [{\citenamefont {Bierwage}\ \emph {et~al.}(2021)\citenamefont
  {Bierwage}, \citenamefont {White},\ and\ \citenamefont
  {Duarte}}]{bierwage2021effect}%
  \BibitemOpen
  \bibfield  {author} {\bibinfo {author} {\bibfnamefont {A.}~\bibnamefont
  {Bierwage}}, \bibinfo {author} {\bibfnamefont {R.~B.}\ \bibnamefont
  {White}},\ and\ \bibinfo {author} {\bibfnamefont {V.~N.}\ \bibnamefont
  {Duarte}},\ }\bibfield  {title} {\bibinfo {title} {On the effect of beating
  during nonlinear frequency chirping},\ }\href
  {https://doi.org/10.1585/pfr.16.1403087} {\bibfield  {journal} {\bibinfo
  {journal} {\href{https://doi.org/10.1585/pfr.16.1403087}{Plasma and Fusion
  Research}}\ }\textbf {\bibinfo {volume} {16}},\ \bibinfo {pages} {1403087}
  (\bibinfo {year} {2021})}\BibitemShut {NoStop}%
\bibitem [{\citenamefont {Krommes}(1986)}]{krommes1986comments}%
  \BibitemOpen
  \bibfield  {author} {\bibinfo {author} {\bibfnamefont {J.~A.}\ \bibnamefont
  {Krommes}},\ }\bibfield  {title} {\bibinfo {title} {Comments on ``{T}heory of
  dissipative density-gradient-driven turbulence in the tokamak edge'''[{P}hys.
  {F}luids 28, 1419 (1985)]},\ }\href {https://doi.org/10.1063/1.865520}
  {\bibfield  {journal} {\bibinfo  {journal}
  {\href{https://doi.org/10.1063/1.865520}{Physics of Fluids}}\ }\textbf
  {\bibinfo {volume} {29}},\ \bibinfo {pages} {2756} (\bibinfo {year}
  {1986})}\BibitemShut {NoStop}%
\bibitem [{\citenamefont {Zhang}(1988)}]{zhang1988comments}%
  \BibitemOpen
  \bibfield  {author} {\bibinfo {author} {\bibfnamefont {C.~F.}\ \bibnamefont
  {Zhang}},\ }\bibfield  {title} {\bibinfo {title} {Comments on clumps theory
  in turbulent plasma},\ }\href
  {https://iopscience.iop.org/article/10.1088/0741-3335/30/7/004} {\bibfield
  {journal} {\bibinfo  {journal}
  {\href{https://iopscience.iop.org/article/10.1088/0741-3335/30/7/004}{Plasma
  Physics and Controlled Fusion}}\ }\textbf {\bibinfo {volume} {30}},\ \bibinfo
  {pages} {853} (\bibinfo {year} {1988})}\BibitemShut {NoStop}%
\bibitem [{\citenamefont {Krommes}(1997)}]{krommes1997clump}%
  \BibitemOpen
  \bibfield  {author} {\bibinfo {author} {\bibfnamefont {J.~A.}\ \bibnamefont
  {Krommes}},\ }\bibfield  {title} {\bibinfo {title} {The clump lifetime
  revisited: Exact calculation of the second-order structure function for a
  model of forced, dissipative turbulence},\ }\href
  {https://doi.org/10.1063/1.872148} {\bibfield  {journal} {\bibinfo  {journal}
  {\href{https://doi.org/10.1063/1.872148}{Physics of Plasmas}}\ }\textbf
  {\bibinfo {volume} {4}},\ \bibinfo {pages} {655} (\bibinfo {year}
  {1997})}\BibitemShut {NoStop}%
\bibitem [{\citenamefont {Diamond}\ \emph {et~al.}(2010)\citenamefont
  {Diamond}, \citenamefont {Itoh},\ and\ \citenamefont
  {Itoh}}]{diamond2010modern}%
  \BibitemOpen
  \bibfield  {author} {\bibinfo {author} {\bibfnamefont {P.~H.}\ \bibnamefont
  {Diamond}}, \bibinfo {author} {\bibfnamefont {S.-I.}\ \bibnamefont {Itoh}},\
  and\ \bibinfo {author} {\bibfnamefont {K.}~\bibnamefont {Itoh}},\ }\href@noop
  {} {\emph {\bibinfo {title} {Modern Plasma Physics: Volume 1, Physical
  Kinetics of Turbulent Plasmas}}}\ (\bibinfo  {publisher} {Cambridge
  University Press},\ \bibinfo {address} {Cambridge},\ \bibinfo {year}
  {2010})\BibitemShut {NoStop}%
\bibitem [{\citenamefont {Kosuga}\ \emph {et~al.}(2017)\citenamefont {Kosuga},
  \citenamefont {Itoh}, \citenamefont {Diamond}, \citenamefont {Itoh},\ and\
  \citenamefont {Lesur}}]{kosuga2017role}%
  \BibitemOpen
  \bibfield  {author} {\bibinfo {author} {\bibfnamefont {Y.}~\bibnamefont
  {Kosuga}}, \bibinfo {author} {\bibfnamefont {S.-I.}\ \bibnamefont {Itoh}},
  \bibinfo {author} {\bibfnamefont {P.~H.}\ \bibnamefont {Diamond}}, \bibinfo
  {author} {\bibfnamefont {K.}~\bibnamefont {Itoh}},\ and\ \bibinfo {author}
  {\bibfnamefont {M.}~\bibnamefont {Lesur}},\ }\bibfield  {title} {\bibinfo
  {title} {Role of phase space structures in collisionless drift wave
  turbulence and impact on transport modeling},\ }\href
  {https://iopscience.iop.org/article/10.1088/1741-4326/57/7/072006/}
  {\bibfield  {journal} {\bibinfo  {journal}
  {\href{https://iopscience.iop.org/article/10.1088/1741-4326/57/7/072006/}{Nuclear
  Fusion}}\ }\textbf {\bibinfo {volume} {57}},\ \bibinfo {pages} {072006}
  (\bibinfo {year} {2017})}\BibitemShut {NoStop}%
\bibitem [{\citenamefont {Lesur}(2020)}]{lesur2020nonlinear}%
  \BibitemOpen
  \bibfield  {author} {\bibinfo {author} {\bibfnamefont {M.}~\bibnamefont
  {Lesur}},\ }\emph {\bibinfo {title} {Nonlinear features of instabilities,
  turbulence and transport in hot plasmas}},\ \href@noop {} {\bibinfo {type}
  {H{D}{R} thesis}},\ \bibinfo  {school} {Universit{\'e} de Lorraine} (\bibinfo
  {year} {2020})\BibitemShut {NoStop}%
\bibitem [{\citenamefont {Berman}\ \emph {et~al.}(1983)\citenamefont {Berman},
  \citenamefont {Tetreault},\ and\ \citenamefont
  {Dupree}}]{berman1983observation}%
  \BibitemOpen
  \bibfield  {author} {\bibinfo {author} {\bibfnamefont {R.~H.}\ \bibnamefont
  {Berman}}, \bibinfo {author} {\bibfnamefont {D.~J.}\ \bibnamefont
  {Tetreault}},\ and\ \bibinfo {author} {\bibfnamefont {T.~H.}\ \bibnamefont
  {Dupree}},\ }\bibfield  {title} {\bibinfo {title} {Observation of
  self-binding turbulent fluctuations in simulation plasma and their relevance
  to plasma kinetic theories},\ }\href {https://doi.org/10.1063/1.864429}
  {\bibfield  {journal} {\bibinfo  {journal}
  {\href{https://doi.org/10.1063/1.864429}{Physics of Fluids}}\ }\textbf
  {\bibinfo {volume} {26}},\ \bibinfo {pages} {2437} (\bibinfo {year}
  {1983})}\BibitemShut {NoStop}%
\bibitem [{\citenamefont {Berman}\ \emph {et~al.}(1985)\citenamefont {Berman},
  \citenamefont {Tetreault},\ and\ \citenamefont
  {Dupree}}]{berman1985simulation}%
  \BibitemOpen
  \bibfield  {author} {\bibinfo {author} {\bibfnamefont {R.~H.}\ \bibnamefont
  {Berman}}, \bibinfo {author} {\bibfnamefont {D.~J.}\ \bibnamefont
  {Tetreault}},\ and\ \bibinfo {author} {\bibfnamefont {T.~H.}\ \bibnamefont
  {Dupree}},\ }\bibfield  {title} {\bibinfo {title} {Simulation of phase space
  hole growth and the development of intermittent plasma turbulence},\ }\href
  {https://doi.org/10.1063/1.865176} {\bibfield  {journal} {\bibinfo  {journal}
  {\href{https://doi.org/10.1063/1.865176}{Physics of Fluids}}\ }\textbf
  {\bibinfo {volume} {28}},\ \bibinfo {pages} {155} (\bibinfo {year}
  {1985})}\BibitemShut {NoStop}%
\bibitem [{\citenamefont {Krommes}\ and\ \citenamefont
  {Hu}(1994)}]{krommes1994role}%
  \BibitemOpen
  \bibfield  {author} {\bibinfo {author} {\bibfnamefont {J.~A.}\ \bibnamefont
  {Krommes}}\ and\ \bibinfo {author} {\bibfnamefont {G.}~\bibnamefont {Hu}},\
  }\bibfield  {title} {\bibinfo {title} {The role of dissipation in the theory
  and simulations of homogeneous plasma turbulence, and resolution of the
  entropy paradox},\ }\href {https://doi.org/10.1063/1.870475} {\bibfield
  {journal} {\bibinfo  {journal}
  {\href{https://doi.org/10.1063/1.870475}{Physics of Plasmas}}\ }\textbf
  {\bibinfo {volume} {1}},\ \bibinfo {pages} {3211} (\bibinfo {year}
  {1994})}\BibitemShut {NoStop}%
\bibitem [{\citenamefont {Abel}\ \emph {et~al.}(2013)\citenamefont {Abel},
  \citenamefont {Plunk}, \citenamefont {Wang}, \citenamefont {Barnes},
  \citenamefont {Cowley}, \citenamefont {Dorland},\ and\ \citenamefont
  {Schekochihin}}]{abel2013multiscale}%
  \BibitemOpen
  \bibfield  {author} {\bibinfo {author} {\bibfnamefont {I.~G.}\ \bibnamefont
  {Abel}}, \bibinfo {author} {\bibfnamefont {G.~G.}\ \bibnamefont {Plunk}},
  \bibinfo {author} {\bibfnamefont {E.}~\bibnamefont {Wang}}, \bibinfo {author}
  {\bibfnamefont {M.}~\bibnamefont {Barnes}}, \bibinfo {author} {\bibfnamefont
  {S.~C.}\ \bibnamefont {Cowley}}, \bibinfo {author} {\bibfnamefont
  {W.}~\bibnamefont {Dorland}},\ and\ \bibinfo {author} {\bibfnamefont {A.~A.}\
  \bibnamefont {Schekochihin}},\ }\bibfield  {title} {\bibinfo {title}
  {Multiscale gyrokinetics for rotating tokamak plasmas: fluctuations,
  transport and energy flows},\ }\href
  {https://doi.org/10.1088/0034-4885/76/11/116201} {\bibfield  {journal}
  {\bibinfo  {journal}
  {\href{https://doi.org/10.1088/0034-4885/76/11/116201}{Reports on Progress in
  Physics}}\ }\textbf {\bibinfo {volume} {76}},\ \bibinfo {pages} {116201}
  (\bibinfo {year} {2013})}\BibitemShut {NoStop}%
\bibitem [{\citenamefont {Dorland}\ and\ \citenamefont
  {Hammett}(1993)}]{dorland1993gyrofluid}%
  \BibitemOpen
  \bibfield  {author} {\bibinfo {author} {\bibfnamefont {W.}~\bibnamefont
  {Dorland}}\ and\ \bibinfo {author} {\bibfnamefont {G.~W.}\ \bibnamefont
  {Hammett}},\ }\bibfield  {title} {\bibinfo {title} {Gyrofluid turbulence
  models with kinetic effects},\ }\href {https://doi.org/10.1063/1.860934}
  {\bibfield  {journal} {\bibinfo  {journal}
  {\href{https://doi.org/10.1063/1.860934}{Physics of Fluids B: Plasma
  Physics}}\ }\textbf {\bibinfo {volume} {5}},\ \bibinfo {pages} {812}
  (\bibinfo {year} {1993})}\BibitemShut {NoStop}%
\bibitem [{\citenamefont {Plunk}\ \emph {et~al.}(2010)\citenamefont {Plunk},
  \citenamefont {Cowley}, \citenamefont {Schekochihin},\ and\ \citenamefont
  {Tatsuno}}]{plunk2010two}%
  \BibitemOpen
  \bibfield  {author} {\bibinfo {author} {\bibfnamefont {G.~G.}\ \bibnamefont
  {Plunk}}, \bibinfo {author} {\bibfnamefont {S.~C.}\ \bibnamefont {Cowley}},
  \bibinfo {author} {\bibfnamefont {A.~A.}\ \bibnamefont {Schekochihin}},\ and\
  \bibinfo {author} {\bibfnamefont {T.}~\bibnamefont {Tatsuno}},\ }\bibfield
  {title} {\bibinfo {title} {Two-dimensional gyrokinetic turbulence},\ }\href
  {https://doi.org/10.1017/S002211201000371X} {\bibfield  {journal} {\bibinfo
  {journal} {\href{https://doi.org/10.1017/S002211201000371X}{Journal of Fluid
  Mechanics}}\ }\textbf {\bibinfo {volume} {664}},\ \bibinfo {pages} {407}
  (\bibinfo {year} {2010})}\BibitemShut {NoStop}%
\bibitem [{\citenamefont {Barnes}\ \emph {et~al.}(2011)\citenamefont {Barnes},
  \citenamefont {Parra},\ and\ \citenamefont
  {Schekochihin}}]{barnes2011critically}%
  \BibitemOpen
  \bibfield  {author} {\bibinfo {author} {\bibfnamefont {M.}~\bibnamefont
  {Barnes}}, \bibinfo {author} {\bibfnamefont {F.~I.}\ \bibnamefont {Parra}},\
  and\ \bibinfo {author} {\bibfnamefont {A.~A.}\ \bibnamefont {Schekochihin}},\
  }\bibfield  {title} {\bibinfo {title} {Critically balanced ion temperature
  gradient turbulence in fusion plasmas},\ }\href
  {https://doi.org/10.1103/PhysRevLett.107.115003} {\bibfield  {journal}
  {\bibinfo  {journal}
  {\href{https://doi.org/10.1103/PhysRevLett.107.115003}{Physical Review
  Letters}}\ }\textbf {\bibinfo {volume} {107}},\ \bibinfo {pages} {115003}
  (\bibinfo {year} {2011})}\BibitemShut {NoStop}%
\bibitem [{\citenamefont {Kunz}\ \emph {et~al.}(2015)\citenamefont {Kunz},
  \citenamefont {Schekochihin}, \citenamefont {Chen}, \citenamefont {Abel},\
  and\ \citenamefont {Cowley}}]{kunz2015inertial}%
  \BibitemOpen
  \bibfield  {author} {\bibinfo {author} {\bibfnamefont {M.~W.}\ \bibnamefont
  {Kunz}}, \bibinfo {author} {\bibfnamefont {A.~A.}\ \bibnamefont
  {Schekochihin}}, \bibinfo {author} {\bibfnamefont {C.~H.~K.}\ \bibnamefont
  {Chen}}, \bibinfo {author} {\bibfnamefont {I.~G.}\ \bibnamefont {Abel}},\
  and\ \bibinfo {author} {\bibfnamefont {S.~C.}\ \bibnamefont {Cowley}},\
  }\bibfield  {title} {\bibinfo {title} {Inertial-range kinetic turbulence in
  pressure-anisotropic astrophysical plasmas},\ }\href
  {https://doi.org/10.1017/S0022377815000811} {\bibfield  {journal} {\bibinfo
  {journal} {\href{https://doi.org/10.1017/S0022377815000811}{Journal of Plasma
  Physics}}\ }\textbf {\bibinfo {volume} {81}},\ \bibinfo {pages} {325810501}
  (\bibinfo {year} {2015})}\BibitemShut {NoStop}%
\bibitem [{\citenamefont {Schekochihin}\ \emph {et~al.}(2016)\citenamefont
  {Schekochihin}, \citenamefont {Parker}, \citenamefont {Highcock},
  \citenamefont {Dellar}, \citenamefont {Dorland},\ and\ \citenamefont
  {Hammett}}]{schekochihin2016phase}%
  \BibitemOpen
  \bibfield  {author} {\bibinfo {author} {\bibfnamefont {A.~A.}\ \bibnamefont
  {Schekochihin}}, \bibinfo {author} {\bibfnamefont {J.~T.}\ \bibnamefont
  {Parker}}, \bibinfo {author} {\bibfnamefont {E.~G.}\ \bibnamefont
  {Highcock}}, \bibinfo {author} {\bibfnamefont {P.~J.}\ \bibnamefont
  {Dellar}}, \bibinfo {author} {\bibfnamefont {W.}~\bibnamefont {Dorland}},\
  and\ \bibinfo {author} {\bibfnamefont {G.~W.}\ \bibnamefont {Hammett}},\
  }\bibfield  {title} {\bibinfo {title} {Phase mixing versus nonlinear
  advection in drift-kinetic plasma turbulence},\ }\href
  {https://doi.org/10.1017/S0022377816000374} {\bibfield  {journal} {\bibinfo
  {journal} {\href{https://doi.org/10.1017/S0022377816000374}{Journal of Plasma
  Physics}}\ }\textbf {\bibinfo {volume} {82}},\ \bibinfo {pages} {905820212}
  (\bibinfo {year} {2016})}\BibitemShut {NoStop}%
\bibitem [{\citenamefont {Kunz}\ \emph {et~al.}(2018)\citenamefont {Kunz},
  \citenamefont {Abel}, \citenamefont {Klein},\ and\ \citenamefont
  {Schekochihin}}]{kunz2018astrophysical}%
  \BibitemOpen
  \bibfield  {author} {\bibinfo {author} {\bibfnamefont {M.~W.}\ \bibnamefont
  {Kunz}}, \bibinfo {author} {\bibfnamefont {I.~G.}\ \bibnamefont {Abel}},
  \bibinfo {author} {\bibfnamefont {K.~G.}\ \bibnamefont {Klein}},\ and\
  \bibinfo {author} {\bibfnamefont {A.~A.}\ \bibnamefont {Schekochihin}},\
  }\bibfield  {title} {\bibinfo {title} {Astrophysical gyrokinetics: turbulence
  in pressure-anisotropic plasmas at ion scales and beyond},\ }\href
  {https://doi.org/10.1017/S0022377818000296} {\bibfield  {journal} {\bibinfo
  {journal} {\href{https://doi.org/10.1017/S0022377818000296}{Journal of Plasma
  Physics}}\ }\textbf {\bibinfo {volume} {84}},\ \bibinfo {pages} {715840201}
  (\bibinfo {year} {2018})}\BibitemShut {NoStop}%
\bibitem [{\citenamefont {Schekochihin}\ \emph {et~al.}(2019)\citenamefont
  {Schekochihin}, \citenamefont {Kawazura},\ and\ \citenamefont
  {Barnes}}]{schekochihin2019constraints}%
  \BibitemOpen
  \bibfield  {author} {\bibinfo {author} {\bibfnamefont {A.~A.}\ \bibnamefont
  {Schekochihin}}, \bibinfo {author} {\bibfnamefont {Y.}~\bibnamefont
  {Kawazura}},\ and\ \bibinfo {author} {\bibfnamefont {M.~A.}\ \bibnamefont
  {Barnes}},\ }\bibfield  {title} {\bibinfo {title} {Constraints on ion versus
  electron heating by plasma turbulence at low beta},\ }\href
  {https://doi.org/10.1017/S0022377819000345} {\bibfield  {journal} {\bibinfo
  {journal} {\href{ttps://doi.org/10.1017/S0022377819000345}{Journal of Plasma
  Physics}}\ }\textbf {\bibinfo {volume} {85}},\ \bibinfo {pages} {905850303}
  (\bibinfo {year} {2019})}\BibitemShut {NoStop}%
\bibitem [{\citenamefont {Tatsuno}\ \emph {et~al.}(2009)\citenamefont
  {Tatsuno}, \citenamefont {Dorland}, \citenamefont {Schekochihin},
  \citenamefont {Plunk}, \citenamefont {Barnes}, \citenamefont {Cowley},\ and\
  \citenamefont {Howes}}]{tatsuno2009nonlinear}%
  \BibitemOpen
  \bibfield  {author} {\bibinfo {author} {\bibfnamefont {T.}~\bibnamefont
  {Tatsuno}}, \bibinfo {author} {\bibfnamefont {W.}~\bibnamefont {Dorland}},
  \bibinfo {author} {\bibfnamefont {A.~A.}\ \bibnamefont {Schekochihin}},
  \bibinfo {author} {\bibfnamefont {G.~G.}\ \bibnamefont {Plunk}}, \bibinfo
  {author} {\bibfnamefont {M.}~\bibnamefont {Barnes}}, \bibinfo {author}
  {\bibfnamefont {S.~C.}\ \bibnamefont {Cowley}},\ and\ \bibinfo {author}
  {\bibfnamefont {G.~G.}\ \bibnamefont {Howes}},\ }\bibfield  {title} {\bibinfo
  {title} {Nonlinear phase mixing and phase-space cascade of entropy in
  gyrokinetic plasma turbulence},\ }\href
  {https://doi.org/10.1103/PhysRevLett.103.015003} {\bibfield  {journal}
  {\bibinfo  {journal}
  {\href{https://doi.org/10.1103/PhysRevLett.103.015003}{Physical Review
  Letters}}\ }\textbf {\bibinfo {volume} {103}},\ \bibinfo {pages} {015003}
  (\bibinfo {year} {2009})}\BibitemShut {NoStop}%
\bibitem [{\citenamefont {Parker}\ \emph {et~al.}(2016)\citenamefont {Parker},
  \citenamefont {Highcock}, \citenamefont {Schekochihin},\ and\ \citenamefont
  {Dellar}}]{parker2016suppression}%
  \BibitemOpen
  \bibfield  {author} {\bibinfo {author} {\bibfnamefont {J.~T.}\ \bibnamefont
  {Parker}}, \bibinfo {author} {\bibfnamefont {E.~G.}\ \bibnamefont
  {Highcock}}, \bibinfo {author} {\bibfnamefont {A.~A.}\ \bibnamefont
  {Schekochihin}},\ and\ \bibinfo {author} {\bibfnamefont {P.~J.}\ \bibnamefont
  {Dellar}},\ }\bibfield  {title} {\bibinfo {title} {Suppression of phase
  mixing in drift-kinetic plasma turbulence},\ }\href
  {https://doi.org/10.1063/1.4958954} {\bibfield  {journal} {\bibinfo
  {journal} {\href{https://doi.org/10.1063/1.4958954}{Physics of Plasmas}}\
  }\textbf {\bibinfo {volume} {23}},\ \bibinfo {pages} {070703} (\bibinfo
  {year} {2016})}\BibitemShut {NoStop}%
\bibitem [{\citenamefont {Pezzi}\ \emph {et~al.}(2018)\citenamefont {Pezzi},
  \citenamefont {Servidio}, \citenamefont {Perrone}, \citenamefont {Valentini},
  \citenamefont {Sorriso-Valvo}, \citenamefont {Greco}, \citenamefont
  {Matthaeus},\ and\ \citenamefont {Veltri}}]{pezzi2018velocity}%
  \BibitemOpen
  \bibfield  {author} {\bibinfo {author} {\bibfnamefont {O.}~\bibnamefont
  {Pezzi}}, \bibinfo {author} {\bibfnamefont {S.}~\bibnamefont {Servidio}},
  \bibinfo {author} {\bibfnamefont {D.}~\bibnamefont {Perrone}}, \bibinfo
  {author} {\bibfnamefont {F.}~\bibnamefont {Valentini}}, \bibinfo {author}
  {\bibfnamefont {L.}~\bibnamefont {Sorriso-Valvo}}, \bibinfo {author}
  {\bibfnamefont {A.}~\bibnamefont {Greco}}, \bibinfo {author} {\bibfnamefont
  {W.}~\bibnamefont {Matthaeus}},\ and\ \bibinfo {author} {\bibfnamefont
  {P.}~\bibnamefont {Veltri}},\ }\bibfield  {title} {\bibinfo {title}
  {Velocity-space cascade in magnetized plasmas: {n}umerical simulations},\
  }\href {https://doi.org/10.1063/1.5027685} {\bibfield  {journal} {\bibinfo
  {journal} {\href{https://doi.org/10.1063/1.5027685}{Physics of Plasmas}}\
  }\textbf {\bibinfo {volume} {25}},\ \bibinfo {pages} {060704} (\bibinfo
  {year} {2018})}\BibitemShut {NoStop}%
\bibitem [{\citenamefont {Kawazura}\ \emph {et~al.}(2019)\citenamefont
  {Kawazura}, \citenamefont {Barnes},\ and\ \citenamefont
  {Schekochihin}}]{kawazura2019thermal}%
  \BibitemOpen
  \bibfield  {author} {\bibinfo {author} {\bibfnamefont {Y.}~\bibnamefont
  {Kawazura}}, \bibinfo {author} {\bibfnamefont {M.}~\bibnamefont {Barnes}},\
  and\ \bibinfo {author} {\bibfnamefont {A.~A.}\ \bibnamefont {Schekochihin}},\
  }\bibfield  {title} {\bibinfo {title} {Thermal disequilibration of ions and
  electrons by collisionless plasma turbulence},\ }\href
  {https://doi.org/10.1073/pnas.1812491116} {\bibfield  {journal} {\bibinfo
  {journal} {\href{https://doi.org/10.1073/pnas.1812491116}{Proceedings of the
  National Academy of Sciences}}\ }\textbf {\bibinfo {volume} {116}},\ \bibinfo
  {pages} {771} (\bibinfo {year} {2019})}\BibitemShut {NoStop}%
\bibitem [{\citenamefont {Meyrand}\ \emph {et~al.}(2019)\citenamefont
  {Meyrand}, \citenamefont {Kanekar}, \citenamefont {Dorland},\ and\
  \citenamefont {Schekochihin}}]{meyrand2019fluidization}%
  \BibitemOpen
  \bibfield  {author} {\bibinfo {author} {\bibfnamefont {R.}~\bibnamefont
  {Meyrand}}, \bibinfo {author} {\bibfnamefont {A.}~\bibnamefont {Kanekar}},
  \bibinfo {author} {\bibfnamefont {W.}~\bibnamefont {Dorland}},\ and\ \bibinfo
  {author} {\bibfnamefont {A.~A.}\ \bibnamefont {Schekochihin}},\ }\bibfield
  {title} {\bibinfo {title} {Fluidization of collisionless plasma turbulence},\
  }\href {https://doi.org/10.1073/pnas.1813913116} {\bibfield  {journal}
  {\bibinfo  {journal}
  {\href{https://doi.org/10.1073/pnas.1813913116}{Proceedings of the National
  Academy of Sciences}}\ }\textbf {\bibinfo {volume} {116}},\ \bibinfo {pages}
  {1185} (\bibinfo {year} {2019})}\BibitemShut {NoStop}%
\bibitem [{\citenamefont {Cerri}\ \emph {et~al.}(2018)\citenamefont {Cerri},
  \citenamefont {Kunz},\ and\ \citenamefont {Califano}}]{cerri2018dual}%
  \BibitemOpen
  \bibfield  {author} {\bibinfo {author} {\bibfnamefont {S.}~\bibnamefont
  {Cerri}}, \bibinfo {author} {\bibfnamefont {M.~W.}\ \bibnamefont {Kunz}},\
  and\ \bibinfo {author} {\bibfnamefont {F.}~\bibnamefont {Califano}},\
  }\bibfield  {title} {\bibinfo {title} {Dual phase-space cascades in {3D}
  hybrid-{V}lasov--{M}axwell turbulence},\ }\href
  {https://doi.org/10.3847/2041-8213/aab557} {\bibfield  {journal} {\bibinfo
  {journal} {\href{https://doi.org/10.3847/2041-8213/aab557}{Astrophysical
  Journal Letters}}\ }\textbf {\bibinfo {volume} {856}},\ \bibinfo {pages}
  {L13} (\bibinfo {year} {2018})}\BibitemShut {NoStop}%
\bibitem [{\citenamefont {Kawazura}\ \emph {et~al.}(2020)\citenamefont
  {Kawazura}, \citenamefont {Schekochihin}, \citenamefont {Barnes},
  \citenamefont {TenBarge}, \citenamefont {Tong}, \citenamefont {Klein},\ and\
  \citenamefont {Dorland}}]{kawazura2020ion}%
  \BibitemOpen
  \bibfield  {author} {\bibinfo {author} {\bibfnamefont {Y.}~\bibnamefont
  {Kawazura}}, \bibinfo {author} {\bibfnamefont {A.~A.}\ \bibnamefont
  {Schekochihin}}, \bibinfo {author} {\bibfnamefont {M.}~\bibnamefont
  {Barnes}}, \bibinfo {author} {\bibfnamefont {J.~M.}\ \bibnamefont
  {TenBarge}}, \bibinfo {author} {\bibfnamefont {Y.}~\bibnamefont {Tong}},
  \bibinfo {author} {\bibfnamefont {K.~G.}\ \bibnamefont {Klein}},\ and\
  \bibinfo {author} {\bibfnamefont {W.}~\bibnamefont {Dorland}},\ }\bibfield
  {title} {\bibinfo {title} {Ion versus electron heating in compressively
  driven astrophysical gyrokinetic turbulence},\ }\href
  {https://doi.org/10.1103/PhysRevX.10.041050} {\bibfield  {journal} {\bibinfo
  {journal} {\href{https://doi.org/10.1103/PhysRevX.10.041050}{Physical Review
  X}}\ }\textbf {\bibinfo {volume} {10}},\ \bibinfo {pages} {041050} (\bibinfo
  {year} {2020})}\BibitemShut {NoStop}%
\bibitem [{\citenamefont {Zhou}\ \emph {et~al.}(2023)\citenamefont {Zhou},
  \citenamefont {Liu},\ and\ \citenamefont {Loureiro}}]{zhou2023electron}%
  \BibitemOpen
  \bibfield  {author} {\bibinfo {author} {\bibfnamefont {M.}~\bibnamefont
  {Zhou}}, \bibinfo {author} {\bibfnamefont {Z.}~\bibnamefont {Liu}},\ and\
  \bibinfo {author} {\bibfnamefont {N.~F.}\ \bibnamefont {Loureiro}},\
  }\bibfield  {title} {\bibinfo {title} {Electron heating in
  kinetic-{A}lfv{\'e}n-wave turbulence},\ }\href
  {https://doi.org/10.1073/pnas.2220927120} {\bibfield  {journal} {\bibinfo
  {journal} {\href{https://doi.org/10.1073/pnas.2220927120}{Proceedings of the
  National Academy of Sciences}}\ }\textbf {\bibinfo {volume} {120}},\ \bibinfo
  {pages} {e2220927120} (\bibinfo {year} {2023})}\BibitemShut {NoStop}%
\bibitem [{\citenamefont {Servidio}\ \emph {et~al.}(2017)\citenamefont
  {Servidio}, \citenamefont {Chasapis}, \citenamefont {Matthaeus},
  \citenamefont {Perrone}, \citenamefont {Valentini}, \citenamefont {Parashar},
  \citenamefont {Veltri}, \citenamefont {Gershman}, \citenamefont {Russell},
  \citenamefont {Giles} \emph {et~al.}}]{servidio2017magnetospheric}%
  \BibitemOpen
  \bibfield  {author} {\bibinfo {author} {\bibfnamefont {S.}~\bibnamefont
  {Servidio}}, \bibinfo {author} {\bibfnamefont {A.}~\bibnamefont {Chasapis}},
  \bibinfo {author} {\bibfnamefont {W.~H.}\ \bibnamefont {Matthaeus}}, \bibinfo
  {author} {\bibfnamefont {D.}~\bibnamefont {Perrone}}, \bibinfo {author}
  {\bibfnamefont {F.}~\bibnamefont {Valentini}}, \bibinfo {author}
  {\bibfnamefont {T.~N.}\ \bibnamefont {Parashar}}, \bibinfo {author}
  {\bibfnamefont {P.}~\bibnamefont {Veltri}}, \bibinfo {author} {\bibfnamefont
  {D.}~\bibnamefont {Gershman}}, \bibinfo {author} {\bibfnamefont {C.~T.}\
  \bibnamefont {Russell}}, \bibinfo {author} {\bibfnamefont {B.}~\bibnamefont
  {Giles}}, \emph {et~al.},\ }\bibfield  {title} {\bibinfo {title}
  {Magnetospheric multiscale observation of plasma velocity-space cascade:
  Hermite representation and theory},\ }\href
  {https://doi.org/10.1103/PhysRevLett.119.205101} {\bibfield  {journal}
  {\bibinfo  {journal}
  {\href{https://doi.org/10.1103/PhysRevLett.119.205101}{Physical Review
  Letters}}\ }\textbf {\bibinfo {volume} {119}},\ \bibinfo {pages} {205101}
  (\bibinfo {year} {2017})}\BibitemShut {NoStop}%
\bibitem [{\citenamefont {Wu}\ \emph {et~al.}(2023)\citenamefont {Wu},
  \citenamefont {He}, \citenamefont {Duan}, \citenamefont {Zhu}, \citenamefont
  {Hou}, \citenamefont {Verscharen}, \citenamefont {Nicolaou}, \citenamefont
  {Owen}, \citenamefont {Fedorov},\ and\ \citenamefont {Louarn}}]{wu2023ion}%
  \BibitemOpen
  \bibfield  {author} {\bibinfo {author} {\bibfnamefont {Z.}~\bibnamefont
  {Wu}}, \bibinfo {author} {\bibfnamefont {J.}~\bibnamefont {He}}, \bibinfo
  {author} {\bibfnamefont {D.}~\bibnamefont {Duan}}, \bibinfo {author}
  {\bibfnamefont {X.}~\bibnamefont {Zhu}}, \bibinfo {author} {\bibfnamefont
  {C.}~\bibnamefont {Hou}}, \bibinfo {author} {\bibfnamefont {D.}~\bibnamefont
  {Verscharen}}, \bibinfo {author} {\bibfnamefont {G.}~\bibnamefont
  {Nicolaou}}, \bibinfo {author} {\bibfnamefont {C.~J.}\ \bibnamefont {Owen}},
  \bibinfo {author} {\bibfnamefont {A.}~\bibnamefont {Fedorov}},\ and\ \bibinfo
  {author} {\bibfnamefont {P.}~\bibnamefont {Louarn}},\ }\bibfield  {title}
  {\bibinfo {title} {Ion energization and thermalization in magnetic
  reconnection exhaust region in the solar wind},\ }\href
  {https://iopscience.iop.org/article/10.3847/1538-4357/accf9b} {\bibfield
  {journal} {\bibinfo  {journal}
  {\href{https://iopscience.iop.org/article/10.3847/1538-4357/accf9b}{The
  Astrophysical Journal}}\ }\textbf {\bibinfo {volume} {951}},\ \bibinfo
  {pages} {98} (\bibinfo {year} {2023})}\BibitemShut {NoStop}%
\bibitem [{\citenamefont {Kawamori}(2013)}]{kawamori2013verification}%
  \BibitemOpen
  \bibfield  {author} {\bibinfo {author} {\bibfnamefont {E.}~\bibnamefont
  {Kawamori}},\ }\bibfield  {title} {\bibinfo {title} {Experimental
  verification of entropy cascade in two-dimensional electrostatic turbulence
  in magnetized plasma},\ }\href
  {https://doi.org/10.1103/PhysRevLett.110.095001} {\bibfield  {journal}
  {\bibinfo  {journal}
  {\href{https://doi.org/10.1103/PhysRevLett.110.095001}{Physical Review
  Letters}}\ }\textbf {\bibinfo {volume} {110}},\ \bibinfo {pages} {095001}
  (\bibinfo {year} {2013})}\BibitemShut {NoStop}%
\bibitem [{\citenamefont {Kawamori}\ and\ \citenamefont
  {Lin}(2022)}]{kawamori2022evidence}%
  \BibitemOpen
  \bibfield  {author} {\bibinfo {author} {\bibfnamefont {E.}~\bibnamefont
  {Kawamori}}\ and\ \bibinfo {author} {\bibfnamefont {Y.-T.}\ \bibnamefont
  {Lin}},\ }\bibfield  {title} {\bibinfo {title} {Evidence of entropy cascade
  in collisionless magnetized plasma turbulence},\ }\href
  {https://www.nature.com/articles/s42005-022-01115-7} {\bibfield  {journal}
  {\bibinfo  {journal}
  {\href{https://www.nature.com/articles/s42005-022-01115-7}{Communications
  Physics}}\ }\textbf {\bibinfo {volume} {5}},\ \bibinfo {pages} {338}
  (\bibinfo {year} {2022})}\BibitemShut {NoStop}%
\bibitem [{\citenamefont {Adkins}\ and\ \citenamefont
  {Schekochihin}(2018)}]{adkins2018solvable}%
  \BibitemOpen
  \bibfield  {author} {\bibinfo {author} {\bibfnamefont {T.}~\bibnamefont
  {Adkins}}\ and\ \bibinfo {author} {\bibfnamefont {A.~A.}\ \bibnamefont
  {Schekochihin}},\ }\bibfield  {title} {\bibinfo {title} {A solvable model of
  {V}lasov-kinetic plasma turbulence in {F}ourier--{H}ermite phase space},\
  }\href {https://doi.org/10.1017/S0022377818000089} {\bibfield  {journal}
  {\bibinfo  {journal}
  {\href{https://doi.org/10.1017/S0022377818000089}{Journal of Plasma
  Physics}}\ }\textbf {\bibinfo {volume} {84}},\ \bibinfo {pages} {905840107}
  (\bibinfo {year} {2018})}\BibitemShut {NoStop}%
\bibitem [{\citenamefont {Celebre}\ \emph {et~al.}(2023)\citenamefont
  {Celebre}, \citenamefont {Servidio},\ and\ \citenamefont
  {Valentini}}]{celebre2023}%
  \BibitemOpen
  \bibfield  {author} {\bibinfo {author} {\bibfnamefont {G.}~\bibnamefont
  {Celebre}}, \bibinfo {author} {\bibfnamefont {S.}~\bibnamefont {Servidio}},\
  and\ \bibinfo {author} {\bibfnamefont {F.}~\bibnamefont {Valentini}},\
  }\bibfield  {title} {\bibinfo {title} {Phase space dynamics of unmagnetized
  plasmas: Collisionless and collisional regimes},\ }\href
  {https://doi.org/10.1063/5.0160549} {\bibfield  {journal} {\bibinfo
  {journal} {\href{https://doi.org/10.1063/5.0160549}{Physics of Plasmas}}\
  }\textbf {\bibinfo {volume} {30}},\ \bibinfo {pages} {092304} (\bibinfo
  {year} {2023})}\BibitemShut {NoStop}%
\bibitem [{\citenamefont {Knorr}(1977)}]{knorr1977time}%
  \BibitemOpen
  \bibfield  {author} {\bibinfo {author} {\bibfnamefont {G.}~\bibnamefont
  {Knorr}},\ }\bibfield  {title} {\bibinfo {title} {Time asymptotic statistics
  of the {V}lasov equation},\ }\href
  {https://doi.org/10.1017/S0022377800020808} {\bibfield  {journal} {\bibinfo
  {journal} {\href{https://doi.org/10.1017/S0022377800020808}{Journal of Plasma
  Physics}}\ }\textbf {\bibinfo {volume} {17}},\ \bibinfo {pages} {553}
  (\bibinfo {year} {1977})}\BibitemShut {NoStop}%
\bibitem [{\citenamefont {Zhdankin}(2022)}]{zhdankin2022generalized}%
  \BibitemOpen
  \bibfield  {author} {\bibinfo {author} {\bibfnamefont {V.}~\bibnamefont
  {Zhdankin}},\ }\bibfield  {title} {\bibinfo {title} {Generalized entropy
  production in collisionless plasma flows and turbulence},\ }\href
  {https://doi.org/10.1103/PhysRevX.12.031011} {\bibfield  {journal} {\bibinfo
  {journal} {\href{https://doi.org/10.1103/PhysRevX.12.031011}{Physical Review
  X}}\ }\textbf {\bibinfo {volume} {12}},\ \bibinfo {pages} {031011} (\bibinfo
  {year} {2022})}\BibitemShut {NoStop}%
\bibitem [{\citenamefont {Kraichnan}(1968)}]{kraichnan1968small}%
  \BibitemOpen
  \bibfield  {author} {\bibinfo {author} {\bibfnamefont {R.~H.}\ \bibnamefont
  {Kraichnan}},\ }\bibfield  {title} {\bibinfo {title} {Small-scale structure
  of a scalar field convected by turbulence},\ }\href
  {https://doi.org/10.1063/1.1692063} {\bibfield  {journal} {\bibinfo
  {journal} {\href{https://doi.org/10.1063/1.1692063}{Physics of Fluids}}\
  }\textbf {\bibinfo {volume} {11}},\ \bibinfo {pages} {945} (\bibinfo {year}
  {1968})}\BibitemShut {NoStop}%
\bibitem [{\citenamefont {Falkovich}\ \emph {et~al.}(2001)\citenamefont
  {Falkovich}, \citenamefont {Gawedzki},\ and\ \citenamefont
  {Vergassola}}]{falkovich2001particles}%
  \BibitemOpen
  \bibfield  {author} {\bibinfo {author} {\bibfnamefont {G.}~\bibnamefont
  {Falkovich}}, \bibinfo {author} {\bibfnamefont {K.}~\bibnamefont
  {Gawedzki}},\ and\ \bibinfo {author} {\bibfnamefont {M.}~\bibnamefont
  {Vergassola}},\ }\bibfield  {title} {\bibinfo {title} {Particles and fields
  in fluid turbulence},\ }\href {https://doi.org/10.1103/RevModPhys.73.913}
  {\bibfield  {journal} {\bibinfo  {journal}
  {\href{https://doi.org/10.1103/RevModPhys.73.913}{Reviews of Modern
  Physics}}\ }\textbf {\bibinfo {volume} {73}},\ \bibinfo {pages} {913}
  (\bibinfo {year} {2001})}\BibitemShut {NoStop}%
\bibitem [{\citenamefont {Kolmogorov}(1941)}]{kolmogorov1941b}%
  \BibitemOpen
  \bibfield  {author} {\bibinfo {author} {\bibfnamefont {A.~N.}\ \bibnamefont
  {Kolmogorov}},\ }\bibfield  {title} {\bibinfo {title} {The local structure of
  turbulence in incompressible viscous fluid at very large {R}eynolds
  numbers},\ }\href@noop {} {\bibfield  {journal} {\bibinfo  {journal} {Doklady
  Akademii Nauk SSSR}\ }\textbf {\bibinfo {volume} {30}},\ \bibinfo {pages}
  {299} (\bibinfo {year} {1941})}\BibitemShut {NoStop}%
\bibitem [{\citenamefont {Obukhov}(1949)}]{obukhov1949structure}%
  \BibitemOpen
  \bibfield  {author} {\bibinfo {author} {\bibfnamefont {A.~M.}\ \bibnamefont
  {Obukhov}},\ }\bibfield  {title} {\bibinfo {title} {Structure of the
  temperature field in a turbulent flow},\ }\href@noop {} {\bibfield  {journal}
  {\bibinfo  {journal} {Izv. Akad. Nauk SSSR, Ser. Geogr. Geofiz}\ }\textbf
  {\bibinfo {volume} {13}},\ \bibinfo {pages} {58} (\bibinfo {year}
  {1949})}\BibitemShut {NoStop}%
\bibitem [{\citenamefont {Corrsin}(1951)}]{corrsin1951spectrum}%
  \BibitemOpen
  \bibfield  {author} {\bibinfo {author} {\bibfnamefont {S.}~\bibnamefont
  {Corrsin}},\ }\bibfield  {title} {\bibinfo {title} {On the spectrum of
  isotropic temperature fluctuations in an isotropic turbulence},\ }\href
  {https://doi.org/10.1063/1.1699986} {\bibfield  {journal} {\bibinfo
  {journal} {\href{https://doi.org/10.1063/1.1699986}{Journal of Applied
  Physics}}\ }\textbf {\bibinfo {volume} {22}},\ \bibinfo {pages} {469}
  (\bibinfo {year} {1951})}\BibitemShut {NoStop}%
\bibitem [{\citenamefont {Frisch}(1995)}]{frisch1995turbulence}%
  \BibitemOpen
  \bibfield  {author} {\bibinfo {author} {\bibfnamefont {U.}~\bibnamefont
  {Frisch}},\ }\href@noop {} {\emph {\bibinfo {title} {Turbulence: The Legacy
  of A. N. Kolmogorov}}}\ (\bibinfo  {publisher} {Cambridge University Press},\
  \bibinfo {address} {Cambridge},\ \bibinfo {year} {1995})\BibitemShut
  {NoStop}%
\bibitem [{\citenamefont {Batchelor}(1959)}]{batchelor1959small}%
  \BibitemOpen
  \bibfield  {author} {\bibinfo {author} {\bibfnamefont {G.~K.}\ \bibnamefont
  {Batchelor}},\ }\bibfield  {title} {\bibinfo {title} {Small-scale variation
  of convected quantities like temperature in turbulent fluid part 1. {G}eneral
  discussion and the case of small conductivity},\ }\href
  {https://doi.org/10.1017/S002211205900009X} {\bibfield  {journal} {\bibinfo
  {journal} {\href{https://doi.org/10.1017/S002211205900009X}{Journal of Fluid
  Mechanics}}\ }\textbf {\bibinfo {volume} {5}},\ \bibinfo {pages} {113}
  (\bibinfo {year} {1959})}\BibitemShut {NoStop}%
\bibitem [{\citenamefont {Goldreich}\ and\ \citenamefont
  {Sridhar}(1995)}]{goldreich1995toward}%
  \BibitemOpen
  \bibfield  {author} {\bibinfo {author} {\bibfnamefont {P.}~\bibnamefont
  {Goldreich}}\ and\ \bibinfo {author} {\bibfnamefont {S.}~\bibnamefont
  {Sridhar}},\ }\bibfield  {title} {\bibinfo {title} {Toward a theory of
  interstellar turbulence. 2. {S}trong {A}lfv{\'e}nic turbulence},\ }\href
  {https://adsabs.harvard.edu/doi/10.1086/175121} {\bibfield  {journal}
  {\bibinfo  {journal}
  {\href{https://adsabs.harvard.edu/doi/10.1086/175121}{Astrophysical
  Journal}}\ }\textbf {\bibinfo {volume} {438}},\ \bibinfo {pages} {763}
  (\bibinfo {year} {1995})}\BibitemShut {NoStop}%
\bibitem [{\citenamefont {Schekochihin}(2022)}]{schekochihin2022mhd}%
  \BibitemOpen
  \bibfield  {author} {\bibinfo {author} {\bibfnamefont {A.~A.}\ \bibnamefont
  {Schekochihin}},\ }\bibfield  {title} {\bibinfo {title} {{MHD} turbulence: a
  biased review},\ }\href {https://doi.org/10.1017/S0022377822000721}
  {\bibfield  {journal} {\bibinfo  {journal}
  {\href{https://doi.org/10.1017/S0022377822000721}{Journal of Plasma
  Physics}}\ }\textbf {\bibinfo {volume} {88}},\ \bibinfo {pages} {155880501}
  (\bibinfo {year} {2022})}\BibitemShut {NoStop}%
\bibitem [{\citenamefont {Sreenivasan}(2019)}]{sreenivasan2019turbulent}%
  \BibitemOpen
  \bibfield  {author} {\bibinfo {author} {\bibfnamefont {K.~R.}\ \bibnamefont
  {Sreenivasan}},\ }\bibfield  {title} {\bibinfo {title} {Turbulent mixing: A
  perspective},\ }\href {https://doi.org/10.1073/pnas.180046311} {\bibfield
  {journal} {\bibinfo  {journal}
  {\href{https://doi.org/10.1073/pnas.180046311}{Proceedings of the National
  Academy of Sciences}}\ }\textbf {\bibinfo {volume} {116}},\ \bibinfo {pages}
  {18175} (\bibinfo {year} {2019})}\BibitemShut {NoStop}%
\bibitem [{\citenamefont {Gould}\ \emph {et~al.}(1967)\citenamefont {Gould},
  \citenamefont {O'{N}eil},\ and\ \citenamefont {Malmberg}}]{gould1967plasma}%
  \BibitemOpen
  \bibfield  {author} {\bibinfo {author} {\bibfnamefont {R.~W.}\ \bibnamefont
  {Gould}}, \bibinfo {author} {\bibfnamefont {T.~M.}\ \bibnamefont
  {O'{N}eil}},\ and\ \bibinfo {author} {\bibfnamefont {J.~H.}\ \bibnamefont
  {Malmberg}},\ }\bibfield  {title} {\bibinfo {title} {Plasma wave echo},\
  }\href {https://doi.org/10.1103/PhysRevLett.19.219} {\bibfield  {journal}
  {\bibinfo  {journal}
  {\href{https://doi.org/10.1103/PhysRevLett.19.219}{Physical Review Letters}}\
  }\textbf {\bibinfo {volume} {19}},\ \bibinfo {pages} {219} (\bibinfo {year}
  {1967})}\BibitemShut {NoStop}%
\bibitem [{\citenamefont {Malmberg}\ \emph {et~al.}(1968)\citenamefont
  {Malmberg}, \citenamefont {Wharton}, \citenamefont {Gould},\ and\
  \citenamefont {O'{N}eil}}]{malmberg1968plasma}%
  \BibitemOpen
  \bibfield  {author} {\bibinfo {author} {\bibfnamefont {J.~H.}\ \bibnamefont
  {Malmberg}}, \bibinfo {author} {\bibfnamefont {C.~B.}\ \bibnamefont
  {Wharton}}, \bibinfo {author} {\bibfnamefont {R.~W.}\ \bibnamefont {Gould}},\
  and\ \bibinfo {author} {\bibfnamefont {T.~M.}\ \bibnamefont {O'{N}eil}},\
  }\bibfield  {title} {\bibinfo {title} {Plasma wave echo experiment},\ }\href
  {https://doi.org/10.1103/PhysRevLett.20.95} {\bibfield  {journal} {\bibinfo
  {journal} {\href{https://doi.org/10.1103/PhysRevLett.20.95}{Physical Review
  Letters}}\ }\textbf {\bibinfo {volume} {20}},\ \bibinfo {pages} {95}
  (\bibinfo {year} {1968})}\BibitemShut {NoStop}%
\bibitem [{\citenamefont {Biglari}\ and\ \citenamefont
  {Diamond}(1988)}]{biglari1988cascade}%
  \BibitemOpen
  \bibfield  {author} {\bibinfo {author} {\bibfnamefont {H.}~\bibnamefont
  {Biglari}}\ and\ \bibinfo {author} {\bibfnamefont {P.~H.}\ \bibnamefont
  {Diamond}},\ }\bibfield  {title} {\bibinfo {title} {Cascade and intermittency
  model for turbulent compressible self-gravitating matter and self-binding
  phase-space density fluctuations},\ }\href
  {https://doi.org/10.1103/PhysRevLett.61.1716} {\bibfield  {journal} {\bibinfo
   {journal} {\href{https://doi.org/10.1103/PhysRevLett.61.1716}{Physical
  Review Letters}}\ }\textbf {\bibinfo {volume} {61}},\ \bibinfo {pages} {1716}
  (\bibinfo {year} {1988})}\BibitemShut {NoStop}%
\bibitem [{\citenamefont {Biglari}\ and\ \citenamefont
  {Diamond}(1989)}]{biglari1989clouds}%
  \BibitemOpen
  \bibfield  {author} {\bibinfo {author} {\bibfnamefont {H.}~\bibnamefont
  {Biglari}}\ and\ \bibinfo {author} {\bibfnamefont {P.~H.}\ \bibnamefont
  {Diamond}},\ }\bibfield  {title} {\bibinfo {title} {Clouds and holes:
  Self-organization in compressible fluid and collisionless plasma
  turbulence},\ }\href {https://doi.org/10.1016/0167-2789(89)90130-9}
  {\bibfield  {journal} {\bibinfo  {journal}
  {\href{https://doi.org/10.1016/0167-2789(89)90130-9}{Physica D: Nonlinear
  Phenomena}}\ }\textbf {\bibinfo {volume} {37}},\ \bibinfo {pages} {206}
  (\bibinfo {year} {1989})}\BibitemShut {NoStop}%
\bibitem [{\citenamefont {Landau}(1965{\natexlab{b}})}]{landau1936transport}%
  \BibitemOpen
  \bibfield  {author} {\bibinfo {author} {\bibfnamefont {L.~D.}\ \bibnamefont
  {Landau}},\ }\bibfield  {title} {\bibinfo {title} {The transport equation in
  the case of {C}oulomb interactions},\ }in\ \href@noop {} {\emph {\bibinfo
  {booktitle} {Collected Papers of L.D. Landau}}},\ \bibinfo {editor} {edited
  by\ \bibinfo {editor} {\bibfnamefont {D.}~\bibnamefont {ter Haar}}}\
  (\bibinfo  {publisher} {Pergamon Press},\ \bibinfo {address} {Oxford},\
  \bibinfo {year} {1965})\ pp.\ \bibinfo {pages} {163--170}\BibitemShut
  {NoStop}%
\bibitem [{\citenamefont {Lenard}(1960)}]{lenard1960bogoliubov}%
  \BibitemOpen
  \bibfield  {author} {\bibinfo {author} {\bibfnamefont {A.}~\bibnamefont
  {Lenard}},\ }\bibfield  {title} {\bibinfo {title} {On {B}ogoliubov's kinetic
  equation for a spatially homogeneous plasma},\ }\href
  {https://doi.org/10.1016/0003-4916(60)90003-8} {\bibfield  {journal}
  {\bibinfo  {journal}
  {\href{https://doi.org/10.1016/0003-4916(60)90003-8}{Annals of Physics}}\
  }\textbf {\bibinfo {volume} {10}},\ \bibinfo {pages} {390} (\bibinfo {year}
  {1960})}\BibitemShut {NoStop}%
\bibitem [{\citenamefont {Balescu}(1960)}]{balescu1960irreversible}%
  \BibitemOpen
  \bibfield  {author} {\bibinfo {author} {\bibfnamefont {R.}~\bibnamefont
  {Balescu}},\ }\bibfield  {title} {\bibinfo {title} {Irreversible processes in
  ionized gases},\ }\href {https://doi.org/10.1063/1.1706002} {\bibfield
  {journal} {\bibinfo  {journal}
  {\href{https://doi.org/10.1063/1.1706002}{Physics of Fluids}}\ }\textbf
  {\bibinfo {volume} {3}},\ \bibinfo {pages} {52} (\bibinfo {year}
  {1960})}\BibitemShut {NoStop}%
\bibitem [{\citenamefont {Bogoliubov}(1946)}]{Bogolyubov1946}%
  \BibitemOpen
  \bibfield  {author} {\bibinfo {author} {\bibfnamefont {N.~N.}\ \bibnamefont
  {Bogoliubov}},\ }\href@noop {} {\emph {\bibinfo {title} {The Problems of
  Dynamic Theory in Statistical Physics}}}\ (\bibinfo  {publisher}
  {Gostechizdat},\ \bibinfo {address} {Moscow},\ \bibinfo {year}
  {1946})\BibitemShut {NoStop}%
\bibitem [{\citenamefont {Oberman}\ \emph {et~al.}(1962)\citenamefont
  {Oberman}, \citenamefont {Ron},\ and\ \citenamefont
  {Dawson}}]{oberman1962high}%
  \BibitemOpen
  \bibfield  {author} {\bibinfo {author} {\bibfnamefont {C.}~\bibnamefont
  {Oberman}}, \bibinfo {author} {\bibfnamefont {A.}~\bibnamefont {Ron}},\ and\
  \bibinfo {author} {\bibfnamefont {J.}~\bibnamefont {Dawson}},\ }\bibfield
  {title} {\bibinfo {title} {High-frequency conductivity of a fully ionized
  plasma},\ }\href {https://doi.org/10.1063/1.1706560} {\bibfield  {journal}
  {\bibinfo  {journal} {\href{https://doi.org/10.1063/1.1706560}{Physics of
  Fluids}}\ }\textbf {\bibinfo {volume} {5}},\ \bibinfo {pages} {1514}
  (\bibinfo {year} {1962})}\BibitemShut {NoStop}%
\bibitem [{\citenamefont {Ye}\ and\ \citenamefont
  {Morrison}(1992)}]{ye1992action}%
  \BibitemOpen
  \bibfield  {author} {\bibinfo {author} {\bibfnamefont {H.}~\bibnamefont
  {Ye}}\ and\ \bibinfo {author} {\bibfnamefont {P.~J.}\ \bibnamefont
  {Morrison}},\ }\bibfield  {title} {\bibinfo {title} {Action principles for
  the {V}lasov equation},\ }\href {https://doi.org/10.1063/1.860231} {\bibfield
   {journal} {\bibinfo  {journal}
  {\href{https://doi.org/10.1063/1.860231}{Physics of Fluids B}}\ }\textbf
  {\bibinfo {volume} {4}},\ \bibinfo {pages} {771} (\bibinfo {year}
  {1992})}\BibitemShut {NoStop}%
\bibitem [{\citenamefont {Schekochihin}(2025)}]{schekochihin2022lectures}%
  \BibitemOpen
  \bibfield  {author} {\bibinfo {author} {\bibfnamefont {A.~A.}\ \bibnamefont
  {Schekochihin}},\ }\bibfield  {title} {\bibinfo {title} {Lectures on kinetic
  theory and magnetohydrodynamics of plasmas},\ }\href
  {http://www-thphys.physics.ox.ac.uk/people/AlexanderSchekochihin/KT/2015/KTLectureNotes.pdf}
  {\bibfield  {journal} {\bibinfo  {journal} {Lecture Notes for the Oxford
  MMathPhys Programme}\ } (\bibinfo {year} {2025})}\BibitemShut {NoStop}%
\bibitem [{\citenamefont {Kraichnan}(1973)}]{kraichnan1973helical}%
  \BibitemOpen
  \bibfield  {author} {\bibinfo {author} {\bibfnamefont {R.~H.}\ \bibnamefont
  {Kraichnan}},\ }\bibfield  {title} {\bibinfo {title} {Helical turbulence and
  absolute equilibrium},\ }\href {https://doi.org/10.1017/S0022112073001837}
  {\bibfield  {journal} {\bibinfo  {journal}
  {\href{https://doi.org/10.1017/S0022112073001837}{Journal of Fluid
  Mechanics}}\ }\textbf {\bibinfo {volume} {59}},\ \bibinfo {pages} {745}
  (\bibinfo {year} {1973})}\BibitemShut {NoStop}%
\bibitem [{\citenamefont {Alexakis}\ and\ \citenamefont
  {Brachet}(2019)}]{alexakis2019thermal}%
  \BibitemOpen
  \bibfield  {author} {\bibinfo {author} {\bibfnamefont {A.}~\bibnamefont
  {Alexakis}}\ and\ \bibinfo {author} {\bibfnamefont {M.-E.}\ \bibnamefont
  {Brachet}},\ }\bibfield  {title} {\bibinfo {title} {On the thermal
  equilibrium state of large-scale flows},\ }\href
  {https://doi.org/10.1017/jfm.2019.394} {\bibfield  {journal} {\bibinfo
  {journal} {\href{https://doi.org/10.1017/jfm.2019.394}{Journal of Fluid
  Mechanics}}\ }\textbf {\bibinfo {volume} {872}},\ \bibinfo {pages} {594}
  (\bibinfo {year} {2019})}\BibitemShut {NoStop}%
\bibitem [{\citenamefont {Hosking}\ and\ \citenamefont
  {Schekochihin}(2023)}]{hosking2023emergence}%
  \BibitemOpen
  \bibfield  {author} {\bibinfo {author} {\bibfnamefont {D.~N.}\ \bibnamefont
  {Hosking}}\ and\ \bibinfo {author} {\bibfnamefont {A.~A.}\ \bibnamefont
  {Schekochihin}},\ }\bibfield  {title} {\bibinfo {title} {Emergence of
  long-range correlations and thermal spectra in forced turbulence},\ }\href
  {https://doi.org/10.1017/jfm.2023.643} {\bibfield  {journal} {\bibinfo
  {journal} {\href{https://doi.org/10.1017/jfm.2023.643}{Journal of Fluid
  Mechanics}}\ }\textbf {\bibinfo {volume} {973}},\ \bibinfo {pages} {A13}
  (\bibinfo {year} {2023})}\BibitemShut {NoStop}%
\bibitem [{Note1()}]{Note1}%
  \BibitemOpen
  \bibinfo {note} {Note that by assuming \protect \eqref {outer_scale_u0}, we
  are ignoring the possibility that the cascade is driven by $\delta \protect
  \! f$ at scales larger than the outer scale $r_0$. If these fluctuations were
  associated with phase-space turbulence at larger scales (e.g., the
  gyrokinetic entropy cascade \cite {schekochihin2008gyrokinetic,
  schekochihin2009astrophysical, tatsuno2009nonlinear, plunk2010two}), they
  would likely have $u_0 < v_{\protect \mathrm {the}}$. How the phase-space
  cascade is modified in the presence of an injection with fine-scale
  velocity-space structure is an interesting question for future work, which,
  along with the other considerations discussed in Section \ref
  {implications_dissipation}, is important for determining how the phase-space
  cascade described in this paper connects to turbulence at scales above
  $r_0$}\BibitemShut {NoStop}%
\bibitem [{\citenamefont {Su}\ and\ \citenamefont
  {Oberman}(1968)}]{su1968collisional}%
  \BibitemOpen
  \bibfield  {author} {\bibinfo {author} {\bibfnamefont {C.}~\bibnamefont
  {Su}}\ and\ \bibinfo {author} {\bibfnamefont {C.}~\bibnamefont {Oberman}},\
  }\bibfield  {title} {\bibinfo {title} {Collisional damping of a plasma
  echo},\ }\href {https://doi.org/10.1103/PhysRevLett.20.427} {\bibfield
  {journal} {\bibinfo  {journal}
  {\href{https://doi.org/10.1103/PhysRevLett.20.427}{Physical Review Letters}}\
  }\textbf {\bibinfo {volume} {20}},\ \bibinfo {pages} {427} (\bibinfo {year}
  {1968})}\BibitemShut {NoStop}%
\bibitem [{\citenamefont {Zocco}\ and\ \citenamefont
  {Schekochihin}(2011)}]{zocco2011reduced}%
  \BibitemOpen
  \bibfield  {author} {\bibinfo {author} {\bibfnamefont {A.}~\bibnamefont
  {Zocco}}\ and\ \bibinfo {author} {\bibfnamefont {A.~A.}\ \bibnamefont
  {Schekochihin}},\ }\bibfield  {title} {\bibinfo {title} {Reduced
  fluid-kinetic equations for low-frequency dynamics, magnetic reconnection,
  and electron heating in low-beta plasmas},\ }\href
  {https://aip.scitation.org/doi/full/10.1063/1.3628639} {\bibfield  {journal}
  {\bibinfo  {journal}
  {\href{https://aip.scitation.org/doi/full/10.1063/1.3628639}{Physics of
  Plasmas}}\ }\textbf {\bibinfo {volume} {18}},\ \bibinfo {pages} {102309}
  (\bibinfo {year} {2011})}\BibitemShut {NoStop}%
\bibitem [{\citenamefont {Kanekar}\ \emph {et~al.}(2015)\citenamefont
  {Kanekar}, \citenamefont {Schekochihin}, \citenamefont {Dorland},\ and\
  \citenamefont {Loureiro}}]{kanekar2015fluctuation}%
  \BibitemOpen
  \bibfield  {author} {\bibinfo {author} {\bibfnamefont {A.}~\bibnamefont
  {Kanekar}}, \bibinfo {author} {\bibfnamefont {A.~A.}\ \bibnamefont
  {Schekochihin}}, \bibinfo {author} {\bibfnamefont {W.}~\bibnamefont
  {Dorland}},\ and\ \bibinfo {author} {\bibfnamefont {N.~F.}\ \bibnamefont
  {Loureiro}},\ }\bibfield  {title} {\bibinfo {title} {Fluctuation-dissipation
  relations for a plasma-kinetic {L}angevin equation},\ }\href
  {https://doi.org/10.1017/S0022377814000622} {\bibfield  {journal} {\bibinfo
  {journal} {\href{https://doi.org/10.1017/S0022377814000622}{Journal of Plasma
  Physics}}\ }\textbf {\bibinfo {volume} {81}},\ \bibinfo {pages} {305810104}
  (\bibinfo {year} {2015})}\BibitemShut {NoStop}%
\bibitem [{\citenamefont {Banik}\ and\ \citenamefont
  {Bhattacharjee}(2024)}]{banik2024relaxation}%
  \BibitemOpen
  \bibfield  {author} {\bibinfo {author} {\bibfnamefont {U.}~\bibnamefont
  {Banik}}\ and\ \bibinfo {author} {\bibfnamefont {A.}~\bibnamefont
  {Bhattacharjee}},\ }\bibfield  {title} {\bibinfo {title} {Relaxation of
  weakly collisional plasma: Continuous spectra, discrete eigenmodes, and the
  decay of echoes},\ }\href {https://doi.org/10.1103/PhysRevE.110.045204}
  {\bibfield  {journal} {\bibinfo  {journal}
  {\href{https://doi.org/10.1103/PhysRevE.110.045204}{Physical Review E}}\
  }\textbf {\bibinfo {volume} {110}},\ \bibinfo {pages} {045204} (\bibinfo
  {year} {2024})}\BibitemShut {NoStop}%
\bibitem [{\citenamefont {Eyink}\ and\ \citenamefont
  {Sreenivasan}(2006)}]{eyink2006onsager}%
  \BibitemOpen
  \bibfield  {author} {\bibinfo {author} {\bibfnamefont {G.~L.}\ \bibnamefont
  {Eyink}}\ and\ \bibinfo {author} {\bibfnamefont {K.~R.}\ \bibnamefont
  {Sreenivasan}},\ }\bibfield  {title} {\bibinfo {title} {Onsager and the
  theory of hydrodynamic turbulence},\ }\href
  {https://doi.org/10.1103/RevModPhys.78.87} {\bibfield  {journal} {\bibinfo
  {journal} {\href{https://doi.org/10.1103/RevModPhys.78.87}{Reviews of Modern
  Physics}}\ }\textbf {\bibinfo {volume} {78}},\ \bibinfo {pages} {87}
  (\bibinfo {year} {2006})}\BibitemShut {NoStop}%
\bibitem [{\citenamefont {Juno}\ \emph {et~al.}(2018)\citenamefont {Juno},
  \citenamefont {Hakim}, \citenamefont {TenBarge}, \citenamefont {Shi},\ and\
  \citenamefont {Dorland}}]{juno2018discontinuous}%
  \BibitemOpen
  \bibfield  {author} {\bibinfo {author} {\bibfnamefont {J.}~\bibnamefont
  {Juno}}, \bibinfo {author} {\bibfnamefont {A.}~\bibnamefont {Hakim}},
  \bibinfo {author} {\bibfnamefont {J.}~\bibnamefont {TenBarge}}, \bibinfo
  {author} {\bibfnamefont {E.}~\bibnamefont {Shi}},\ and\ \bibinfo {author}
  {\bibfnamefont {W.}~\bibnamefont {Dorland}},\ }\bibfield  {title} {\bibinfo
  {title} {Discontinuous {G}alerkin algorithms for fully kinetic plasmas},\
  }\href {https://doi.org/10.1016/j.jcp.2017.10.009} {\bibfield  {journal}
  {\bibinfo  {journal}
  {\href{https://doi.org/10.1016/j.jcp.2017.10.009}{Journal of Computational
  Physics}}\ }\textbf {\bibinfo {volume} {353}},\ \bibinfo {pages} {110}
  (\bibinfo {year} {2018})}\BibitemShut {NoStop}%
\bibitem [{\citenamefont {Hakim}\ and\ \citenamefont {Juno}(2020)}]{9355299}%
  \BibitemOpen
  \bibfield  {author} {\bibinfo {author} {\bibfnamefont {A.}~\bibnamefont
  {Hakim}}\ and\ \bibinfo {author} {\bibfnamefont {J.}~\bibnamefont {Juno}},\
  }\bibfield  {title} {\bibinfo {title} {Alias-free, matrix-free, and
  quadrature-free discontinuous {G}alerkin algorithms for (plasma) kinetic
  equations},\ }in\ \href {https://doi.org/10.1109/SC41405.2020.00077} {\emph
  {\bibinfo {booktitle} {SC20: International Conference for High Performance
  Computing, Networking, Storage and Analysis}}}\ (\bibinfo  {publisher}
  {IEEE},\ \bibinfo {year} {2020})\ pp.\ \bibinfo {pages} {1--15}\BibitemShut
  {NoStop}%
\bibitem [{\citenamefont {Dougherty}(1964)}]{dougherty1964model}%
  \BibitemOpen
  \bibfield  {author} {\bibinfo {author} {\bibfnamefont {J.~P.}\ \bibnamefont
  {Dougherty}},\ }\bibfield  {title} {\bibinfo {title} {Model {F}okker-{P}lanck
  equation for a plasma and its solution},\ }\href
  {https://doi.org/10.1063/1.2746779} {\bibfield  {journal} {\bibinfo
  {journal} {\href{https://doi.org/10.1063/1.2746779}{Physics of Fluids}}\
  }\textbf {\bibinfo {volume} {7}},\ \bibinfo {pages} {1788} (\bibinfo {year}
  {1964})}\BibitemShut {NoStop}%
\bibitem [{\citenamefont {Hakim}\ \emph {et~al.}(2020)\citenamefont {Hakim},
  \citenamefont {Francisquez}, \citenamefont {Juno},\ and\ \citenamefont
  {Hammett}}]{hakim2020conservative}%
  \BibitemOpen
  \bibfield  {author} {\bibinfo {author} {\bibfnamefont {A.}~\bibnamefont
  {Hakim}}, \bibinfo {author} {\bibfnamefont {M.}~\bibnamefont {Francisquez}},
  \bibinfo {author} {\bibfnamefont {J.}~\bibnamefont {Juno}},\ and\ \bibinfo
  {author} {\bibfnamefont {G.~W.}\ \bibnamefont {Hammett}},\ }\bibfield
  {title} {\bibinfo {title} {Conservative discontinuous {G}alerkin schemes for
  nonlinear {D}ougherty--{F}okker--{P}lanck collision operators},\ }\href
  {https://doi.org/10.1017/S0022377820000586} {\bibfield  {journal} {\bibinfo
  {journal} {\href{https://doi.org/10.1017/S0022377820000586}{Journal of Plasma
  Physics}}\ }\textbf {\bibinfo {volume} {86}},\ \bibinfo {pages} {905860403}
  (\bibinfo {year} {2020})}\BibitemShut {NoStop}%
\bibitem [{Note2()}]{Note2}%
  \BibitemOpen
  \bibinfo {note} {Note that $v_{\protect \mathrm {th}}$ in \protect \eqref
  {col_op_FP} is the thermal velocity of the full distribution function $f$
  and, therefore, depends not just on time but also on space. Throughout the
  paper, we use $v_{\protect \mathrm {the}}$ to denote the thermal velocity of
  $f_0$, which is generally time-dependent but independent of space. However,
  for spatially homogeneous turbulence, we expect $v_{\protect \mathrm {the}}
  \sim v_{\protect \mathrm {th}}$, so this distinction, while important
  for~\protect \eqref {col_op_FP} to satisfy exactly its necessary conservation
  laws, is unimportant, e.g., for dimensional estimates.}\BibitemShut {Stop}%
\bibitem [{Note3()}]{Note3}%
  \BibitemOpen
  \bibinfo {note} {We solve Ampère's law for the electric field rather than
  Poisson's equation purely out of convenience of the simulation code that we
  are using. We ensure that the total volume-averaged current is zero at all
  time steps, which enforces that the volume-averaged electric field is zero at
  all time steps. Then, in a periodic box, the 1D-1V Vlasov-Ampère system is
  equivalent to the 1D-1V Vlasov--Poisson system.}\BibitemShut {Stop}%
\bibitem [{\citenamefont {Kloeden}\ and\ \citenamefont
  {Platen}(2013)}]{kloeden2013stochastic}%
  \BibitemOpen
  \bibfield  {author} {\bibinfo {author} {\bibfnamefont {P.~E.}\ \bibnamefont
  {Kloeden}}\ and\ \bibinfo {author} {\bibfnamefont {E.}~\bibnamefont
  {Platen}},\ }\href@noop {} {\emph {\bibinfo {title} {Numerical Solution of
  Stochastic Differential Equations}}},\ Stochastic Modelling and Applied
  Probability\ (\bibinfo  {publisher} {Springer Berlin},\ \bibinfo {address}
  {Heidelberg},\ \bibinfo {year} {2013})\BibitemShut {NoStop}%
\bibitem [{\citenamefont {Yeung}\ \emph {et~al.}(2002)\citenamefont {Yeung},
  \citenamefont {Xu},\ and\ \citenamefont {Sreenivasan}}]{yeung2002schmidt}%
  \BibitemOpen
  \bibfield  {author} {\bibinfo {author} {\bibfnamefont {P.~K.}\ \bibnamefont
  {Yeung}}, \bibinfo {author} {\bibfnamefont {S.}~\bibnamefont {Xu}},\ and\
  \bibinfo {author} {\bibfnamefont {K.~R.}\ \bibnamefont {Sreenivasan}},\
  }\bibfield  {title} {\bibinfo {title} {Schmidt number effects on turbulent
  transport with uniform mean scalar gradient},\ }\href
  {https://doi.org/10.1063/1.1517298} {\bibfield  {journal} {\bibinfo
  {journal} {\href{https://doi.org/10.1063/1.1517298}{Physics of Fluids}}\
  }\textbf {\bibinfo {volume} {14}},\ \bibinfo {pages} {4178} (\bibinfo {year}
  {2002})}\BibitemShut {NoStop}%
\bibitem [{\citenamefont {Donzis}\ \emph {et~al.}(2010)\citenamefont {Donzis},
  \citenamefont {Sreenivasan},\ and\ \citenamefont
  {Yeung}}]{donzis2010batchelor}%
  \BibitemOpen
  \bibfield  {author} {\bibinfo {author} {\bibfnamefont {D.~A.}\ \bibnamefont
  {Donzis}}, \bibinfo {author} {\bibfnamefont {K.~R.}\ \bibnamefont
  {Sreenivasan}},\ and\ \bibinfo {author} {\bibfnamefont {P.~K.}\ \bibnamefont
  {Yeung}},\ }\bibfield  {title} {\bibinfo {title} {The {B}atchelor spectrum
  for mixing of passive scalars in isotropic turbulence},\ }\href
  {https://link.springer.com/article/10.1007/s10494-010-9271-6} {\bibfield
  {journal} {\bibinfo  {journal}
  {\href{https://link.springer.com/article/10.1007/s10494-010-9271-6}{Flow,
  Turbulence and Combustion}}\ }\textbf {\bibinfo {volume} {85}},\ \bibinfo
  {pages} {549} (\bibinfo {year} {2010})}\BibitemShut {NoStop}%
\bibitem [{\citenamefont {Schekochihin}\ \emph {et~al.}(2004)\citenamefont
  {Schekochihin}, \citenamefont {Haynes},\ and\ \citenamefont
  {Cowley}}]{schekochihin2004diffusion}%
  \BibitemOpen
  \bibfield  {author} {\bibinfo {author} {\bibfnamefont {A.~A.}\ \bibnamefont
  {Schekochihin}}, \bibinfo {author} {\bibfnamefont {P.~H.}\ \bibnamefont
  {Haynes}},\ and\ \bibinfo {author} {\bibfnamefont {S.~C.}\ \bibnamefont
  {Cowley}},\ }\bibfield  {title} {\bibinfo {title} {Diffusion of passive
  scalar in a finite-scale random flow},\ }\href
  {https://doi.org/10.1103/PhysRevE.70.046304} {\bibfield  {journal} {\bibinfo
  {journal} {\href{https://doi.org/10.1103/PhysRevE.70.046304}{Physical Review
  E}}\ }\textbf {\bibinfo {volume} {70}},\ \bibinfo {pages} {046304} (\bibinfo
  {year} {2004})}\BibitemShut {NoStop}%
\bibitem [{\citenamefont {Kadomtsev}\ and\ \citenamefont
  {Pogutse}(1970)}]{kadomtsev1970collisionless}%
  \BibitemOpen
  \bibfield  {author} {\bibinfo {author} {\bibfnamefont {B.~B.}\ \bibnamefont
  {Kadomtsev}}\ and\ \bibinfo {author} {\bibfnamefont {O.~P.}\ \bibnamefont
  {Pogutse}},\ }\bibfield  {title} {\bibinfo {title} {Collisionless relaxation
  in systems with {C}oulomb interactions},\ }\href
  {https://doi.org/10.1103/PhysRevLett.25.1155} {\bibfield  {journal} {\bibinfo
   {journal} {\href{https://doi.org/10.1103/PhysRevLett.25.1155}{Physical
  Review Letters}}\ }\textbf {\bibinfo {volume} {25}},\ \bibinfo {pages} {1155}
  (\bibinfo {year} {1970})}\BibitemShut {NoStop}%
\bibitem [{\citenamefont {Ewart}\ \emph {et~al.}(2022)\citenamefont {Ewart},
  \citenamefont {Brown}, \citenamefont {Adkins},\ and\ \citenamefont
  {Schekochihin}}]{ewart2022}%
  \BibitemOpen
  \bibfield  {author} {\bibinfo {author} {\bibfnamefont {R.~J.}\ \bibnamefont
  {Ewart}}, \bibinfo {author} {\bibfnamefont {A.}~\bibnamefont {Brown}},
  \bibinfo {author} {\bibfnamefont {T.}~\bibnamefont {Adkins}},\ and\ \bibinfo
  {author} {\bibfnamefont {A.~A.}\ \bibnamefont {Schekochihin}},\ }\bibfield
  {title} {\bibinfo {title} {Collisionless relaxation of a {L}ynden-{B}ell
  plasma},\ }\href {https://doi.org/10.1017/S0022377822000782} {\bibfield
  {journal} {\bibinfo  {journal}
  {\href{https://doi.org/10.1017/S0022377822000782}{Journal of Plasma
  Physics}}\ }\textbf {\bibinfo {volume} {88}},\ \bibinfo {pages} {925880501}
  (\bibinfo {year} {2022})}\BibitemShut {NoStop}%
\bibitem [{\citenamefont {Ewart}\ \emph {et~al.}(2023)\citenamefont {Ewart},
  \citenamefont {Nastac},\ and\ \citenamefont
  {Schekochihin}}]{ewart2023nonthermal}%
  \BibitemOpen
  \bibfield  {author} {\bibinfo {author} {\bibfnamefont {R.~J.}\ \bibnamefont
  {Ewart}}, \bibinfo {author} {\bibfnamefont {M.~L.}\ \bibnamefont {Nastac}},\
  and\ \bibinfo {author} {\bibfnamefont {A.~A.}\ \bibnamefont {Schekochihin}},\
  }\bibfield  {title} {\bibinfo {title} {Non-thermal particle acceleration and
  power-law tails via relaxation to universal {L}ynden-{B}ell equilibria},\
  }\href {https://doi.org/10.1017/S0022377823000983} {\bibfield  {journal}
  {\bibinfo  {journal}
  {\href{https://doi.org/10.1017/S0022377823000983}{Journal of Plasma
  Physics}}\ }\textbf {\bibinfo {volume} {89}},\ \bibinfo {pages} {905890516}
  (\bibinfo {year} {2023})}\BibitemShut {NoStop}%
\bibitem [{\citenamefont {Ewart}\ \emph {et~al.}(2024)\citenamefont {Ewart},
  \citenamefont {Nastac}, \citenamefont {Bilbao}, \citenamefont {Silva},
  \citenamefont {Silva},\ and\ \citenamefont
  {Schekochihin}}]{ewart2024relaxation}%
  \BibitemOpen
  \bibfield  {author} {\bibinfo {author} {\bibfnamefont {R.~J.}\ \bibnamefont
  {Ewart}}, \bibinfo {author} {\bibfnamefont {M.~L.}\ \bibnamefont {Nastac}},
  \bibinfo {author} {\bibfnamefont {P.~J.}\ \bibnamefont {Bilbao}}, \bibinfo
  {author} {\bibfnamefont {T.}~\bibnamefont {Silva}}, \bibinfo {author}
  {\bibfnamefont {L.~O.}\ \bibnamefont {Silva}},\ and\ \bibinfo {author}
  {\bibfnamefont {A.~A.}\ \bibnamefont {Schekochihin}},\ }\bibfield  {title}
  {\bibinfo {title} {Relaxation to universal non-{M}axwellian equilibria in a
  collisionless plasma},\ }\href {https://arxiv.org/abs/2409.01742} {\bibfield
  {journal} {\bibinfo  {journal}
  {\href{https://arxiv.org/abs/2409.01742}{arXiv:2409.01742}}\ } (\bibinfo
  {year} {2024})}\BibitemShut {NoStop}%
\bibitem [{\citenamefont {Sturrock}(1966)}]{sturrock1966stochastic}%
  \BibitemOpen
  \bibfield  {author} {\bibinfo {author} {\bibfnamefont {P.~A.}\ \bibnamefont
  {Sturrock}},\ }\bibfield  {title} {\bibinfo {title} {Stochastic
  acceleration},\ }\href {https://doi.org/10.1103/PhysRev.141.186} {\bibfield
  {journal} {\bibinfo  {journal}
  {\href{https://doi.org/10.1103/PhysRev.141.186}{Physical Review}}\ }\textbf
  {\bibinfo {volume} {141}},\ \bibinfo {pages} {186} (\bibinfo {year}
  {1966})}\BibitemShut {NoStop}%
\bibitem [{\citenamefont {Chandran}\ \emph {et~al.}(2010)\citenamefont
  {Chandran}, \citenamefont {Li}, \citenamefont {Rogers}, \citenamefont
  {Quataert},\ and\ \citenamefont {Germaschewski}}]{chandran2010perpendicular}%
  \BibitemOpen
  \bibfield  {author} {\bibinfo {author} {\bibfnamefont {B.~D.~G.}\
  \bibnamefont {Chandran}}, \bibinfo {author} {\bibfnamefont {B.}~\bibnamefont
  {Li}}, \bibinfo {author} {\bibfnamefont {B.~N.}\ \bibnamefont {Rogers}},
  \bibinfo {author} {\bibfnamefont {E.}~\bibnamefont {Quataert}},\ and\
  \bibinfo {author} {\bibfnamefont {K.}~\bibnamefont {Germaschewski}},\
  }\bibfield  {title} {\bibinfo {title} {Perpendicular ion heating by
  low-frequency {A}lfv{\'e}n-wave turbulence in the solar wind},\ }\href
  {https://doi.org/10.1088/0004-637X/720/1/503} {\bibfield  {journal} {\bibinfo
   {journal} {\href{https://doi.org/10.1088/0004-637X/720/1/503}{Astrophysical
  Journal}}\ }\textbf {\bibinfo {volume} {720}},\ \bibinfo {pages} {503}
  (\bibinfo {year} {2010})}\BibitemShut {NoStop}%
\bibitem [{\citenamefont {Verscharen}\ \emph {et~al.}(2019)\citenamefont
  {Verscharen}, \citenamefont {Klein},\ and\ \citenamefont
  {Maruca}}]{verscharen2019multi}%
  \BibitemOpen
  \bibfield  {author} {\bibinfo {author} {\bibfnamefont {D.}~\bibnamefont
  {Verscharen}}, \bibinfo {author} {\bibfnamefont {K.~G.}\ \bibnamefont
  {Klein}},\ and\ \bibinfo {author} {\bibfnamefont {B.~A.}\ \bibnamefont
  {Maruca}},\ }\bibfield  {title} {\bibinfo {title} {The multi-scale nature of
  the solar wind},\ }\href {https://doi.org/10.1007/s41116-019-0021-0}
  {\bibfield  {journal} {\bibinfo  {journal}
  {\href{https://doi.org/10.1007/s41116-019-0021-0}{Living Reviews in Solar
  Physics}}\ }\textbf {\bibinfo {volume} {16}},\ \bibinfo {pages} {5} (\bibinfo
  {year} {2019})}\BibitemShut {NoStop}%
\bibitem [{\citenamefont {Cerri}\ \emph {et~al.}(2021)\citenamefont {Cerri},
  \citenamefont {Arzamasskiy},\ and\ \citenamefont
  {Kunz}}]{cerri2021stochastic}%
  \BibitemOpen
  \bibfield  {author} {\bibinfo {author} {\bibfnamefont {S.~S.}\ \bibnamefont
  {Cerri}}, \bibinfo {author} {\bibfnamefont {L.}~\bibnamefont {Arzamasskiy}},\
  and\ \bibinfo {author} {\bibfnamefont {M.~W.}\ \bibnamefont {Kunz}},\
  }\bibfield  {title} {\bibinfo {title} {On stochastic heating and its
  phase-space signatures in low-beta kinetic turbulence},\ }\href
  {https://doi.org/10.3847/1538-4357/abfbde} {\bibfield  {journal} {\bibinfo
  {journal} {\href{https://doi.org/10.3847/1538-4357/abfbde}{Astrophysical
  Journal}}\ }\textbf {\bibinfo {volume} {916}},\ \bibinfo {pages} {120}
  (\bibinfo {year} {2021})}\BibitemShut {NoStop}%
\bibitem [{\citenamefont {Banik}\ \emph {et~al.}(2024)\citenamefont {Banik},
  \citenamefont {Bhattacharjee},\ and\ \citenamefont
  {Sengupta}}]{banik2024universal}%
  \BibitemOpen
  \bibfield  {author} {\bibinfo {author} {\bibfnamefont {U.}~\bibnamefont
  {Banik}}, \bibinfo {author} {\bibfnamefont {A.}~\bibnamefont
  {Bhattacharjee}},\ and\ \bibinfo {author} {\bibfnamefont {W.}~\bibnamefont
  {Sengupta}},\ }\bibfield  {title} {\bibinfo {title} {Universal nonthermal
  power-law distribution functions from the self-consistent evolution of
  collisionless electrostatic plasmas},\ }\href
  {https://iopscience.iop.org/article/10.3847/1538-4357/ad91a1} {\bibfield
  {journal} {\bibinfo  {journal}
  {\href{https://iopscience.iop.org/article/10.3847/1538-4357/ad91a1}{The
  Astrophysical Journal}}\ }\textbf {\bibinfo {volume} {977}},\ \bibinfo
  {pages} {91} (\bibinfo {year} {2024})}\BibitemShut {NoStop}%
\bibitem [{Note4()}]{Note4}%
  \BibitemOpen
  \bibinfo {note} {Implicit in using $\delta C_2$ as a proxy for the amplitude
  of $\delta \protect \! f$ is that the logarithmic dependence of \protect
  \eqref {delta_C2_amp} on $\nu $ is order-unity. For finite $\nu $, this
  approximation is reasonable and essentially captures how the amplitude of
  $\delta \protect \! f$ depends on the flux $\varepsilon $.}\BibitemShut
  {Stop}%
\bibitem [{\citenamefont {Morrison}\ and\ \citenamefont
  {Shadwick}(2008)}]{morrison2008fluctuation}%
  \BibitemOpen
  \bibfield  {author} {\bibinfo {author} {\bibfnamefont {P.~J.}\ \bibnamefont
  {Morrison}}\ and\ \bibinfo {author} {\bibfnamefont {B.~A.}\ \bibnamefont
  {Shadwick}},\ }\bibfield  {title} {\bibinfo {title} {On the fluctuation
  spectrum of plasma},\ }\href {https://doi.org/10.1016/j.cnsns.2007.04.005}
  {\bibfield  {journal} {\bibinfo  {journal}
  {\href{https://doi.org/10.1016/j.cnsns.2007.04.005}{Communications in
  Nonlinear Science and Numerical Simulation}}\ }\textbf {\bibinfo {volume}
  {13}},\ \bibinfo {pages} {130} (\bibinfo {year} {2008})}\BibitemShut
  {NoStop}%
\bibitem [{\citenamefont {Fouvry}(2022)}]{fouvry2022kinetic}%
  \BibitemOpen
  \bibfield  {author} {\bibinfo {author} {\bibfnamefont {J.-B.}\ \bibnamefont
  {Fouvry}},\ }\bibfield  {title} {\bibinfo {title} {Kinetic theory of
  one-dimensional inhomogeneous long-range interacting {N}-body systems at
  order $1/{N}^2$ without collective effects},\ }\href
  {https://doi.org/10.1103/PhysRevE.106.054123} {\bibfield  {journal} {\bibinfo
   {journal} {\href{https://doi.org/10.1103/PhysRevE.106.054123}{Physical
  Review E}}\ }\textbf {\bibinfo {volume} {106}},\ \bibinfo {pages} {054123}
  (\bibinfo {year} {2022})}\BibitemShut {NoStop}%
\bibitem [{\citenamefont {Birdsall}\ and\ \citenamefont
  {Langdon}(2004)}]{birdsall2018plasma}%
  \BibitemOpen
  \bibfield  {author} {\bibinfo {author} {\bibfnamefont {C.~K.}\ \bibnamefont
  {Birdsall}}\ and\ \bibinfo {author} {\bibfnamefont {A.~B.}\ \bibnamefont
  {Langdon}},\ }\href@noop {} {\emph {\bibinfo {title} {Plasma Physics via
  Computer Simulation}}}\ (\bibinfo  {publisher} {Taylor \& Francis Group},\
  \bibinfo {address} {New York},\ \bibinfo {year} {2004})\BibitemShut {NoStop}%
\bibitem [{\citenamefont {Zakharov}(1972)}]{zakharov1972collapse}%
  \BibitemOpen
  \bibfield  {author} {\bibinfo {author} {\bibfnamefont {V.~E.}\ \bibnamefont
  {Zakharov}},\ }\bibfield  {title} {\bibinfo {title} {Collapse of {L}angmuir
  waves},\ }\href@noop {} {\bibfield  {journal} {\bibinfo  {journal} {Soviet
  Physics Journal of Experimental and Theoretical Physics}\ }\textbf {\bibinfo
  {volume} {35}},\ \bibinfo {pages} {908} (\bibinfo {year} {1972})}\BibitemShut
  {NoStop}%
\bibitem [{\citenamefont {Goldman}(1984)}]{goldman1984strong}%
  \BibitemOpen
  \bibfield  {author} {\bibinfo {author} {\bibfnamefont {M.~V.}\ \bibnamefont
  {Goldman}},\ }\bibfield  {title} {\bibinfo {title} {Strong turbulence of
  plasma waves},\ }\href {https://doi.org/10.1103/RevModPhys.56.709} {\bibfield
   {journal} {\bibinfo  {journal}
  {\href{https://doi.org/10.1103/RevModPhys.56.709}{Reviews of Modern
  Physics}}\ }\textbf {\bibinfo {volume} {56}},\ \bibinfo {pages} {709}
  (\bibinfo {year} {1984})}\BibitemShut {NoStop}%
\bibitem [{\citenamefont {Robinson}(1997)}]{robinson1997nonlinear}%
  \BibitemOpen
  \bibfield  {author} {\bibinfo {author} {\bibfnamefont {P.}~\bibnamefont
  {Robinson}},\ }\bibfield  {title} {\bibinfo {title} {Nonlinear wave collapse
  and strong turbulence},\ }\href {https://doi.org/10.1103/RevModPhys.69.507}
  {\bibfield  {journal} {\bibinfo  {journal}
  {\href{https://doi.org/10.1103/RevModPhys.69.507}{Reviews of Modern
  Physics}}\ }\textbf {\bibinfo {volume} {69}},\ \bibinfo {pages} {507}
  (\bibinfo {year} {1997})}\BibitemShut {NoStop}%
\bibitem [{\citenamefont {Che}\ \emph {et~al.}(2017)\citenamefont {Che},
  \citenamefont {Goldstein}, \citenamefont {Diamond},\ and\ \citenamefont
  {Sagdeev}}]{che2017electron}%
  \BibitemOpen
  \bibfield  {author} {\bibinfo {author} {\bibfnamefont {H.}~\bibnamefont
  {Che}}, \bibinfo {author} {\bibfnamefont {M.~L.}\ \bibnamefont {Goldstein}},
  \bibinfo {author} {\bibfnamefont {P.~H.}\ \bibnamefont {Diamond}},\ and\
  \bibinfo {author} {\bibfnamefont {R.~Z.}\ \bibnamefont {Sagdeev}},\
  }\bibfield  {title} {\bibinfo {title} {How electron two-stream instability
  drives cyclic {L}angmuir collapse and continuous coherent emission},\ }\href
  {https://doi.org/10.1073/pnas.1614055114} {\bibfield  {journal} {\bibinfo
  {journal} {\href{https://doi.org/10.1073/pnas.1614055114}{Proceedings of the
  National Academy of Sciences}}\ }\textbf {\bibinfo {volume} {114}},\ \bibinfo
  {pages} {1502} (\bibinfo {year} {2017})}\BibitemShut {NoStop}%
\bibitem [{\citenamefont {Reid}\ and\ \citenamefont
  {Kontar}(2021)}]{reid2021fine}%
  \BibitemOpen
  \bibfield  {author} {\bibinfo {author} {\bibfnamefont {H.~A.~S.}\
  \bibnamefont {Reid}}\ and\ \bibinfo {author} {\bibfnamefont {E.~P.}\
  \bibnamefont {Kontar}},\ }\bibfield  {title} {\bibinfo {title} {Fine
  structure of type {III} solar radio bursts from {L}angmuir wave motion in
  turbulent plasma},\ }\href
  {https://www.nature.com/articles/s41550-021-01370-8} {\bibfield  {journal}
  {\bibinfo  {journal}
  {\href{https://www.nature.com/articles/s41550-021-01370-8}{Nature
  Astronomy}}\ }\textbf {\bibinfo {volume} {5}},\ \bibinfo {pages} {796}
  (\bibinfo {year} {2021})}\BibitemShut {NoStop}%
\bibitem [{\citenamefont {Thorne}\ and\ \citenamefont
  {Blandford}(2017)}]{thorne2017modern}%
  \BibitemOpen
  \bibfield  {author} {\bibinfo {author} {\bibfnamefont {K.~S.}\ \bibnamefont
  {Thorne}}\ and\ \bibinfo {author} {\bibfnamefont {R.~D.}\ \bibnamefont
  {Blandford}},\ }\href@noop {} {\emph {\bibinfo {title} {Modern Classical
  Physics: Optics, Fluids, Plasmas, Elasticity, Relativity, and Statistical
  Physics}}}\ (\bibinfo  {publisher} {Princeton University Press},\ \bibinfo
  {address} {Princeton},\ \bibinfo {year} {2017})\BibitemShut {NoStop}%
\bibitem [{\citenamefont {Quataert}\ and\ \citenamefont
  {Gruzinov}(1999)}]{quataert1999turbulence}%
  \BibitemOpen
  \bibfield  {author} {\bibinfo {author} {\bibfnamefont {E.}~\bibnamefont
  {Quataert}}\ and\ \bibinfo {author} {\bibfnamefont {A.}~\bibnamefont
  {Gruzinov}},\ }\bibfield  {title} {\bibinfo {title} {Turbulence and particle
  heating in advection-dominated accretion flows},\ }\href
  {https://doi.org/10.1086/307423} {\bibfield  {journal} {\bibinfo  {journal}
  {\href{https://doi.org/10.1086/307423}{Astrophysical Journal}}\ }\textbf
  {\bibinfo {volume} {520}},\ \bibinfo {pages} {248} (\bibinfo {year}
  {1999})}\BibitemShut {NoStop}%
\bibitem [{\citenamefont {De~Pontieu}\ \emph {et~al.}(2007)\citenamefont
  {De~Pontieu}, \citenamefont {McIntosh}, \citenamefont {Carlsson},
  \citenamefont {Hansteen}, \citenamefont {Tarbell}, \citenamefont {Schrijver},
  \citenamefont {Title}, \citenamefont {Shine}, \citenamefont {Tsuneta},
  \citenamefont {Katsukawa} \emph {et~al.}}]{de2007chromospheric}%
  \BibitemOpen
  \bibfield  {author} {\bibinfo {author} {\bibfnamefont {B.}~\bibnamefont
  {De~Pontieu}}, \bibinfo {author} {\bibfnamefont {S.}~\bibnamefont
  {McIntosh}}, \bibinfo {author} {\bibfnamefont {M.}~\bibnamefont {Carlsson}},
  \bibinfo {author} {\bibfnamefont {V.}~\bibnamefont {Hansteen}}, \bibinfo
  {author} {\bibfnamefont {T.}~\bibnamefont {Tarbell}}, \bibinfo {author}
  {\bibfnamefont {C.}~\bibnamefont {Schrijver}}, \bibinfo {author}
  {\bibfnamefont {A.}~\bibnamefont {Title}}, \bibinfo {author} {\bibfnamefont
  {R.}~\bibnamefont {Shine}}, \bibinfo {author} {\bibfnamefont
  {S.}~\bibnamefont {Tsuneta}}, \bibinfo {author} {\bibfnamefont
  {Y.}~\bibnamefont {Katsukawa}}, \emph {et~al.},\ }\bibfield  {title}
  {\bibinfo {title} {Chromospheric {A}lfv{\'e}nic waves strong enough to power
  the solar wind},\ }\href
  {https://www.science.org/doi/full/10.1126/science.1151747} {\bibfield
  {journal} {\bibinfo  {journal}
  {\href{https://www.science.org/doi/full/10.1126/science.1151747}{Science}}\
  }\textbf {\bibinfo {volume} {318}},\ \bibinfo {pages} {1574} (\bibinfo {year}
  {2007})}\BibitemShut {NoStop}%
\bibitem [{\citenamefont {Chen}(2016)}]{chen2016recent}%
  \BibitemOpen
  \bibfield  {author} {\bibinfo {author} {\bibfnamefont {C.~H.~K.}\
  \bibnamefont {Chen}},\ }\bibfield  {title} {\bibinfo {title} {Recent progress
  in astrophysical plasma turbulence from solar wind observations},\ }\href
  {https://doi.org/10.1017/S0022377816001124} {\bibfield  {journal} {\bibinfo
  {journal} {\href{https://doi.org/10.1017/S0022377816001124}{Journal of Plasma
  Physics}}\ }\textbf {\bibinfo {volume} {82}},\ \bibinfo {pages} {535820602}
  (\bibinfo {year} {2016})}\BibitemShut {NoStop}%
\bibitem [{\citenamefont {Howes}\ \emph {et~al.}(2008)\citenamefont {Howes},
  \citenamefont {Cowley}, \citenamefont {Dorland}, \citenamefont {Hammett},
  \citenamefont {Quataert},\ and\ \citenamefont
  {Schekochihin}}]{howes2008model}%
  \BibitemOpen
  \bibfield  {author} {\bibinfo {author} {\bibfnamefont {G.~G.}\ \bibnamefont
  {Howes}}, \bibinfo {author} {\bibfnamefont {S.~C.}\ \bibnamefont {Cowley}},
  \bibinfo {author} {\bibfnamefont {W.}~\bibnamefont {Dorland}}, \bibinfo
  {author} {\bibfnamefont {G.~W.}\ \bibnamefont {Hammett}}, \bibinfo {author}
  {\bibfnamefont {E.}~\bibnamefont {Quataert}},\ and\ \bibinfo {author}
  {\bibfnamefont {A.~A.}\ \bibnamefont {Schekochihin}},\ }\bibfield  {title}
  {\bibinfo {title} {A model of turbulence in magnetized plasmas: Implications
  for the dissipation range in the solar wind},\ }\href
  {https://doi.org/10.1029/2007JA012665} {\bibfield  {journal} {\bibinfo
  {journal} {\href{https://doi.org/10.1029/2007JA012665}{Journal of Geophysical
  Research: Space Physics}}\ }\textbf {\bibinfo {volume} {113}} (\bibinfo
  {year} {2008})}\BibitemShut {NoStop}%
\bibitem [{\citenamefont {Bandak}\ \emph {et~al.}(2022)\citenamefont {Bandak},
  \citenamefont {Goldenfeld}, \citenamefont {Mailybaev},\ and\ \citenamefont
  {Eyink}}]{bandak2022dissipation}%
  \BibitemOpen
  \bibfield  {author} {\bibinfo {author} {\bibfnamefont {D.}~\bibnamefont
  {Bandak}}, \bibinfo {author} {\bibfnamefont {N.}~\bibnamefont {Goldenfeld}},
  \bibinfo {author} {\bibfnamefont {A.~A.}\ \bibnamefont {Mailybaev}},\ and\
  \bibinfo {author} {\bibfnamefont {G.}~\bibnamefont {Eyink}},\ }\bibfield
  {title} {\bibinfo {title} {Dissipation-range fluid turbulence and thermal
  noise},\ }\href {https://doi.org/10.1103/PhysRevE.105.065113} {\bibfield
  {journal} {\bibinfo  {journal}
  {\href{https://doi.org/10.1103/PhysRevE.105.065113}{Physical Review E}}\
  }\textbf {\bibinfo {volume} {105}},\ \bibinfo {pages} {065113} (\bibinfo
  {year} {2022})}\BibitemShut {NoStop}%
\bibitem [{\citenamefont {McMullen}\ \emph {et~al.}(2022)\citenamefont
  {McMullen}, \citenamefont {Krygier}, \citenamefont {Torczynski},\ and\
  \citenamefont {Gallis}}]{mcmullen2022navier}%
  \BibitemOpen
  \bibfield  {author} {\bibinfo {author} {\bibfnamefont {R.~M.}\ \bibnamefont
  {McMullen}}, \bibinfo {author} {\bibfnamefont {M.~C.}\ \bibnamefont
  {Krygier}}, \bibinfo {author} {\bibfnamefont {J.~R.}\ \bibnamefont
  {Torczynski}},\ and\ \bibinfo {author} {\bibfnamefont {M.~A.}\ \bibnamefont
  {Gallis}},\ }\bibfield  {title} {\bibinfo {title} {{N}avier-{S}tokes
  equations do not describe the smallest scales of turbulence in gases},\
  }\href {https://doi.org/10.1103/PhysRevLett.128.114501} {\bibfield  {journal}
  {\bibinfo  {journal}
  {\href{https://doi.org/10.1103/PhysRevLett.128.114501}{Physical Review
  Letters}}\ }\textbf {\bibinfo {volume} {128}},\ \bibinfo {pages} {114501}
  (\bibinfo {year} {2022})}\BibitemShut {NoStop}%
\bibitem [{\citenamefont {Bell}\ \emph {et~al.}(2022)\citenamefont {Bell},
  \citenamefont {Nonaka}, \citenamefont {Garcia},\ and\ \citenamefont
  {Eyink}}]{bell2022thermal}%
  \BibitemOpen
  \bibfield  {author} {\bibinfo {author} {\bibfnamefont {J.~B.}\ \bibnamefont
  {Bell}}, \bibinfo {author} {\bibfnamefont {A.}~\bibnamefont {Nonaka}},
  \bibinfo {author} {\bibfnamefont {A.~L.}\ \bibnamefont {Garcia}},\ and\
  \bibinfo {author} {\bibfnamefont {G.}~\bibnamefont {Eyink}},\ }\bibfield
  {title} {\bibinfo {title} {Thermal fluctuations in the dissipation range of
  homogeneous isotropic turbulence},\ }\href
  {https://doi.org/10.1017/jfm.2022.188} {\bibfield  {journal} {\bibinfo
  {journal} {\href{https://doi.org/10.1017/jfm.2022.188}{Journal of Fluid
  Mechanics}}\ }\textbf {\bibinfo {volume} {939}},\ \bibinfo {pages} {A12}
  (\bibinfo {year} {2022})}\BibitemShut {NoStop}%
\bibitem [{\citenamefont {Dupree}(1982)}]{dupree1982theory}%
  \BibitemOpen
  \bibfield  {author} {\bibinfo {author} {\bibfnamefont {T.~H.}\ \bibnamefont
  {Dupree}},\ }\bibfield  {title} {\bibinfo {title} {Theory of phase-space
  density holes},\ }\href {https://doi.org/10.1063/1.863734} {\bibfield
  {journal} {\bibinfo  {journal}
  {\href{https://doi.org/10.1063/1.863734}{Physics of Fluids}}\ }\textbf
  {\bibinfo {volume} {25}},\ \bibinfo {pages} {277} (\bibinfo {year}
  {1982})}\BibitemShut {NoStop}%
\bibitem [{\citenamefont {Dupree}(1983)}]{dupree1983growth}%
  \BibitemOpen
  \bibfield  {author} {\bibinfo {author} {\bibfnamefont {T.~H.}\ \bibnamefont
  {Dupree}},\ }\bibfield  {title} {\bibinfo {title} {Growth of phase-space
  density holes},\ }\href {https://doi.org/10.1063/1.864430} {\bibfield
  {journal} {\bibinfo  {journal}
  {\href{https://doi.org/10.1063/1.864430}{Physics of Fluids}}\ }\textbf
  {\bibinfo {volume} {26}},\ \bibinfo {pages} {2460} (\bibinfo {year}
  {1983})}\BibitemShut {NoStop}%
\bibitem [{\citenamefont {Bernstein}\ \emph {et~al.}(1957)\citenamefont
  {Bernstein}, \citenamefont {Greene},\ and\ \citenamefont
  {Kruskal}}]{bernstein1957exact}%
  \BibitemOpen
  \bibfield  {author} {\bibinfo {author} {\bibfnamefont {I.~B.}\ \bibnamefont
  {Bernstein}}, \bibinfo {author} {\bibfnamefont {J.~M.}\ \bibnamefont
  {Greene}},\ and\ \bibinfo {author} {\bibfnamefont {M.~D.}\ \bibnamefont
  {Kruskal}},\ }\bibfield  {title} {\bibinfo {title} {Exact nonlinear plasma
  oscillations},\ }\href {https://doi.org/10.1103/PhysRev.108.546} {\bibfield
  {journal} {\bibinfo  {journal}
  {\href{https://doi.org/10.1103/PhysRev.108.546}{Physical Review}}\ }\textbf
  {\bibinfo {volume} {108}},\ \bibinfo {pages} {546} (\bibinfo {year}
  {1957})}\BibitemShut {NoStop}%
\bibitem [{\citenamefont {Roberts}\ and\ \citenamefont
  {Berk}(1967)}]{roberts1967nonlinear}%
  \BibitemOpen
  \bibfield  {author} {\bibinfo {author} {\bibfnamefont {K.~V.}\ \bibnamefont
  {Roberts}}\ and\ \bibinfo {author} {\bibfnamefont {H.~L.}\ \bibnamefont
  {Berk}},\ }\bibfield  {title} {\bibinfo {title} {Nonlinear evolution of a
  two-stream instability},\ }\href {https://doi.org/10.1103/PhysRevLett.19.297}
  {\bibfield  {journal} {\bibinfo  {journal}
  {\href{https://doi.org/10.1103/PhysRevLett.19.297}{Physical Review Letters}}\
  }\textbf {\bibinfo {volume} {19}},\ \bibinfo {pages} {297} (\bibinfo {year}
  {1967})}\BibitemShut {NoStop}%
\bibitem [{\citenamefont {Morse}\ and\ \citenamefont
  {Nielson}(1969)}]{morse1969one}%
  \BibitemOpen
  \bibfield  {author} {\bibinfo {author} {\bibfnamefont {R.~L.}\ \bibnamefont
  {Morse}}\ and\ \bibinfo {author} {\bibfnamefont {C.~W.}\ \bibnamefont
  {Nielson}},\ }\bibfield  {title} {\bibinfo {title} {One-, two-, and
  three-dimensional numerical simulation of two-beam plasmas},\ }\href
  {https://doi.org/10.1103/PhysRevLett.23.1087} {\bibfield  {journal} {\bibinfo
   {journal} {\href{https://doi.org/10.1103/PhysRevLett.23.1087}{Physical
  Review Letters}}\ }\textbf {\bibinfo {volume} {23}},\ \bibinfo {pages} {1087}
  (\bibinfo {year} {1969})}\BibitemShut {NoStop}%
\bibitem [{\citenamefont {Berk}\ \emph {et~al.}(1970)\citenamefont {Berk},
  \citenamefont {Nielsen},\ and\ \citenamefont {Roberts}}]{berk1970phase}%
  \BibitemOpen
  \bibfield  {author} {\bibinfo {author} {\bibfnamefont {H.~L.}\ \bibnamefont
  {Berk}}, \bibinfo {author} {\bibfnamefont {C.~E.}\ \bibnamefont {Nielsen}},\
  and\ \bibinfo {author} {\bibfnamefont {K.~V.}\ \bibnamefont {Roberts}},\
  }\bibfield  {title} {\bibinfo {title} {Phase space hydrodynamics of
  equivalent nonlinear systems: Experimental and computational observations},\
  }\href {https://doi.org/10.1063/1.1693039} {\bibfield  {journal} {\bibinfo
  {journal} {\href{https://doi.org/10.1063/1.1693039}{Physics of Fluids}}\
  }\textbf {\bibinfo {volume} {13}},\ \bibinfo {pages} {980} (\bibinfo {year}
  {1970})}\BibitemShut {NoStop}%
\bibitem [{\citenamefont {Schamel}(1986)}]{schamel1986electron}%
  \BibitemOpen
  \bibfield  {author} {\bibinfo {author} {\bibfnamefont {H.}~\bibnamefont
  {Schamel}},\ }\bibfield  {title} {\bibinfo {title} {Electron holes, ion holes
  and double layers: Electrostatic phase space structures in theory and
  experiment},\ }\href {https://doi.org/10.1016/0370-1573(86)90043-8}
  {\bibfield  {journal} {\bibinfo  {journal}
  {\href{https://doi.org/10.1016/0370-1573(86)90043-8}{Physics Reports}}\
  }\textbf {\bibinfo {volume} {140}},\ \bibinfo {pages} {161} (\bibinfo {year}
  {1986})}\BibitemShut {NoStop}%
\bibitem [{\citenamefont {Ghizzo}\ \emph {et~al.}(1988)\citenamefont {Ghizzo},
  \citenamefont {Izrar}, \citenamefont {Bertrand}, \citenamefont {Fijalkow},
  \citenamefont {Feix},\ and\ \citenamefont {Shoucri}}]{ghizzo1988stability}%
  \BibitemOpen
  \bibfield  {author} {\bibinfo {author} {\bibfnamefont {A.}~\bibnamefont
  {Ghizzo}}, \bibinfo {author} {\bibfnamefont {B.}~\bibnamefont {Izrar}},
  \bibinfo {author} {\bibfnamefont {P.}~\bibnamefont {Bertrand}}, \bibinfo
  {author} {\bibfnamefont {E.}~\bibnamefont {Fijalkow}}, \bibinfo {author}
  {\bibfnamefont {M.~R.}\ \bibnamefont {Feix}},\ and\ \bibinfo {author}
  {\bibfnamefont {M.}~\bibnamefont {Shoucri}},\ }\bibfield  {title} {\bibinfo
  {title} {Stability of {B}ernstein--{G}reene--{K}ruskal plasma equilibria.
  numerical experiments over a long time},\ }\href
  {https://doi.org/10.1063/1.866579} {\bibfield  {journal} {\bibinfo  {journal}
  {\href{https://doi.org/10.1063/1.866579}{Physics of Fluids}}\ }\textbf
  {\bibinfo {volume} {31}},\ \bibinfo {pages} {72} (\bibinfo {year}
  {1988})}\BibitemShut {NoStop}%
\bibitem [{\citenamefont {Hutchinson}(2017)}]{hutchinson2017electron}%
  \BibitemOpen
  \bibfield  {author} {\bibinfo {author} {\bibfnamefont {I.~H.}\ \bibnamefont
  {Hutchinson}},\ }\bibfield  {title} {\bibinfo {title} {Electron holes in
  phase space: What they are and why they matter},\ }\href
  {https://doi.org/10.1063/1.4976854} {\bibfield  {journal} {\bibinfo
  {journal} {\href{https://doi.org/10.1063/1.4976854}{Physics of Plasmas}}\
  }\textbf {\bibinfo {volume} {24}} (\bibinfo {year} {2017})}\BibitemShut
  {NoStop}%
\bibitem [{\citenamefont {Hutchinson}(2024)}]{hutchinson2024kinetic}%
  \BibitemOpen
  \bibfield  {author} {\bibinfo {author} {\bibfnamefont {I.}~\bibnamefont
  {Hutchinson}},\ }\bibfield  {title} {\bibinfo {title} {Kinetic solitary
  electrostatic structures in collisionless plasma: {P}hase-space holes},\
  }\href {https://doi.org/10.1103/RevModPhys.96.045007} {\bibfield  {journal}
  {\bibinfo  {journal}
  {\href{https://doi.org/10.1103/RevModPhys.96.045007}{Reviews of Modern
  Physics}}\ }\textbf {\bibinfo {volume} {96}},\ \bibinfo {pages} {045007}
  (\bibinfo {year} {2024})}\BibitemShut {NoStop}%
\bibitem [{\citenamefont {Galeotti}\ and\ \citenamefont
  {Califano}(2005)}]{galeotti2005asymptotic}%
  \BibitemOpen
  \bibfield  {author} {\bibinfo {author} {\bibfnamefont {L.}~\bibnamefont
  {Galeotti}}\ and\ \bibinfo {author} {\bibfnamefont {F.}~\bibnamefont
  {Califano}},\ }\bibfield  {title} {\bibinfo {title} {Asymptotic evolution of
  weakly collisional {V}lasov-{P}oisson plasmas},\ }\href
  {https://doi.org/10.1103/PhysRevLett.95.015002} {\bibfield  {journal}
  {\bibinfo  {journal}
  {\href{https://doi.org/10.1103/PhysRevLett.95.015002}{Physical Review
  Letters}}\ }\textbf {\bibinfo {volume} {95}},\ \bibinfo {pages} {015002}
  (\bibinfo {year} {2005})}\BibitemShut {NoStop}%
\bibitem [{\citenamefont {Califano}\ \emph {et~al.}(2006)\citenamefont
  {Califano}, \citenamefont {Galeotti},\ and\ \citenamefont
  {Mangeney}}]{califano2006vlasov}%
  \BibitemOpen
  \bibfield  {author} {\bibinfo {author} {\bibfnamefont {F.}~\bibnamefont
  {Califano}}, \bibinfo {author} {\bibfnamefont {L.}~\bibnamefont {Galeotti}},\
  and\ \bibinfo {author} {\bibfnamefont {A.}~\bibnamefont {Mangeney}},\
  }\bibfield  {title} {\bibinfo {title} {The {V}lasov-{P}oisson model and the
  validity of a numerical approach},\ }\href
  {https://doi.org/10.1063/1.2215596} {\bibfield  {journal} {\bibinfo
  {journal} {\href{https://doi.org/10.1063/1.2215596}{Physics of Plasmas}}\
  }\textbf {\bibinfo {volume} {13}} (\bibinfo {year} {2006})}\BibitemShut
  {NoStop}%
\bibitem [{\citenamefont {Carril}\ \emph {et~al.}(2023)\citenamefont {Carril},
  \citenamefont {Gidi}, \citenamefont {Navarro},\ and\ \citenamefont
  {Araneda}}]{carril2023formation}%
  \BibitemOpen
  \bibfield  {author} {\bibinfo {author} {\bibfnamefont {H.~A.}\ \bibnamefont
  {Carril}}, \bibinfo {author} {\bibfnamefont {J.~A.}\ \bibnamefont {Gidi}},
  \bibinfo {author} {\bibfnamefont {R.~E.}\ \bibnamefont {Navarro}},\ and\
  \bibinfo {author} {\bibfnamefont {J.~A.}\ \bibnamefont {Araneda}},\
  }\bibfield  {title} {\bibinfo {title} {Formation of multiple {BGK}-like
  structures in the time-asymptotic state of collisionless {V}lasov-{P}oisson
  plasmas},\ }\href {https://doi.org/10.1103/PhysRevE.107.065203} {\bibfield
  {journal} {\bibinfo  {journal}
  {\href{https://doi.org/10.1103/PhysRevE.107.065203}{Physical Review E}}\
  }\textbf {\bibinfo {volume} {107}},\ \bibinfo {pages} {065203} (\bibinfo
  {year} {2023})}\BibitemShut {NoStop}%
\bibitem [{\citenamefont {Boltzmann}(1896)}]{Boltzmann}%
  \BibitemOpen
  \bibfield  {author} {\bibinfo {author} {\bibfnamefont {L.}~\bibnamefont
  {Boltzmann}},\ }\href@noop {} {\emph {\bibinfo {title} {Vorlesugnen \"uber
  Gastheorie}}}\ (\bibinfo  {publisher} {J. A. Barth},\ \bibinfo {address}
  {Leipzig},\ \bibinfo {year} {1896})\BibitemShut {NoStop}%
\bibitem [{\citenamefont {Davidson}(2015)}]{davidson2015turbulence}%
  \BibitemOpen
  \bibfield  {author} {\bibinfo {author} {\bibfnamefont {P.}~\bibnamefont
  {Davidson}},\ }\href@noop {} {\emph {\bibinfo {title} {Turbulence: {A}n
  {I}ntroduction for {S}cientists and {E}ngineers}}}\ (\bibinfo  {publisher}
  {Oxford University Press},\ \bibinfo {address} {Oxford},\ \bibinfo {year}
  {2015})\BibitemShut {NoStop}%
\bibitem [{\citenamefont {Lynden-Bell}(1967)}]{lynden1967statistical}%
  \BibitemOpen
  \bibfield  {author} {\bibinfo {author} {\bibfnamefont {D.}~\bibnamefont
  {Lynden-Bell}},\ }\bibfield  {title} {\bibinfo {title} {Statistical mechanics
  of violent relaxation in stellar systems},\ }\href
  {https://doi.org/10.1093/mnras/136.1.101} {\bibfield  {journal} {\bibinfo
  {journal} {\href{https://doi.org/10.1093/mnras/136.1.101}{Monthly Notices of
  the Royal Astronomical Society}}\ }\textbf {\bibinfo {volume} {136}},\
  \bibinfo {pages} {101} (\bibinfo {year} {1967})}\BibitemShut {NoStop}%
\bibitem [{\citenamefont {Chavanis}(2022)}]{chavanis2021kinetic}%
  \BibitemOpen
  \bibfield  {author} {\bibinfo {author} {\bibfnamefont {P.-H.}\ \bibnamefont
  {Chavanis}},\ }\bibfield  {title} {\bibinfo {title} {Kinetic theory of
  collisionless relaxation for systems with long-range interactions},\ }\href
  {https://doi.org/10.1016/j.physa.2022.128089} {\bibfield  {journal} {\bibinfo
   {journal} {\href{https://doi.org/10.1016/j.physa.2022.128089}{Physica A:
  Statistical Mechanics and its Applications}}\ }\textbf {\bibinfo {volume}
  {606}},\ \bibinfo {pages} {128089} (\bibinfo {year} {2022})}\BibitemShut
  {NoStop}%
\bibitem [{Note5()}]{Note5}%
  \BibitemOpen
  \bibinfo {note} {An alternative explanation for the power-law tail in energy
  with exponent $-2$ could be that the phase-space holes effectively act as a
  large-scale forcing at $k_{\protect \mathrm {f}} \lambda _{\protect \mathrm
  {De}} \gtrsim 1$. According to the quasilinear theory of \cite
  {banik2024universal}, who considered the $k_{\protect \mathrm {f}} \lambda
  _{\protect \mathrm {De}} \ll 1$ limit of the dressed diffusion equation
  \protect \eqref {QL_evol_general}, such forcing dynamically generates the
  power-law tail, akin to how the flat top \protect \eqref {flat-top} in
  Section~\ref {thermo_turb} emerges in the $k_{\protect \mathrm {f}} \lambda
  _{\protect \mathrm {De}} \gg 1$ regime. If the mean distribution function in
  the simulations of \cite {ewart2024relaxation} is indeed governed by
  quasilinear theory, it must also be reconciled with how this quasilinear
  behavior is consistent with the presence of phase-space turbulence (cf. the
  discussion at the end of Section \ref
  {Stochastic_heating_numerics}).}\BibitemShut {Stop}%
\bibitem [{\citenamefont {Marino}\ and\ \citenamefont
  {Sorriso-Valvo}(2023)}]{marino2023scaling}%
  \BibitemOpen
  \bibfield  {author} {\bibinfo {author} {\bibfnamefont {R.}~\bibnamefont
  {Marino}}\ and\ \bibinfo {author} {\bibfnamefont {L.}~\bibnamefont
  {Sorriso-Valvo}},\ }\bibfield  {title} {\bibinfo {title} {Scaling laws for
  the energy transfer in space plasma turbulence},\ }\href
  {https://www.sciencedirect.com/science/article/pii/S0370157322003969}
  {\bibfield  {journal} {\bibinfo  {journal}
  {\href{https://www.sciencedirect.com/science/article/pii/S0370157322003969}{Physics
  Reports}}\ }\textbf {\bibinfo {volume} {1006}},\ \bibinfo {pages} {1}
  (\bibinfo {year} {2023})}\BibitemShut {NoStop}%
\bibitem [{\citenamefont {Binney}\ and\ \citenamefont
  {Tremaine}(2011)}]{binney2011galactic}%
  \BibitemOpen
  \bibfield  {author} {\bibinfo {author} {\bibfnamefont {J.}~\bibnamefont
  {Binney}}\ and\ \bibinfo {author} {\bibfnamefont {S.}~\bibnamefont
  {Tremaine}},\ }\href@noop {} {\emph {\bibinfo {title} {Galactic Dynamics}}}\
  (\bibinfo  {publisher} {Princeton University Press},\ \bibinfo {address}
  {Princeton, NJ},\ \bibinfo {year} {2011})\BibitemShut {NoStop}%
\bibitem [{\citenamefont {Rampf}(2021)}]{rampf2021cosmological}%
  \BibitemOpen
  \bibfield  {author} {\bibinfo {author} {\bibfnamefont {C.}~\bibnamefont
  {Rampf}},\ }\bibfield  {title} {\bibinfo {title} {Cosmological
  {V}lasov--{P}oisson equations for dark matter: Recent developments and
  connections to selected plasma problems},\ }\href
  {https://link.springer.com/article/10.1007/s41614-021-00055-z} {\bibfield
  {journal} {\bibinfo  {journal}
  {\href{https://link.springer.com/article/10.1007/s41614-021-00055-z}{Reviews
  of Modern Plasma Physics}}\ }\textbf {\bibinfo {volume} {5}},\ \bibinfo
  {pages} {10} (\bibinfo {year} {2021})}\BibitemShut {NoStop}%
\bibitem [{\citenamefont {Hamilton}\ and\ \citenamefont
  {Fouvry}(2024)}]{hamilton2024kinetic}%
  \BibitemOpen
  \bibfield  {author} {\bibinfo {author} {\bibfnamefont {C.}~\bibnamefont
  {Hamilton}}\ and\ \bibinfo {author} {\bibfnamefont {J.-B.}\ \bibnamefont
  {Fouvry}},\ }\bibfield  {title} {\bibinfo {title} {Kinetic theory of stellar
  systems: {A} tutorial},\ }\href {https://doi.org/10.1063/5.0204214}
  {\bibfield  {journal} {\bibinfo  {journal}
  {\href{https://doi.org/10.1063/5.0204214}{Physics of Plasmas}}\ }\textbf
  {\bibinfo {volume} {31}},\ \bibinfo {pages} {120901} (\bibinfo {year}
  {2024})}\BibitemShut {NoStop}%
\bibitem [{\citenamefont {Ginat}\ \emph {et~al.}(2025)\citenamefont {Ginat},
  \citenamefont {Nastac}, \citenamefont {Ewart}, \citenamefont {Konrad},
  \citenamefont {Bartelmann},\ and\ \citenamefont
  {Schekochihin}}]{ginat2024cosmological}%
  \BibitemOpen
  \bibfield  {author} {\bibinfo {author} {\bibfnamefont {Y.~B.}\ \bibnamefont
  {Ginat}}, \bibinfo {author} {\bibfnamefont {M.~L.}\ \bibnamefont {Nastac}},
  \bibinfo {author} {\bibfnamefont {R.~J.}\ \bibnamefont {Ewart}}, \bibinfo
  {author} {\bibfnamefont {S.}~\bibnamefont {Konrad}}, \bibinfo {author}
  {\bibfnamefont {M.}~\bibnamefont {Bartelmann}},\ and\ \bibinfo {author}
  {\bibfnamefont {A.~A.}\ \bibnamefont {Schekochihin}},\ }\bibfield  {title}
  {\bibinfo {title} {Gravitational turbulence: the small-scale limit of the
  cold-dark-matter power spectrum},\ }\href {https://arxiv.org/abs/2501.01524}
  {\bibfield  {journal} {\bibinfo  {journal}
  {\href{https://arxiv.org/abs/2501.01524}{arXiv:2501.01524}}\ } (\bibinfo
  {year} {2025})}\BibitemShut {NoStop}%
\bibitem [{\citenamefont {Eyink}(1995)}]{eyink1995besov}%
  \BibitemOpen
  \bibfield  {author} {\bibinfo {author} {\bibfnamefont {G.~L.}\ \bibnamefont
  {Eyink}},\ }\bibfield  {title} {\bibinfo {title} {Besov spaces and the
  multifractal hypothesis},\ }\href {https://doi.org/10.1007/BF02183353}
  {\bibfield  {journal} {\bibinfo  {journal}
  {\href{https://doi.org/10.1007/BF02183353}{Journal of Statistical Physics}}\
  }\textbf {\bibinfo {volume} {78}},\ \bibinfo {pages} {353} (\bibinfo {year}
  {1995})}\BibitemShut {NoStop}%
\bibitem [{\citenamefont {Cho}\ and\ \citenamefont
  {Lazarian}(2009)}]{cho2009simulations}%
  \BibitemOpen
  \bibfield  {author} {\bibinfo {author} {\bibfnamefont {J.}~\bibnamefont
  {Cho}}\ and\ \bibinfo {author} {\bibfnamefont {A.}~\bibnamefont {Lazarian}},\
  }\bibfield  {title} {\bibinfo {title} {Simulations of electron
  magnetohydrodynamic turbulence},\ }\href
  {https://iopscience.iop.org/article/10.1088/0004-637X/701/1/236} {\bibfield
  {journal} {\bibinfo  {journal}
  {\href{https://iopscience.iop.org/article/10.1088/0004-637X/701/1/236}{The
  Astrophysical Journal}}\ }\textbf {\bibinfo {volume} {701}},\ \bibinfo
  {pages} {236} (\bibinfo {year} {2009})}\BibitemShut {NoStop}%
\bibitem [{\citenamefont {Schekochihin}\ \emph {et~al.}(2002)\citenamefont
  {Schekochihin}, \citenamefont {Boldyrev},\ and\ \citenamefont
  {Kulsrud}}]{schekochihin2002spectra}%
  \BibitemOpen
  \bibfield  {author} {\bibinfo {author} {\bibfnamefont {A.~A.}\ \bibnamefont
  {Schekochihin}}, \bibinfo {author} {\bibfnamefont {S.~A.}\ \bibnamefont
  {Boldyrev}},\ and\ \bibinfo {author} {\bibfnamefont {R.~M.}\ \bibnamefont
  {Kulsrud}},\ }\bibfield  {title} {\bibinfo {title} {Spectra and growth rates
  of fluctuating magnetic fields in the kinematic dynamo theory with large
  magnetic {P}randtl numbers},\ }\href
  {https://iopscience.iop.org/article/10.1086/338697/} {\bibfield  {journal}
  {\bibinfo  {journal}
  {\href{https://iopscience.iop.org/article/10.1086/338697}{The Astrophysical
  Journal}}\ }\textbf {\bibinfo {volume} {567}},\ \bibinfo {pages} {828}
  (\bibinfo {year} {2002})}\BibitemShut {NoStop}%
\bibitem [{\citenamefont {Furutsu}(1964)}]{furutsu1964statistical}%
  \BibitemOpen
  \bibfield  {author} {\bibinfo {author} {\bibfnamefont {K.}~\bibnamefont
  {Furutsu}},\ }\href@noop {} {\emph {\bibinfo {title} {On the Statistical
  Theory of Electromagnetic Waves in a Fluctuating Medium (II)}}}\ (\bibinfo
  {publisher} {U.S. National Bureau of Standards},\ \bibinfo {address}
  {Washington, D.C.},\ \bibinfo {year} {1964})\BibitemShut {NoStop}%
\bibitem [{\citenamefont {Novikov}(1965)}]{novikov1965functionals}%
  \BibitemOpen
  \bibfield  {author} {\bibinfo {author} {\bibfnamefont {E.~A.}\ \bibnamefont
  {Novikov}},\ }\bibfield  {title} {\bibinfo {title} {Functionals and the
  random-force method in turbulence theory},\ }\href@noop {} {\bibfield
  {journal} {\bibinfo  {journal} {Soviet Physics Journal of Experimental and
  Theoretical Physics}\ }\textbf {\bibinfo {volume} {20}},\ \bibinfo {pages}
  {1290} (\bibinfo {year} {1965})}\BibitemShut {NoStop}%
\bibitem [{\citenamefont {Suzuki}(1984)}]{suzuki1984scaling}%
  \BibitemOpen
  \bibfield  {author} {\bibinfo {author} {\bibfnamefont {M.}~\bibnamefont
  {Suzuki}},\ }\bibfield  {title} {\bibinfo {title} {Scaling property of the
  relative diffusion of charged particles in turbulent electric fields. i},\
  }\href {https://doi.org/10.1143/PTP.71.267} {\bibfield  {journal} {\bibinfo
  {journal} {\href{https://doi.org/10.1143/PTP.71.267}{Progress of Theoretical
  Physics}}\ }\textbf {\bibinfo {volume} {71}},\ \bibinfo {pages} {267}
  (\bibinfo {year} {1984})}\BibitemShut {NoStop}%
\bibitem [{\citenamefont {P{\'e}cseli}(1990)}]{pecseli1990phase}%
  \BibitemOpen
  \bibfield  {author} {\bibinfo {author} {\bibfnamefont {H.}~\bibnamefont
  {P{\'e}cseli}},\ }\bibfield  {title} {\bibinfo {title} {Phase space diffusion
  in turbulent plasmas},\ }\href
  {https://iopscience.iop.org/article/10.1088/0031-8949/1990/T30/021}
  {\bibfield  {journal} {\bibinfo  {journal}
  {\href{https://iopscience.iop.org/article/10.1088/0031-8949/1990/T30/021}{Physica
  Scripta}}\ }\textbf {\bibinfo {volume} {1990}},\ \bibinfo {pages} {159}
  (\bibinfo {year} {1990})}\BibitemShut {NoStop}%
\bibitem [{\citenamefont {Cardy}\ \emph {et~al.}(2008)\citenamefont {Cardy},
  \citenamefont {Falkovich},\ and\ \citenamefont {Gawedzki}}]{cardy2008non}%
  \BibitemOpen
  \bibfield  {author} {\bibinfo {author} {\bibfnamefont {J.}~\bibnamefont
  {Cardy}}, \bibinfo {author} {\bibfnamefont {G.}~\bibnamefont {Falkovich}},\
  and\ \bibinfo {author} {\bibfnamefont {K.}~\bibnamefont {Gawedzki}},\
  }\href@noop {} {\emph {\bibinfo {title} {Non-equilibrium {S}tatistical
  {M}echanics and {T}urbulence}}}\ (\bibinfo  {publisher} {Cambridge University
  Press},\ \bibinfo {address} {Cambridge},\ \bibinfo {year} {2008})\BibitemShut
  {NoStop}%
\bibitem [{\citenamefont {Chertkov}\ \emph {et~al.}(1995)\citenamefont
  {Chertkov}, \citenamefont {Falkovich}, \citenamefont {Kolokolov},\ and\
  \citenamefont {Lebedev}}]{chertkov1995}%
  \BibitemOpen
  \bibfield  {author} {\bibinfo {author} {\bibfnamefont {M.}~\bibnamefont
  {Chertkov}}, \bibinfo {author} {\bibfnamefont {G.}~\bibnamefont {Falkovich}},
  \bibinfo {author} {\bibfnamefont {I.}~\bibnamefont {Kolokolov}},\ and\
  \bibinfo {author} {\bibfnamefont {V.}~\bibnamefont {Lebedev}},\ }\bibfield
  {title} {\bibinfo {title} {Statistics of a passive scalar advected by a
  large-scale two-dimensional velocity field: Analytic solution},\ }\href
  {https://doi.org/10.1103/PhysRevE.51.5609} {\bibfield  {journal} {\bibinfo
  {journal} {\href{https://link.aps.org/doi/10.1103/PhysRevE.51.5609}{Physical
  Review E}}\ }\textbf {\bibinfo {volume} {51}},\ \bibinfo {pages} {5609}
  (\bibinfo {year} {1995})}\BibitemShut {NoStop}%
\bibitem [{\citenamefont {Balkovsky}\ and\ \citenamefont
  {Fouxon}(1999)}]{balkovsky1999universal}%
  \BibitemOpen
  \bibfield  {author} {\bibinfo {author} {\bibfnamefont {E.}~\bibnamefont
  {Balkovsky}}\ and\ \bibinfo {author} {\bibfnamefont {A.}~\bibnamefont
  {Fouxon}},\ }\bibfield  {title} {\bibinfo {title} {Universal long-time
  properties of {L}agrangian statistics in the {B}atchelor regime and their
  application to the passive scalar problem},\ }\href
  {https://doi.org/10.1103/PhysRevE.60.4164} {\bibfield  {journal} {\bibinfo
  {journal} {\href{https://doi.org/10.1103/PhysRevE.60.4164}{Physical Review
  E}}\ }\textbf {\bibinfo {volume} {60}},\ \bibinfo {pages} {4164} (\bibinfo
  {year} {1999})}\BibitemShut {NoStop}%
\bibitem [{\citenamefont {Furstenberg}\ and\ \citenamefont
  {H.~Kesten}(1960)}]{furstenberg1960products}%
  \BibitemOpen
  \bibfield  {author} {\bibinfo {author} {\bibfnamefont {H.}~\bibnamefont
  {Furstenberg}}\ and\ \bibinfo {author} {\bibfnamefont {H.}~\bibnamefont
  {H.~Kesten}},\ }\bibfield  {title} {\bibinfo {title} {Products of random
  matrices},\ }\href@noop {} {\bibfield  {journal} {\bibinfo  {journal} {The
  Annals of Mathematical Statistics}\ }\textbf {\bibinfo {volume} {31}},\
  \bibinfo {pages} {457} (\bibinfo {year} {1960})}\BibitemShut {NoStop}%
\bibitem [{\citenamefont {Oseledec}(1968)}]{oseledec1968multiplicative}%
  \BibitemOpen
  \bibfield  {author} {\bibinfo {author} {\bibfnamefont {V.~I.}\ \bibnamefont
  {Oseledec}},\ }\bibfield  {title} {\bibinfo {title} {A multiplicative ergodic
  theorem: {L}yapunov characteristic numbers for dynamical systems},\
  }\href@noop {} {\bibfield  {journal} {\bibinfo  {journal} {Transactions of
  the Moscow Mathematical Society}\ }\textbf {\bibinfo {volume} {19}},\
  \bibinfo {pages} {197} (\bibinfo {year} {1968})}\BibitemShut {NoStop}%
\bibitem [{\citenamefont {Chavanis}(2023)}]{chavanis2023secular}%
  \BibitemOpen
  \bibfield  {author} {\bibinfo {author} {\bibfnamefont {P.-H.}\ \bibnamefont
  {Chavanis}},\ }\bibfield  {title} {\bibinfo {title} {The secular dressed
  diffusion equation},\ }\href {https://doi.org/10.3390/universe9020068}
  {\bibfield  {journal} {\bibinfo  {journal}
  {\href{https://doi.org/10.3390/universe9020068}{Universe}}\ }\textbf
  {\bibinfo {volume} {9}},\ \bibinfo {pages} {68} (\bibinfo {year}
  {2023})}\BibitemShut {NoStop}%
\bibitem [{\citenamefont {Vanden~Eijnden}(1997)}]{vanden1997some}%
  \BibitemOpen
  \bibfield  {author} {\bibinfo {author} {\bibfnamefont {E.}~\bibnamefont
  {Vanden~Eijnden}},\ }\bibfield  {title} {\bibinfo {title} {Some remarks on
  the quasilinear treatment of the stochastic acceleration problem},\ }\href
  {https://doi.org/10.1063/1.872548} {\bibfield  {journal} {\bibinfo  {journal}
  {\href{https://doi.org/10.1063/1.872548}{Physics of Plasmas}}\ }\textbf
  {\bibinfo {volume} {4}},\ \bibinfo {pages} {1486} (\bibinfo {year}
  {1997})}\BibitemShut {NoStop}%
\bibitem [{\citenamefont {Jeans}(1902)}]{jeans1902stability}%
  \BibitemOpen
  \bibfield  {author} {\bibinfo {author} {\bibfnamefont {J.~H.}\ \bibnamefont
  {Jeans}},\ }\bibfield  {title} {\bibinfo {title} {I. {T}he stability of a
  spherical nebula},\ }\href {https://doi.org/10.1098/rsta.1902.0012}
  {\bibfield  {journal} {\bibinfo  {journal}
  {\href{https://doi.org/10.1098/rsta.1902.0012}{Philosophical Transactions of
  the Royal Society A}}\ }\textbf {\bibinfo {volume} {199}},\ \bibinfo {pages}
  {1} (\bibinfo {year} {1902})}\BibitemShut {NoStop}%
\bibitem [{\citenamefont {Ng}\ and\ \citenamefont
  {Bhattacharjee}(2005)}]{ng2005bernstein}%
  \BibitemOpen
  \bibfield  {author} {\bibinfo {author} {\bibfnamefont {C.~S.}\ \bibnamefont
  {Ng}}\ and\ \bibinfo {author} {\bibfnamefont {A.}~\bibnamefont
  {Bhattacharjee}},\ }\bibfield  {title} {\bibinfo {title}
  {{B}ernstein-{G}reene-{K}ruskal modes in a three-dimensional plasma},\ }\href
  {https://doi.org/10.1103/PhysRevLett.95.245004} {\bibfield  {journal}
  {\bibinfo  {journal}
  {\href{https://doi.org/10.1103/PhysRevLett.95.245004}{Physical Review
  Letters}}\ }\textbf {\bibinfo {volume} {95}},\ \bibinfo {pages} {245004}
  (\bibinfo {year} {2005})}\BibitemShut {NoStop}%
\bibitem [{\citenamefont {Ng}\ \emph {et~al.}(2006)\citenamefont {Ng},
  \citenamefont {Bhattacharjee},\ and\ \citenamefont {Skiff}}]{ng2006weakly}%
  \BibitemOpen
  \bibfield  {author} {\bibinfo {author} {\bibfnamefont {C.~S.}\ \bibnamefont
  {Ng}}, \bibinfo {author} {\bibfnamefont {A.}~\bibnamefont {Bhattacharjee}},\
  and\ \bibinfo {author} {\bibfnamefont {F.}~\bibnamefont {Skiff}},\ }\bibfield
   {title} {\bibinfo {title} {Weakly collisional {L}andau damping and
  three-dimensional {B}ernstein-{G}reene-{K}ruskal modes: New results on old
  problems},\ }\href {https://doi.org/10.1063/1.2186187} {\bibfield  {journal}
  {\bibinfo  {journal} {\href{https://doi.org/10.1063/1.2186187}{Physics of
  Plasmas}}\ }\textbf {\bibinfo {volume} {13}} (\bibinfo {year}
  {2006})}\BibitemShut {NoStop}%
\end{thebibliography}%

\end{document}